\documentclass[12pt,aps,prd,groupedaddress,floatfix,nofootinbib]{revtex4-1}
\makeatletter
\@ifundefined{@parse@version@dash}{%
\def\@parse@version#1{\@parse@version@0#1}
\def\@parse@version@#1/#2/#3#4#5\@nil{%
\@parse@version@dash#1-#2-#3#4\@nil}
\def\@parse@version@dash#1-#2-#3#4#5\@nil{%
\if\relax#2\relax\else#1\fi#2#3#4 }
}{}
\makeatother

\usepackage{graphicx}
\usepackage{dcolumn}
\usepackage{bm}
\usepackage{lineno}
\usepackage[normalem]{ulem}
\usepackage{hyperref}

\usepackage{color}
\usepackage{wrapfig}

\newcommand{\ISU}{Department of Physics and Astronomy, Iowa State University, Ames IA 50011}
\newcommand{\BNL}{Department of Physics, Brookhaven National Laboratory, Upton, NY 11793}
\newcommand{\kent}{Department of Physics, Kent State University, Kent OH 44242 USA}

\newcommand{\UIC}{University of Illinois at Chicago, Chicago, IL 60607}

\newcommand{\WSU}{Department of Physics and Astronomy, Wayne State University, Detroit, Michigan 48201, USA}
\newcommand{\UIUC}{University of Illinois at Urbana-Champaign, Urbana, IL 61801}

\newcommand{\NCSU}{North Carolina State University, Raleigh, NC 27695}
\newcommand{\INT}{Institute for Nuclear Theory, University of Washington, Seattle, WA 98195, USA}
\newcommand{\SBU}{Stony Brook University, Stony Brook, NY 11794}

\newcommand{\UH}{Department of Physics, University of Houston, Houston, TX 77204, USA}
\newcommand{\CIT}{Department of Physics, California Institute of Technology, Pasadena, California 91125, USA}
\newcommand{\CITLab}{LIGO Laboratory, California Institute of Technology, Pasadena, California 91125, USA}
\newcommand{\LANL}{Theoretical Division, Los Alamos National Laboratory, Los Alamos, NM 87545, USA}
\newcommand{\WASHU}{Washington University in St.~Louis, St.~Louis, MO 63130, USA}
\newcommand{\MD}{University of Maryland, Department of Astronomy and Joint Space-Science Institute, University of Maryland, College Park, MD 20742}
\newcommand{\MDastro}{University of Maryland, Department of Astronomy, College Park, MD 20742}
\newcommand{\MDphys}{Department of Physics, Maryland Center for Fundamental Physics, and NSF Institute for Robust Quantum Simulation, University of Maryland, College Park, MD 20742}
\newcommand{\MIT}{Center for Theoretical Physics, Massachusetts Institute of Technology, Cambridge, MA 02139}
\newcommand{\PEPPU}{Natural Science Division, Pepperdine University, Malibu, CA 90263, USA}
\newcommand{\PUCTS}{Princeton Center for Theoretical Science, Princeton University, Princeton, NJ 08544, USA}
\newcommand{\PUGI}{Princeton Gravity Initiative, Princeton University, Princeton, NJ 08544, USA}
\newcommand{\IAS}{School of Natural Sciences, Institute for Advanced Study, Princeton, NJ 08540, USA}
\newcommand{\UB}{Fakult{\"a}t f{\"u}r Physik, Universit{\"a}t Bielefeld, D-33615 Bielefeld, Germany}
\newcommand{\TAMUC}{Texas A\& M University-Commerce, Commerce, TX 75429, USA}
\newcommand{\OSU}{Department of Physics, The Ohio State University, Columbus, OH 43210-1117, USA}
\newcommand{\GSI}{GSI Helmholtz Centre for Heavy-ion Research, Planckstr. 1, 64291 Darmstadt, Germany}
\newcommand{\UNCCH}{Department of Physics and Astronomy, The  University of North Carolina at Chapel Hill, Chapel Hill, NC, 27599}
\newcommand{\IGC}{Institute for Gravitation and the Cosmos, The Pennsylvania State University, University Park, PA 16802, USA}
\newcommand{\PSUP}{Department of Physics, The Pennsylvania State University, University Park, PA 16802, USA}
\newcommand{\PSUAA}{Department of Astronomy \& Astrophysics, The Pennsylvania State University, University Park, PA 16802, USA}
\newcommand{\UiS}{Faculty of Science and Technology, University of Stavanger, 4036 Stavanger, Norway}
\newcommand{\OHIOU}{Department of Physics and Astronomy and Institute of Nuclear and Particle Physics, Ohio University, Athens, OH 45701, USA}
\newcommand{\FRIB}{Facility for Rare Isotope Beams and Department of Physics and Astronomy, Michigan State University, East Lansing, MI 48824, USA}
\newcommand{\CSUF}{Nicholas and Lee Begovich Center for Gravitational Wave Physics and Astronomy, California State University Fullerton, CA 92831, USA}
\newcommand{\GSFC}{Astroparticle Physics Laboratory, NASA/GSFC, Greenbelt, MD 20771}
\newcommand{\CRESST}{Center for Research and Exploration in Space Science and Technology, NASA/GSFC, Greenbelt, MD 20771}
\newcommand{\UFABC}{Center for Mathematics, Computation and Cognition, UFABC, Santo Andre, 09210-170, Brazil}
\newcommand{\UFRJ}{Instituto de Física, Universidade Federal do Rio de Janeiro, Caixa Postal 68528, 21941-972, Rio de Janeiro, RJ, Brazil}
\newcommand{\giessen}{Institute for Theoretical Physics, Justus Liebig University Giessen, Heinrich-Buff-Ring 16, 35392 Giessen, Germany}
\newcommand{\hfhf}{Helmholtz Research Academy Hesse for FAIR (HFHF), Campus Giessen, 35392 Giessen, Germany}
\newcommand{\tamu}{Department of Physics and Astronomy and Cyclotron Institute, Texas A\&M University, College Station, TX 77843, USA}
\newcommand{\UTK}{Department of Physics and Astronomy and University of Tennessee, Knoxville, Knoxville, TN 37996, USA}
\newcommand{\ORNL}{Physics Division, Oak Ridge National Laboratory, Oak Ridge, TN 37830, USA}
\newcommand{\UM}{School of Physics and Astronomy, University of Minnesota, Minneapolis, Minnesota 55455 USA}
\newcommand{\LIGOlabMIT}{LIGO Laboratory, Massachusetts Institute of Technology, 185 Albany St, Cambridge, MA 02139, USA}
\newcommand{\MKI}{Department of Physics and Kavli Institute for Astrophysics and Space Research, Massachusetts Institute of Technology,  77 Massachusetts Ave, Cambridge, MA 02139, USA}
\newcommand{\ANL}{Physics Division, Argonne National Laboratory, Lemont, IL 60439, USA}
\newcommand{\PSU}{Pennsylvania State University, Department of Physics, University Park, Pennsylvania 16802, USA}
\newcommand{\BUW}{Department of Physics, Wuppertal University, Gaussstr. 20, D-42119, Wuppertal, Germany}
\newcommand{\FZJ}{Juelich Supercomputing Centre, Forschungszentrum Juelich, D-52425 Juelich, Germany}
\newcommand{\ELTE}{Eotvos Lorand University, Institute for Theoretical Physics,
H-1117, Budapest, Hungary}
\newcommand{\UCSD}{Physics Department, UCSD, San Diego, CA 92093, USA}
\newcommand{\IU}{Physics Department, Indiana University, Bloomington, IN 47405, USA}
\newcommand{\CITA}{Canadian Institute for Theoretical Astrophysics, University of Toronto, Toronto, Ontario M5S 3H8, Canada}
\usepackage[dvipsnames]{xcolor}

\newcommand{\snn} {\sqrt{s_{_{\rm NN}}}}

\newcommand{\txt}[1]{\textrm{#1}}

\begin{document}


\title{Long Range Plan: Dense matter theory for heavy-ion collisions and neutron stars
\vspace{1cm}
}

\def\authspc{\vspace{-0.2cm}}
\def\affspc{\vspace{-0.2cm}}

\author{\authspc Alessandro Lovato}\affiliation{\affspc\ANL}
\author{\authspc Travis Dore}\affiliation{\UB}
\author{\authspc Robert D. Pisarski and Bjoern Schenke} \affiliation{\affspc\BNL}
\author{\authspc Katerina Chatziioannou}\affiliation{\affspc\CIT}\affiliation{\affspc\CITLab}
\author{\authspc Jocelyn S. Read}\affiliation{\CSUF}
\author{\authspc Philippe Landry}\affiliation{\CITA}
\author{\authspc Pawel Danielewicz, Dean Lee, Scott Pratt}\affiliation{\FRIB}
\author{\authspc Fabian Rennecke}\affiliation{\affspc\giessen}\affiliation{\affspc\hfhf}
\author{\authspc Hannah Elfner}\affiliation{\affspc \GSI}
\author{\authspc Veronica Dexheimer, Rajesh Kumar, Michael Strickland}\affiliation{\kent}
\author{\authspc Johannes Jahan, Claudia Ratti and Volodymyr Vovchenko}\affiliation{\affspc\UH}
\author{\authspc Mikhail Stephanov}\affiliation{\affspc \UIC}
\author{\authspc Dekrayat Almaalol, Gordon Baym, Mauricio Hippert, Jacquelyn Noronha-Hostler, Jorge Noronha,  Enrico Speranza, and Nicol\'as Yunes}\affiliation{\affspc\UIUC}
\author{\authspc Chuck J. Horowitz}\affiliation{\IU}
\author{\authspc Steven P. Harris, Larry McLerran, Sanjay Reddy, Agnieszka Sorensen}\affiliation{\affspc\INT}
\author{\authspc Srimoyee Sen}\affiliation{\affspc\ISU}
\author{\authspc Stefano Gandolfi and Ingo Tews}\affiliation{\affspc\LANL}
\author{\authspc M. Coleman Miller}\affiliation{\affspc\MD}
\author{\authspc Cecilia Chirenti}\affiliation{\affspc \MDastro}\affiliation{\affspc \GSFC}\affiliation{\affspc \CRESST}\affiliation{\affspc \UFABC}
\author{\authspc Zohreh Davoudi}\affiliation{\affspc\MDphys}
\author{\authspc Jamie M. Karthein and Krishna Rajagopal}\affiliation{\affspc \MIT}
\author{\authspc Salvatore Vitale}\affiliation{\MKI }\affiliation{\LIGOlabMIT}
\author{\authspc Joseph Kapusta}\affiliation{\UM}
\author{\authspc G\"ok\c ce Ba\c sar}\affiliation{\affspc \UNCCH}
\author{\authspc Thomas Schaefer and Vladimir Skokov}\affiliation{\affspc\NCSU}
\author{\authspc Ulrich Heinz}\affiliation{\affspc\OSU}
\author{\authspc Christian Drischler, Daniel R.~Phillips, Madappa Prakash}\affiliation{\OHIOU}
\author{\authspc Zoltan Fodor}\affiliation{\PSU}\affiliation{\BUW}\affiliation{\FZJ}\affiliation{\ELTE}\affiliation{\UCSD}
\author{\authspc David Radice}\affiliation{\affspc\IGC}\affiliation{\affspc\PSUP}\affiliation{\affspc\PSUAA}
\author{\authspc Christopher Plumberg} \affiliation{\affspc \PEPPU}
\author{\authspc Elias R. Most, Carolyn A. Raithel}\affiliation{\affspc\PUCTS}\affiliation{\affspc\PUGI}\affiliation{\affspc\IAS}
\author{\authspc Eduardo S. Fraga}\affiliation{\UFRJ}
\author{\authspc Aleksi Kurkela}\affiliation{\UiS}
\author{\authspc James M. Lattimer}\affiliation{\affspc \SBU}
\author{\authspc Andrew W. Steiner}\affiliation{\UTK}\affiliation{\ORNL}
\author{\authspc Jeremy W. Holt}\affiliation{\tamu}
\author{\authspc Bao-An Li}\affiliation{\affspc\TAMUC}

\author{\authspc Chun Shen}\affiliation{\affspc\WSU}

\author{\authspc Mark Alford, Alexander Haber, Saori Pastore,  Maria Piarulli}\affiliation{\affspc\WASHU}

\date{\today}

\renewcommand\abstractname{{\bf Executive Summary}}
\begin{abstract}
\begin{center}
{\bf Executive Summary}
\end{center}

Since the release of the 2015 Long Range Plan in Nuclear Physics, major events
have occurred that reshaped our understanding of quantum chromodynamics (QCD) 
and nuclear matter at large densities, in and out of equilibrium. The US nuclear
community has an opportunity to capitalize on advances in 
astrophysical observations and nuclear experiments and engage in
an interdisciplinary effort in the theory of dense baryonic matter that  
connects low- and high-energy nuclear physics, astrophysics, gravitational waves physics, and data science.

\begin{itemize}
\item {\bf Now is the time to pursue dense matter studies:} Over the 
past decade we have seen the first detection of a gravitational wave 
signal from a binary neutron star merger, together with its electromagnetic counterparts~\cite{TheLIGOScientific:2017qsa}, and the radius measurement of a two-solar-mass neutron 
star~\cite{Riley:2021pdl,Miller:2021qha,Fonseca:2021wxt} from NASA's NICER mission.  
New experimental results possibly indicative of a QCD critical 
point~\cite{Adamczewski-Musch:2019byl,HADES:2019auv,STAR:2020tga,HADES:2020wpc,STAR:2021rls} from heavy ion collisions at STAR and HADES and new inference of the neutron-skin thickness has led to  (although with significant uncertainties) the extraction of the slope of the symmetry energy, 
from PREX-II~\cite{PREX:2021umo} and CREX data~\cite{CREX:2022kgg}. These results
cover a wide range of scales, from single nuclei to neutron stars, and 
physical probes, from gravitational waves to high-energy particles, 
and call for an interdisciplinary effort to unravel the properties of 
strongly interacting matter. Soon we expect to see new 
experimental information on neutron-rich atomic nuclei and the equation 
of state EOS (including new kinds of neutron skin thickness measurements using mirror nuclei~\cite{Pineda:2021shy}) from FRIB, as well as from  heavy-ion collisions at RHIC, SPS, 
and FAIR, the possible detection of multiple neutron star mergers per year 
by the LIGO-Virgo-KAGRA detectors with electromagnetic counterparts
\cite{Patricelli:2022hhr,Colombo:2022zzp}, and new data by the NICER mission for 
PSR  J0437.  Interpreting this new experimental and observational data fully will require significant coordinated efforts in dense matter theory beyond current levels and improved coordination among theorists, experimentalists, and observers.

\item {\bf Interpreting Beam Energy Scan II data:} Results from phase II of 
the RHIC Beam Energy Scan program (BESII) are anticipated over the coming year or two, since data-taking 
concluded in 2021. A central goal of BESII is to measure the beam 
energy dependence of fluctuation observables and identify a possible  
critical point in the QCD phase diagram,  addressing one of the big open questions
in the field~\cite{Busza:2018rrf}. Significant theoretical efforts are needed to reliably interpret
the results. This includes
tools to study the non-equilibrium evolution of non-Gaussian fluctuations, 
inclusion of strangeness neutrality and electric charge diffusion in
hydrodynamic models, and an EOS and dynamic tools to 
study the meta-stable regime. All of these tools, together with the
already developed   Beam Energy Scan Theory (BEST) Collaboration \cite{An:2021wof} framework, have to be integrated in a 
Bayesian analysis framework. Only then will the BESII data 
allow us to either confirm the discovery of a critical point in the phase 
diagram and pinpoint its location,  
or place constraints on the location of any critical point by excluding 
its presence in the regime explored in BESII.
Either outcome is important for  understanding the  QCD phase diagram. 
A critical point discovery would imply that the transition at high baryon density, $n_B$, is  discontinuous and would motivate a program for exploring 
consequences of a coexistence region.

\item {\bf Uncertainty quantification in chiral effective field theory:}
At low to moderate $n_B$ and temperatures $T$, the dense-matter EOS can be computed using nuclear many-body methods that use as input nuclear interactions derived from chiral effective field theory  ($\chi$EFT)~\cite{Epelbaum:2008ga,Machleidt:2011zz}. 
These microscopic calculations of dense matter have become more sophisticated, with different computational methods agreeing to good accuracy~\cite{Hebeler:2009iv,Coraggio:2012ca,Hagen:2013yba,Carbone:2014mja,Lynn:2019rdt,Piarulli:2019pfq,Huth:2020ozf,Drischler:2021kxf,Lovato:2022apd}. 
Present theoretical uncertainties are dominated by the nuclear interactions employed in these calculations. 
$\chi$EFT enables us to estimate these uncertainties as it is based on a systematic momentum expansion. 
The use of Bayesian tools has enabled tremendous progress in the rigorous quantification of EFT uncertainties over the past 7 years~\cite{Furnstahl:2014xsa,Wesolowski:2015fqa,Wesolowski:2018lzj,Lim:2018bkq,Melendez:2019izc,Drischler:2020hwi,Drischler:2020yad,Elhatisari:2022qfr}. 
However, there remain several problems pertinent to the theory of dense QCD matter where model uncertainties need to be assessed. 
Further methodological developments and software tools are needed to achieve this goal~\cite{Phillips:2020dmw}. 
Several studies have indicated that $\chi$EFT calculations of neutron-rich dense matter might be valid up to twice the nuclear saturation density ($n_{sat}=0.16$ fm$^{-3}$)~\cite{Tews:2018kmu,Drischler:2020hwi,Drischler:2020yad}, but it is still unclear where and how $\chi$EFT breaks down.
For studies of neutron-star mergers, it is crucial to access dense matter at finite temperatures~\cite{Wellenhofer:2015qba,Carbone:2019pkr,Keller:2020qhx}.
Finite temperatures and the addition of protons might influence the breakdown of the theory in dense matter.
The possibility of a large neutron skin in ${}^{208}$Pb~\cite{Reed:2021nqk}  might lead to tension with $\chi$EFT calculations~\cite{Hu:2021trw} should future experiments confirm its central value~\cite{PREX:2021umo} with improved precision~\cite{Essick:2021ezp}.
However, the relatively small neutron skin in $^{48}$Ca inferred by the CREX measurements~\cite{CREX:2022kgg} is more in line with $\chi$EFT calculations~\cite{Hagen:2015yea}. 
Grounding the EOS at low $n_B$ in reliable nuclear-theory calculations, including lattice QCD~\cite{Drischler:2019xuo}, is extremely important, given the upcoming data from observations and experiments. 
    
\item {\bf Connecting laboratory experiments to astrophysics:}  With the 
upcoming BESII data, low-energy FRIB data,
observational data of neutron stars and their mergers, the advances of 
lattice QCD, perturbative QCD (pQCD), and $\chi$EFT, and previous knowledge of the liquid-gas phase 
transition, we will have disconnected regions in the QCD phase diagram that 
will need to be connected through effective models. In equilibrium, 
multidimensional EOS, in terms of $T$ and different 
chemical potentials $\mu$, must be flexible enough to allow studies of parameter space 
to quantify uncertainties and reproduce all known constraints. They will 
need to provide particle composition, necessary for calculations of in and out 
of equilibrium properties, such as strangeness production, neutrino emissivity, and bulk viscosity. Additionally, dynamical models are needed to make 
direct connection to experimental heavy-ion data and post-merger signals that 
will require stable and causal equations of motion that can handle multiple 
conserved charges, microscopic models of transport coefficients, 
state-of-the-art hadronic transport codes, and proper handling of neutrino
interactions. These challenges require interdisciplinary collaborations to 
share knowledge and develop open-source tools that can be applied within and across the
different communities. 

\item {\bf Leveraging novel simulation and computation technologies:} Hamiltonian simulations performed on quantum simulators and computers has the promise of enabling efficient first-principles simulation of dense matter, in and out of equilibrium, by avoiding the sign problem in current Monte Carlo based methods. Quantum information tools provide novel probes of quantum state of matter, phases and phase transitions, and equilibration and thermalization processes. Over the next decade, the community will identify computational problems that can benefit from quantum advantage, develop efficient algorithms to access them in a quantum-simulating devise, and  engage in implementation and co-design efforts involving quantum technologists to gain a deeper understanding of gauge-theory dynamics even on near-term noisy quantum simulators and computers.
\end{itemize}
\end{abstract}

\maketitle
\tableofcontents

\section{Overview}

\begin{figure}
    \centering
    \includegraphics[width=0.7\linewidth]{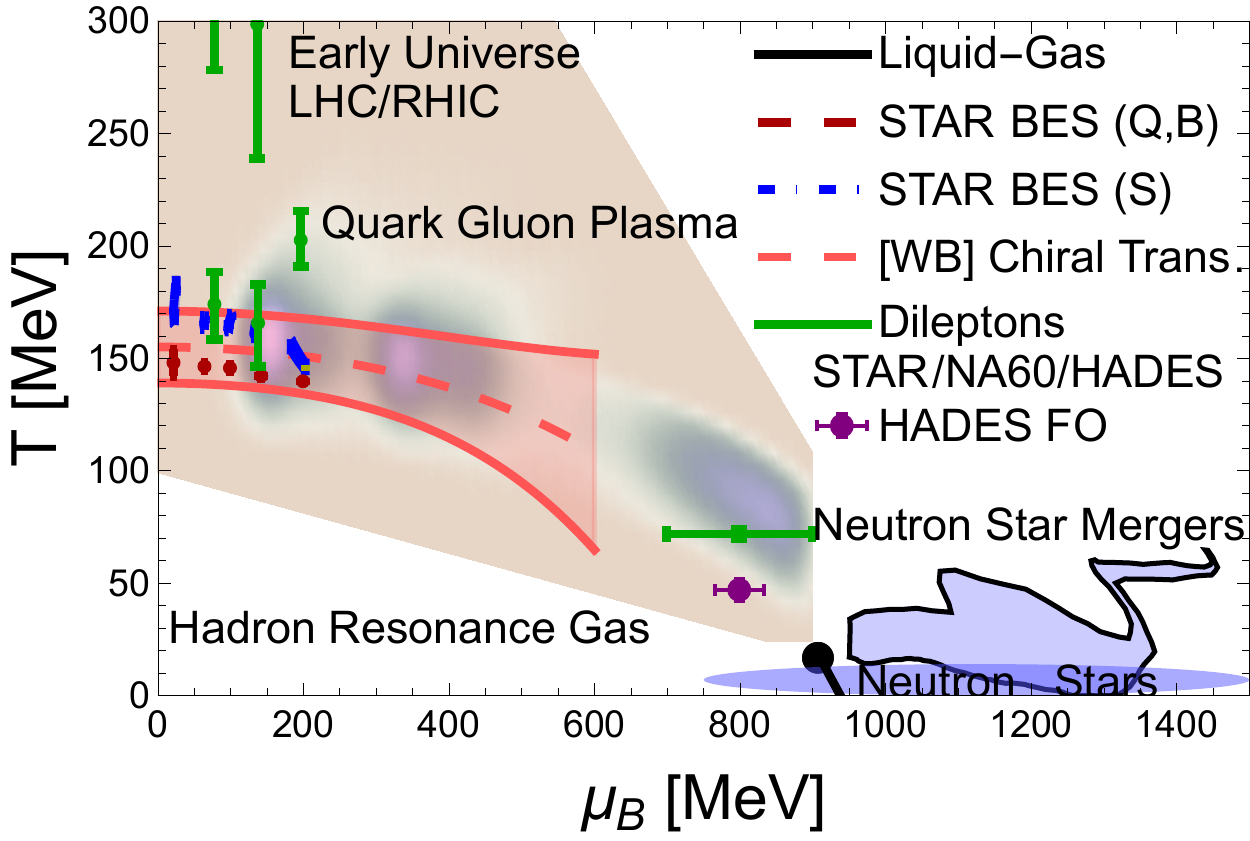}
    \caption{QCD phase diagram with the latest interpretation of experimental data and dynamical simulations of regions probed by various systems.  The zero baryon chemical potential axis follows the trajectory of the early universe and coincides with the Large Hadron Collider (LHC). RHIC, through the STAR BESII and fixed target (FXT) programs, explores the brown shaded region as inferred from relativistic viscous hydrodynamic simulations  \cite{Shen:2022oyg} for $\sqrt{s_{NN}}=3,\, 7.7,\, 27$ GeV. Estimates for neutron star mergers comes from numerical relativity simulations  \cite{Most:2018eaw} and the $T=0$ neutron star  range comes from various EOS estimates. Net-proton, K, $\pi$ data from STAR were used to extract $T,\mu_B$ of light particles at freeze-out  (red)  and  net-K, $\Lambda$ for strange particles  (blue) from \cite{Alba:2020jir}. Dilepton measurements of temperatures within the quark-gluon plasma phase from STAR, NA60 \cite{NA60:2008dcb}, and HADES \cite{HADES:2019auv} are shown in green. Thermal model fits to HADES particle yields provide a freeze-out estimate for $\sqrt{s_{NN}}=2.4$ GeV \cite{Harabasz:2020sei} (some ambiguity still exists \cite{Motornenko:2021nds}), shown in maroon. The nuclear liquid-gas phase transition is based on experimental data from \cite{Elliott:2013pna} and the $\mu_B$ estimate from \cite{Vovchenko:2016rkn}.  The chiral transition (light red) comes from lattice QCD calculations  \cite{Borsanyi:2020fev}. The possible QCD critical point and the associated 1st-order phase transition line are not shown due to uncertainty, which STAR BESII and FXT aim to reduce. }
    \label{fig:PhaseDiagram}
\end{figure}

 Understanding the behavior of dense baryonic matter is one of the
central problems in nuclear physics. Dense matter in this context
is any nuclear system that contains a net excess of quarks over 
anti-quarks, either in the form of neutrons and protons, as 
baryonic resonances, or as quark matter. Nuclear physicists address 
fundamental questions about the properties of dense matter. How does dense
matter respond to compression and heating, that is, what is the 
EOS? Does it undergo phase transformations, possibly
to states which exhibit macroscopic order, such as superfluidity,
superconductivity, or crystalline phases? How does dense baryonic matter emit 
radiation, in the form of light, neutrinos, or gravitational 
waves? How does it conduct heat or electric current? 

 Information about dense matter comes from an unusually 
broad array of sources, ranging from astrophysical objects
such as neutron stars and neutron star mergers, to 
laboratory experiments involving collisions of heavy ions,
or electro-weak probes of nuclei. Theorists working on dense
matter employ effective theories of nuclear forces, perturbative
QCD (pQCD) calculations, numerical calculations in lattice QCD, 
and microscopic approaches such as kinetic theory  and 
(general) relativistic fluid dynamics. They work with data
scientists and experts in machine learning.

 Since the last long-range plan \cite{Aprahamian:2015qub}, there have been several 
important developments, including new multi-messenger observations of 
 neutron star mergers, X-ray observations of neutron star radii, radio pulsar timing observations of potentially ultraheavy neutron stars in excess of $2M_\odot$, parity-violating electron scattering measurements of the neutron skin thicknesses of Ca$^{48}$ and Pb$^{208}$, and the completion of the analysis
of data from phase I of RHIC's Beam Energy Scan and the taking of data from its phase II. Constraints on neutron
star masses and radii will dramatically improve in the 
foreseeable future, and many additional mass measurements are on the horizon. New kinds of neutron star radius measurements, such as from measurements of moments of inertia, are possible.  Also, data from BES II will become 
available, and a future CBM program at GSI will provide
very high luminosity data in the energy range studied by 
the STAR fixed target program \cite{Almaalol:2022xwv}.
The interpretation of these observational and experimental 
results relies on advances in nuclear theory.  There is a unique opportunity to join together different sub-fields of nuclear theory (low and high energy) and external fields, such as astrophysics and gravitational wave physics, to  study dense matter at large $n_B$, beginning below  $n_{sat}$  and ranging up to many times $n_{sat}$.
In the following sections, we review the status of nuclear theory across heavy-ions, lattice QCD, nuclear astrophysics, and gravitational waves (as it pertains to neutron stars) and discuss the outlook and challenges for the next 10 years.

\section{EOS at large densities}

The sign problem prevents direct first principles lattice QCD calculations of the QCD EOS at $\mu_B\neq 0$ \cite{deForcrand:2009zkb}. Theorists continue to 
explore ideas, such as the complex Langevin method \cite{Damgaard:1987rr}, or relatively new ideas such as path integrals 
on Lefschetz thimbles \cite{Alexandru:2020wrj},
etc. to overcome the sign problem. 
Functional methods offer the possibility to study QCD directly at finite $n_B$ in the continuum, making recent progress on the description of the phase diagram \cite{Fischer:2018sdj,Fu:2019hdw,Gao:2020fbl,Gunkel:2021oya}.
It is difficult to predict future advances, but a new opportunity is provided by progress
in quantum information science and the availability of noisy, 
intermediate scale quantum computers and quantum simulators~\cite{NSAC-QIS-2019-QuantumInformationScience,davoudi2022quantum,catterall2022report,humble2022snowmass}. 
This will enable exploratory studies of the Hamiltonian evolution
of systems with non-zero quark density, see Sec.~\ref{sec:QIS} for further discussions.

In the meantime, expansion schemes can reach out to finite $n_B$, where the series coefficients can be computed without a sign problem. Notably these schemes are sensitive to thermodynamic singularities of the EOS in the complex $\mu_B$ plane, whose locations are sensitive to whether there is a critical point in the phase diagram \cite{Fisher:1974series}. In principle, they encode information about the possible phase transitions (crossover, first order, etc.) in the phase diagram. The nature of the corresponding phase transitions, whether they are a cross-over or a first-order that ends at a critical point, is of great interest to the nuclear physics community because this may have observable signatures in heavy-ion collisions and neutron star mergers. In practice, only a limited number of terms are available in such expansions \cite{Ratti:2018ksb,Borsanyi:2021sxv,Bollweg:2022rps}. However, combined with the universal nature of the complex singularities \cite{An:2017brc,Mukherjee:2019eou,Connelly:2020gwa}, these expansion schemes can still be used to extract valuable information regarding the conjectured critical point and the critical contribution to the EOS in its vicinity \cite{Ratti:2018ksb,Basar:2021hdf,Basar:2021gyi}.
Beyond the range covered by the existing expansion schemes, effective models are required to extend the $n_B$ coverage. At very low $T$ but around $1-2\ n_{sat}$, it is also possible to calculate the EOS using $\chi$EFT, that is relevant for neutron stars, and at extremely large $T$ or $n_B$ one should approach pQCD calculations \cite{Komoltsev:2021jzg}. However, these points in the phase diagram need to be connected through effective theories.  Entirely new phases of matter may exist  that we have yet to discover such as quarkyonic matter, color superconducting states or condensates of pions or kaons. 
The EOS provides a starting point to discover these new stages of matter and phase transitions, but much is not yet understood and significant theoretical development is needed over the next decade to make advances in the future.

The EOS describes the properties of dense matter in equilibrium.  However, the systems that probe this dense matter are initiated (heavy ion collisions) or driven (neutron star mergers; heavy ion collisions that pass near a critical point or through a first order transition) sufficiently far out-of-equilibrium that an understanding of matter in equilibrium (the EOS) is only a starting point for understanding the physics. Subsequent sections we shall describe progress and opportunities in describing the relevant out-of-equilibrium dynamics for heavy ion collisions and neutron star mergers, including both modeling of far-from-equilibrium dynamical evolution and transport coefficients that describe how a perturbed system returns to equilibrium. These transport coefficients should have the same degrees-of-freedom as the EOS.  For instance, in heavy-ion collisions that probe QCD time scales (too short lived to experience weak decays), the transport coefficients should be calculated directly from QCD.  That is currently not possible (although quantum computing algorithms promise an interesting future possibility), so various effective models are employed instead. In neutron star mergers, where time-scales are twenty orders of magnitude longer, the relevant out-of-equilibrium effects are governed by (slow) weak interactions that seek to maintain chemical equilibrium and that yield a significant bulk viscosity when they cannot.

\subsection{Lattice QCD}
The challenge for hot and dense lattice QCD in the near future is to extend the coverage of the phase diagram to large $\mu_B$, reaching towards the neutron star mergers. The EOS of QCD at $\mu_B=0$ for a system of 2+1 flavors has been known from first principles to a few
percent for nearly a decade \cite{Borsanyi:2010cj,Borsanyi:2013bia,Bazavov:2014pvz}, which exhibits a smooth crossover at $T\simeq 156$ MeV \cite{Aoki:2006we} with the thermodynamics of the high-$T$ phase being well described by resummed pQCD calculations \cite{Andersen:2011sf,Strickland:2014zka,Ghiglieri:2020dpq,Haque:2020eyj}.  Extracting the EOS (and other properties) of QCD at finite
$\mu_B$ from regular Monte Carlo simulations is complicated due to the  sign problem. Several indirect methods have been proposed to study QCD
matter at small $\mu_B$. A conceptually straightforward method is
Taylor expansion, where high statistics simulations at $\mu_B=0$ are used to extract
$\mu_B$-derivatives \cite{Allton:2005gk,Bazavov:2017dus,Bazavov:2020bjn}. An
alternative extrapolation method uses analytic continuation from imaginary $\mu_B$
\cite{deForcrand:2002hgr,DElia:2002tig,Wu:2006su,Conradi:2007be,deForcrand:2008vr,DElia:2009pdy,Borsanyi:2018grb}. Reweighting of the
configurations to finite $\mu_B$ was also discussed 
\cite{Fodor:2001au,Fodor:2001pe,Csikor:2004ik,Fodor:2004nz}, in
combination with density of state methods \cite{Fodor:2007vv,Alexandru:2014hga}
or canonical ensembles \cite{Alexandru:2005ix,Kratochvila:2005mk,Ejiri:2008xt}.

A recently introduced \cite{Giordano:2020roi} and tested \cite{Borsanyi:2021hbk}
reweighting scheme shifts the starting point of the extrapolation to an
ensemble much closer to the finite-$\mu_B$ system than those used at $\mu_B=0$
in the original works. This significantly reduces the overlap problem
and mitigates the sign problem. Because this method is computationally very demanding and, it is currently limited to very small lattices.
Direct methods, such as the complex Langevin equation, promise access to high
$n_B$ at a lower cost. Gauge cooling \cite{Seiler:2012wz} allowed the
generalization to dynamical QCD \cite{Sexty:2013ica}, and
the EOS was addressed \cite{Sexty:2019vqx}. A more effective
dynamical stabilization method was suggested \cite{Attanasio:2018rtq}
that enabled low-$T$, high $\mu_B$ simulations \cite{Attanasio:2022mjd},
pointing to the maturity of this technique to be used soon at physical conditions.

A full study of the phase diagram can only be
pursued using different techniques in their respective region of validity. Overlapping regions allow for cross-checks,
which are crucial to high-$\mu_B$ simulations. Another key
future challenge involves obtaining the full 4D nature of the QCD phase diagram.  One requires $T,~\mu_B,~\mu_S,~\mu_Q$  in order to explore strangeness and electric charge fluctuations that occur in event-by-event fluctuations of the initial conditions in heavy-ion collisions (also due to gluon splittings into quark anti-quark pairs).

\subsection{Many-body theory and chiral effective field theory}

Many-body theory has made significant advances in predicting the nuclear EOS with quantified uncertainties based on microscopic two- and three-nucleon interactions consistently derived from $\chi$EFT~\cite{Hebeler:2015hla,Lynn:2019rdt,Stroberg:2019mxo,Hergert:2020bxy,Drischler:2021kxf}. 
Nuclear observables can be computed using different many-body frameworks to estimate method uncertainties, while systematic order-by-order calculations have enabled us to estimate the uncertainties due to truncating the chiral expansion at a finite order (so-called EFT truncation errors).
Quantifying and propagating these theoretical uncertainties is crucial for meaningful comparisons between competing nuclear theory predictions and/or constraints from nuclear experiments and neutron-star observations. Bayesian methods facilitate such comparisons in a statistically rigorous way to take full advantage of the rich empirical constraints on the nuclear EOS anticipated in the next 15 years.
Tremendous progress has been made in the past few years in predicting properties of nuclei and calculating the nuclear matter EOS using Quantum Monte Carlo (QMC) methods~\cite{Carlson_2015,Gandolfi:2020pbj}. Those nonperturbative methods, e.g. the auxiliary field Diffusion Monte Carlo (AFDMC), are based on imaginary-time propagation of a many-body wave function, and are used to extract ground-state properties of many-body nuclear system.

Recent QMC calculations take as input nuclear interactions derived within $\chi$EFT~\cite{Epelbaum:2008ga,Piarulli:2019cqu}.
The latter provides a way to systematically construct nuclear interactions by choosing the relevant effective degrees of freedom, like nucleons (N's), $\pi$'s, or $\Delta$'s. All two-nucleon and multi-nucleon interaction terms are expanded in powers of nuclear momenta over a breakdown scale. At each order in the EFT expansion, several diagrams enter the description of the interaction. 
For example, at leading-order (LO) only the one-$\pi$ exchange between two nucleons as well as a momentum-independent contact interaction contribute. At next-to-leading-order (NLO), two-$\pi$ exchanges are included as well as momentum-dependent contact interactions, and similarly, more involved terms appear at higher orders. The various coupling constants are determined from fits to experimental data, e.g, the $\pi$-N couplings are fit to $\pi$-N scattering and low energy constants describing short-range interactions are fit to scattering data. The advantage of interactions obtained within $\chi$EFT is that many-body interactions, like three-N forces, are naturally emerging within the chiral expansion. Another important feature of $\chi$EFT is that theoretical uncertainties from the truncation of the expansion can be estimated. Several techniques have been introduced to estimate these systematic (or theoretical) uncertainties~\cite{Epelbaum:2014efa,Drischler:2020hwi}. QMC calculations based on $\chi$EFT Hamiltonians of binding energies, radii, and electroweak transitions of nuclei up to $A=16$~\cite{Gezerlis_2013,Gezerlis_2014,Lynn_2017,Tews_2021,Piarulli:2014bda,Piarulli:2016vel,Baroni:2018fdn} are in very good agreement with experimental data~\cite{Lonardoni_2018,Lonardoni_2020,Piarulli:2017dwd,King:2020wmp}. 
QMC methods were also used to calculate the EOS of matter up to $\sim 2\ n_{sat}$~\cite{Lynn:2015jua,Tews:2018kmu,Piarulli:2019pfq,Lovato:2022apd}. The calculated EOS include estimates of systematic truncation uncertainties, and are commonly used to constrain properties of neutron stars~\cite{Tews:2018kmu,Al_Mamun_2021,Dietrich:2020efo}.

The past decade has seen a renaissance for many-body perturbation theory (MBPT) calculations in nuclear physics~\cite{Drischler:2021kxf,Tichai:2020dna}.
The similarity renormalization group~\cite{Bogner:2009bt} has enabled us to systematically lower the resolution scale of chiral interactions, making them softer to accelerate the MBPT convergence.
Furthermore, recent advances in automatic diagram generation~\cite{Arthuis:2020tjz} combined with automatic code generation~\cite{Drischler:2017wtt} and high-performance computing have led to a fully automated approach to MBPT calculations in nuclear physics~\cite{Drischler:2021kxf}, in which chiral two- and multi-nucleon forces can be included to high orders in the chiral and MBPT expansions. 
MBPT has been demonstrated to be a computationally efficient and versatile tool for
studying the nuclear EOS as a function of $n_B$, isospin asymmetry, and $T$~\cite{Keller:2020qhx} and their implications for the neutron star structure~\cite{Drischler:2021kxf}. 
In particular, MBPT allows us to compute neutron-star (i.e., $\beta$-equilibrated) matter directly rather than by interpolation, providing important insights into isospin asymmetry expansions of the low-density nuclear EOS~\cite{Drischler:2013iza,Drischler:2015eba,Wellenhofer:2015qba,Wellenhofer:2017qla,Wen:2020nqs,Somasundaram:2020chb}.
But MBPT's versatility goes well beyond the nuclear EOS and includes MBPT calculations of the linear response and transport coefficients that could fuel more accurate numerical simulations of supernovae and neutron-star mergers~\cite{Du:2021rhq}.
Global optical potentials with quantified uncertainties have been derived from MBPT calculations of infinite matter based on chiral NN and 3N interactions~\cite{Whitehead:2020wwb, Holt:2022piv}.
These microscopic optical potentials will play a crucial role in the FRIB era to interpret nuclear reaction experiments to constrain the nuclear EOS and beyond~\cite{Hebborn:2022vzm}.

High-quality astrophysical simulations of neutron-star mergers are key to extracting vital information from multimessenger observations and require robust input for the EOS and neutrino opacities with quantified theoretical uncertainties. 
Future $\chi$EFT-based calculations 
will have to be devoted to further improving the nuclear Hamiltonians to reduce the  uncertainties. 
The constrained EOS around nuclear densities will be crucial to understand the ``low-$n_B$'' limit of dense QCD.
Such calculations are needed at finite $T$ as well as arbitrary neutron-proton asymmetries.
$\chi$EFT is valid for nucleon momenta below $\approx 500-600$~MeV, roughly translating into densities below $\sim 2\ n_{sat}$~\cite{Tews:2018kmu,Drischler:2020hwi}, but the exact breakdown density and the way $\chi$EFT breaks down in dense matter is unknown.
In addition, the convergence pattern of the nuclear EOS at finite proton fractions and $T$ needs to be investigated more closely with rigorous statistical analysis.
These questions will need to be answered to make reliable connections to laboratory experiments and astrophysical observations.

\subsection{pQCD}
Due to the asymptotic freedom of QCD, various quantities including the EOS can be calculated directly from the Lagrangian of QCD at extremely high $n_B$ and/or $T$ using resummed perturbation theory. The first results for the EOS came more than four decades ago for massless quarks \cite{Freedman:1976ub,Baluni:1977ms,Blaizot:2000fc,Fraga:2001id,Vuorinen:2003fs} and then including strange quark mass \cite{Freedman:1977gz,Farhi:1984qu,Fraga:2004gz,Kurkela:2009gj}. In the past years there has been significant activity to improve the perturbative calculations of the EOS, generalizing the Next-to-Next-to-Leading-Order (N2LO) calculations \cite{Kurkela:2009gj} to cover all $n_B$ and $T$ \cite{Kurkela:2016was}, including the effects arising from the strange quark mass \cite{Gorda:2021gha}. Partial N3LO results at $T=0$ and zero quark masses have been obtained \cite{Gorda:2018gpy,Gorda:2021znl}. Resummation schemes that go beyond strict weak coupling expansion have been explored \cite{Andersen:2002jz,Fujimoto:2020tjc,Fernandez:2021jfr}

At low $T$, the current state-of-the-art results become quantitatively reliable at  $n_B \gtrsim 40\ n_{sat}$ \cite{Kurkela:2014vha}. While these densities are well above $n_B$ reached in stable neutron stars, it has been recently demonstrated that the knowledge of this high-$n_B$ limit constrains the EoS at neutron-star densities \cite{Kurkela:2014vha,Komoltsev:2021jzg}. 
These constraints arise from the requirement that the low-$n_B$ EoS be connected to the high-density limit such that the EoS is mechanically stable, causal, and thermodynamically consistent at all $n_B$. Further improvements of the perturbative results, in particular the completion of the N3LO computation, will offer even stronger constraints to the EOS at neutron-star densities. 
The impact of these constraints due to the nontrivial interplay between  perturbative constraints and astrophysical observations have been discussed recently~\cite{Kurkela:2014vha,Gorda:2022jvk,Somasundaram:2022ztm,Ecker:2022dlg,Altiparmak:2022bke}, and the pQCD input might lead to softening of the EoS at the highest $n_B$ reached in neutron stars.

\subsection{Effective models}

Beyond the current reach of $\chi$EFT and other ab-initio theories,  effective models provide an alternative way to acquire insight about the QCD EOS at large $n_B$ and finite $T$. A subset of these models begin with lattice QCD results and reconstruct an EOS with additional physics added at large $\mu_B$. An example of this, is \cite{Parotto:2018pwx} where they constructed an EOS that respects what is known from lattice QCD at $\mu_B=0$ and at the same time has a critical point in the appropriate 3D Ising model university class placed at a point in the $(T,\mu_B)$ phase diagram of your choosing, and with the mapping of the axes of the universal Ising model into the QCD phase diagram also of your choosing. This construction has also been extended to incorporate the constraint of strangeness neutrality~\cite{Karthein:2021nxe}. This constructed EoS can then be used in heavy-ion dynamical simulations~\cite{Shen:2020gef,An:2021wof,Dore:2022qyz}, with the goal of incorporating calculations of the critical fluctuations near a critical point in such hydrodynamic modeling. By so doing, the resulting fluctuations of observed particle multiplicities can be computed and compared to measurements anticipated from phase II of the RHIC BES program. Thus, constraints can be placed on the choices made in constructing the EoS, including in particular the location of a possible critical point in the QCD phase diagram.
Given the presence of multiple baryon number, electric charge, and strangeness (BSQ) conserved charges in heavy-ion collisions, this framework for incorporating a critical point must also be developed to include the electric charge and strangeness chemical potentials, $\mu_Q$ and $\mu_S$.

Holographic models tuned to fit lattice QCD results at $n_B=0$ have been useful to investigate the equilibrium and transport properties of strongly-coupled matter near the critical point \cite{Rougemont:2017tlu,Critelli:2017oub,Grefa:2021qvt,Grefa:2022sav}.  Relativistic mean field models can be applied to any energy regime, as long as they contain the appropriate degrees of freedom, namely a variety of baryons, mesons, and deconfined quarks \cite{Dexheimer:2009hi,Motornenko:2019arp}. Effective models provide particle population and can incorporate various strangeness and isospin densities.
Modern Bayesian tools enable us to determine uncertainties in, and correlations between, model parameters. This information can be straightforwardly propagated to model predictions thereby quantifying the uncertainty within a particular model. Quantifying the uncertainty of a particular model is a harder problem, although Bayesian Model Mixing~\cite{bandframework,Phillips:2020dmw} shows promise for assessing and controlling it.
Effective models can be constrained by lattice QCD, pQCD, and $\chi$EFT and be fitted to reproduce nuclear saturation properties, heavy-ion collision data, and neutron star observational data. 
To make meaningful comparisons between heavy-ion collisions and neutron star mergers, which overlap in properties such as entropy per baryon and $n_B$ \cite{Most:2022wgo}, they need to contain  chiral symmetry restoration and an appropriate treatment for in medium masses of mesons. One unresolved complication is that the condensation of mesons needs to be properly accounted for in the case of mergers. 

Developments will be driven by upcoming data and new data analyses. 
Once we have a realistic continuous description connecting the QCD phase diagram, developments in e.g.~astrophysics will immediately reflect on constraints for matter created in heavy-ion collisions and vice-versa. In particular, the confirmation of the QC critical point would focus astrophysics efforts onto looking for signals of quark deconfinement in neutron stars \cite{Alford:2013aca,Benic:2014jia,Dexheimer:2014pea}, their births \cite{Sagert:2008ka,Fischer:2017lag}, and their mergers \cite{Most:2018eaw,Bauswein:2018bma}.

\subsection{Transport coefficients}

In the finite $T$ and small $n_B$ regime probed by high-energy heavy-ion collisions, one can already calculate the EOS directly from lattice QCD calculations. 
There, the primary uncertainty lies in both the correct description of the initial state and the transport coefficients needed for relativistic viscous hydrodynamic simulations. To overcome this issue, groups perform large scale Bayesian analyses at $\mu_B=0$ \cite{Bernhard:2019bmu} to extract these properties and these transport coefficients can in turn be used to make predictions for future runs.   
However, at large $n_B$ this is no longer possible because not only are the initial conditions and transport coefficients not yet known, but also the EOS is unknown.  Furthermore, the transport coefficients are sensitive to phase transitions \cite{Soloveva:2019xph,Grefa:2022sav} and vary as functions of  $\left\{T,\mu_B,\mu_S,\mu_Q\right\}$ \cite{Rose:2017bjz,McLaughlin:2021dph} and, therefore, also require multidimensional tables, just like the EOS. Microscopic models and effective theories are required to determine reasonable functional forms of these transport coefficients to guide future statistical studies. This picture is further complicated by the BSQ diffusion matrix that is relevant at large $n_B$ \cite{Greif:2017byw} and second-order transport coefficients that provide stability to far-from-equilibrium systems and couple various dissipative currents \cite{Almaalol:2022pjc}, but we have very little knowledge about their values at finite $n_B$.  Thus, significant theoretical development is required before a full scale Bayesian analysis can be performed at these $n_B$. Potentially new data science techniques will be necessary because of the much larger scale of free parameters at large $n_B$. 

Analogously, similar challenges exist for neutron star mergers \cite{Alford:2017rxf,Most:2021zvc,Most:2022yhe}. In that case, finite $T$ effects are not negligible, at which point weak interactions become important. While these can be computed as effective reaction rates \cite{Yakovlev:2004iq}, their corresponding impact on the hydrodynamic system can be understood in terms of transport coefficients. These must be determined taking into account a variety of out-of-equilibrium processes, which depend on the specific phase and particle content of the matter which, in turn, vary with $n_B$. Current calculations have been limited to very simple forms of npe$\mu$ matter \cite{Alford:2019qtm,Alford:2019kdw,Alford:2020lla,Alford:2021lpp,Alford:2022ufz} and hyperons \cite{Alford:2020pld}, with uncertainties concerning the relevant thermodynamic conditions in the neutrino trapped regime \cite{Perego:2019adq,Zappa:2022rpd}. 

\subsection{New phases of matter}

There are many reasons to believe that the phase structure 
of dense QCD is rich. Beyond a couple of times $n_{sat}$, new degrees of freedom (other than nucleons) are expected to populate dense matter. These are hyperons, spin 3/2 resonances, negative parity states, and a myriad of mesons. Their appearance softens the EOS, by opening new fermionic channels or by condensing bosons. There may also be a hadronic phase with restored chiral symmetry, quarkyonic matter, with mesons, baryons, and glueballs \cite{McLerran:2007qj,McLerran:2018hbz,Pisarski:2018bct,Pisarski:2020dnx,Lajer:2021kcz}. Eventually, at larger $n_B$ and $T$, hadrons start to overlap, and a description based on quarks and gluons is necessary \cite{Baym:2017whm}. We anticipate a regime of color superconducting quark matter phases \cite{Alford:2007xm} at extremely high $n_B$, which itself is predicted to have several phases at various locations on the phase diagram. These phases may be inhomogeneous \cite{Kojo:2009ha,Kojo:2010fe,Kojo:2011cn,Pisarski:2018bct,Pisarski:2020dnx,Lajer:2021kcz,Pisarski:2021qof} and many of them superfluids \cite{Page:2010aw,Hippert:2021gfs,Manuel:2007pz}, including the CFL phase \cite{Alford:1998mk,Alford:2004hz,Alford:2007xm}, which is favored at the highest $\mu_B$'s. Among the color superconducting phases, there may be some that are crystalline \cite{Alford:2000ze,Bowers:2002xr,Mannarelli:2007bs} and some in which chiral symmetry is broken \cite{Alford:1998mk,Bedaque:2001je}.

In quark matter phases it is expected that the EoS becomes approximately conformal leading to a characteristic softening \cite{Annala:2019puf, Fujimoto:2022ohj} that may be used as an indicator of the active degrees of freedom and the phase of the dense matter. It has been shown (through nuclear-gravity collaborations) that these exotic phases produce signatures related to the characteristic  bump in the speed of sound $c_s^2$ (as a function of $n_B$) of dense matter or a first-order phase transition, that could be observed in the near future in neutron stars mergers \cite{Tan:2021nat} from the binary love relation.
If these phases of matter could be accessed by heavy-ion collisions there are likely other unique signatures that have yet to be explored.  Further study and collaborations with gravitational wave physicists and astrophysicists will be necessary to find new observable signatures in multi-messenger astrophysics.

\section{Heavy-ion collisions}\label{sec:heavyions}

 A central goal of the RHIC BES program is the discovery of a 
possible critical point or a first-order phase
transition in the QCD phase diagram. Such a discovery
would be of central importance to our understanding
of QCD, as it would represent the first clear distinction
between quark-gluon  and hadronic matter.  Matter produced at the top RHIC beam energies ($\sqrt{s_{NN}}$) is characterized
by a very low  $\mu_B<25$ MeV, and can be well described by the standard model of heavy-ion collisions: event-by-event initial conditions, relativistic viscous hydrodynamics, and hadronic interactions \cite{Alba:2017hhe,Bernhard:2019bmu, Schenke:2020mbo, JETSCAPE:2020mzn, Nijs:2020ors}. 
At lower $\sqrt{s_{NN}}$, more of the initial baryons of 
the colliding nuclei are stopped, and higher $\mu_B$ are reached. For these $\sqrt{s_{NN}}$, significant upgrades must be made to high-energy heavy-ion collision simulations to handle baryon stopping, large $n_B$, new transport coefficients, etc. Fundamental questions exist on the boundaries of applicability of hydrodynamics vs.\ hadron transport around extremely far-from-equilibrium, short-lived, droplets of QGP.   

Phases I and II of the RHIC BES explore matter up to $\mu_B\simeq 750$ MeV. While data from  BES I have
been analyzed, data from BES II, which will add significantly
more statistics at low $\sqrt{s_{NN}}$, have not been published yet.  The main observables that have been explored are fluctuations of particle multiplicities \cite{Stephanov:1999zu,Hatta:2003wn,Stephanov:2008qz,Athanasiou:2010kw,Stephanov:2011pb,Brewer:2018abr,Mroczek:2020rpm}, which are predicted to show an enhancement in the critical region. Data from BES I provide intriguing hints of non-monotonic behavior
in higher order cumulants of the net-proton distribution as a function of the beam energy~\cite{STAR:2020tga}. 
There are also indications for an excess of the variance of the proton number relative to the non-critical baseline in the regime $\snn \lesssim 20$~GeV if the expected effects of baryon conservation and repulsive interactions are taken into account \cite{Vovchenko:2021kxx}.
Data from BES II, as
well as a suitable theoretical analysis framework, will be required 
to draw firm conclusions. A central challenge, and, hence, a central element of the theoretical analysis framework, arises from the fact that, because of critical slowing down, the fluctuations that develop in a heavy ion collision that passes near a QCD critical point do not have the time to grow as they would in thermal equilibrium~\cite{Berdnikov:1999ph}. This means that calculations of these fluctuations (see below) done for the purpose of estimating the magnitude of fluctuations in particle multiplicities to be measured in BES II must be done out-of-equilibrium, and must be done self-consistently with the hydrodynamic calculation of the expansion and cooling of the droplet of QGP. Even if it were possible to extrapolate lattice calculations of equilibrium susceptibilities to the $\mu_B$ of interest, this would not suffice: if BES II collisions pass near a critical point, because of critical slowing down the measurements of fluctuation observables will need to be compared to dynamical
calculations of critical fluctuations that are far-from-equilibrium.
Finally, we note that future STAR BES II results will be significantly extended by CBM at 
FAIR \cite{Almaalol:2022xwv}.

Independent of the possible existence of a critical point,  heavy-ion collisions data at low $\sqrt{s_{NN}}$ will shed light on other properties of dense QCD matter, with far-reaching consequences. While collisions at very high $\sqrt{s_{NN}}$ explore a regime  similar to that in the early universe, characterized by high $T$ and low $\mu_B$, collisions at low $\sqrt{s_{NN}}$ create systems closely resembling those encountered in neutron star mergers \cite{HADES:2019auv}, with low $T$ and high $\mu_B$. This opens up the possibility of new connections between nuclear and astrophysics. 
Can we constrain the EOS beyond the regime accessible by lattice QCD? 
Can heavy-ion collisions be used to determine the EOS in the regime probed by neutron star mergers? How does the EOS evolve from the small-$\mu_B$ regime, characterized by a reduction of  $c_s^2$ in the crossover region, to the high
$\mu_B$ regime, where neutron star data indicate a rapid rise in $c_s^2$? What are the degrees of freedom responsible for this  behavior? How do the strong interactions depend on density?
Does the QGP still look like a nearly perfect fluid at large $\mu_B$?
Studies of these questions will spur fruitful interdisciplinary work that will advance
fluid dynamic and transport theories in both astrophysics and heavy-ions.

\subsection{Relativistic viscous fluid dynamics} \label{sec:hic_visc}

The fluid dynamic framework at high-energies is significantly simplier than what is required at low energies (and as such, many high-energy hydrodynamic codes are not fully prepared to be ran at low energies yet).
Extending fluid dynamics to low $\sqrt{s_{NN}}$, with possible critical fluctuations or a metastable region, involves a number of challenges.

First, given the role of higher order cumulants as an experimental signature of the critical point, it is important to include the 
evolution of non-Gaussian cumulants of the order parameter.  A framework for computing
the Gaussian component of critical fluctuations self-consistently with hydrodynamics, called HYDRO+ \cite{Stephanov:2017ghc}, has been developed and used to model heavy ion collisions near a critical point~\cite{Rajagopal:2019xwg,Du:2020bxp}, and in Ref.~\cite{Pradeep:2022mkf} 
these out-of-equilibrium fluctuations described in concert with the hydrodynamic evolution of the droplet of matter produced
in a heavy ion collision have been ``particlized'', namely converted into cumulants of proton multiplicities that can subsequently be measured.
Generalization to non-critical fluctuations in relativistic fluids has also been introduced for Gaussian fluctuations \cite{An:2019osr,An:2019csj}.
A similar framework for non-Gaussian fluctuations will be crucial for direct comparison of hydrodynamic models that incorporate a 
critical point with BES II data. An important first step towards this goal has been reported in \cite{An:2020vri}.  
Relativistic viscous fluid dynamics frameworks also need to be extended to incorporate the metastable regime since fluid elements may pass across the first-order phase transition. This will require deriving the equations of motion that can handle a mestable regime (e.g. incorporating chiral symmetery restoration), demonstrating that they are stable \cite{Almaalol:2022pjc}, and ensuring that the hydrodynamic codes can handle numerical challenges passing across a first-order line.

Second, at finite $\mu_B$ the fluid dynamic description has to 
incorporate the dynamics of all (BSQ) conserved charges. This implies hydrodynamics codes must be compatible with 4D EOS tables \cite{Noronha-Hostler:2019ayj,Monnai:2019hkn}, transport coefficients that vary in a 4D space which may have sharp features due to phase transitions, and equations of motion that contain the full matrix of diffusion coefficients (with all appropriate coupling terms to other dissipative currents). 

Third, initial conditions have been updated to incorporate baryon stopping \cite{Shen:2017bsr} or gluon splittings into BSQ densities \cite{Martinez:2019jbu,Carzon:2019qja}, but a single model that incorporates both does not yet exist.  Because the EOS varies with electric charge, isopin effects should also be studied in these initial conditions, which may be non-trivial depending on the type of nuclei (e.g., nuclei with neutron skin). Out-of-equilibrium effects (i.e., the full $T^{\mu\nu}$ and $J^{\mu}$ for each conserved charge) should also be considered, which is possible within transport models such as SMASH \cite{Inghirami:2022afu}, but has not yet been considered in alternative approaches.

Fourth, fundamental questions still remain on the proper way to convert the fluid-like QGP into particles.  The typical approach is to use a  Cooper-Frye approach that constructs a distribution function $f=f_{eq}+\delta f$, where $f_{eq}$ is the equilibrium contribution and $\delta f$ contributions appear for each dissipative current (assumed to be small comparatively). Due to the role the shear and bulk viscosities, $\eta$ and $\zeta$, and diffusion play at large $\mu_B$, one may reach a regime where $\delta f>f_{eq}$ and this approach must be revisited \cite{Pratt:2010jt,Alqahtani:2016rth,Alqahtani:2017tnq,Alqahtani:2017mhy,McNelis:2021acu}.  Additionally, even if the correct distribution function is known, one must ensure that BSQ charges are conserved in a single event \cite{Oliinychenko:2019zfk,Vovchenko:2022syc}. Finally, with the measurement of global $\Lambda$ polarization by the STAR collaboration \cite{STAR:2017ckg}, the development of consistent theories of spin hydrodynamics that can be used in simulations has become an important problem in the field. Despite the intense interest displayed by the community (see, for instance, \cite{Florkowski:2017ruc,Hattori:2019lfp,Bhadury:2020puc,Weickgenannt:2020aaf,Hongo:2021ona,Weickgenannt:2022zxs}), many questions still remain concerning the formulation of consistent (causal and stable) theories of viscous spin hydrodynamics and  their implementation in heavy-ion simulations.

This list indisputably goes beyond what can be achieved by individual investigators.  A~collaborative approach to tackling these challenges
was initiated by the BEST Collaboration.
The BEST Collaboration achieved significant progress reported in \cite{An:2021wof}, 
but the work still remains to be done to enable quantitative comparison to heavy-ion collision experiments and to locate the critical point. Further efforts are also needed to make direct connections to the physics of the neutron star mergers.

\subsection{Hadronic transport} 

Hadronic transport simulations have been used extensively to study heavy-ion collisions at very low to intermediate beam energies ($\snn \approx 1.9 ~ \txt{GeV} $ to $\snn \approx 8.0 ~ \txt{GeV}$), where the far-from-equilibrium dynamics of the system questions suitability of fluid dynamical approaches. In addition to providing a means of studying non-equilibrium evolution, hadronic transport naturally includes baryon, strangeness, and charge diffusion,
as well as describes effects due to the interplay between the evolving collision zone and the spectators, which are crucial for a correct description of, e.g., flow observables. 

By comparing the simulations with experimental data, hadronic transport can be used to extract the EOS and in-medium properties of nuclear matter at finite $T$ and large $n_B$. The single-nucleon mean-field potential, including its density-, isospin-, and momentum-dependence, can be extracted by comparisons of several observables, including the directed and elliptical flow. In particular, the difference between proton and neutron flows and the spectrum ratio of energetic neutrons over protons can be used to constrain the symmetry energy at both sub- and suprasaturation densities \cite{Li:2000bj,Li:2014oda,Colonna:2020euy, Xu:2019hqg,Russotto:2011hq, Cozma:2011nr, Giordano:2010pv} (see also Sec.\ \ref{sec:SymExpan}). Meson yields in inelastic nucleon-nucleon reactions during heavy-ion collisions can also provide information about the isospin dependence of the EOS and regions of the phase diagram probed in the collisions \cite{Colonna:2020euy,Fuchs:2000kp,Li:2002qx,Xiao:2008vm,Yong:2022pyb,SRIT:2021gcy}.

While hadronic transport studies have been successful in gaining understanding of the dynamics of heavy-ion collisions, precise quantitative statements about the dense nuclear matter EOS are not yet possible. Significant differences exist between EOS extracted using various theoretical fits to heavy-ion data \cite{LeFevre:2015paj,Nara:2020ztb,Danielewicz:2002pu}, and there is a considerable tension between analyses that imply a soft EOS of symmetric nuclear matter at $n_B\sim 3$--$4\ n_{\rm{sat}}$ \cite{Danielewicz:2002pu, Oliinychenko:2022uvy} and neutron star studies that point towards a very stiff asymmetric EOS in the same $n_B$ range \cite{Bedaque:2014sqa,Tews:2018kmu,Fujimoto:2019hxv,Tan:2021ahl,Marczenko:2022jhl}. Further development is needed to resolve these issues.

 Improvements may include maximally flexible parametrizations of nucleon interactions that would enable description of multiple non-trivial features of the EOS, such as, e.g., a phase transition at high $T$ and high $n_B$ or a particular behavior of $c_s^2$ as a function of $n_B$. Such a flexible parametrization of the symmetric EOS based on relativistically covariant vector-type interactions was recently developed \cite{Sorensen:2020ygf, Oliinychenko:2022uvy}. However, similarly versatile isospin-dependent and momentum-dependent EOS with the corresponding single-nucleon potential is still needed. Improvements may include, e.g., generalizing parametrizations \cite{Cai:2022grw} based on Gogny-like energy density functionals (which encapsulate main features of experimentally measured nucleon-nucleon short-range correlations and optical potentials at $n_{\rm{sat}}$ \cite{Hen:2016kwk,Li:2018lpy}) or using parametrizations based on scalar-type interactions \cite{Sorensen:2021zxd}. The incorporation of mesonic degrees of freedom among species explicitly affected by mean-field potentials may also lead to more realistic simulations. Recent increases in available computational power have rendered these possible code refinements accessible in realistic implementations.

In parallel to these possible advances, an effort should be made to reduce theoretical uncertainties coming from model assumptions and details of implementation. This could be achieved through systematic comparisons between different hadronic transport codes. Such studies could be done within or modeled after the work of the Transport Model Evaluation Project Collaboration, which to date provided several benchmark results as well as recommendations for improvements (see \cite{TMEP:2022xjg} for a review).

\subsection{Far-from-equilibrium relativistic fluids}

Large deviations from equilibrium are found in ultrarelativistic heavy-ion collisions, especially at the early stages of the evolution \cite{Niemi:2014wta,Noronha-Hostler:2015coa}. Recent experimental findings in small systems such as $p$+A collisions indicate potential fluid-like behavior (akin to A+A systems), and motivated a deep investigation towards the foundations of hydrodynamic behavior \cite{PHENIX:2018lia}. This has lead to new insights into what constitutes relativistic hydrodynamics and its possible extension to the far-from-equilibrium regime \cite{Martinez:2010sc,Florkowski:2010cf,Alqahtani:2017mhy,Romatschke:2017ejr,Berges:2020fwq}. Analogously, developments in the far-from-equilibrium regime have also been made in strong coupling \cite{Chesler:2010bi,Heller:2011ju,Heller:2012km,vanderSchee:2013pia,Heller:2013oxa,Casalderrey-Solana:2013aba,Chesler:2013lia,Chesler:2015wra,Keegan:2015avk,Spalinski:2017mel}, and kinetic theory \cite{Denicol:2014xca,Denicol:2014tha,Kurkela:2015qoa,Keegan:2015avk,Bazow:2015dha,Denicol:2016bjh,Heller:2016rtz,Romatschke:2017vte,Strickland:2018ayk,Kurkela:2018wud,Kurkela:2018vqr,Kurkela:2019set,Strickland:2019hff,Denicol:2019lio,Almaalol:2020rnu,Jaiswal:2022udf,Alalawi:2022pmg,Mullins:2022fbx}. Theoretical developments include (but are not limited to)  hydrodynamic attractors \cite{Heller:2015dha,Romatschke:2017vte,Florkowski:2017olj,Strickland:2017kux,Jaiswal:2022udf}, anisotropic hydrodynamics \cite{Martinez:2010sc,Florkowski:2010cf,Alqahtani:2017mhy}, the formulation of new causal and stable (general-)relativistic viscous hydrodynamic theories \cite{Bemfica:2017wps,Kovtun:2019hdm,Bemfica:2019knx,Hoult:2020eho,Bemfica:2020zjp,Noronha:2021syv}, and the proof that causality is a necessary condition for thermodynamic stability in relativistic fluids \cite{Gavassino:2021owo}. Furthermore, the means by which rapid longitudinal expansion can drive kinetic theory 
into the ground state of an effective Hamiltonian whose subsequent adiabatic evolution later hydrodynamizes has been also been considered~\cite{Brewer:2019oha,Brewer:2022vkq}.  Thus, it may be possible to extend hydrodynamical treatments to the far-from-equilibrium conditions that are generated in low-energy heavy-ion collisions.

In heavy-ion simulations when hydrodynamics is initialized, the system can be so far-from-equilibrium that the standard 2nd-order hydrodynamic formulations \cite{Israel:1979wp,Denicol:2012cn} can display unphysical, acausal behavior in simulations \cite{Chiu:2021muk,Plumberg:2021bme}. At the same time, the hydrodynamic paradigm has been enormously successfully at predicting experimental observables at the percent level \cite{Niemi:2015voa,Noronha-Hostler:2015uye} and describing a menagerie of observables \cite{Alba:2017hhe,Schenke:2020mbo,Hirvonen:2022xfv}. Thus, there is a tension between foundational physical principles and successful phenomenology. Mastering the far-from-equilibrium regime of relativistic fluids is crucial not only for accurate simulations, but also for understanding why hydrodynamics is applicable to the extremely short-lived and very rapidly expanding systems formed in heavy-ion collisions. 

Much less is known about the applicability of hydrodynamics at finite $n_B$. Far-from-equilibrium constraints, such as  \cite{Bemfica:2019cop,Bemfica:2020xym} for $\mu_B=0$, are not known. Following \cite{Almaalol:2022pjc}, further work on the causality and stability properties of hydrodynamics with multiple conserved charges (with and without the inclusion of critical phenomena) is needed. Large deviations from equilibrium may be expected in the vicinity of the QCD critical point depending on the structure and properties of the critical region \cite{Martinez:2019bsn,Monnai:2016kud}. The initial conditions may also be far from equilibrium which may complicate comparisons to lattice QCD (since these systems will not conserve entropy) as well as comparisons between events at similar beam energy since the exact phase diagram trajectory the system takes is heavily dependent on initial conditions \cite{Dore:2020jye,Dore:2022qyz,Chattopadhyay:2022sxk}. Thus, the inclusion of out-of-equilibrium effects is necessary for systematic comparisons between the matter formed in low energy heavy-ion collisions (RHIC, HADES, and FAIR) and  neutron star mergers.

\subsection{Bayesian analyses }

Bayesian analyses of LHC and highest-energy RHIC collisions have been proven capable of constraining the EOS of baryon-free matter at high $T$ \cite{Pratt:2015zsa}. For similar success in constraining the baryon- and isospin-dependence of the EOS, low $\sqrt{s_{NN}}$ simulations must first reckon with  several practical and conceptual challenges. Once these upgrades are incorporated to heavy-ion simulations, it will then be possible to apply a Bayesian analysis framework that treats aspects of the EOS, transport coefficients and pre-equilibrium dynamics as  free parameters to be determined by comparisons with experimental data. Additionally, one could apply model selection techniques to determine if a specific $\sqrt{s_{NN}}$ can be best described by hydrodynamic or hadron transport simulations \cite{JETSCAPE:2020mzn}. In such an approach, the EOS and transport coefficients could be extracted from experimental data, and possibly even determine the location of the QCD critical point or first-order phase transition line.

\section{Neutron stars}

\begin{figure}
    \begin{tabular}{cc}
       \includegraphics[width=0.5\linewidth]{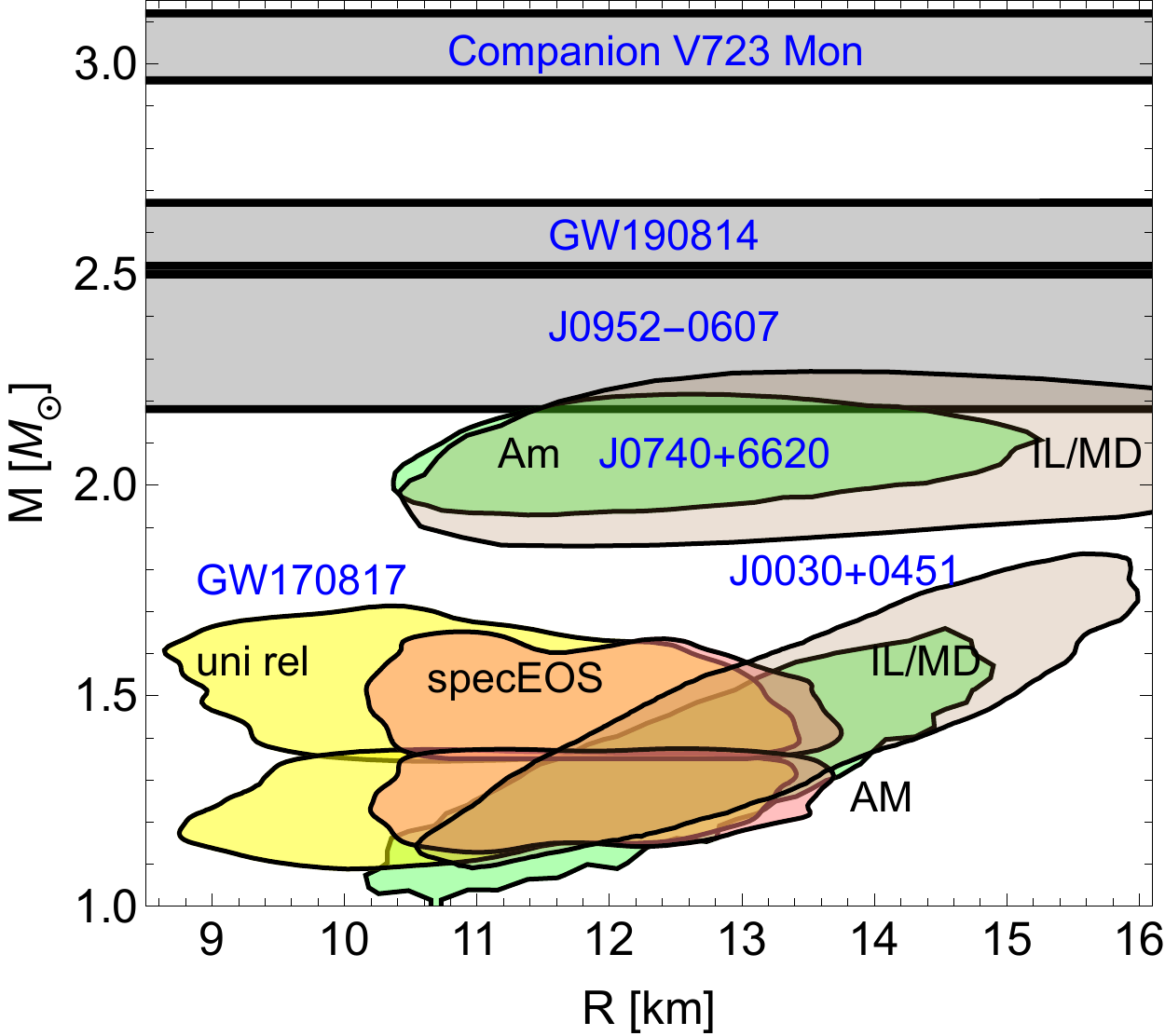}   &  \includegraphics[width=0.47\linewidth]{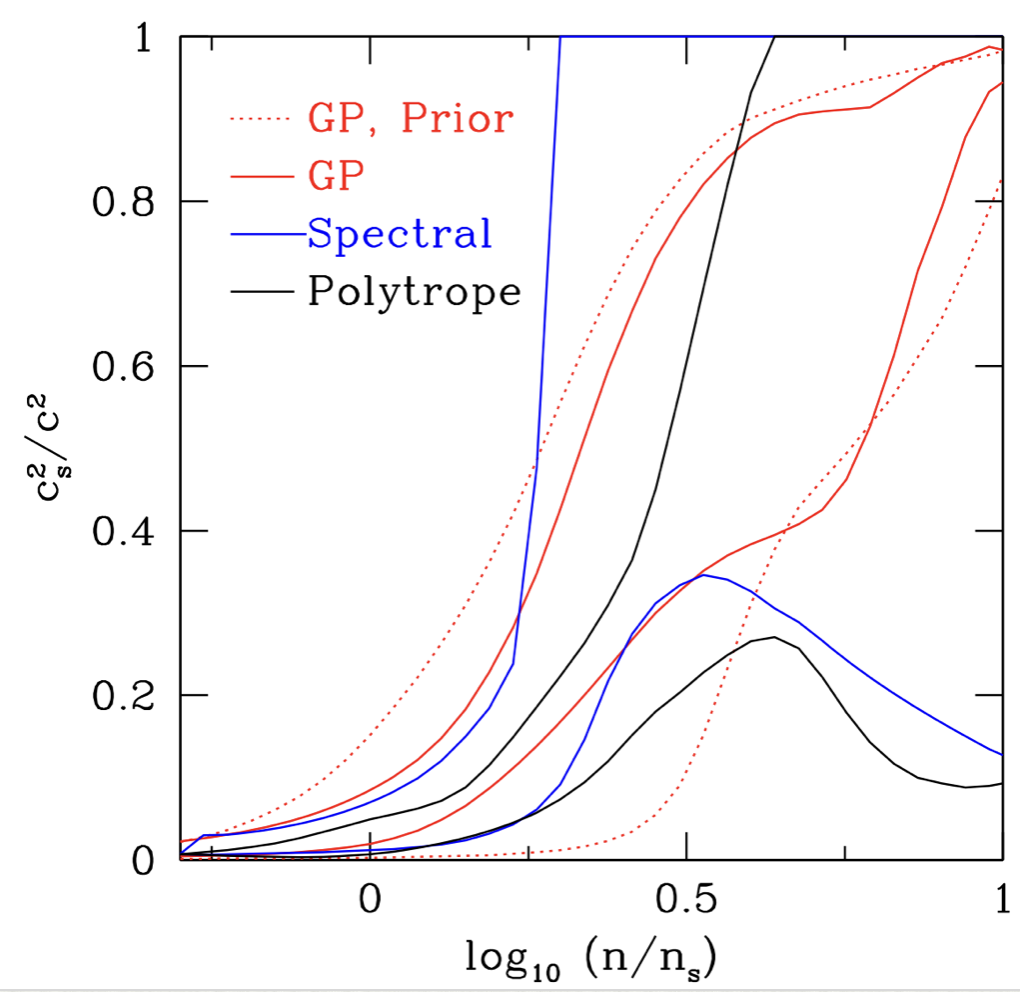}
    \end{tabular}
   
    \caption{Left: Constraints at the $2\sigma$ confidence level from LIGO/Virgo GW170817 \cite{TheLIGOScientific:2017qsa}, NICER J0030, J0740 \cite{Riley:2021pdl,Miller:2021qha,Riley:2019yda,Miller:2019cac}.  Potential neutron stars candidates GW190814, J0952-0607, and V723 Mon are included, but still under contentious debate. Figure adapted from \cite{Tan:2021ahl}. Right: Extraction of  $c_s^2$ vs. $n_B$ from GW170817,  J0030, J0740 data using three different functional forms for the EOS where the upper and lower bounds of the posteriors are color coded in red for Gaussian Processes, blue for spectral EOS, and black for polytropes. Figure from \cite{Miller:2021qha}}
    \label{fig:MRconstraints}
\end{figure}

Recent gravitational-wave, X-ray and radio observations have the potential to revolutionize our understanding of nuclear physics at high $n_B$ and low $T$. The advanced LIGO and Virgo detectors have already observed the inspiral of a neutron star binary in 2017 (GW170817), another possible neutron star binary in 2019 (GW190425), and the possible merger of two neutron-star/black-hole binaries in 2020 (GW200105 and GW200115). Measurements of the maximum mass ($M_{max}$), mass-radius ($M-R$) relation, and tidal deformability ($\Lambda$) of neutron stars can be used to infer the $T=0$ EOS at intermediate and large $n_B$. Current electromagnetic and gravitational constraints are shown in Fig.~\ref{fig:MRconstraints}.  

\begin{figure}
    \centering
    \includegraphics[width=\linewidth]{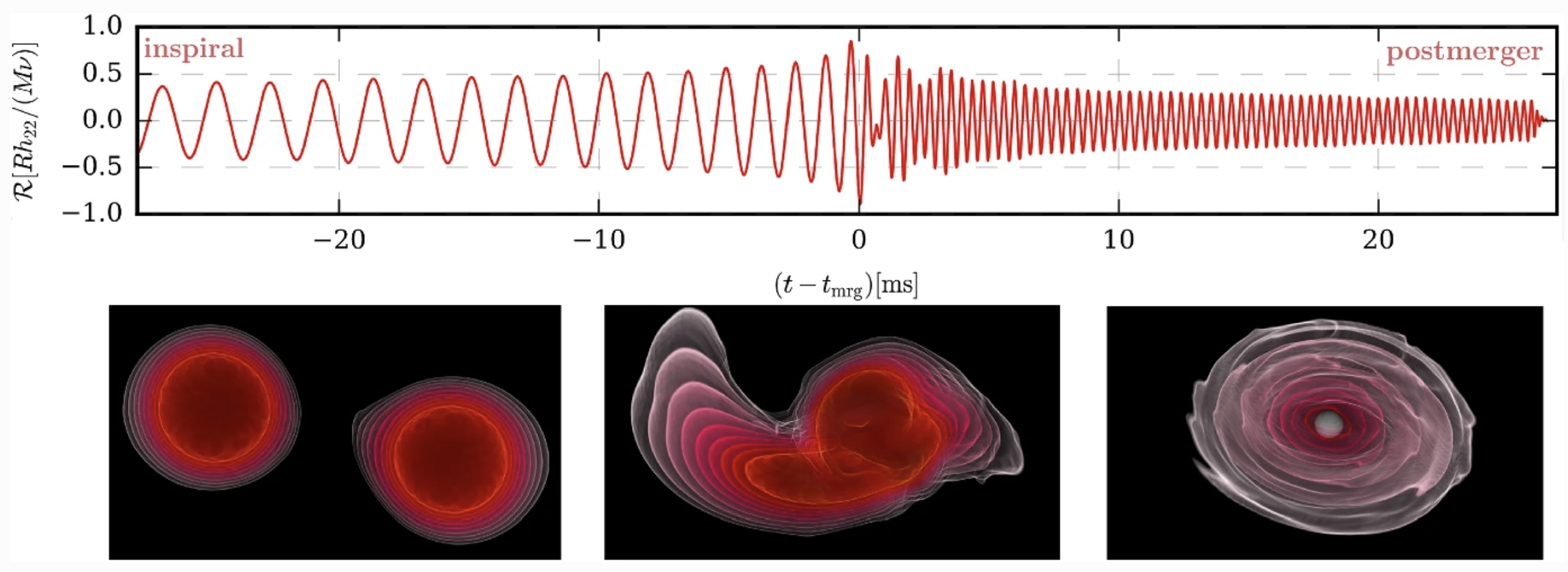}
    \caption{Phases of the coalescence of binary neutron stars. During the inspiral phase (before touching), the neutron stars are deformed due to their gravitational pull and one can extract the tidal deformability $\Lambda$. The inspiral is the only phase currently measurable from gravitational waves. The post-merger is the point where the neutron stars touch, finite T and $Y_Q$ play a role, and the gravitational waves from this phase may be measurable in O5 or  3G~\cite{Torres-Rivas:2018svp}. The remnant (the final remaining object) is either a hypermassive neutron star or black hole.  Figure taken from \cite{Dietrich:2020eud}.}
    \label{fig:NSM}
\end{figure}

During its fourth (O4) and fifth (O5) observing runs (to occur in 2023 and c. 2025), the LIGO/Virgo/KAGRA detector network is expected to detect many tens of events including neutron stars. In the third-generation (3G) era, circa late 2030s, Cosmic Explorer in the United States~\cite{Reitze:2019iox}, the Einstein Telescope in Europe~\cite{Punturo:2010zz} and NEMO in Australia~\cite{Ackley:2020atn} are expected to observe 10s to 100,000s of neutron star events per year. Similarly, the NICER telescope has announced radius constraints for 2 pulsars to date, and observations will continue at the minimum of 3 more years, if not longer.    
These observations will not only be numerous, but they will  come with increased sensitivity, i.e., it will be possible to pull new physics out of the data. While GW170817 could  place an upper limit on the tidal deformability $\Lambda$ \cite{LIGOScientific:2018hze}, the error bars were large. If the same event were detected in $O5$ or in the 3G era, the $\Lambda$ is expected to be measured $\sim$ 6 times and $\sim 30$ times more accurately \cite{Carson:2019rjx}, respectively. Quantum squeezing implemented during O3 and frequency-dependent squeezing under implementation O4 will yield further improvements in the high-frequency sensitivity of the detector. Such improvement in the $\gtrsim 700$\,Hz sensitivity opens up the possibility of detecting the gravitational waves emitted during and after the merger of neutron stars (see Fig.\ \ref{fig:NSM}). The merger phase  has the potential of carrying information about possible phase transitions at high $n_B$ or $T$~\cite{Most:2018eaw,Bauswein:2018bma}. It has been suggested that out-of-equilibrium effects can also play a crucial role in the post-merger \cite{Most:2021zvc, Hammond:2022uua, Most:2022yhe}, which could lead to a whole new field of research and strong synergies with heavy-ion collisions. However, simulations with out-of-equilibrium neutrino-radiation hydrodynamics suggest that out-of-equilibrium effects are too small to be resolved in current simulations \cite{Zappa:2022rpd}.

\subsection{X-Ray observations}

The radii of neutron stars, if measured with precision and reliability, provide data sensitive to the EOS at a few times $n_{sat}$.  Early attempts at radius measurements relied on time-integrated measures (such as the average flux and spectrum, supplemented by a distance estimate to the measured neutron stars) and have been found to be susceptible to systematic error.  For example, data can be equally well-fit assuming hydrogen atmospheres or helium atmospheres, but the inferred radii can differ by $\sim 50$\%,  many times the formal statistical uncertainty \cite{2013ApJ...764..145C,2012MNRAS.423.1556S}.
In contrast, the X-ray data obtained using NASA's NICER mission times photons to better than 100 nanoseconds, such that it is possible to obtain spectra as a function of rotational phase.  This extra time dimension in the data renders radius estimates much more robust against systematic errors.  

To date, NICER-based radius measurements have been published for two neutron stars \cite{Riley:2021pdl,Miller:2021qha,Riley:2019yda,Miller:2019cac}.  A measurement for another high-signal neutron star is expected within the next several months, and measurements will also be reported for three additional neutron stars with weaker signals.  It is expected that by the end of the NICER mission it will at least double the total exposure on each of these neutron stars, which will lead to an expected reduction in the current $\sim 1-2$~km radius uncertainties by a factor of $\sqrt{2}$, and possibly more depending on the total exposure.

\subsection{Inspiral of neutron stars}

Ground-based detectors observe gravitational waves from the coalescence of two neutron stars for minutes, depending on their low frequency sensitivity. Though the neutron star structure does not affect the majority of the signal (modulo potential resonance effects), this long inspiral can be used to measure the chirp mass of the system to astonishing relative accuracy ${\cal{O}}(10^{-4})$~\cite{Farr:2015lna}. The mass ratio of the system is less well measured and correlated with the spin, making the precise determination of the individual masses challenging~\cite{Ng:2018neg}. In the coming years, as more systems are observed, information about the population distribution of merging neutron star spins can help alleviate this degeneracy and obtain more constraining individual mass measurements. Regardless, from the chirp mass alone, it is inferred that GW190425 was a compact binary whose primary had mass $\gtrsim 1.7M_{\odot}$.  If, indeed, the primary was a neutron star, this would mark a novel kind of astrophysical system compared to the known galactic binary neutron star systems \cite{LIGOScientific:2020aai} that have a mass distribution characterized by $M=1.35\pm0.1M_\odot$~\cite{Lattimer:2012nd}.

During the final tens of milliseconds, tidal interactions between the two coalescing neutron stars result in a speed up of the inspiral, which manifests as a shift in the phase of the waveform. The tidal deformability $\Lambda$ quantifies this effect and it is a strong function of the neutron star mass and radius, scaling as a high negative power of the stellar compactness $C=G M/(c^2 R)$. Due to this scaling with compactness, tidal interactions between massive neutron stars -such as GW190425- are undetectable at current sensitivity, and the interpretation of some binaries is based purely on the inferred masses. 
Notable among them is GW190814 \cite{LIGOScientific:2020zkf}, with a secondary mass at $2.6M_{\odot}$, whose nature is an unsettled debate.   Improved sensitivity will allow us to probe potential tidal interactions of heavier binaries, with third generation detectors able to reliably infer the nature of objects up to $\sim 2M_{\odot}$ from the inspiral signal alone. At the same time, the improved sensitivity and the large number of detections could result in binary tidal deformability measurements down to $\sim 100$ during the fifth observing run and $\sim 20$ for next generation detectors~\cite{Chatziioannou:2021tdi}.

Besides overall constraints on $\Lambda$ (which can be translated to constraints on the radius and combined/contrasted with other probes, e.g., \cite{Annala:2017llu,Bauswein:2017vtn,De:2018uhw,Most:2018hfd,Raithel:2018ncd,LIGOScientific:2018cki,Raithel:2020vvg,Dietrich:2020efo}), inspiral signals can probe the entire $\Lambda-m$ relation. As mentioned above, such constraints become intrinsically weaker as $m$ increases. However, $\sim 100$ events (expected circa the fifth observing run) will suffice for the identification of (or a constraint on) a sudden softening in $\Lambda-m$ at low $n_B$/mass scales around $1.4M{\odot}$~\cite{Chatziioannou:2019yko,Pang:2020ilf}.   

\subsection{Potential ultraheavy neutron stars}

In the last years, there have been potential observations of ultraheavy neutron stars ($M\gtrsim 2.1 M_{\odot}$).  These compact objects would fall into the mass gap between the largest well-measured neutron star masses and what is anticipated to be the lightest black hole $M_{BH}^{min}\gtrsim 3M_{\odot}$.  As previously mentioned, GW190814 included a compact object of $2.6M_{\odot}$, where only the mass alone could be confidently measured; the tidal deformability for such a heavy compact object, if it was a neutron star, would be well below $\Lambda \sim 2-20$ \cite{Tan:2020ics,Tan:2021ahl}, which, in turn, is below the values of $\Lambda = \mathcal{O}(100)$ that can be extracted with current detectors \cite{Carson:2019rjx}. Not long after GW190814 was observed, the companion to the red giant V723 Mon was estimated to have a mass of $M\geq 2.91\pm0.08$~M$_\odot $ \cite{Jayasinghe:2021uqb}, although astrophysical and modeling systematics make this measurement less robust than the GW190814 one; furthermore, the companion could be a black hole. In addition, several ``black widow" pulsars are known (so-named because their primaries may be consuming their sub-stellar companions, leading to a high mass), the first known of which is PSR1957+20 with an estimated mass of $2.40\pm0.12M_\odot$~\cite{vanKerkwijk:2010mt}.  In just the last months, a new black widow pulsar PSR J0952-0607 was announced, having a mass of $M=2.35\pm 0.17 M_{\odot}$ \cite{Romani:2022jhd} and also having the second fastest known spin rate (707 Hz) of any pulsar. Some of these ultraheavy neutron star candidates are shown as gray bands in Fig.\ \ref{fig:MRconstraints} because there is no known information about their sizes. Physics inferred from these stars comes with several caveats, which makes it difficult to determine today what an accurate minimum for the maximum mass of a neutron star is.  
However, if even one of these objects were to be unambiguously confirmed as an ultraheavy neutron star, it would have strong implications for the EOS \cite{Tan:2020ics,Tan:2021ahl,Tews:2020ylw}.

In contrast, from the well-measured mass of GW170817 and inferences from its electromagnetic signature that it promptly (within a second) collapsed into a black hole, it has been estimated that the neutron star maximum mass is no larger than $2.2-2.3M_\odot$ (see \cite{Margalit:2017dij,Rezzolla:2017aly,Shibata:2019ctb}). Future mergers with accompanying electromagnetic signals and additional pulsar detections will refine this estimate.  Besides its importance to the properties of the neutron star EOS, the neutron star maximum mass will help determine if future gravitational wave merger detections involve neutron stars or black holes.

\subsection{Merging neutron stars and multi-messenger signals}

Different from the inspiral, where the orbiting neutron stars are essentially at $T\simeq 0\, \rm MeV$, the collision of two neutron stars will probe the hot and dense phase of nuclear matter (e.g., \cite{Bauswein:2010dn,Kastaun:2016yaf,Hanauske:2016gia,Perego:2019adq,Endrizzi:2019trv,Raithel:2021hye}). 
Shock heating during merger will generate temperatures up to $T\simeq 50\, \rm MeV$, whereas the potentially metastable neutron star remnant will probe  $n_B\sim 2-10\, n_{sat}$  \cite{Radice:2020ddv,Most:2019onn}. Despite the obvious difference in isospin, the conditions probed in the merger are similar to those expected in low-energy heavy-ion collisions \cite{Most:2022wgo} and see Fig.\ \ref{fig:PhaseDiagram}.
Hyperons may be copiously present at those densities \cite{Sekiguchi:2011mc,Radice:2016rys}, and the phase transition (at high $n_B$ and $T$) to deconfined quark matter can occur \cite{Most:2018eaw,Bauswein:2018bma,Most:2019onn,Weih:2019xvw,Blacker:2020nlq,Liebling:2020dhf,Prakash:2021wpz}.
Accessing this information via future gravitational waves and indirectly through electromagnetic counterparts \cite{LIGOScientific:2017ync,Kasliwal:2017ngb,Fujimoto:2022xhv}, would offer the tantalizing prospect of obtaining novel insights into dense matter.

Several ways of potentially leveraging this information have been suggested.
The most direct way of probing hot and dense matter would be via the high-frequency part of the gravitational-wave signal, e.g., through next-generation gravitational-wave detectors such as Cosmic Explorer \cite{Reitze:2019iox}, Einstein Telescope \cite{Punturo:2010zz} or NEMO \cite{Ackley:2020atn}, see also,\cite{Bose:2017jvk,Breschi:2022ens,Breschi:2022xnc,Wijngaarden:2022sah,Puecher:2022oiz}. This kilohertz signal is sourced by oscillations of the merger remnant, which has been shown to be correlated with the underlying EOS \cite{Bauswein:2011tp,Bernuzzi:2012ci,Takami:2014zpa}. Since they are sourced by the densest part of the star they would offer a pristine probe of the conditions inside neutron stars, which correlate with the EOS. \cite{Breschi:2022ens,Raithel:2022orm}. Finite-$T$ effects \cite{Bauswein:2010dn,Raithel:2021hye}, and especially exotic degrees of freedom \cite{Sekiguchi:2011mc,Radice:2016rys,Most:2018eaw,Bauswein:2018bma} can lead to distinct shifts in the post-merger spectrum. 
Neutron-rich matter ejected from the system will eventually undergo r-process nucleosynthesis~\cite{Lattimer:1977igd},  powering electromagnetic afterglows \cite{Metzger:2019zeh}. Since the amount and composition of this material are to some extent dependent on the underlying (finite-$T$) EOS, constraints on the electromagnetic emission can provide additional constraints on the dense matter phase \cite{Radice:2017lry,Coughlin:2018miv,Kiuchi:2019lls,Fujimoto:2022xhv}. Such inferences are naturally prone to uncertainties in the numerical modeling and underlying nuclear physics at low $n_B$ (reaction rates, opacities, etc.). 

The absence or presence of certain counterparts can provide more direct constraints on the densest states of nuclear matter present in the most massive neutron stars. 
Rapidly spinning neutron stars can support between $20-30\,\%$ higher masses than the maximum mass, $M_{\rm max}$ of non-rotating neutron stars \cite{Breu:2016ufb,Bozzola:2019tit,Koliogiannis:2019rvh,Annala:2021gom}. The merger of massive enough neutron stars will eventually result in direct black hole formation \cite{Baiotti:2019sew}. This can be accompanied with a reduction in available remnant baryon matter, leading to a potential suppression of electromagnetic counterparts \cite{Ruiz:2017inq,Bernuzzi:2020txg}. The threshold binary mass, at which a merger remnant would immediately collapse into a black hole, has been shown to be correlated with $M_{\rm max}$ and their compactness, offering an indirect probe of dense matter  \cite{Bauswein:2013jpa,Koppel:2019pys,Bauswein:2020aag,Tootle:2021umi,Kashyap:2021wzs,Kolsch:2021lub,Perego:2021mkd}.
Determining the outcome of a merger in terms of delayed black hole formation also promises to lead to a number of constraints on the neutron star $M_{\rm max}$ \cite{Radice:2017lry,Margalit:2017dij, Rezzolla:2017aly, Ruiz:2017due, Shibata:2019ctb}.

\subsection{Beyond the standard model}

Because neutron stars couple all four fundamental forces of nature (strong, weak, electromagnetism, and gravity), it is natural to question the influence of beyond standard model (BSM) physics on neutron stars. A number of groups have been incorporating effects of dark matter accumulating within neutron stars through dark matter capture \cite{Ellis:2018bkr,Deliyergiyev:2019vti,Kain:2021hpk,Rutherford:2022xeb} or consider the potential of dark compact objects \cite{Horowitz:2019aim}. Another dark matter candidate, mirror matter, is an exact copy of QCD but with heavy quark masses, which would lead to mirror neutron stars (dark stars with much smaller mass/radii than standard model neutron stars) \cite{Hippert:2021fch}.   Other types of BSM model physics such as modified gravity \cite{Yagi:2013awa} or by adding a QCD axion to their EOS \cite{Lopes:2022efy} have also been explored. If  BSM particles couple to the standard model (beyond merely gravitationally), then they can be efficiently produced inside of a neutron star core \cite{Raffelt:1996wa,Fortin:2021cog}.  If the coupling between the BSM particle and the standard model is sufficiently weak, then the neutron star will emit BSM particles, altering how it cools over time \cite{Sedrakian:2015krq,Hamaguchi:2018oqw}, and perhaps generating an additional electromagnetic signal \cite{Fortin:2021cog,Buschmann:2019pfp,Giangrandi:2022wht}.  

Neutron star mergers have become a focus for discovering BSM physics \cite{Fortin:2021cog,Baryakhtar:2022hbu}.  Large quantities of accumulated dark matter (interacting purely gravitationally) have been shown in numerical simulations to modify post-merger dynamics \cite{Ellis:2017jgp,Bezares:2019jcb}.  Ultralight axions coupled to standard model particles can modify the neutron star inspiral phase prior to merger \cite{Zhang:2021mks}.  BSM particles that couple to nucleons can, as with neutron stars, be produced in the merger remnant and either enhance transport inside the remnant or be emitted, cooling down the remnant \cite{Dev:2021kje,Harris:2020qim} and possibly generating an additional electromagnetic signal with their subsequent decay \cite{Diamond:2021ekg}.  More precise the knowledge of the standard model EOS greatly improves the chance for BSM physics to be found as well. 

\section{Merging heavy-ions and neutron stars}\label{sec:HI_and_NSM}

With the arrival of better constraints from  neutron stars and heavy-ions, groups are beginning to attempt to make direct comparisons between both systems.  However, major theoretical roadblocks still remain for these comparisons that require large scale theoretical effort. Below, we detail challenges to be tackled over the next decade.

\subsection{Connecting symmetric and asymmetric nuclear matter}\label{sec:SymExpan}

One inherent difference between matter created in heavy-ion collisions and inside the core of neutron stars, is their isospin content. Presently, heavy-ions collisions rely on stable nuclei such that the proton number $Z$ to nucleon number $A$ from the initial state is approximately $Y_Q\equiv Z/A\sim 0.5$ i.e. ``symmetric nuclear matter", where the fraction $Y_Q=n_Q/n_B$ is preserved throughout the evolution even though $n_Q$ and $n_B$ is distributed across a wide range of hadrons (e.g. $\pi^\pm$ or $\Sigma^\pm$) due to baryon number and electric charge conservation, not just in protons and neutrons. The heaviest nuclei such as $^{238}U$ may reach $Y_Q\sim 0.38$, which brings $Y_Q$ closer to neutron stars (but not quite as low). 

In contrast, the process of neutronization that takes place in supernova explosions leaves neutron star cores with a disproportionate amount of neutrons, which remains even after merging events \cite{Most:2019onn}, and neutron stars are made of what is known as ``asymmetric nuclear matter". The strongly interacting charge fraction $Y_Q$ refers to not only the amount of protons, but also hyperons and quarks. The charge fraction\footnote{Note, the net total charge in stars is, of course, zero but that is due to the additional contribution from leptons, which is not included in $Y_Q$.} is significantly smaller, $Y_Q\leq 0.2$,  but sometimes orders of magnitude smaller than that.  

Nuclear symmetry energy, $E_{sym}$, measures the energy cost to make nuclear systems more neutron rich. It is approximately the energy difference between pure neutron matter $Y_Q\rightarrow 0$ and symmetric nuclear matter $Y_Q=0.5$. Many interesting issues regarding its physical origins (e.g., the spin-isospin dependence of three-body and tensor forces, as well as the resulting strong isospin-dependence of nucleon-nucleon short-range correlations), $n_B$ dependence, as well as impact on properties of nuclei, neutron stars and their collisions remain to be addressed (see \cite{Li:2014oda}).
One must rely on the symmetry energy expansion at $n_{sat}$, where only the first two coefficients: the magnitude $E_{sym}(n_{sat})$ and slope $L(n_{sat})$, currently have significant experimental and observational constraints (and $L$ is still hotly debated, see Fig.~5 of \cite{Li:2019xxz}). It may be possible to determine neutron skin thickness from high-energy heavy-ion collisions \cite{Xu:2022ikx} or measurements of mirror nuclei at GSI or FRIB~\cite{Pineda:2021shy}, which, when  compared to neutron star $\Lambda$ or radii measured in neutron star mergers  \cite{Raithel:2019ejc,Li:2021thg,Essick:2021kjb,Most:2021ktk}, could add a new piece to the puzzle. Moreover, heavy-ion reactions induced by high-energy rare isotopes \cite{LCK08} at FAIR and the planned 400 MeV/nucleon upgrade of FRIB will provide new opportunities to constrain more tightly the symmetry energy, especially around $2\ n_{sat}$ \cite{FRIB400,Huth:2021bsp}.

Deeper questions remain because strange particles are not relevant at $n_{sat}$ but appear at higher $n_B$.  Therefore, a Taylor expansion scheme around $n_{sat}$ may not correctly describe their behavior \cite{Aryal:2020ocm,Costa:2020dgc}.  Quarks would be relevant at even higher $n_B$ and would require an entirely new approach as well.  Furthermore, a Taylor expansion would most likely break down if a phase transition is approached and in that case it is unlikely that  symmetry energy expansion can correctly connect the EOS across all $n_B$.  Note that the deconfinement phase transition is sensitive to $Y_Q$, especially because of the sensitivity of the hyperons \cite{Aryal:2020ocm}. Finally, if structure in $c_s^2$ as a function of $n_B$ (such as a bump, as seen in the right panel of Fig. \ref{fig:MRconstraints}) is observationally verified  from neutron stars in the near future, one would have to understand its signature in heavy-ion collisions as well (or determine how a $c_s^2$ bump varies with $Y_Q$).  Thus, many questions still remain. 

\subsection{Signatures of phase transitions}

If either the QCD critical point or a first-order phase transition into quarks is found, it will be clear that a real phase transition exists between quarks and hadrons.  Active searches are underway for this transition in low energy heavy-ion collisions, neutron star mergers, and for neutron stars (where the $T=0$ limit is applicable).  Signs of a phase transition would present differently in each limit.  In heavy-ion collisions, the signatures of the QCD critical point and first-order phase transition were discussed already in Sec.\ \ref{sec:heavyions}.  However, in nuclear astrophysics very specific signatures may appear (and others may yet to be discovered through further collaborations between nuclear and gravitational wave physicists).
At $T=0$  a first-order phase transition may lead to ``mass twins" that appear as a second stable branch in the mass-radius sequence. These are neutron stars with approximately the same mass but distinguishable radii \cite{Alford:2013aca,Benic:2014jia,Dexheimer:2014pea}. A merger involving twin stars could produce unique structure in the binary Love relations detectable in the near future by LIGO and Virgo \cite{Tan:2021nat}.
During the merger itself, finite $T$ are reached and, therefore, numerical relativity simulations that necessitate a multidimensional phase diagram are required.  The signatures of a first-order phase transition then would show up in the post-merger peak frequency that will be possible to measure in O4 or beyond. There one could find the modification of the gravitational wave form \cite{Most:2018eaw} and the peak frequency \cite{Bauswein:2018bma} of post merger signals

A word of caution is needed: we previously noted that in Sec.\ \ref{sec:SymExpan} that heavy-ion collisions and neutron stars are not on the same axis of the QCD phase diagram due to their differences in electric charge/isospin and strangeness \cite{Aryal:2020ocm}.  Therefore, it is possible that the nature of a phase transition may change between these two limits.  In fact, for the liquid-gas phase transition, while it is observable for symmetric nuclear matter, it  may vanish for asymmetric nuclear matter \cite{Qian:2000fq}.  There is another possibility of a second critical point at low $T$, such that it would be probed by neutron star mergers \cite{Hatsuda:2006ps}.  That would imply that a first-order phase transition may exist at finite $T$ but it would not reach the $T\rightarrow 0$ limit. 

\subsection{Statistical analyses}

It is clear that there is enormous potential through cross-talk and collaboration between the various communities working\footnote{Across nuclear theory this is a regular challenge so a Computational Nuclear Physics 
and AI/ML Town Hall was held to address these issues.} on dense QCD and nuclear matter.  The logical next step is to conduct joint statistical analyses to extract the EOS across various regions of the QCD phase diagram, using nuclear experimental data, theoretical constraints from lattice QCD and $\chi$EFT, and astrophysical constraints from NICER and gravitational wave detectors. 

Such an endeavor would be far from a trivial task because upgrades are needed in nearly all theoretical frameworks, as detailed in previous sections.  However, uncertainty quantification and data analysis presents its own challenges as well.  On the data side alone, Bayesian analysis~ \cite{Danielewicz:2021vqq,dagostini_multidimensional_1995} may reveal information that was thought to be inaccessible, but useful for theory, such as triple-differential distributions tied to the reaction plane at the energies where the reaction-plane effects are still pronounced.  At the interface between experiment and theory, as a further example, to determine the EOS from data one must determine reasonable functional forms of the EOS.  Already in the simplest case of neutron stars at $T=0$ this has proven challenging because microscopic models of the EOS often contain more structure than available methods like polytropes and spectral EOS are able to handle \cite{Tan:2020ics}.  New ideas are needed such as Gaussian processes \cite{Landry:2020vaw,Gorda:2022jvk}, neural networks \cite{Fujimoto:2019hxv,Han:2022sxt} or other new methods not yet identified.  Additionally, if one wants to consider data from heavy-ion collisions or post-merger data then one also requires a multidimensional EOS in a reasonable functional form, which presents an entirely new challenge.  Furthermore, heavy-ion collisions also require functional forms of the transport coefficients as well.  Thus, a large scale statistical analysis to extract the dense matter EOS quickly becomes a problem that is significantly too large for a single group to handle and requires expertise across a variety of fields and large amounts of computational power. 

\subsection{Magnetic fields}

As in other properties, such as $n_B$ and $T$, there can be overlap in values of extremely high magnetic fields produced in heavy-ion collisions and inside young neutron stars or when they merge. Magnetic fields with poloidal components of $10^{15}$ G \cite{haensel2006neutron} on the surface and $10^{16}$ G \cite{Makishima:2014dua,DallOsso:2018dos} inside have been  found in magnetars, neutron stars in which the magnetic-field decay powers the emission of high-energy electromagnetic radiation. The poloidal field was shown to increase quadratically with the chemical potential \cite{Dexheimer:2016yqu}, reaching around one order of magnitude larger in the stellar center than on the surface. Nevertheless, values of $10^{18}$ G are allowed by the virial theorem in the center of isolated stars \cite{1991ApJ...383..745L,Bonazzola:1993zz}.
During the merger of two neutron stars, a small scale turbulent dynamo operating at the shear layer between the two merging stars can significantly amplify the magnetic field strength \cite{Price:2006fi}, potentially in excess of $10^{16-17}\, \rm G$ \cite{Giacomazzo:2014qba,Kiuchi:2015sga,Kiuchi:2017zzg}. Due to the computational challenge of fully resolving this effect numerically \cite{Palenzuela:2021gdo,Aguilera-Miret:2021fre}, novel approaches to study this regime will be needed.

\subsection{Chemical equilibration processes in neutron star mergers}

Chemical equilibrium refers to the establishment of steady-state particle fractions, like the proton fraction, strangeness fraction or lepton species fractions. In dense matter this is usually achieved via weak interactions\footnote{When additional particle degrees of freedom enter the ground state, combined weak/strong and purely strong interactions become important in chemical equilibration, for example in hyperonic matter \cite{Alford:2020pld} and matter containing a thermal $\pi^-$ population \cite{Fore:2019wib}.}, e.g. Urca processes \cite{Yakovlev:2000jp}.  In neutron star mergers (and in some oscillations of isolated neutron stars) rapid changes in $T$ and $n_B$ 
\cite{Kastaun:2016yaf,Perego:2019adq,Alford:2017rxf} will drive fluid elements out of chemical equilibrium, and weak interactions will relax them back to equilibrium on a timescale that can be comparable to that of the fluid flow dynamics, namely tens of milliseconds \cite{Alford:2017rxf,Arras:2018fxj}. This means that chemical equilibration may influence the fluid dynamics enough to create signatures in the post-merger gravitational wave emission \cite{Hammond:2022uua,Most:2022yhe}. However, quantifying the impact of these effects will require simulations 
that include sufficiently accurate rates for all the relevant chemical equilibration processes, and for faster rates may require resolutions substantially higher than those currently employed in production numerical relativity simulations \cite{Zappa:2022rpd}. 

One specific mechanism is bulk viscosity, which 
achieves a resonant maximum when the chemical relaxation time is comparable to the fluid dynamical timescale.
This provides a connection \cite{Gavassino:2020kwo} to similar phenomena present in heavy ion collisions (see Sec.\ref{sec:hic_visc}) where bulk viscosity also peaks when the fluid dynamical and relaxation timescales become
comparable~\cite{Stephanov:2017ghc,Martinez:2019bsn}. In heavy ion collisions these are strong interaction timescales, but the close analogy between the two scenarios motivates the further development of viscous fluid dynamics \cite{Pandya:2022pif,Pandya:2022sff,Camelio:2022ljs,Camelio:2022fds} and out-of-equilibrium \cite{Foucart:2016rxm,Radice:2021jtw} simulation techniques in general relativity.  In heavy ion collisions, the fluid dynamical time-scales are of order several fm$/c$, and critical slowing down near a critical point yields strong interaction dynamics on comparable time-scales (which are slow by the standards of the strong interactions); in neutron star mergers both the fluid dynamical time-scales and the time-scales for the weak-interaction processes that control chemical equilibration are of order milliseconds.  In both cases, because the two relevant timescales are comparable to each other (even though they differ by twenty orders of magnitude between one case and the other) non-equilibrium dynamics becomes important and in both cases the bulk viscosity is enhanced as the relevant equilibration processes lag behind the fluid evolution.

The rate of chemical equilibration will depend on basic properties of the dense matter (making  it a potential probe of exotic phases) and also on astrophysical conditions such as the
thermodynamic conditions probed in the merger \cite{Perego:2019adq,Endrizzi:2019trv} and the degree to which neutrinos are trapped \cite{Perego:2019adq,Zappa:2022rpd}. Clarifying this point not only requires consistent neutrino transport models \cite{Foucart:2016rxm,Foucart:2020qjb,Radice:2021jtw}, but also advances in nuclear theory, including the modeling of more exotic matter like hyperonic or (color-superconducting) quark matter as well as the interfaces and conversion between different phases of matter.

\section{Exploratory directions: Quantum information science for dense-matter theory
\label{sec:QIS}}
First-principles calculations of dense matter, in and out of equilibrium, using the method of lattice QCD, has faced significant computational challenges to date, as mentioned before. The conventional lattice-QCD method relies on Monte Carlo sampling of quantum configurations contributing to QCD correlation functions, which is enabled by Wick rotating to imaginary time. Introducing a quark chemical potential, and/or keeping the Minkowski signature of spacetime leads to an oscillatory sampling weight, hence a sign problem, which is argued to be NP-hard~\cite{troyer2005computational}. An alternative approach is the Hamiltonian simulation of such systems but the classical algorithms scale exponentially with the system size and are not feasible. While for low-dimensional quantum many-body systems, including lattice gauge theories, efficient Hamiltonian-simulation techniques such as tensor networks have been developed and successfully applied~\cite{banuls2020review,meurice2020tensor,meurice2022tensor}, such methods are generally limited when the entanglement grows rapidly. Quantum simulators and quantum computers are the most natural simulating platform for implementing Hamiltonian dynamics, and can perform the real-time dynamics of matter efficiently, i.e., with resources that scale only polynomially in the system size~\cite{lloyd1996universal}. With rapid advances in quantum hardware, algorithms, and theory, and significant investment by technology companies and the government sector in the U.S. and worldwide in quantum information sciences, the era of quantum-enhanced computing appears to be in reach. The next decade, therefore, marks a critical time for the nuclear-physics community during which the computational problems that can benefit from quantum advantage need to be identified, efficient algorithms to access them in a quantum-simulating devise, in a digital or analog form, need to be devised, and co-design efforts involving quantum technologists need to be formed to potentially impact the development of hardware, as have been articulated in recent community whitepapers and reports~\cite{NSAC-QIS-2019-QuantumInformationScience,davoudi2022quantum,catterall2022report,humble2022snowmass}. 

When it comes to physics of dense matter, in and out of equilibrium, there are a number of problems in which quantum simulation can provide a way forward. A direct simulation of the collision of particles, as in lepton, hadron, and ion colliders, is possible in a quantum computer. Quantum algorithms exist for evaluating the full S-matrix in a quantum field theory~\cite{jordan2011quantum}, and for real-time evolution of gauge theories~\cite{shaw2020quantum,Byrnes:2005qx,ciavarella2021trailhead,Lamm:2019bik,kan2021lattice}, but current resource estimates are substantial and, hence, beyond the limits of current and near-future quantum computers~\cite{banuls2020simulating,klco2022standard}. An alternate path in the short term is to resort to the same effective theories currently in play in modeling scattering process, in which the amplitudes are factorized into short-distance perturbative contributions and long-distance non-perturbative matrix elements. A quantum computer can then provide the non-perturbative input into the long-distance observables, such as parton distribution functions~\cite{lamm2020parton,echevarria2021quantum,kreshchuk2021simulating,li2021partonic,perez2021determining}, hadronic tensor~\cite{lamm2020parton}, transport coefficient~\cite{cohen2021quantum}, and jet and parton-shower functions~\cite{deJong:2020riy,nachman2021quantum,Yao:2022eqm}, at a lower computational cost compared to a quantum simulation involving the entire collision. The evaluation of the low-order transport coefficients for use in relativistic hydrodynamics is among quantities that may be accessed in shorter term, but the qubit and gate estimates are still substantial for a proper QCD computation~\cite{cohen2021quantum,kan2021lattice}. Such simulations, for example, need to include the non-trivial task of preparing initial hadronic states on a quantum computer. Thermal states at arbitrary matter density can also be efficiently prepared in a quantum computer using a variety of methods, and the phase diagram of quantum field theories can, in principle, be explored~\cite{czajka2022quantum,davoudi2022toward}.

Quantum simulation, given its ability to track the dynamics of the system in real time, can shed light on the approach to equilibrium in strongly-interacting systems~\cite{riera2012thermalization,deJong:2021wsd}. It is argued that quantum entanglement, beyond just the interactions in the system, can be the key in addressing the question of why the strongly-interacting matter thermalizes in a remarkably short period of time, contrary to naive scaling arguments~\cite{kharzeev2005color,kharzeev2017deep,baker2018thermal,berges2021qcd}. Hence, it is interesting to explore the implications of the hypothesized \emph{eigenstate thermalization hypothesis}~\cite{deutsch1991quantum,srednicki1994chaos} which aims to explain how systems which are initially prepared in far-from-equilibrium pure states can evolve to a state which appears to be in thermal equilibrium. The use of quantum computing is necessary in studying the real-time evolution of entanglement in strongly interacting QCD matter. Furthermore, it is known that entanglement measures, such as entanglement spectrum~\cite{li2008entanglement,amico2008entanglement,eisert2010colloquium}, hold key insights about the stages of equilibration and thermalization of a quantum many-body system. For example, the distribution of level spacing in the entanglement spectrum is shown to reveal whether the system thermalizes, and the evolution of entanglement-spectral gaps can provide a signature of phase transitions~\cite{geraedts2016many, kaufman2016quantum, yang2015two,Mueller:2021gxd}. Thermalization and other related questions in gauge theories are starting to be explored using entanglement measures~\cite{Mueller:2021gxd,halimeh2022robust,banerjee2021quantum,brenes2018many,dalmonte2022entanglement} but without quantum simulators and quantum computers of significant capacity and capability, they cannot be studied directly for QCD~\cite{kokail2021entanglement,pichler2016measurement,yirka2021qubit,kokail2021quantum}.  

Finally, it is important for the QCD community to identify a set of near-term problems that even a noisy intermediate-scale quantum computer can perform reliably, and provide at least qualitative understanding of the salient features of gauge-theory dynamics out of equilibrium. Quench experiments may reveal interesting out-of-equilibrium features of the physical system~\cite{mitra2018quantum,canovi2011quantum}, including dynamical quantum phase transitions~\cite{heyl2019dynamical,zhang2017observation,jurcevic2017direct,nie2019experimental} that are shown to be relevant in gauge theories as well~\cite{zache2019dynamical,mueller2022quantum,van2022anatomy}. These experiments are relatively straightforward to set up given an easily preparable initial state~\cite{bernien2017probing,zhang2017observation}, and may, therefore, be the first playground for studies of non-equilibrium physics of gauge theories.



\bibliography{inspire,NOTinspire} 

\begin{thebibliography}{549}%
\makeatletter
\providecommand \@ifxundefined [1]{%
 \@ifx{#1\undefined}
}%
\providecommand \@ifnum [1]{%
 \ifnum #1\expandafter \@firstoftwo
 \else \expandafter \@secondoftwo
 \fi
}%
\providecommand \@ifx [1]{%
 \ifx #1\expandafter \@firstoftwo
 \else \expandafter \@secondoftwo
 \fi
}%
\providecommand \natexlab [1]{#1}%
\providecommand \enquote  [1]{``#1''}%
\providecommand \bibnamefont  [1]{#1}%
\providecommand \bibfnamefont [1]{#1}%
\providecommand \citenamefont [1]{#1}%
\providecommand \href@noop [0]{\@secondoftwo}%
\providecommand \href [0]{\begingroup \@sanitize@url \@href}%
\providecommand \@href[1]{\@@startlink{#1}\@@href}%
\providecommand \@@href[1]{\endgroup#1\@@endlink}%
\providecommand \@sanitize@url [0]{\catcode `\\12\catcode `\$12\catcode
  `\&12\catcode `\#12\catcode `\^12\catcode `\_12\catcode `\%12\relax}%
\providecommand \@@startlink[1]{}%
\providecommand \@@endlink[0]{}%
\providecommand \url  [0]{\begingroup\@sanitize@url \@url }%
\providecommand \@url [1]{\endgroup\@href {#1}{\urlprefix }}%
\providecommand \urlprefix  [0]{URL }%
\providecommand \Eprint [0]{\href }%
\providecommand \doibase [0]{http://dx.doi.org/}%
\providecommand \selectlanguage [0]{\@gobble}%
\providecommand \bibinfo  [0]{\@secondoftwo}%
\providecommand \bibfield  [0]{\@secondoftwo}%
\providecommand \translation [1]{[#1]}%
\providecommand \BibitemOpen [0]{}%
\providecommand \bibitemStop [0]{}%
\providecommand \bibitemNoStop [0]{.\EOS\space}%
\providecommand \EOS [0]{\spacefactor3000\relax}%
\providecommand \BibitemShut  [1]{\csname bibitem#1\endcsname}%
\let\auto@bib@innerbib\@empty
\bibitem [{\citenamefont {Abbott}\ \emph
  {et~al.}(2017{\natexlab{a}})\citenamefont {Abbott} \emph
  {et~al.}}]{TheLIGOScientific:2017qsa}%
  \BibitemOpen
  \bibfield  {author} {\bibinfo {author} {\bibfnamefont {B.~P.}\ \bibnamefont
  {Abbott}} \emph {et~al.} (\bibinfo {collaboration} {LIGO Scientific,
  Virgo}),\ }\href {\doibase 10.1103/PhysRevLett.119.161101} {\bibfield
  {journal} {\bibinfo  {journal} {Phys. Rev. Lett.}\ }\textbf {\bibinfo
  {volume} {119}},\ \bibinfo {pages} {161101} (\bibinfo {year}
  {2017}{\natexlab{a}})},\ \Eprint {http://arxiv.org/abs/1710.05832}
  {arXiv:1710.05832 [gr-qc]} \BibitemShut {NoStop}%
\bibitem [{\citenamefont {Riley}\ \emph {et~al.}(2021)\citenamefont {Riley}
  \emph {et~al.}}]{Riley:2021pdl}%
  \BibitemOpen
  \bibfield  {author} {\bibinfo {author} {\bibfnamefont {T.~E.}\ \bibnamefont
  {Riley}} \emph {et~al.},\ }\href {\doibase 10.3847/2041-8213/ac0a81}
  {\bibfield  {journal} {\bibinfo  {journal} {Astrophys. J. Lett.}\ }\textbf
  {\bibinfo {volume} {918}},\ \bibinfo {pages} {L27} (\bibinfo {year}
  {2021})},\ \Eprint {http://arxiv.org/abs/2105.06980} {arXiv:2105.06980
  [astro-ph.HE]} \BibitemShut {NoStop}%
\bibitem [{\citenamefont {Miller}\ \emph {et~al.}(2021)\citenamefont {Miller}
  \emph {et~al.}}]{Miller:2021qha}%
  \BibitemOpen
  \bibfield  {author} {\bibinfo {author} {\bibfnamefont {M.~C.}\ \bibnamefont
  {Miller}} \emph {et~al.},\ }\href {\doibase 10.3847/2041-8213/ac089b}
  {\bibfield  {journal} {\bibinfo  {journal} {Astrophys. J. Lett.}\ }\textbf
  {\bibinfo {volume} {918}},\ \bibinfo {pages} {L28} (\bibinfo {year}
  {2021})},\ \Eprint {http://arxiv.org/abs/2105.06979} {arXiv:2105.06979
  [astro-ph.HE]} \BibitemShut {NoStop}%
\bibitem [{\citenamefont {Fonseca}\ \emph {et~al.}(2021)\citenamefont {Fonseca}
  \emph {et~al.}}]{Fonseca:2021wxt}%
  \BibitemOpen
  \bibfield  {author} {\bibinfo {author} {\bibfnamefont {E.}~\bibnamefont
  {Fonseca}} \emph {et~al.},\ }\href {\doibase 10.3847/2041-8213/ac03b8}
  {\bibfield  {journal} {\bibinfo  {journal} {Astrophys. J. Lett.}\ }\textbf
  {\bibinfo {volume} {915}},\ \bibinfo {pages} {L12} (\bibinfo {year}
  {2021})},\ \Eprint {http://arxiv.org/abs/2104.00880} {arXiv:2104.00880
  [astro-ph.HE]} \BibitemShut {NoStop}%
\bibitem [{\citenamefont {Adamczewski-Musch}\ \emph
  {et~al.}(2019{\natexlab{a}})\citenamefont {Adamczewski-Musch} \emph
  {et~al.}}]{Adamczewski-Musch:2019byl}%
  \BibitemOpen
  \bibfield  {author} {\bibinfo {author} {\bibfnamefont {J.}~\bibnamefont
  {Adamczewski-Musch}} \emph {et~al.} (\bibinfo {collaboration} {HADES}),\
  }\href {\doibase 10.1038/s41567-019-0583-8} {\bibfield  {journal} {\bibinfo
  {journal} {Nature Phys.}\ }\textbf {\bibinfo {volume} {15}},\ \bibinfo
  {pages} {1040} (\bibinfo {year} {2019}{\natexlab{a}})}\BibitemShut {NoStop}%
\bibitem [{\citenamefont {Adamczewski-Musch}\ \emph
  {et~al.}(2019{\natexlab{b}})\citenamefont {Adamczewski-Musch} \emph
  {et~al.}}]{HADES:2019auv}%
  \BibitemOpen
  \bibfield  {author} {\bibinfo {author} {\bibfnamefont {J.}~\bibnamefont
  {Adamczewski-Musch}} \emph {et~al.} (\bibinfo {collaboration} {HADES}),\
  }\href {\doibase 10.1038/s41567-019-0583-8} {\bibfield  {journal} {\bibinfo
  {journal} {Nature Phys.}\ }\textbf {\bibinfo {volume} {15}},\ \bibinfo
  {pages} {1040} (\bibinfo {year} {2019}{\natexlab{b}})}\BibitemShut {NoStop}%
\bibitem [{\citenamefont {Adam}\ \emph {et~al.}(2021)\citenamefont {Adam} \emph
  {et~al.}}]{STAR:2020tga}%
  \BibitemOpen
  \bibfield  {author} {\bibinfo {author} {\bibfnamefont {J.}~\bibnamefont
  {Adam}} \emph {et~al.} (\bibinfo {collaboration} {STAR}),\ }\href {\doibase
  10.1103/PhysRevLett.126.092301} {\bibfield  {journal} {\bibinfo  {journal}
  {Phys. Rev. Lett.}\ }\textbf {\bibinfo {volume} {126}},\ \bibinfo {pages}
  {092301} (\bibinfo {year} {2021})},\ \Eprint
  {http://arxiv.org/abs/2001.02852} {arXiv:2001.02852 [nucl-ex]} \BibitemShut
  {NoStop}%
\bibitem [{\citenamefont {Adamczewski-Musch}\ \emph {et~al.}(2020)\citenamefont
  {Adamczewski-Musch} \emph {et~al.}}]{HADES:2020wpc}%
  \BibitemOpen
  \bibfield  {author} {\bibinfo {author} {\bibfnamefont {J.}~\bibnamefont
  {Adamczewski-Musch}} \emph {et~al.} (\bibinfo {collaboration} {HADES}),\
  }\href {\doibase 10.1103/PhysRevC.102.024914} {\bibfield  {journal} {\bibinfo
   {journal} {Phys. Rev. C}\ }\textbf {\bibinfo {volume} {102}},\ \bibinfo
  {pages} {024914} (\bibinfo {year} {2020})},\ \Eprint
  {http://arxiv.org/abs/2002.08701} {arXiv:2002.08701 [nucl-ex]} \BibitemShut
  {NoStop}%
\bibitem [{\citenamefont {Abdallah}\ \emph {et~al.}(2021)\citenamefont
  {Abdallah} \emph {et~al.}}]{STAR:2021rls}%
  \BibitemOpen
  \bibfield  {author} {\bibinfo {author} {\bibfnamefont {M.}~\bibnamefont
  {Abdallah}} \emph {et~al.} (\bibinfo {collaboration} {STAR}),\ }\href
  {\doibase 10.1103/PhysRevLett.127.262301} {\bibfield  {journal} {\bibinfo
  {journal} {Phys. Rev. Lett.}\ }\textbf {\bibinfo {volume} {127}},\ \bibinfo
  {pages} {262301} (\bibinfo {year} {2021})},\ \Eprint
  {http://arxiv.org/abs/2105.14698} {arXiv:2105.14698 [nucl-ex]} \BibitemShut
  {NoStop}%
\bibitem [{\citenamefont {Adhikari}\ \emph {et~al.}(2021)\citenamefont
  {Adhikari} \emph {et~al.}}]{PREX:2021umo}%
  \BibitemOpen
  \bibfield  {author} {\bibinfo {author} {\bibfnamefont {D.}~\bibnamefont
  {Adhikari}} \emph {et~al.} (\bibinfo {collaboration} {PREX}),\ }\href
  {\doibase 10.1103/PhysRevLett.126.172502} {\bibfield  {journal} {\bibinfo
  {journal} {Phys. Rev. Lett.}\ }\textbf {\bibinfo {volume} {126}},\ \bibinfo
  {pages} {172502} (\bibinfo {year} {2021})},\ \Eprint
  {http://arxiv.org/abs/2102.10767} {arXiv:2102.10767 [nucl-ex]} \BibitemShut
  {NoStop}%
\bibitem [{\citenamefont {Adhikari}\ \emph {et~al.}(2022)\citenamefont
  {Adhikari} \emph {et~al.}}]{CREX:2022kgg}%
  \BibitemOpen
  \bibfield  {author} {\bibinfo {author} {\bibfnamefont {D.}~\bibnamefont
  {Adhikari}} \emph {et~al.} (\bibinfo {collaboration} {CREX}),\ }\href
  {\doibase 10.1103/PhysRevLett.129.042501} {\bibfield  {journal} {\bibinfo
  {journal} {Phys. Rev. Lett.}\ }\textbf {\bibinfo {volume} {129}},\ \bibinfo
  {pages} {042501} (\bibinfo {year} {2022})},\ \Eprint
  {http://arxiv.org/abs/2205.11593} {arXiv:2205.11593 [nucl-ex]} \BibitemShut
  {NoStop}%
\bibitem [{\citenamefont {Pineda}\ \emph {et~al.}(2021)\citenamefont {Pineda}
  \emph {et~al.}}]{Pineda:2021shy}%
  \BibitemOpen
  \bibfield  {author} {\bibinfo {author} {\bibfnamefont {S.~V.}\ \bibnamefont
  {Pineda}} \emph {et~al.},\ }\href {\doibase 10.1103/PhysRevLett.127.182503}
  {\bibfield  {journal} {\bibinfo  {journal} {Phys. Rev. Lett.}\ }\textbf
  {\bibinfo {volume} {127}},\ \bibinfo {pages} {182503} (\bibinfo {year}
  {2021})},\ \Eprint {http://arxiv.org/abs/2106.10378} {arXiv:2106.10378
  [nucl-ex]} \BibitemShut {NoStop}%
\bibitem [{\citenamefont {Patricelli}\ \emph {et~al.}(2022)\citenamefont
  {Patricelli}, \citenamefont {Bernardini}, \citenamefont {Mapelli},
  \citenamefont {D'Avanzo}, \citenamefont {Santoliquido}, \citenamefont
  {Cella}, \citenamefont {Razzano},\ and\ \citenamefont
  {Cuoco}}]{Patricelli:2022hhr}%
  \BibitemOpen
  \bibfield  {author} {\bibinfo {author} {\bibfnamefont {B.}~\bibnamefont
  {Patricelli}}, \bibinfo {author} {\bibfnamefont {M.~G.}\ \bibnamefont
  {Bernardini}}, \bibinfo {author} {\bibfnamefont {M.}~\bibnamefont {Mapelli}},
  \bibinfo {author} {\bibfnamefont {P.}~\bibnamefont {D'Avanzo}}, \bibinfo
  {author} {\bibfnamefont {F.}~\bibnamefont {Santoliquido}}, \bibinfo {author}
  {\bibfnamefont {G.}~\bibnamefont {Cella}}, \bibinfo {author} {\bibfnamefont
  {M.}~\bibnamefont {Razzano}}, \ and\ \bibinfo {author} {\bibfnamefont
  {E.}~\bibnamefont {Cuoco}},\ }\href {\doibase 10.1093/mnras/stac1167}
  {\bibfield  {journal} {\bibinfo  {journal} {Mon. Not. Roy. Astron. Soc.}\
  }\textbf {\bibinfo {volume} {513}},\ \bibinfo {pages} {4159} (\bibinfo {year}
  {2022})},\ \bibinfo {note} {[Erratum: Mon.Not.Roy.Astron.Soc. 514, 3395
  (2022)]},\ \Eprint {http://arxiv.org/abs/2204.12504} {arXiv:2204.12504
  [astro-ph.HE]} \BibitemShut {NoStop}%
\bibitem [{\citenamefont {Colombo}\ \emph {et~al.}(2022)\citenamefont
  {Colombo}, \citenamefont {Salafia}, \citenamefont {Gabrielli}, \citenamefont
  {Ghirlanda}, \citenamefont {Giacomazzo}, \citenamefont {Perego},\ and\
  \citenamefont {Colpi}}]{Colombo:2022zzp}%
  \BibitemOpen
  \bibfield  {author} {\bibinfo {author} {\bibfnamefont {A.}~\bibnamefont
  {Colombo}}, \bibinfo {author} {\bibfnamefont {O.~S.}\ \bibnamefont
  {Salafia}}, \bibinfo {author} {\bibfnamefont {F.}~\bibnamefont {Gabrielli}},
  \bibinfo {author} {\bibfnamefont {G.}~\bibnamefont {Ghirlanda}}, \bibinfo
  {author} {\bibfnamefont {B.}~\bibnamefont {Giacomazzo}}, \bibinfo {author}
  {\bibfnamefont {A.}~\bibnamefont {Perego}}, \ and\ \bibinfo {author}
  {\bibfnamefont {M.}~\bibnamefont {Colpi}},\ }\href {\doibase
  10.3847/1538-4357/ac8d00} {\bibfield  {journal} {\bibinfo  {journal}
  {Astrophys. J.}\ }\textbf {\bibinfo {volume} {937}},\ \bibinfo {pages} {79}
  (\bibinfo {year} {2022})},\ \Eprint {http://arxiv.org/abs/2204.07592}
  {arXiv:2204.07592 [astro-ph.HE]} \BibitemShut {NoStop}%
\bibitem [{\citenamefont {Busza}\ \emph {et~al.}(2018)\citenamefont {Busza},
  \citenamefont {Rajagopal},\ and\ \citenamefont {van~der
  Schee}}]{Busza:2018rrf}%
  \BibitemOpen
  \bibfield  {author} {\bibinfo {author} {\bibfnamefont {W.}~\bibnamefont
  {Busza}}, \bibinfo {author} {\bibfnamefont {K.}~\bibnamefont {Rajagopal}}, \
  and\ \bibinfo {author} {\bibfnamefont {W.}~\bibnamefont {van~der Schee}},\
  }\href {\doibase 10.1146/annurev-nucl-101917-020852} {\bibfield  {journal}
  {\bibinfo  {journal} {Ann. Rev. Nucl. Part. Sci.}\ }\textbf {\bibinfo
  {volume} {68}},\ \bibinfo {pages} {339} (\bibinfo {year} {2018})},\ \Eprint
  {http://arxiv.org/abs/1802.04801} {arXiv:1802.04801 [hep-ph]} \BibitemShut
  {NoStop}%
\bibitem [{\citenamefont {An}\ \emph {et~al.}(2022)\citenamefont {An} \emph
  {et~al.}}]{An:2021wof}%
  \BibitemOpen
  \bibfield  {author} {\bibinfo {author} {\bibfnamefont {X.}~\bibnamefont {An}}
  \emph {et~al.},\ }\href {\doibase 10.1016/j.nuclphysa.2021.122343} {\bibfield
   {journal} {\bibinfo  {journal} {Nucl. Phys. A}\ }\textbf {\bibinfo {volume}
  {1017}},\ \bibinfo {pages} {122343} (\bibinfo {year} {2022})},\ \Eprint
  {http://arxiv.org/abs/2108.13867} {arXiv:2108.13867 [nucl-th]} \BibitemShut
  {NoStop}%
\bibitem [{\citenamefont {Epelbaum}\ \emph {et~al.}(2009)\citenamefont
  {Epelbaum}, \citenamefont {Hammer},\ and\ \citenamefont
  {Meissner}}]{Epelbaum:2008ga}%
  \BibitemOpen
  \bibfield  {author} {\bibinfo {author} {\bibfnamefont {E.}~\bibnamefont
  {Epelbaum}}, \bibinfo {author} {\bibfnamefont {H.-W.}\ \bibnamefont
  {Hammer}}, \ and\ \bibinfo {author} {\bibfnamefont {U.-G.}\ \bibnamefont
  {Meissner}},\ }\href {\doibase 10.1103/RevModPhys.81.1773} {\bibfield
  {journal} {\bibinfo  {journal} {Rev. Mod. Phys.}\ }\textbf {\bibinfo {volume}
  {81}},\ \bibinfo {pages} {1773} (\bibinfo {year} {2009})},\ \Eprint
  {http://arxiv.org/abs/0811.1338} {arXiv:0811.1338 [nucl-th]} \BibitemShut
  {NoStop}%
\bibitem [{\citenamefont {Machleidt}\ and\ \citenamefont
  {Entem}(2011)}]{Machleidt:2011zz}%
  \BibitemOpen
  \bibfield  {author} {\bibinfo {author} {\bibfnamefont {R.}~\bibnamefont
  {Machleidt}}\ and\ \bibinfo {author} {\bibfnamefont {D.~R.}\ \bibnamefont
  {Entem}},\ }\href {\doibase 10.1016/j.physrep.2011.02.001} {\bibfield
  {journal} {\bibinfo  {journal} {Phys. Rept.}\ }\textbf {\bibinfo {volume}
  {503}},\ \bibinfo {pages} {1} (\bibinfo {year} {2011})},\ \Eprint
  {http://arxiv.org/abs/1105.2919} {arXiv:1105.2919 [nucl-th]} \BibitemShut
  {NoStop}%
\bibitem [{\citenamefont {Hebeler}\ and\ \citenamefont
  {Schwenk}(2010)}]{Hebeler:2009iv}%
  \BibitemOpen
  \bibfield  {author} {\bibinfo {author} {\bibfnamefont {K.}~\bibnamefont
  {Hebeler}}\ and\ \bibinfo {author} {\bibfnamefont {A.}~\bibnamefont
  {Schwenk}},\ }\href {\doibase 10.1103/PhysRevC.82.014314} {\bibfield
  {journal} {\bibinfo  {journal} {Phys. Rev. C}\ }\textbf {\bibinfo {volume}
  {82}},\ \bibinfo {pages} {014314} (\bibinfo {year} {2010})},\ \Eprint
  {http://arxiv.org/abs/0911.0483} {arXiv:0911.0483 [nucl-th]} \BibitemShut
  {NoStop}%
\bibitem [{\citenamefont {Coraggio}\ \emph {et~al.}(2013)\citenamefont
  {Coraggio}, \citenamefont {Holt}, \citenamefont {Itaco}, \citenamefont
  {Machleidt},\ and\ \citenamefont {Sammarruca}}]{Coraggio:2012ca}%
  \BibitemOpen
  \bibfield  {author} {\bibinfo {author} {\bibfnamefont {L.}~\bibnamefont
  {Coraggio}}, \bibinfo {author} {\bibfnamefont {J.~W.}\ \bibnamefont {Holt}},
  \bibinfo {author} {\bibfnamefont {N.}~\bibnamefont {Itaco}}, \bibinfo
  {author} {\bibfnamefont {R.}~\bibnamefont {Machleidt}}, \ and\ \bibinfo
  {author} {\bibfnamefont {F.}~\bibnamefont {Sammarruca}},\ }\href {\doibase
  10.1103/PhysRevC.87.014322} {\bibfield  {journal} {\bibinfo  {journal} {Phys.
  Rev. C}\ }\textbf {\bibinfo {volume} {87}},\ \bibinfo {pages} {014322}
  (\bibinfo {year} {2013})},\ \Eprint {http://arxiv.org/abs/1209.5537}
  {arXiv:1209.5537 [nucl-th]} \BibitemShut {NoStop}%
\bibitem [{\citenamefont {Hagen}\ \emph {et~al.}(2014)\citenamefont {Hagen},
  \citenamefont {Papenbrock}, \citenamefont {Ekstr\"om}, \citenamefont {Wendt},
  \citenamefont {Baardsen}, \citenamefont {Gandolfi}, \citenamefont
  {Hjorth-Jensen},\ and\ \citenamefont {Horowitz}}]{Hagen:2013yba}%
  \BibitemOpen
  \bibfield  {author} {\bibinfo {author} {\bibfnamefont {G.}~\bibnamefont
  {Hagen}}, \bibinfo {author} {\bibfnamefont {T.}~\bibnamefont {Papenbrock}},
  \bibinfo {author} {\bibfnamefont {A.}~\bibnamefont {Ekstr\"om}}, \bibinfo
  {author} {\bibfnamefont {K.~A.}\ \bibnamefont {Wendt}}, \bibinfo {author}
  {\bibfnamefont {G.}~\bibnamefont {Baardsen}}, \bibinfo {author}
  {\bibfnamefont {S.}~\bibnamefont {Gandolfi}}, \bibinfo {author}
  {\bibfnamefont {M.}~\bibnamefont {Hjorth-Jensen}}, \ and\ \bibinfo {author}
  {\bibfnamefont {C.~J.}\ \bibnamefont {Horowitz}},\ }\href {\doibase
  10.1103/PhysRevC.89.014319} {\bibfield  {journal} {\bibinfo  {journal} {Phys.
  Rev. C}\ }\textbf {\bibinfo {volume} {89}},\ \bibinfo {pages} {014319}
  (\bibinfo {year} {2014})},\ \Eprint {http://arxiv.org/abs/1311.2925}
  {arXiv:1311.2925 [nucl-th]} \BibitemShut {NoStop}%
\bibitem [{\citenamefont {Carbone}\ \emph {et~al.}(2014)\citenamefont
  {Carbone}, \citenamefont {Rios},\ and\ \citenamefont
  {Polls}}]{Carbone:2014mja}%
  \BibitemOpen
  \bibfield  {author} {\bibinfo {author} {\bibfnamefont {A.}~\bibnamefont
  {Carbone}}, \bibinfo {author} {\bibfnamefont {A.}~\bibnamefont {Rios}}, \
  and\ \bibinfo {author} {\bibfnamefont {A.}~\bibnamefont {Polls}},\ }\href
  {\doibase 10.1103/PhysRevC.90.054322} {\bibfield  {journal} {\bibinfo
  {journal} {Phys. Rev. C}\ }\textbf {\bibinfo {volume} {90}},\ \bibinfo
  {pages} {054322} (\bibinfo {year} {2014})},\ \Eprint
  {http://arxiv.org/abs/1408.0717} {arXiv:1408.0717 [nucl-th]} \BibitemShut
  {NoStop}%
\bibitem [{\citenamefont {Lynn}\ \emph {et~al.}(2019)\citenamefont {Lynn},
  \citenamefont {Tews}, \citenamefont {Gandolfi},\ and\ \citenamefont
  {Lovato}}]{Lynn:2019rdt}%
  \BibitemOpen
  \bibfield  {author} {\bibinfo {author} {\bibfnamefont {J.~E.}\ \bibnamefont
  {Lynn}}, \bibinfo {author} {\bibfnamefont {I.}~\bibnamefont {Tews}}, \bibinfo
  {author} {\bibfnamefont {S.}~\bibnamefont {Gandolfi}}, \ and\ \bibinfo
  {author} {\bibfnamefont {A.}~\bibnamefont {Lovato}},\ }\href {\doibase
  10.1146/annurev-nucl-101918-023600} {\bibfield  {journal} {\bibinfo
  {journal} {Ann. Rev. Nucl. Part. Sci.}\ }\textbf {\bibinfo {volume} {69}},\
  \bibinfo {pages} {279} (\bibinfo {year} {2019})},\ \Eprint
  {http://arxiv.org/abs/1901.04868} {arXiv:1901.04868 [nucl-th]} \BibitemShut
  {NoStop}%
\bibitem [{\citenamefont {Piarulli}\ \emph {et~al.}(2020)\citenamefont
  {Piarulli}, \citenamefont {Bombaci}, \citenamefont {Logoteta}, \citenamefont
  {Lovato},\ and\ \citenamefont {Wiringa}}]{Piarulli:2019pfq}%
  \BibitemOpen
  \bibfield  {author} {\bibinfo {author} {\bibfnamefont {M.}~\bibnamefont
  {Piarulli}}, \bibinfo {author} {\bibfnamefont {I.}~\bibnamefont {Bombaci}},
  \bibinfo {author} {\bibfnamefont {D.}~\bibnamefont {Logoteta}}, \bibinfo
  {author} {\bibfnamefont {A.}~\bibnamefont {Lovato}}, \ and\ \bibinfo {author}
  {\bibfnamefont {R.~B.}\ \bibnamefont {Wiringa}},\ }\href {\doibase
  10.1103/PhysRevC.101.045801} {\bibfield  {journal} {\bibinfo  {journal}
  {Phys. Rev. C}\ }\textbf {\bibinfo {volume} {101}},\ \bibinfo {pages}
  {045801} (\bibinfo {year} {2020})},\ \Eprint
  {http://arxiv.org/abs/1908.04426} {arXiv:1908.04426 [nucl-th]} \BibitemShut
  {NoStop}%
\bibitem [{\citenamefont {Huth}\ \emph {et~al.}(2021)\citenamefont {Huth},
  \citenamefont {Wellenhofer},\ and\ \citenamefont {Schwenk}}]{Huth:2020ozf}%
  \BibitemOpen
  \bibfield  {author} {\bibinfo {author} {\bibfnamefont {S.}~\bibnamefont
  {Huth}}, \bibinfo {author} {\bibfnamefont {C.}~\bibnamefont {Wellenhofer}}, \
  and\ \bibinfo {author} {\bibfnamefont {A.}~\bibnamefont {Schwenk}},\ }\href
  {\doibase 10.1103/PhysRevC.103.025803} {\bibfield  {journal} {\bibinfo
  {journal} {Phys. Rev. C}\ }\textbf {\bibinfo {volume} {103}},\ \bibinfo
  {pages} {025803} (\bibinfo {year} {2021})},\ \Eprint
  {http://arxiv.org/abs/2009.08885} {arXiv:2009.08885 [nucl-th]} \BibitemShut
  {NoStop}%
\bibitem [{\citenamefont {Drischler}\ \emph
  {et~al.}(2021{\natexlab{a}})\citenamefont {Drischler}, \citenamefont {Holt},\
  and\ \citenamefont {Wellenhofer}}]{Drischler:2021kxf}%
  \BibitemOpen
  \bibfield  {author} {\bibinfo {author} {\bibfnamefont {C.}~\bibnamefont
  {Drischler}}, \bibinfo {author} {\bibfnamefont {J.~W.}\ \bibnamefont {Holt}},
  \ and\ \bibinfo {author} {\bibfnamefont {C.}~\bibnamefont {Wellenhofer}},\
  }\href {\doibase 10.1146/annurev-nucl-102419-041903} {\bibfield  {journal}
  {\bibinfo  {journal} {Ann. Rev. Nucl. Part. Sci.}\ }\textbf {\bibinfo
  {volume} {71}},\ \bibinfo {pages} {403} (\bibinfo {year}
  {2021}{\natexlab{a}})},\ \Eprint {http://arxiv.org/abs/2101.01709}
  {arXiv:2101.01709 [nucl-th]} \BibitemShut {NoStop}%
\bibitem [{\citenamefont {Lovato}\ \emph {et~al.}(2022)\citenamefont {Lovato},
  \citenamefont {Bombaci}, \citenamefont {Logoteta}, \citenamefont {Piarulli},\
  and\ \citenamefont {Wiringa}}]{Lovato:2022apd}%
  \BibitemOpen
  \bibfield  {author} {\bibinfo {author} {\bibfnamefont {A.}~\bibnamefont
  {Lovato}}, \bibinfo {author} {\bibfnamefont {I.}~\bibnamefont {Bombaci}},
  \bibinfo {author} {\bibfnamefont {D.}~\bibnamefont {Logoteta}}, \bibinfo
  {author} {\bibfnamefont {M.}~\bibnamefont {Piarulli}}, \ and\ \bibinfo
  {author} {\bibfnamefont {R.~B.}\ \bibnamefont {Wiringa}},\ }\href {\doibase
  10.1103/PhysRevC.105.055808} {\bibfield  {journal} {\bibinfo  {journal}
  {Phys. Rev. C}\ }\textbf {\bibinfo {volume} {105}},\ \bibinfo {pages}
  {055808} (\bibinfo {year} {2022})},\ \Eprint
  {http://arxiv.org/abs/2202.10293} {arXiv:2202.10293 [nucl-th]} \BibitemShut
  {NoStop}%
\bibitem [{\citenamefont {Furnstahl}\ \emph {et~al.}(2015)\citenamefont
  {Furnstahl}, \citenamefont {Phillips},\ and\ \citenamefont
  {Wesolowski}}]{Furnstahl:2014xsa}%
  \BibitemOpen
  \bibfield  {author} {\bibinfo {author} {\bibfnamefont {R.~J.}\ \bibnamefont
  {Furnstahl}}, \bibinfo {author} {\bibfnamefont {D.~R.}\ \bibnamefont
  {Phillips}}, \ and\ \bibinfo {author} {\bibfnamefont {S.}~\bibnamefont
  {Wesolowski}},\ }\href {\doibase 10.1088/0954-3899/42/3/034028} {\bibfield
  {journal} {\bibinfo  {journal} {J. Phys. G}\ }\textbf {\bibinfo {volume}
  {42}},\ \bibinfo {pages} {034028} (\bibinfo {year} {2015})},\ \Eprint
  {http://arxiv.org/abs/1407.0657} {arXiv:1407.0657 [nucl-th]} \BibitemShut
  {NoStop}%
\bibitem [{\citenamefont {Wesolowski}\ \emph {et~al.}(2016)\citenamefont
  {Wesolowski}, \citenamefont {Klco}, \citenamefont {Furnstahl}, \citenamefont
  {Phillips},\ and\ \citenamefont {Thapaliya}}]{Wesolowski:2015fqa}%
  \BibitemOpen
  \bibfield  {author} {\bibinfo {author} {\bibfnamefont {S.}~\bibnamefont
  {Wesolowski}}, \bibinfo {author} {\bibfnamefont {N.}~\bibnamefont {Klco}},
  \bibinfo {author} {\bibfnamefont {R.~J.}\ \bibnamefont {Furnstahl}}, \bibinfo
  {author} {\bibfnamefont {D.~R.}\ \bibnamefont {Phillips}}, \ and\ \bibinfo
  {author} {\bibfnamefont {A.}~\bibnamefont {Thapaliya}},\ }\href {\doibase
  10.1088/0954-3899/43/7/074001} {\bibfield  {journal} {\bibinfo  {journal} {J.
  Phys. G}\ }\textbf {\bibinfo {volume} {43}},\ \bibinfo {pages} {074001}
  (\bibinfo {year} {2016})},\ \Eprint {http://arxiv.org/abs/1511.03618}
  {arXiv:1511.03618 [nucl-th]} \BibitemShut {NoStop}%
\bibitem [{\citenamefont {Wesolowski}\ \emph {et~al.}(2019)\citenamefont
  {Wesolowski}, \citenamefont {Furnstahl}, \citenamefont {Melendez},\ and\
  \citenamefont {Phillips}}]{Wesolowski:2018lzj}%
  \BibitemOpen
  \bibfield  {author} {\bibinfo {author} {\bibfnamefont {S.}~\bibnamefont
  {Wesolowski}}, \bibinfo {author} {\bibfnamefont {R.~J.}\ \bibnamefont
  {Furnstahl}}, \bibinfo {author} {\bibfnamefont {J.~A.}\ \bibnamefont
  {Melendez}}, \ and\ \bibinfo {author} {\bibfnamefont {D.~R.}\ \bibnamefont
  {Phillips}},\ }\href {\doibase 10.1088/1361-6471/aaf5fc} {\bibfield
  {journal} {\bibinfo  {journal} {J. Phys. G}\ }\textbf {\bibinfo {volume}
  {46}},\ \bibinfo {pages} {045102} (\bibinfo {year} {2019})},\ \Eprint
  {http://arxiv.org/abs/1808.08211} {arXiv:1808.08211 [nucl-th]} \BibitemShut
  {NoStop}%
\bibitem [{\citenamefont {Lim}\ and\ \citenamefont {Holt}(2018)}]{Lim:2018bkq}%
  \BibitemOpen
  \bibfield  {author} {\bibinfo {author} {\bibfnamefont {Y.}~\bibnamefont
  {Lim}}\ and\ \bibinfo {author} {\bibfnamefont {J.~W.}\ \bibnamefont {Holt}},\
  }\href {\doibase 10.1103/PhysRevLett.121.062701} {\bibfield  {journal}
  {\bibinfo  {journal} {Phys. Rev. Lett.}\ }\textbf {\bibinfo {volume} {121}},\
  \bibinfo {pages} {062701} (\bibinfo {year} {2018})},\ \Eprint
  {http://arxiv.org/abs/1803.02803} {arXiv:1803.02803 [nucl-th]} \BibitemShut
  {NoStop}%
\bibitem [{\citenamefont {Melendez}\ \emph {et~al.}(2019)\citenamefont
  {Melendez}, \citenamefont {Furnstahl}, \citenamefont {Phillips},
  \citenamefont {Pratola},\ and\ \citenamefont
  {Wesolowski}}]{Melendez:2019izc}%
  \BibitemOpen
  \bibfield  {author} {\bibinfo {author} {\bibfnamefont {J.~A.}\ \bibnamefont
  {Melendez}}, \bibinfo {author} {\bibfnamefont {R.~J.}\ \bibnamefont
  {Furnstahl}}, \bibinfo {author} {\bibfnamefont {D.~R.}\ \bibnamefont
  {Phillips}}, \bibinfo {author} {\bibfnamefont {M.~T.}\ \bibnamefont
  {Pratola}}, \ and\ \bibinfo {author} {\bibfnamefont {S.}~\bibnamefont
  {Wesolowski}},\ }\href {\doibase 10.1103/PhysRevC.100.044001} {\bibfield
  {journal} {\bibinfo  {journal} {Phys. Rev. C}\ }\textbf {\bibinfo {volume}
  {100}},\ \bibinfo {pages} {044001} (\bibinfo {year} {2019})},\ \Eprint
  {http://arxiv.org/abs/1904.10581} {arXiv:1904.10581 [nucl-th]} \BibitemShut
  {NoStop}%
\bibitem [{\citenamefont {Drischler}\ \emph
  {et~al.}(2020{\natexlab{a}})\citenamefont {Drischler}, \citenamefont
  {Furnstahl}, \citenamefont {Melendez},\ and\ \citenamefont
  {Phillips}}]{Drischler:2020hwi}%
  \BibitemOpen
  \bibfield  {author} {\bibinfo {author} {\bibfnamefont {C.}~\bibnamefont
  {Drischler}}, \bibinfo {author} {\bibfnamefont {R.~J.}\ \bibnamefont
  {Furnstahl}}, \bibinfo {author} {\bibfnamefont {J.~A.}\ \bibnamefont
  {Melendez}}, \ and\ \bibinfo {author} {\bibfnamefont {D.~R.}\ \bibnamefont
  {Phillips}},\ }\href {\doibase 10.1103/PhysRevLett.125.202702} {\bibfield
  {journal} {\bibinfo  {journal} {Phys. Rev. Lett.}\ }\textbf {\bibinfo
  {volume} {125}},\ \bibinfo {pages} {202702} (\bibinfo {year}
  {2020}{\natexlab{a}})},\ \Eprint {http://arxiv.org/abs/2004.07232}
  {arXiv:2004.07232 [nucl-th]} \BibitemShut {NoStop}%
\bibitem [{\citenamefont {Drischler}\ \emph
  {et~al.}(2020{\natexlab{b}})\citenamefont {Drischler}, \citenamefont
  {Melendez}, \citenamefont {Furnstahl},\ and\ \citenamefont
  {Phillips}}]{Drischler:2020yad}%
  \BibitemOpen
  \bibfield  {author} {\bibinfo {author} {\bibfnamefont {C.}~\bibnamefont
  {Drischler}}, \bibinfo {author} {\bibfnamefont {J.~A.}\ \bibnamefont
  {Melendez}}, \bibinfo {author} {\bibfnamefont {R.~J.}\ \bibnamefont
  {Furnstahl}}, \ and\ \bibinfo {author} {\bibfnamefont {D.~R.}\ \bibnamefont
  {Phillips}},\ }\href {\doibase 10.1103/PhysRevC.102.054315} {\bibfield
  {journal} {\bibinfo  {journal} {Phys. Rev. C}\ }\textbf {\bibinfo {volume}
  {102}},\ \bibinfo {pages} {054315} (\bibinfo {year} {2020}{\natexlab{b}})},\
  \Eprint {http://arxiv.org/abs/2004.07805} {arXiv:2004.07805 [nucl-th]}
  \BibitemShut {NoStop}%
\bibitem [{\citenamefont {Elhatisari}\ \emph {et~al.}(2022)\citenamefont
  {Elhatisari} \emph {et~al.}}]{Elhatisari:2022qfr}%
  \BibitemOpen
  \bibfield  {author} {\bibinfo {author} {\bibfnamefont {S.}~\bibnamefont
  {Elhatisari}} \emph {et~al.},\ }\href@noop {} {\  (\bibinfo {year} {2022})},\
  \Eprint {http://arxiv.org/abs/2210.17488} {arXiv:2210.17488 [nucl-th]}
  \BibitemShut {NoStop}%
\bibitem [{\citenamefont {Phillips}\ \emph {et~al.}(2021)\citenamefont
  {Phillips} \emph {et~al.}}]{Phillips:2020dmw}%
  \BibitemOpen
  \bibfield  {author} {\bibinfo {author} {\bibfnamefont {D.~R.}\ \bibnamefont
  {Phillips}} \emph {et~al.},\ }\href {\doibase 10.1088/1361-6471/abf1df}
  {\bibfield  {journal} {\bibinfo  {journal} {J. Phys. G}\ }\textbf {\bibinfo
  {volume} {48}},\ \bibinfo {pages} {072001} (\bibinfo {year} {2021})},\
  \Eprint {http://arxiv.org/abs/2012.07704} {arXiv:2012.07704 [nucl-th]}
  \BibitemShut {NoStop}%
\bibitem [{\citenamefont {Tews}\ \emph {et~al.}(2018)\citenamefont {Tews},
  \citenamefont {Carlson}, \citenamefont {Gandolfi},\ and\ \citenamefont
  {Reddy}}]{Tews:2018kmu}%
  \BibitemOpen
  \bibfield  {author} {\bibinfo {author} {\bibfnamefont {I.}~\bibnamefont
  {Tews}}, \bibinfo {author} {\bibfnamefont {J.}~\bibnamefont {Carlson}},
  \bibinfo {author} {\bibfnamefont {S.}~\bibnamefont {Gandolfi}}, \ and\
  \bibinfo {author} {\bibfnamefont {S.}~\bibnamefont {Reddy}},\ }\href
  {\doibase 10.3847/1538-4357/aac267} {\bibfield  {journal} {\bibinfo
  {journal} {Astrophys. J.}\ }\textbf {\bibinfo {volume} {860}},\ \bibinfo
  {pages} {149} (\bibinfo {year} {2018})},\ \Eprint
  {http://arxiv.org/abs/1801.01923} {arXiv:1801.01923 [nucl-th]} \BibitemShut
  {NoStop}%
\bibitem [{\citenamefont {Wellenhofer}\ \emph {et~al.}(2015)\citenamefont
  {Wellenhofer}, \citenamefont {Holt},\ and\ \citenamefont
  {Kaiser}}]{Wellenhofer:2015qba}%
  \BibitemOpen
  \bibfield  {author} {\bibinfo {author} {\bibfnamefont {C.}~\bibnamefont
  {Wellenhofer}}, \bibinfo {author} {\bibfnamefont {J.~W.}\ \bibnamefont
  {Holt}}, \ and\ \bibinfo {author} {\bibfnamefont {N.}~\bibnamefont
  {Kaiser}},\ }\href {\doibase 10.1103/PhysRevC.92.015801} {\bibfield
  {journal} {\bibinfo  {journal} {Phys. Rev. C}\ }\textbf {\bibinfo {volume}
  {92}},\ \bibinfo {pages} {015801} (\bibinfo {year} {2015})},\ \Eprint
  {http://arxiv.org/abs/1504.00177} {arXiv:1504.00177 [nucl-th]} \BibitemShut
  {NoStop}%
\bibitem [{\citenamefont {Carbone}\ and\ \citenamefont
  {Schwenk}(2019)}]{Carbone:2019pkr}%
  \BibitemOpen
  \bibfield  {author} {\bibinfo {author} {\bibfnamefont {A.}~\bibnamefont
  {Carbone}}\ and\ \bibinfo {author} {\bibfnamefont {A.}~\bibnamefont
  {Schwenk}},\ }\href {\doibase 10.1103/PhysRevC.100.025805} {\bibfield
  {journal} {\bibinfo  {journal} {Phys. Rev. C}\ }\textbf {\bibinfo {volume}
  {100}},\ \bibinfo {pages} {025805} (\bibinfo {year} {2019})},\ \Eprint
  {http://arxiv.org/abs/1904.00924} {arXiv:1904.00924 [nucl-th]} \BibitemShut
  {NoStop}%
\bibitem [{\citenamefont {Keller}\ \emph {et~al.}(2021)\citenamefont {Keller},
  \citenamefont {Wellenhofer}, \citenamefont {Hebeler},\ and\ \citenamefont
  {Schwenk}}]{Keller:2020qhx}%
  \BibitemOpen
  \bibfield  {author} {\bibinfo {author} {\bibfnamefont {J.}~\bibnamefont
  {Keller}}, \bibinfo {author} {\bibfnamefont {C.}~\bibnamefont {Wellenhofer}},
  \bibinfo {author} {\bibfnamefont {K.}~\bibnamefont {Hebeler}}, \ and\
  \bibinfo {author} {\bibfnamefont {A.}~\bibnamefont {Schwenk}},\ }\href
  {\doibase 10.1103/PhysRevC.103.055806} {\bibfield  {journal} {\bibinfo
  {journal} {Phys. Rev. C}\ }\textbf {\bibinfo {volume} {103}},\ \bibinfo
  {pages} {055806} (\bibinfo {year} {2021})},\ \Eprint
  {http://arxiv.org/abs/2011.05855} {arXiv:2011.05855 [nucl-th]} \BibitemShut
  {NoStop}%
\bibitem [{\citenamefont {Reed}\ \emph {et~al.}(2021)\citenamefont {Reed},
  \citenamefont {Fattoyev}, \citenamefont {Horowitz},\ and\ \citenamefont
  {Piekarewicz}}]{Reed:2021nqk}%
  \BibitemOpen
  \bibfield  {author} {\bibinfo {author} {\bibfnamefont {B.~T.}\ \bibnamefont
  {Reed}}, \bibinfo {author} {\bibfnamefont {F.~J.}\ \bibnamefont {Fattoyev}},
  \bibinfo {author} {\bibfnamefont {C.~J.}\ \bibnamefont {Horowitz}}, \ and\
  \bibinfo {author} {\bibfnamefont {J.}~\bibnamefont {Piekarewicz}},\ }\href
  {\doibase 10.1103/PhysRevLett.126.172503} {\bibfield  {journal} {\bibinfo
  {journal} {Phys. Rev. Lett.}\ }\textbf {\bibinfo {volume} {126}},\ \bibinfo
  {pages} {172503} (\bibinfo {year} {2021})},\ \Eprint
  {http://arxiv.org/abs/2101.03193} {arXiv:2101.03193 [nucl-th]} \BibitemShut
  {NoStop}%
\bibitem [{\citenamefont {Hu}\ \emph {et~al.}(2022)\citenamefont {Hu} \emph
  {et~al.}}]{Hu:2021trw}%
  \BibitemOpen
  \bibfield  {author} {\bibinfo {author} {\bibfnamefont {B.}~\bibnamefont {Hu}}
  \emph {et~al.},\ }\href {\doibase 10.1038/s41567-022-01715-8} {\bibfield
  {journal} {\bibinfo  {journal} {Nature Phys.}\ }\textbf {\bibinfo {volume}
  {18}},\ \bibinfo {pages} {1196} (\bibinfo {year} {2022})},\ \Eprint
  {http://arxiv.org/abs/2112.01125} {arXiv:2112.01125 [nucl-th]} \BibitemShut
  {NoStop}%
\bibitem [{\citenamefont {Essick}\ \emph
  {et~al.}(2021{\natexlab{a}})\citenamefont {Essick}, \citenamefont {Landry},
  \citenamefont {Schwenk},\ and\ \citenamefont {Tews}}]{Essick:2021ezp}%
  \BibitemOpen
  \bibfield  {author} {\bibinfo {author} {\bibfnamefont {R.}~\bibnamefont
  {Essick}}, \bibinfo {author} {\bibfnamefont {P.}~\bibnamefont {Landry}},
  \bibinfo {author} {\bibfnamefont {A.}~\bibnamefont {Schwenk}}, \ and\
  \bibinfo {author} {\bibfnamefont {I.}~\bibnamefont {Tews}},\ }\href {\doibase
  10.1103/PhysRevC.104.065804} {\bibfield  {journal} {\bibinfo  {journal}
  {Phys. Rev. C}\ }\textbf {\bibinfo {volume} {104}},\ \bibinfo {pages}
  {065804} (\bibinfo {year} {2021}{\natexlab{a}})},\ \Eprint
  {http://arxiv.org/abs/2107.05528} {arXiv:2107.05528 [nucl-th]} \BibitemShut
  {NoStop}%
\bibitem [{\citenamefont {Hagen}\ \emph {et~al.}(2015)\citenamefont {Hagen}
  \emph {et~al.}}]{Hagen:2015yea}%
  \BibitemOpen
  \bibfield  {author} {\bibinfo {author} {\bibfnamefont {G.}~\bibnamefont
  {Hagen}} \emph {et~al.},\ }\href {\doibase 10.1038/nphys3529} {\bibfield
  {journal} {\bibinfo  {journal} {Nature Phys.}\ }\textbf {\bibinfo {volume}
  {12}},\ \bibinfo {pages} {186} (\bibinfo {year} {2015})},\ \Eprint
  {http://arxiv.org/abs/1509.07169} {arXiv:1509.07169 [nucl-th]} \BibitemShut
  {NoStop}%
\bibitem [{\citenamefont {Drischler}\ \emph
  {et~al.}(2021{\natexlab{b}})\citenamefont {Drischler}, \citenamefont
  {Haxton}, \citenamefont {McElvain}, \citenamefont {Mereghetti}, \citenamefont
  {Nicholson}, \citenamefont {Vranas},\ and\ \citenamefont
  {Walker-Loud}}]{Drischler:2019xuo}%
  \BibitemOpen
  \bibfield  {author} {\bibinfo {author} {\bibfnamefont {C.}~\bibnamefont
  {Drischler}}, \bibinfo {author} {\bibfnamefont {W.}~\bibnamefont {Haxton}},
  \bibinfo {author} {\bibfnamefont {K.}~\bibnamefont {McElvain}}, \bibinfo
  {author} {\bibfnamefont {E.}~\bibnamefont {Mereghetti}}, \bibinfo {author}
  {\bibfnamefont {A.}~\bibnamefont {Nicholson}}, \bibinfo {author}
  {\bibfnamefont {P.}~\bibnamefont {Vranas}}, \ and\ \bibinfo {author}
  {\bibfnamefont {A.}~\bibnamefont {Walker-Loud}},\ }\href {\doibase
  10.1016/j.ppnp.2021.103888} {\bibfield  {journal} {\bibinfo  {journal} {Prog.
  Part. Nucl. Phys.}\ }\textbf {\bibinfo {volume} {121}},\ \bibinfo {pages}
  {103888} (\bibinfo {year} {2021}{\natexlab{b}})},\ \Eprint
  {http://arxiv.org/abs/1910.07961} {arXiv:1910.07961 [nucl-th]} \BibitemShut
  {NoStop}%
\bibitem [{\citenamefont {Shen}\ and\ \citenamefont
  {Schenke}(2022)}]{Shen:2022oyg}%
  \BibitemOpen
  \bibfield  {author} {\bibinfo {author} {\bibfnamefont {C.}~\bibnamefont
  {Shen}}\ and\ \bibinfo {author} {\bibfnamefont {B.}~\bibnamefont {Schenke}},\
  }\href {\doibase 10.1103/PhysRevC.105.064905} {\bibfield  {journal} {\bibinfo
   {journal} {Phys. Rev. C}\ }\textbf {\bibinfo {volume} {105}},\ \bibinfo
  {pages} {064905} (\bibinfo {year} {2022})},\ \Eprint
  {http://arxiv.org/abs/2203.04685} {arXiv:2203.04685 [nucl-th]} \BibitemShut
  {NoStop}%
\bibitem [{\citenamefont {Most}\ \emph {et~al.}(2019)\citenamefont {Most},
  \citenamefont {Papenfort}, \citenamefont {Dexheimer}, \citenamefont
  {Hanauske}, \citenamefont {Schramm}, \citenamefont {St\"ocker},\ and\
  \citenamefont {Rezzolla}}]{Most:2018eaw}%
  \BibitemOpen
  \bibfield  {author} {\bibinfo {author} {\bibfnamefont {E.~R.}\ \bibnamefont
  {Most}}, \bibinfo {author} {\bibfnamefont {L.~J.}\ \bibnamefont {Papenfort}},
  \bibinfo {author} {\bibfnamefont {V.}~\bibnamefont {Dexheimer}}, \bibinfo
  {author} {\bibfnamefont {M.}~\bibnamefont {Hanauske}}, \bibinfo {author}
  {\bibfnamefont {S.}~\bibnamefont {Schramm}}, \bibinfo {author} {\bibfnamefont
  {H.}~\bibnamefont {St\"ocker}}, \ and\ \bibinfo {author} {\bibfnamefont
  {L.}~\bibnamefont {Rezzolla}},\ }\href {\doibase
  10.1103/PhysRevLett.122.061101} {\bibfield  {journal} {\bibinfo  {journal}
  {Phys. Rev. Lett.}\ }\textbf {\bibinfo {volume} {122}},\ \bibinfo {pages}
  {061101} (\bibinfo {year} {2019})},\ \Eprint
  {http://arxiv.org/abs/1807.03684} {arXiv:1807.03684 [astro-ph.HE]}
  \BibitemShut {NoStop}%
\bibitem [{\citenamefont {Alba}\ \emph {et~al.}(2020)\citenamefont {Alba},
  \citenamefont {Sarti}, \citenamefont {Noronha-Hostler}, \citenamefont
  {Parotto}, \citenamefont {Portillo-Vazquez}, \citenamefont {Ratti},\ and\
  \citenamefont {Stafford}}]{Alba:2020jir}%
  \BibitemOpen
  \bibfield  {author} {\bibinfo {author} {\bibfnamefont {P.}~\bibnamefont
  {Alba}}, \bibinfo {author} {\bibfnamefont {V.~M.}\ \bibnamefont {Sarti}},
  \bibinfo {author} {\bibfnamefont {J.}~\bibnamefont {Noronha-Hostler}},
  \bibinfo {author} {\bibfnamefont {P.}~\bibnamefont {Parotto}}, \bibinfo
  {author} {\bibfnamefont {I.}~\bibnamefont {Portillo-Vazquez}}, \bibinfo
  {author} {\bibfnamefont {C.}~\bibnamefont {Ratti}}, \ and\ \bibinfo {author}
  {\bibfnamefont {J.~M.}\ \bibnamefont {Stafford}},\ }\href {\doibase
  10.1103/PhysRevC.101.054905} {\bibfield  {journal} {\bibinfo  {journal}
  {Phys. Rev. C}\ }\textbf {\bibinfo {volume} {101}},\ \bibinfo {pages}
  {054905} (\bibinfo {year} {2020})},\ \Eprint
  {http://arxiv.org/abs/2002.12395} {arXiv:2002.12395 [hep-ph]} \BibitemShut
  {NoStop}%
\bibitem [{\citenamefont {Arnaldi}\ \emph {et~al.}(2009)\citenamefont {Arnaldi}
  \emph {et~al.}}]{NA60:2008dcb}%
  \BibitemOpen
  \bibfield  {author} {\bibinfo {author} {\bibfnamefont {R.}~\bibnamefont
  {Arnaldi}} \emph {et~al.} (\bibinfo {collaboration} {NA60}),\ }\href
  {\doibase 10.1140/epjc/s10052-008-0857-2} {\bibfield  {journal} {\bibinfo
  {journal} {Eur. Phys. J. C}\ }\textbf {\bibinfo {volume} {59}},\ \bibinfo
  {pages} {607} (\bibinfo {year} {2009})},\ \Eprint
  {http://arxiv.org/abs/0810.3204} {arXiv:0810.3204 [nucl-ex]} \BibitemShut
  {NoStop}%
\bibitem [{\citenamefont {Harabasz}\ \emph {et~al.}(2020)\citenamefont
  {Harabasz}, \citenamefont {Florkowski}, \citenamefont {Galatyuk},
  \citenamefont {Ma~Lgorzata~Gumberidze}, \citenamefont {Ryblewski},
  \citenamefont {Salabura},\ and\ \citenamefont {Stroth}}]{Harabasz:2020sei}%
  \BibitemOpen
  \bibfield  {author} {\bibinfo {author} {\bibfnamefont {S.}~\bibnamefont
  {Harabasz}}, \bibinfo {author} {\bibfnamefont {W.}~\bibnamefont
  {Florkowski}}, \bibinfo {author} {\bibfnamefont {T.}~\bibnamefont
  {Galatyuk}}, \bibinfo {author} {\bibfnamefont {t.}~\bibnamefont
  {Ma~Lgorzata~Gumberidze}}, \bibinfo {author} {\bibfnamefont {R.}~\bibnamefont
  {Ryblewski}}, \bibinfo {author} {\bibfnamefont {P.}~\bibnamefont {Salabura}},
  \ and\ \bibinfo {author} {\bibfnamefont {J.}~\bibnamefont {Stroth}},\ }\href
  {\doibase 10.1103/PhysRevC.102.054903} {\bibfield  {journal} {\bibinfo
  {journal} {Phys. Rev. C}\ }\textbf {\bibinfo {volume} {102}},\ \bibinfo
  {pages} {054903} (\bibinfo {year} {2020})},\ \Eprint
  {http://arxiv.org/abs/2003.12992} {arXiv:2003.12992 [nucl-th]} \BibitemShut
  {NoStop}%
\bibitem [{\citenamefont {Motornenko}\ \emph {et~al.}(2021)\citenamefont
  {Motornenko}, \citenamefont {Steinheimer}, \citenamefont {Vovchenko},
  \citenamefont {Stock},\ and\ \citenamefont {Stoecker}}]{Motornenko:2021nds}%
  \BibitemOpen
  \bibfield  {author} {\bibinfo {author} {\bibfnamefont {A.}~\bibnamefont
  {Motornenko}}, \bibinfo {author} {\bibfnamefont {J.}~\bibnamefont
  {Steinheimer}}, \bibinfo {author} {\bibfnamefont {V.}~\bibnamefont
  {Vovchenko}}, \bibinfo {author} {\bibfnamefont {R.}~\bibnamefont {Stock}}, \
  and\ \bibinfo {author} {\bibfnamefont {H.}~\bibnamefont {Stoecker}},\ }\href
  {\doibase 10.1016/j.physletb.2021.136703} {\bibfield  {journal} {\bibinfo
  {journal} {Phys. Lett. B}\ }\textbf {\bibinfo {volume} {822}},\ \bibinfo
  {pages} {136703} (\bibinfo {year} {2021})},\ \Eprint
  {http://arxiv.org/abs/2104.06036} {arXiv:2104.06036 [hep-ph]} \BibitemShut
  {NoStop}%
\bibitem [{\citenamefont {Elliott}\ \emph {et~al.}(2013)\citenamefont
  {Elliott}, \citenamefont {Lake}, \citenamefont {Moretto},\ and\ \citenamefont
  {Phair}}]{Elliott:2013pna}%
  \BibitemOpen
  \bibfield  {author} {\bibinfo {author} {\bibfnamefont {J.~B.}\ \bibnamefont
  {Elliott}}, \bibinfo {author} {\bibfnamefont {P.~T.}\ \bibnamefont {Lake}},
  \bibinfo {author} {\bibfnamefont {L.~G.}\ \bibnamefont {Moretto}}, \ and\
  \bibinfo {author} {\bibfnamefont {L.}~\bibnamefont {Phair}},\ }\href
  {\doibase 10.1103/PhysRevC.87.054622} {\bibfield  {journal} {\bibinfo
  {journal} {Phys. Rev. C}\ }\textbf {\bibinfo {volume} {87}},\ \bibinfo
  {pages} {054622} (\bibinfo {year} {2013})}\BibitemShut {NoStop}%
\bibitem [{\citenamefont {Vovchenko}\ \emph {et~al.}(2017)\citenamefont
  {Vovchenko}, \citenamefont {Gorenstein},\ and\ \citenamefont
  {Stoecker}}]{Vovchenko:2016rkn}%
  \BibitemOpen
  \bibfield  {author} {\bibinfo {author} {\bibfnamefont {V.}~\bibnamefont
  {Vovchenko}}, \bibinfo {author} {\bibfnamefont {M.~I.}\ \bibnamefont
  {Gorenstein}}, \ and\ \bibinfo {author} {\bibfnamefont {H.}~\bibnamefont
  {Stoecker}},\ }\href {\doibase 10.1103/PhysRevLett.118.182301} {\bibfield
  {journal} {\bibinfo  {journal} {Phys. Rev. Lett.}\ }\textbf {\bibinfo
  {volume} {118}},\ \bibinfo {pages} {182301} (\bibinfo {year} {2017})},\
  \Eprint {http://arxiv.org/abs/1609.03975} {arXiv:1609.03975 [hep-ph]}
  \BibitemShut {NoStop}%
\bibitem [{\citenamefont {Borsanyi}\ \emph {et~al.}(2020)\citenamefont
  {Borsanyi}, \citenamefont {Fodor}, \citenamefont {Guenther}, \citenamefont
  {Kara}, \citenamefont {Katz}, \citenamefont {Parotto}, \citenamefont
  {Pasztor}, \citenamefont {Ratti},\ and\ \citenamefont
  {Szabo}}]{Borsanyi:2020fev}%
  \BibitemOpen
  \bibfield  {author} {\bibinfo {author} {\bibfnamefont {S.}~\bibnamefont
  {Borsanyi}}, \bibinfo {author} {\bibfnamefont {Z.}~\bibnamefont {Fodor}},
  \bibinfo {author} {\bibfnamefont {J.~N.}\ \bibnamefont {Guenther}}, \bibinfo
  {author} {\bibfnamefont {R.}~\bibnamefont {Kara}}, \bibinfo {author}
  {\bibfnamefont {S.~D.}\ \bibnamefont {Katz}}, \bibinfo {author}
  {\bibfnamefont {P.}~\bibnamefont {Parotto}}, \bibinfo {author} {\bibfnamefont
  {A.}~\bibnamefont {Pasztor}}, \bibinfo {author} {\bibfnamefont
  {C.}~\bibnamefont {Ratti}}, \ and\ \bibinfo {author} {\bibfnamefont {K.~K.}\
  \bibnamefont {Szabo}},\ }\href {\doibase 10.1103/PhysRevLett.125.052001}
  {\bibfield  {journal} {\bibinfo  {journal} {Phys. Rev. Lett.}\ }\textbf
  {\bibinfo {volume} {125}},\ \bibinfo {pages} {052001} (\bibinfo {year}
  {2020})},\ \Eprint {http://arxiv.org/abs/2002.02821} {arXiv:2002.02821
  [hep-lat]} \BibitemShut {NoStop}%
\bibitem [{\citenamefont {Aprahamian}\ \emph {et~al.}(2015)\citenamefont
  {Aprahamian} \emph {et~al.}}]{Aprahamian:2015qub}%
  \BibitemOpen
  \bibfield  {author} {\bibinfo {author} {\bibfnamefont {A.}~\bibnamefont
  {Aprahamian}} \emph {et~al.},\ }\href@noop {} {\  (\bibinfo {year}
  {2015})}\BibitemShut {NoStop}%
\bibitem [{\citenamefont {Almaalol}\ \emph
  {et~al.}(2022{\natexlab{a}})\citenamefont {Almaalol} \emph
  {et~al.}}]{Almaalol:2022xwv}%
  \BibitemOpen
  \bibfield  {author} {\bibinfo {author} {\bibfnamefont {D.}~\bibnamefont
  {Almaalol}} \emph {et~al.},\ }\href@noop {} {\  (\bibinfo {year}
  {2022}{\natexlab{a}})},\ \Eprint {http://arxiv.org/abs/2209.05009}
  {arXiv:2209.05009 [nucl-ex]} \BibitemShut {NoStop}%
\bibitem [{\citenamefont {de~Forcrand}(2009)}]{deForcrand:2009zkb}%
  \BibitemOpen
  \bibfield  {author} {\bibinfo {author} {\bibfnamefont {P.}~\bibnamefont
  {de~Forcrand}},\ }\href {\doibase 10.22323/1.091.0010} {\bibfield  {journal}
  {\bibinfo  {journal} {PoS}\ }\textbf {\bibinfo {volume} {LAT2009}},\ \bibinfo
  {pages} {010} (\bibinfo {year} {2009})},\ \Eprint
  {http://arxiv.org/abs/1005.0539} {arXiv:1005.0539 [hep-lat]} \BibitemShut
  {NoStop}%
\bibitem [{\citenamefont {Damgaard}\ and\ \citenamefont
  {Huffel}(1987)}]{Damgaard:1987rr}%
  \BibitemOpen
  \bibfield  {author} {\bibinfo {author} {\bibfnamefont {P.~H.}\ \bibnamefont
  {Damgaard}}\ and\ \bibinfo {author} {\bibfnamefont {H.}~\bibnamefont
  {Huffel}},\ }\href {\doibase 10.1016/0370-1573(87)90144-X} {\bibfield
  {journal} {\bibinfo  {journal} {Phys. Rept.}\ }\textbf {\bibinfo {volume}
  {152}},\ \bibinfo {pages} {227} (\bibinfo {year} {1987})}\BibitemShut
  {NoStop}%
\bibitem [{\citenamefont {Alexandru}\ \emph {et~al.}(2022)\citenamefont
  {Alexandru}, \citenamefont {Basar}, \citenamefont {Bedaque},\ and\
  \citenamefont {Warrington}}]{Alexandru:2020wrj}%
  \BibitemOpen
  \bibfield  {author} {\bibinfo {author} {\bibfnamefont {A.}~\bibnamefont
  {Alexandru}}, \bibinfo {author} {\bibfnamefont {G.}~\bibnamefont {Basar}},
  \bibinfo {author} {\bibfnamefont {P.~F.}\ \bibnamefont {Bedaque}}, \ and\
  \bibinfo {author} {\bibfnamefont {N.~C.}\ \bibnamefont {Warrington}},\ }\href
  {\doibase 10.1103/RevModPhys.94.015006} {\bibfield  {journal} {\bibinfo
  {journal} {Rev. Mod. Phys.}\ }\textbf {\bibinfo {volume} {94}},\ \bibinfo
  {pages} {015006} (\bibinfo {year} {2022})},\ \Eprint
  {http://arxiv.org/abs/2007.05436} {arXiv:2007.05436 [hep-lat]} \BibitemShut
  {NoStop}%
\bibitem [{\citenamefont {Fischer}(2019)}]{Fischer:2018sdj}%
  \BibitemOpen
  \bibfield  {author} {\bibinfo {author} {\bibfnamefont {C.~S.}\ \bibnamefont
  {Fischer}},\ }\href {\doibase 10.1016/j.ppnp.2019.01.002} {\bibfield
  {journal} {\bibinfo  {journal} {Prog. Part. Nucl. Phys.}\ }\textbf {\bibinfo
  {volume} {105}},\ \bibinfo {pages} {1} (\bibinfo {year} {2019})},\ \Eprint
  {http://arxiv.org/abs/1810.12938} {arXiv:1810.12938 [hep-ph]} \BibitemShut
  {NoStop}%
\bibitem [{\citenamefont {Fu}\ \emph {et~al.}(2020)\citenamefont {Fu},
  \citenamefont {Pawlowski},\ and\ \citenamefont {Rennecke}}]{Fu:2019hdw}%
  \BibitemOpen
  \bibfield  {author} {\bibinfo {author} {\bibfnamefont {W.-j.}\ \bibnamefont
  {Fu}}, \bibinfo {author} {\bibfnamefont {J.~M.}\ \bibnamefont {Pawlowski}}, \
  and\ \bibinfo {author} {\bibfnamefont {F.}~\bibnamefont {Rennecke}},\ }\href
  {\doibase 10.1103/PhysRevD.101.054032} {\bibfield  {journal} {\bibinfo
  {journal} {Phys. Rev. D}\ }\textbf {\bibinfo {volume} {101}},\ \bibinfo
  {pages} {054032} (\bibinfo {year} {2020})},\ \Eprint
  {http://arxiv.org/abs/1909.02991} {arXiv:1909.02991 [hep-ph]} \BibitemShut
  {NoStop}%
\bibitem [{\citenamefont {Gao}\ and\ \citenamefont
  {Pawlowski}(2021)}]{Gao:2020fbl}%
  \BibitemOpen
  \bibfield  {author} {\bibinfo {author} {\bibfnamefont {F.}~\bibnamefont
  {Gao}}\ and\ \bibinfo {author} {\bibfnamefont {J.~M.}\ \bibnamefont
  {Pawlowski}},\ }\href {\doibase 10.1016/j.physletb.2021.136584} {\bibfield
  {journal} {\bibinfo  {journal} {Phys. Lett. B}\ }\textbf {\bibinfo {volume}
  {820}},\ \bibinfo {pages} {136584} (\bibinfo {year} {2021})},\ \Eprint
  {http://arxiv.org/abs/2010.13705} {arXiv:2010.13705 [hep-ph]} \BibitemShut
  {NoStop}%
\bibitem [{\citenamefont {Gunkel}\ and\ \citenamefont
  {Fischer}(2021)}]{Gunkel:2021oya}%
  \BibitemOpen
  \bibfield  {author} {\bibinfo {author} {\bibfnamefont {P.~J.}\ \bibnamefont
  {Gunkel}}\ and\ \bibinfo {author} {\bibfnamefont {C.~S.}\ \bibnamefont
  {Fischer}},\ }\href {\doibase 10.1103/PhysRevD.104.054022} {\bibfield
  {journal} {\bibinfo  {journal} {Phys. Rev. D}\ }\textbf {\bibinfo {volume}
  {104}},\ \bibinfo {pages} {054022} (\bibinfo {year} {2021})},\ \Eprint
  {http://arxiv.org/abs/2106.08356} {arXiv:2106.08356 [hep-ph]} \BibitemShut
  {NoStop}%
\bibitem [{NSA(2019)}]{NSAC-QIS-2019-QuantumInformationScience}%
  \BibitemOpen
  \href {https://science.osti.gov/-/media/np/pdf/Reports/NSAC_QIS_Report.pdf}
  {\emph {\bibinfo {title} {Nuclear {Physics} and {Quantum} {Information}
  {Science}: {Report} by the {NSAC} {QIS} {Subcommittee}}}},\ \bibinfo {type}
  {Tech. Rep.}\ (\bibinfo  {institution} {NSF \& DOE Office of Science},\
  \bibinfo {year} {2019})\BibitemShut {NoStop}%
\bibitem [{\citenamefont {Davoudi}\ \emph
  {et~al.}(2022{\natexlab{a}})\citenamefont {Davoudi}, \citenamefont
  {Balantekin}, \citenamefont {Bhattacharya}, \citenamefont {Carena},
  \citenamefont {de~Jong}, \citenamefont {Draper}, \citenamefont {El-Khadra},
  \citenamefont {Gemelke}, \citenamefont {Hanada}, \citenamefont {Kharzeev}
  \emph {et~al.}}]{davoudi2022quantum}%
  \BibitemOpen
  \bibfield  {author} {\bibinfo {author} {\bibfnamefont {C.~W.~B.}\
  \bibnamefont {Davoudi}}, \bibinfo {author} {\bibfnamefont {A.}~\bibnamefont
  {Balantekin}}, \bibinfo {author} {\bibfnamefont {T.}~\bibnamefont
  {Bhattacharya}}, \bibinfo {author} {\bibfnamefont {M.}~\bibnamefont
  {Carena}}, \bibinfo {author} {\bibfnamefont {W.~A.}\ \bibnamefont {de~Jong}},
  \bibinfo {author} {\bibfnamefont {P.}~\bibnamefont {Draper}}, \bibinfo
  {author} {\bibfnamefont {A.}~\bibnamefont {El-Khadra}}, \bibinfo {author}
  {\bibfnamefont {N.}~\bibnamefont {Gemelke}}, \bibinfo {author} {\bibfnamefont
  {M.}~\bibnamefont {Hanada}}, \bibinfo {author} {\bibfnamefont
  {D.}~\bibnamefont {Kharzeev}},  \emph {et~al.},\ }\href@noop {} {\bibfield
  {journal} {\bibinfo  {journal} {arXiv preprint arXiv:2204.03381}\ } (\bibinfo
  {year} {2022}{\natexlab{a}})}\BibitemShut {NoStop}%
\bibitem [{\citenamefont {Catterall}\ \emph {et~al.}(2022)\citenamefont
  {Catterall}, \citenamefont {Harnik}, \citenamefont {Hubeny}, \citenamefont
  {Bauer}, \citenamefont {Berlin}, \citenamefont {Davoudi}, \citenamefont
  {Faulkner}, \citenamefont {Hartman}, \citenamefont {Headrick}, \citenamefont
  {Kahn} \emph {et~al.}}]{catterall2022report}%
  \BibitemOpen
  \bibfield  {author} {\bibinfo {author} {\bibfnamefont {S.}~\bibnamefont
  {Catterall}}, \bibinfo {author} {\bibfnamefont {R.}~\bibnamefont {Harnik}},
  \bibinfo {author} {\bibfnamefont {V.~E.}\ \bibnamefont {Hubeny}}, \bibinfo
  {author} {\bibfnamefont {C.~W.}\ \bibnamefont {Bauer}}, \bibinfo {author}
  {\bibfnamefont {A.}~\bibnamefont {Berlin}}, \bibinfo {author} {\bibfnamefont
  {Z.}~\bibnamefont {Davoudi}}, \bibinfo {author} {\bibfnamefont
  {T.}~\bibnamefont {Faulkner}}, \bibinfo {author} {\bibfnamefont
  {T.}~\bibnamefont {Hartman}}, \bibinfo {author} {\bibfnamefont
  {M.}~\bibnamefont {Headrick}}, \bibinfo {author} {\bibfnamefont {Y.~F.}\
  \bibnamefont {Kahn}},  \emph {et~al.},\ }\href@noop {} {\bibfield  {journal}
  {\bibinfo  {journal} {arXiv preprint arXiv:2209.14839}\ } (\bibinfo {year}
  {2022})}\BibitemShut {NoStop}%
\bibitem [{\citenamefont {Humble}\ \emph {et~al.}(2022)\citenamefont {Humble},
  \citenamefont {Perdue},\ and\ \citenamefont {Savage}}]{humble2022snowmass}%
  \BibitemOpen
  \bibfield  {author} {\bibinfo {author} {\bibfnamefont {T.~S.}\ \bibnamefont
  {Humble}}, \bibinfo {author} {\bibfnamefont {G.~N.}\ \bibnamefont {Perdue}},
  \ and\ \bibinfo {author} {\bibfnamefont {M.~J.}\ \bibnamefont {Savage}},\
  }\href@noop {} {\bibfield  {journal} {\bibinfo  {journal} {arXiv preprint
  arXiv:2209.06786}\ } (\bibinfo {year} {2022})}\BibitemShut {NoStop}%
\bibitem [{\citenamefont {Fisher}(1974)}]{Fisher:1974series}%
  \BibitemOpen
  \bibfield  {author} {\bibinfo {author} {\bibfnamefont {M.~E.}\ \bibnamefont
  {Fisher}},\ }\href@noop {} {\bibfield  {journal} {\bibinfo  {journal} {Rocky
  Mountain Journal of Mathematics}\ }\textbf {\bibinfo {volume} {4}},\ \bibinfo
  {pages} {181} (\bibinfo {year} {1974})}\BibitemShut {NoStop}%
\bibitem [{\citenamefont {Ratti}(2018)}]{Ratti:2018ksb}%
  \BibitemOpen
  \bibfield  {author} {\bibinfo {author} {\bibfnamefont {C.}~\bibnamefont
  {Ratti}},\ }\href {\doibase 10.1088/1361-6633/aabb97} {\bibfield  {journal}
  {\bibinfo  {journal} {Rept. Prog. Phys.}\ }\textbf {\bibinfo {volume} {81}},\
  \bibinfo {pages} {084301} (\bibinfo {year} {2018})},\ \Eprint
  {http://arxiv.org/abs/1804.07810} {arXiv:1804.07810 [hep-lat]} \BibitemShut
  {NoStop}%
\bibitem [{\citenamefont {Bors\'anyi}\ \emph {et~al.}(2021)\citenamefont
  {Bors\'anyi}, \citenamefont {Fodor}, \citenamefont {Guenther}, \citenamefont
  {Kara}, \citenamefont {Katz}, \citenamefont {Parotto}, \citenamefont
  {P\'asztor}, \citenamefont {Ratti},\ and\ \citenamefont
  {Szab\'o}}]{Borsanyi:2021sxv}%
  \BibitemOpen
  \bibfield  {author} {\bibinfo {author} {\bibfnamefont {S.}~\bibnamefont
  {Bors\'anyi}}, \bibinfo {author} {\bibfnamefont {Z.}~\bibnamefont {Fodor}},
  \bibinfo {author} {\bibfnamefont {J.~N.}\ \bibnamefont {Guenther}}, \bibinfo
  {author} {\bibfnamefont {R.}~\bibnamefont {Kara}}, \bibinfo {author}
  {\bibfnamefont {S.~D.}\ \bibnamefont {Katz}}, \bibinfo {author}
  {\bibfnamefont {P.}~\bibnamefont {Parotto}}, \bibinfo {author} {\bibfnamefont
  {A.}~\bibnamefont {P\'asztor}}, \bibinfo {author} {\bibfnamefont
  {C.}~\bibnamefont {Ratti}}, \ and\ \bibinfo {author} {\bibfnamefont {K.~K.}\
  \bibnamefont {Szab\'o}},\ }\href {\doibase 10.1103/PhysRevLett.126.232001}
  {\bibfield  {journal} {\bibinfo  {journal} {Phys. Rev. Lett.}\ }\textbf
  {\bibinfo {volume} {126}},\ \bibinfo {pages} {232001} (\bibinfo {year}
  {2021})},\ \Eprint {http://arxiv.org/abs/2102.06660} {arXiv:2102.06660
  [hep-lat]} \BibitemShut {NoStop}%
\bibitem [{\citenamefont {Bollweg}\ \emph {et~al.}(2022)\citenamefont
  {Bollweg}, \citenamefont {Goswami}, \citenamefont {Kaczmarek}, \citenamefont
  {Karsch}, \citenamefont {Mukherjee}, \citenamefont {Petreczky}, \citenamefont
  {Schmidt},\ and\ \citenamefont {Scior}}]{Bollweg:2022rps}%
  \BibitemOpen
  \bibfield  {author} {\bibinfo {author} {\bibfnamefont {D.}~\bibnamefont
  {Bollweg}}, \bibinfo {author} {\bibfnamefont {J.}~\bibnamefont {Goswami}},
  \bibinfo {author} {\bibfnamefont {O.}~\bibnamefont {Kaczmarek}}, \bibinfo
  {author} {\bibfnamefont {F.}~\bibnamefont {Karsch}}, \bibinfo {author}
  {\bibfnamefont {S.}~\bibnamefont {Mukherjee}}, \bibinfo {author}
  {\bibfnamefont {P.}~\bibnamefont {Petreczky}}, \bibinfo {author}
  {\bibfnamefont {C.}~\bibnamefont {Schmidt}}, \ and\ \bibinfo {author}
  {\bibfnamefont {P.}~\bibnamefont {Scior}} (\bibinfo {collaboration}
  {HotQCD}),\ }\href {\doibase 10.1103/PhysRevD.105.074511} {\bibfield
  {journal} {\bibinfo  {journal} {Phys. Rev. D}\ }\textbf {\bibinfo {volume}
  {105}},\ \bibinfo {pages} {074511} (\bibinfo {year} {2022})},\ \Eprint
  {http://arxiv.org/abs/2202.09184} {arXiv:2202.09184 [hep-lat]} \BibitemShut
  {NoStop}%
\bibitem [{\citenamefont {An}\ \emph {et~al.}(2018)\citenamefont {An},
  \citenamefont {Mesterh\'azy},\ and\ \citenamefont {Stephanov}}]{An:2017brc}%
  \BibitemOpen
  \bibfield  {author} {\bibinfo {author} {\bibfnamefont {X.}~\bibnamefont
  {An}}, \bibinfo {author} {\bibfnamefont {D.}~\bibnamefont {Mesterh\'azy}}, \
  and\ \bibinfo {author} {\bibfnamefont {M.~A.}\ \bibnamefont {Stephanov}},\
  }\href {\doibase 10.1088/1742-5468/aaac4a} {\bibfield  {journal} {\bibinfo
  {journal} {J. Stat. Mech.}\ }\textbf {\bibinfo {volume} {1803}},\ \bibinfo
  {pages} {033207} (\bibinfo {year} {2018})},\ \Eprint
  {http://arxiv.org/abs/1707.06447} {arXiv:1707.06447 [hep-th]} \BibitemShut
  {NoStop}%
\bibitem [{\citenamefont {Mukherjee}\ and\ \citenamefont
  {Skokov}(2021)}]{Mukherjee:2019eou}%
  \BibitemOpen
  \bibfield  {author} {\bibinfo {author} {\bibfnamefont {S.}~\bibnamefont
  {Mukherjee}}\ and\ \bibinfo {author} {\bibfnamefont {V.}~\bibnamefont
  {Skokov}},\ }\href {\doibase 10.1103/PhysRevD.103.L071501} {\bibfield
  {journal} {\bibinfo  {journal} {Phys. Rev. D}\ }\textbf {\bibinfo {volume}
  {103}},\ \bibinfo {pages} {L071501} (\bibinfo {year} {2021})},\ \Eprint
  {http://arxiv.org/abs/1909.04639} {arXiv:1909.04639 [hep-ph]} \BibitemShut
  {NoStop}%
\bibitem [{\citenamefont {Connelly}\ \emph {et~al.}(2020)\citenamefont
  {Connelly}, \citenamefont {Johnson}, \citenamefont {Rennecke},\ and\
  \citenamefont {Skokov}}]{Connelly:2020gwa}%
  \BibitemOpen
  \bibfield  {author} {\bibinfo {author} {\bibfnamefont {A.}~\bibnamefont
  {Connelly}}, \bibinfo {author} {\bibfnamefont {G.}~\bibnamefont {Johnson}},
  \bibinfo {author} {\bibfnamefont {F.}~\bibnamefont {Rennecke}}, \ and\
  \bibinfo {author} {\bibfnamefont {V.}~\bibnamefont {Skokov}},\ }\href
  {\doibase 10.1103/PhysRevLett.125.191602} {\bibfield  {journal} {\bibinfo
  {journal} {Phys. Rev. Lett.}\ }\textbf {\bibinfo {volume} {125}},\ \bibinfo
  {pages} {191602} (\bibinfo {year} {2020})},\ \Eprint
  {http://arxiv.org/abs/2006.12541} {arXiv:2006.12541 [cond-mat.stat-mech]}
  \BibitemShut {NoStop}%
\bibitem [{\citenamefont {Basar}(2021)}]{Basar:2021hdf}%
  \BibitemOpen
  \bibfield  {author} {\bibinfo {author} {\bibfnamefont {G.}~\bibnamefont
  {Basar}},\ }\href {\doibase 10.1103/PhysRevLett.127.171603} {\bibfield
  {journal} {\bibinfo  {journal} {Phys. Rev. Lett.}\ }\textbf {\bibinfo
  {volume} {127}},\ \bibinfo {pages} {171603} (\bibinfo {year} {2021})},\
  \Eprint {http://arxiv.org/abs/2105.08080} {arXiv:2105.08080 [hep-th]}
  \BibitemShut {NoStop}%
\bibitem [{\citenamefont {Basar}\ \emph {et~al.}(2022)\citenamefont {Basar},
  \citenamefont {Dunne},\ and\ \citenamefont {Yin}}]{Basar:2021gyi}%
  \BibitemOpen
  \bibfield  {author} {\bibinfo {author} {\bibfnamefont {G.}~\bibnamefont
  {Basar}}, \bibinfo {author} {\bibfnamefont {G.~V.}\ \bibnamefont {Dunne}}, \
  and\ \bibinfo {author} {\bibfnamefont {Z.}~\bibnamefont {Yin}},\ }\href
  {\doibase 10.1103/PhysRevD.105.105002} {\bibfield  {journal} {\bibinfo
  {journal} {Phys. Rev. D}\ }\textbf {\bibinfo {volume} {105}},\ \bibinfo
  {pages} {105002} (\bibinfo {year} {2022})},\ \Eprint
  {http://arxiv.org/abs/2112.14269} {arXiv:2112.14269 [hep-th]} \BibitemShut
  {NoStop}%
\bibitem [{\citenamefont {Komoltsev}\ and\ \citenamefont
  {Kurkela}(2022)}]{Komoltsev:2021jzg}%
  \BibitemOpen
  \bibfield  {author} {\bibinfo {author} {\bibfnamefont {O.}~\bibnamefont
  {Komoltsev}}\ and\ \bibinfo {author} {\bibfnamefont {A.}~\bibnamefont
  {Kurkela}},\ }\href {\doibase 10.1103/PhysRevLett.128.202701} {\bibfield
  {journal} {\bibinfo  {journal} {Phys. Rev. Lett.}\ }\textbf {\bibinfo
  {volume} {128}},\ \bibinfo {pages} {202701} (\bibinfo {year} {2022})},\
  \Eprint {http://arxiv.org/abs/2111.05350} {arXiv:2111.05350 [nucl-th]}
  \BibitemShut {NoStop}%
\bibitem [{\citenamefont {Borsanyi}\ \emph {et~al.}(2010)\citenamefont
  {Borsanyi}, \citenamefont {Endrodi}, \citenamefont {Fodor}, \citenamefont
  {Jakovac}, \citenamefont {Katz}, \citenamefont {Krieg}, \citenamefont
  {Ratti},\ and\ \citenamefont {Szabo}}]{Borsanyi:2010cj}%
  \BibitemOpen
  \bibfield  {author} {\bibinfo {author} {\bibfnamefont {S.}~\bibnamefont
  {Borsanyi}}, \bibinfo {author} {\bibfnamefont {G.}~\bibnamefont {Endrodi}},
  \bibinfo {author} {\bibfnamefont {Z.}~\bibnamefont {Fodor}}, \bibinfo
  {author} {\bibfnamefont {A.}~\bibnamefont {Jakovac}}, \bibinfo {author}
  {\bibfnamefont {S.~D.}\ \bibnamefont {Katz}}, \bibinfo {author}
  {\bibfnamefont {S.}~\bibnamefont {Krieg}}, \bibinfo {author} {\bibfnamefont
  {C.}~\bibnamefont {Ratti}}, \ and\ \bibinfo {author} {\bibfnamefont {K.~K.}\
  \bibnamefont {Szabo}},\ }\href {\doibase 10.1007/JHEP11(2010)077} {\bibfield
  {journal} {\bibinfo  {journal} {JHEP}\ }\textbf {\bibinfo {volume} {11}},\
  \bibinfo {pages} {077} (\bibinfo {year} {2010})},\ \Eprint
  {http://arxiv.org/abs/1007.2580} {arXiv:1007.2580 [hep-lat]} \BibitemShut
  {NoStop}%
\bibitem [{\citenamefont {Borsanyi}\ \emph {et~al.}(2014)\citenamefont
  {Borsanyi}, \citenamefont {Fodor}, \citenamefont {Hoelbling}, \citenamefont
  {Katz}, \citenamefont {Krieg},\ and\ \citenamefont
  {Szabo}}]{Borsanyi:2013bia}%
  \BibitemOpen
  \bibfield  {author} {\bibinfo {author} {\bibfnamefont {S.}~\bibnamefont
  {Borsanyi}}, \bibinfo {author} {\bibfnamefont {Z.}~\bibnamefont {Fodor}},
  \bibinfo {author} {\bibfnamefont {C.}~\bibnamefont {Hoelbling}}, \bibinfo
  {author} {\bibfnamefont {S.~D.}\ \bibnamefont {Katz}}, \bibinfo {author}
  {\bibfnamefont {S.}~\bibnamefont {Krieg}}, \ and\ \bibinfo {author}
  {\bibfnamefont {K.~K.}\ \bibnamefont {Szabo}},\ }\href {\doibase
  10.1016/j.physletb.2014.01.007} {\bibfield  {journal} {\bibinfo  {journal}
  {Phys. Lett. B}\ }\textbf {\bibinfo {volume} {730}},\ \bibinfo {pages} {99}
  (\bibinfo {year} {2014})},\ \Eprint {http://arxiv.org/abs/1309.5258}
  {arXiv:1309.5258 [hep-lat]} \BibitemShut {NoStop}%
\bibitem [{\citenamefont {Bazavov}\ \emph {et~al.}(2014)\citenamefont {Bazavov}
  \emph {et~al.}}]{Bazavov:2014pvz}%
  \BibitemOpen
  \bibfield  {author} {\bibinfo {author} {\bibfnamefont {A.}~\bibnamefont
  {Bazavov}} \emph {et~al.} (\bibinfo {collaboration} {HotQCD}),\ }\href
  {\doibase 10.1103/PhysRevD.90.094503} {\bibfield  {journal} {\bibinfo
  {journal} {Phys. Rev. D}\ }\textbf {\bibinfo {volume} {90}},\ \bibinfo
  {pages} {094503} (\bibinfo {year} {2014})},\ \Eprint
  {http://arxiv.org/abs/1407.6387} {arXiv:1407.6387 [hep-lat]} \BibitemShut
  {NoStop}%
\bibitem [{\citenamefont {Aoki}\ \emph {et~al.}(2006)\citenamefont {Aoki},
  \citenamefont {Endrodi}, \citenamefont {Fodor}, \citenamefont {Katz},\ and\
  \citenamefont {Szabo}}]{Aoki:2006we}%
  \BibitemOpen
  \bibfield  {author} {\bibinfo {author} {\bibfnamefont {Y.}~\bibnamefont
  {Aoki}}, \bibinfo {author} {\bibfnamefont {G.}~\bibnamefont {Endrodi}},
  \bibinfo {author} {\bibfnamefont {Z.}~\bibnamefont {Fodor}}, \bibinfo
  {author} {\bibfnamefont {S.~D.}\ \bibnamefont {Katz}}, \ and\ \bibinfo
  {author} {\bibfnamefont {K.~K.}\ \bibnamefont {Szabo}},\ }\href {\doibase
  10.1038/nature05120} {\bibfield  {journal} {\bibinfo  {journal} {Nature}\
  }\textbf {\bibinfo {volume} {443}},\ \bibinfo {pages} {675} (\bibinfo {year}
  {2006})},\ \Eprint {http://arxiv.org/abs/hep-lat/0611014}
  {arXiv:hep-lat/0611014} \BibitemShut {NoStop}%
\bibitem [{\citenamefont {Andersen}\ \emph {et~al.}(2011)\citenamefont
  {Andersen}, \citenamefont {Leganger}, \citenamefont {Strickland},\ and\
  \citenamefont {Su}}]{Andersen:2011sf}%
  \BibitemOpen
  \bibfield  {author} {\bibinfo {author} {\bibfnamefont {J.~O.}\ \bibnamefont
  {Andersen}}, \bibinfo {author} {\bibfnamefont {L.~E.}\ \bibnamefont
  {Leganger}}, \bibinfo {author} {\bibfnamefont {M.}~\bibnamefont
  {Strickland}}, \ and\ \bibinfo {author} {\bibfnamefont {N.}~\bibnamefont
  {Su}},\ }\href {\doibase 10.1007/JHEP08(2011)053} {\bibfield  {journal}
  {\bibinfo  {journal} {JHEP}\ }\textbf {\bibinfo {volume} {08}},\ \bibinfo
  {pages} {053} (\bibinfo {year} {2011})},\ \Eprint
  {http://arxiv.org/abs/1103.2528} {arXiv:1103.2528 [hep-ph]} \BibitemShut
  {NoStop}%
\bibitem [{\citenamefont {Strickland}\ \emph {et~al.}(2014)\citenamefont
  {Strickland}, \citenamefont {Andersen}, \citenamefont {Bandyopadhyay},
  \citenamefont {Haque}, \citenamefont {Mustafa},\ and\ \citenamefont
  {Su}}]{Strickland:2014zka}%
  \BibitemOpen
  \bibfield  {author} {\bibinfo {author} {\bibfnamefont {M.}~\bibnamefont
  {Strickland}}, \bibinfo {author} {\bibfnamefont {J.~O.}\ \bibnamefont
  {Andersen}}, \bibinfo {author} {\bibfnamefont {A.}~\bibnamefont
  {Bandyopadhyay}}, \bibinfo {author} {\bibfnamefont {N.}~\bibnamefont
  {Haque}}, \bibinfo {author} {\bibfnamefont {M.~G.}\ \bibnamefont {Mustafa}},
  \ and\ \bibinfo {author} {\bibfnamefont {N.}~\bibnamefont {Su}},\ }\href
  {\doibase 10.1016/j.nuclphysa.2014.09.022} {\bibfield  {journal} {\bibinfo
  {journal} {Nucl. Phys. A}\ }\textbf {\bibinfo {volume} {931}},\ \bibinfo
  {pages} {841} (\bibinfo {year} {2014})},\ \Eprint
  {http://arxiv.org/abs/1407.3671} {arXiv:1407.3671 [hep-ph]} \BibitemShut
  {NoStop}%
\bibitem [{\citenamefont {Ghiglieri}\ \emph {et~al.}(2020)\citenamefont
  {Ghiglieri}, \citenamefont {Kurkela}, \citenamefont {Strickland},\ and\
  \citenamefont {Vuorinen}}]{Ghiglieri:2020dpq}%
  \BibitemOpen
  \bibfield  {author} {\bibinfo {author} {\bibfnamefont {J.}~\bibnamefont
  {Ghiglieri}}, \bibinfo {author} {\bibfnamefont {A.}~\bibnamefont {Kurkela}},
  \bibinfo {author} {\bibfnamefont {M.}~\bibnamefont {Strickland}}, \ and\
  \bibinfo {author} {\bibfnamefont {A.}~\bibnamefont {Vuorinen}},\ }\href
  {\doibase 10.1016/j.physrep.2020.07.004} {\bibfield  {journal} {\bibinfo
  {journal} {Phys. Rept.}\ }\textbf {\bibinfo {volume} {880}},\ \bibinfo
  {pages} {1} (\bibinfo {year} {2020})},\ \Eprint
  {http://arxiv.org/abs/2002.10188} {arXiv:2002.10188 [hep-ph]} \BibitemShut
  {NoStop}%
\bibitem [{\citenamefont {Haque}\ and\ \citenamefont
  {Strickland}(2021)}]{Haque:2020eyj}%
  \BibitemOpen
  \bibfield  {author} {\bibinfo {author} {\bibfnamefont {N.}~\bibnamefont
  {Haque}}\ and\ \bibinfo {author} {\bibfnamefont {M.}~\bibnamefont
  {Strickland}},\ }\href {\doibase 10.1103/PhysRevC.103.L031901} {\bibfield
  {journal} {\bibinfo  {journal} {Phys. Rev. C}\ }\textbf {\bibinfo {volume}
  {103}},\ \bibinfo {pages} {031901} (\bibinfo {year} {2021})},\ \Eprint
  {http://arxiv.org/abs/2011.06938} {arXiv:2011.06938 [hep-ph]} \BibitemShut
  {NoStop}%
\bibitem [{\citenamefont {Allton}\ \emph {et~al.}(2005)\citenamefont {Allton},
  \citenamefont {Doring}, \citenamefont {Ejiri}, \citenamefont {Hands},
  \citenamefont {Kaczmarek}, \citenamefont {Karsch}, \citenamefont {Laermann},\
  and\ \citenamefont {Redlich}}]{Allton:2005gk}%
  \BibitemOpen
  \bibfield  {author} {\bibinfo {author} {\bibfnamefont {C.~R.}\ \bibnamefont
  {Allton}}, \bibinfo {author} {\bibfnamefont {M.}~\bibnamefont {Doring}},
  \bibinfo {author} {\bibfnamefont {S.}~\bibnamefont {Ejiri}}, \bibinfo
  {author} {\bibfnamefont {S.~J.}\ \bibnamefont {Hands}}, \bibinfo {author}
  {\bibfnamefont {O.}~\bibnamefont {Kaczmarek}}, \bibinfo {author}
  {\bibfnamefont {F.}~\bibnamefont {Karsch}}, \bibinfo {author} {\bibfnamefont
  {E.}~\bibnamefont {Laermann}}, \ and\ \bibinfo {author} {\bibfnamefont
  {K.}~\bibnamefont {Redlich}},\ }\href {\doibase 10.1103/PhysRevD.71.054508}
  {\bibfield  {journal} {\bibinfo  {journal} {Phys. Rev. D}\ }\textbf {\bibinfo
  {volume} {71}},\ \bibinfo {pages} {054508} (\bibinfo {year} {2005})},\
  \Eprint {http://arxiv.org/abs/hep-lat/0501030} {arXiv:hep-lat/0501030}
  \BibitemShut {NoStop}%
\bibitem [{\citenamefont {Bazavov}\ \emph {et~al.}(2017)\citenamefont {Bazavov}
  \emph {et~al.}}]{Bazavov:2017dus}%
  \BibitemOpen
  \bibfield  {author} {\bibinfo {author} {\bibfnamefont {A.}~\bibnamefont
  {Bazavov}} \emph {et~al.},\ }\href {\doibase 10.1103/PhysRevD.95.054504}
  {\bibfield  {journal} {\bibinfo  {journal} {Phys. Rev. D}\ }\textbf {\bibinfo
  {volume} {95}},\ \bibinfo {pages} {054504} (\bibinfo {year} {2017})},\
  \Eprint {http://arxiv.org/abs/1701.04325} {arXiv:1701.04325 [hep-lat]}
  \BibitemShut {NoStop}%
\bibitem [{\citenamefont {Bazavov}\ \emph {et~al.}(2020)\citenamefont {Bazavov}
  \emph {et~al.}}]{Bazavov:2020bjn}%
  \BibitemOpen
  \bibfield  {author} {\bibinfo {author} {\bibfnamefont {A.}~\bibnamefont
  {Bazavov}} \emph {et~al.},\ }\href {\doibase 10.1103/PhysRevD.101.074502}
  {\bibfield  {journal} {\bibinfo  {journal} {Phys. Rev. D}\ }\textbf {\bibinfo
  {volume} {101}},\ \bibinfo {pages} {074502} (\bibinfo {year} {2020})},\
  \Eprint {http://arxiv.org/abs/2001.08530} {arXiv:2001.08530 [hep-lat]}
  \BibitemShut {NoStop}%
\bibitem [{\citenamefont {de~Forcrand}\ and\ \citenamefont
  {Philipsen}(2002)}]{deForcrand:2002hgr}%
  \BibitemOpen
  \bibfield  {author} {\bibinfo {author} {\bibfnamefont {P.}~\bibnamefont
  {de~Forcrand}}\ and\ \bibinfo {author} {\bibfnamefont {O.}~\bibnamefont
  {Philipsen}},\ }\href {\doibase 10.1016/S0550-3213(02)00626-0} {\bibfield
  {journal} {\bibinfo  {journal} {Nucl. Phys. B}\ }\textbf {\bibinfo {volume}
  {642}},\ \bibinfo {pages} {290} (\bibinfo {year} {2002})},\ \Eprint
  {http://arxiv.org/abs/hep-lat/0205016} {arXiv:hep-lat/0205016} \BibitemShut
  {NoStop}%
\bibitem [{\citenamefont {D'Elia}\ and\ \citenamefont
  {Lombardo}(2003)}]{DElia:2002tig}%
  \BibitemOpen
  \bibfield  {author} {\bibinfo {author} {\bibfnamefont {M.}~\bibnamefont
  {D'Elia}}\ and\ \bibinfo {author} {\bibfnamefont {M.-P.}\ \bibnamefont
  {Lombardo}},\ }\href {\doibase 10.1103/PhysRevD.67.014505} {\bibfield
  {journal} {\bibinfo  {journal} {Phys. Rev. D}\ }\textbf {\bibinfo {volume}
  {67}},\ \bibinfo {pages} {014505} (\bibinfo {year} {2003})},\ \Eprint
  {http://arxiv.org/abs/hep-lat/0209146} {arXiv:hep-lat/0209146} \BibitemShut
  {NoStop}%
\bibitem [{\citenamefont {Wu}\ \emph {et~al.}(2007)\citenamefont {Wu},
  \citenamefont {Luo},\ and\ \citenamefont {Chen}}]{Wu:2006su}%
  \BibitemOpen
  \bibfield  {author} {\bibinfo {author} {\bibfnamefont {L.-K.}\ \bibnamefont
  {Wu}}, \bibinfo {author} {\bibfnamefont {X.-Q.}\ \bibnamefont {Luo}}, \ and\
  \bibinfo {author} {\bibfnamefont {H.-S.}\ \bibnamefont {Chen}},\ }\href
  {\doibase 10.1103/PhysRevD.76.034505} {\bibfield  {journal} {\bibinfo
  {journal} {Phys. Rev. D}\ }\textbf {\bibinfo {volume} {76}},\ \bibinfo
  {pages} {034505} (\bibinfo {year} {2007})},\ \Eprint
  {http://arxiv.org/abs/hep-lat/0611035} {arXiv:hep-lat/0611035} \BibitemShut
  {NoStop}%
\bibitem [{\citenamefont {Conradi}\ and\ \citenamefont
  {D'Elia}(2007)}]{Conradi:2007be}%
  \BibitemOpen
  \bibfield  {author} {\bibinfo {author} {\bibfnamefont {S.}~\bibnamefont
  {Conradi}}\ and\ \bibinfo {author} {\bibfnamefont {M.}~\bibnamefont
  {D'Elia}},\ }\href {\doibase 10.1103/PhysRevD.76.074501} {\bibfield
  {journal} {\bibinfo  {journal} {Phys. Rev. D}\ }\textbf {\bibinfo {volume}
  {76}},\ \bibinfo {pages} {074501} (\bibinfo {year} {2007})},\ \Eprint
  {http://arxiv.org/abs/0707.1987} {arXiv:0707.1987 [hep-lat]} \BibitemShut
  {NoStop}%
\bibitem [{\citenamefont {de~Forcrand}\ and\ \citenamefont
  {Philipsen}(2008)}]{deForcrand:2008vr}%
  \BibitemOpen
  \bibfield  {author} {\bibinfo {author} {\bibfnamefont {P.}~\bibnamefont
  {de~Forcrand}}\ and\ \bibinfo {author} {\bibfnamefont {O.}~\bibnamefont
  {Philipsen}},\ }\href {\doibase 10.1088/1126-6708/2008/11/012} {\bibfield
  {journal} {\bibinfo  {journal} {JHEP}\ }\textbf {\bibinfo {volume} {11}},\
  \bibinfo {pages} {012} (\bibinfo {year} {2008})},\ \Eprint
  {http://arxiv.org/abs/0808.1096} {arXiv:0808.1096 [hep-lat]} \BibitemShut
  {NoStop}%
\bibitem [{\citenamefont {D'Elia}\ and\ \citenamefont
  {Sanfilippo}(2009)}]{DElia:2009pdy}%
  \BibitemOpen
  \bibfield  {author} {\bibinfo {author} {\bibfnamefont {M.}~\bibnamefont
  {D'Elia}}\ and\ \bibinfo {author} {\bibfnamefont {F.}~\bibnamefont
  {Sanfilippo}},\ }\href {\doibase 10.1103/PhysRevD.80.014502} {\bibfield
  {journal} {\bibinfo  {journal} {Phys. Rev. D}\ }\textbf {\bibinfo {volume}
  {80}},\ \bibinfo {pages} {014502} (\bibinfo {year} {2009})},\ \Eprint
  {http://arxiv.org/abs/0904.1400} {arXiv:0904.1400 [hep-lat]} \BibitemShut
  {NoStop}%
\bibitem [{\citenamefont {Borsanyi}\ \emph {et~al.}(2018)\citenamefont
  {Borsanyi}, \citenamefont {Fodor}, \citenamefont {Guenther}, \citenamefont
  {Katz}, \citenamefont {Szabo}, \citenamefont {Pasztor}, \citenamefont
  {Portillo},\ and\ \citenamefont {Ratti}}]{Borsanyi:2018grb}%
  \BibitemOpen
  \bibfield  {author} {\bibinfo {author} {\bibfnamefont {S.}~\bibnamefont
  {Borsanyi}}, \bibinfo {author} {\bibfnamefont {Z.}~\bibnamefont {Fodor}},
  \bibinfo {author} {\bibfnamefont {J.~N.}\ \bibnamefont {Guenther}}, \bibinfo
  {author} {\bibfnamefont {S.~K.}\ \bibnamefont {Katz}}, \bibinfo {author}
  {\bibfnamefont {K.~K.}\ \bibnamefont {Szabo}}, \bibinfo {author}
  {\bibfnamefont {A.}~\bibnamefont {Pasztor}}, \bibinfo {author} {\bibfnamefont
  {I.}~\bibnamefont {Portillo}}, \ and\ \bibinfo {author} {\bibfnamefont
  {C.}~\bibnamefont {Ratti}},\ }\href {\doibase 10.1007/JHEP10(2018)205}
  {\bibfield  {journal} {\bibinfo  {journal} {JHEP}\ }\textbf {\bibinfo
  {volume} {10}},\ \bibinfo {pages} {205} (\bibinfo {year} {2018})},\ \Eprint
  {http://arxiv.org/abs/1805.04445} {arXiv:1805.04445 [hep-lat]} \BibitemShut
  {NoStop}%
\bibitem [{\citenamefont {Fodor}\ and\ \citenamefont
  {Katz}(2002{\natexlab{a}})}]{Fodor:2001au}%
  \BibitemOpen
  \bibfield  {author} {\bibinfo {author} {\bibfnamefont {Z.}~\bibnamefont
  {Fodor}}\ and\ \bibinfo {author} {\bibfnamefont {S.~D.}\ \bibnamefont
  {Katz}},\ }\href {\doibase 10.1016/S0370-2693(02)01583-6} {\bibfield
  {journal} {\bibinfo  {journal} {Phys. Lett. B}\ }\textbf {\bibinfo {volume}
  {534}},\ \bibinfo {pages} {87} (\bibinfo {year} {2002}{\natexlab{a}})},\
  \Eprint {http://arxiv.org/abs/hep-lat/0104001} {arXiv:hep-lat/0104001}
  \BibitemShut {NoStop}%
\bibitem [{\citenamefont {Fodor}\ and\ \citenamefont
  {Katz}(2002{\natexlab{b}})}]{Fodor:2001pe}%
  \BibitemOpen
  \bibfield  {author} {\bibinfo {author} {\bibfnamefont {Z.}~\bibnamefont
  {Fodor}}\ and\ \bibinfo {author} {\bibfnamefont {S.~D.}\ \bibnamefont
  {Katz}},\ }\href {\doibase 10.1088/1126-6708/2002/03/014} {\bibfield
  {journal} {\bibinfo  {journal} {JHEP}\ }\textbf {\bibinfo {volume} {03}},\
  \bibinfo {pages} {014} (\bibinfo {year} {2002}{\natexlab{b}})},\ \Eprint
  {http://arxiv.org/abs/hep-lat/0106002} {arXiv:hep-lat/0106002} \BibitemShut
  {NoStop}%
\bibitem [{\citenamefont {Csikor}\ \emph {et~al.}(2004)\citenamefont {Csikor},
  \citenamefont {Egri}, \citenamefont {Fodor}, \citenamefont {Katz},
  \citenamefont {Szabo},\ and\ \citenamefont {Toth}}]{Csikor:2004ik}%
  \BibitemOpen
  \bibfield  {author} {\bibinfo {author} {\bibfnamefont {F.}~\bibnamefont
  {Csikor}}, \bibinfo {author} {\bibfnamefont {G.~I.}\ \bibnamefont {Egri}},
  \bibinfo {author} {\bibfnamefont {Z.}~\bibnamefont {Fodor}}, \bibinfo
  {author} {\bibfnamefont {S.~D.}\ \bibnamefont {Katz}}, \bibinfo {author}
  {\bibfnamefont {K.~K.}\ \bibnamefont {Szabo}}, \ and\ \bibinfo {author}
  {\bibfnamefont {A.~I.}\ \bibnamefont {Toth}},\ }\href {\doibase
  10.1088/1126-6708/2004/05/046} {\bibfield  {journal} {\bibinfo  {journal}
  {JHEP}\ }\textbf {\bibinfo {volume} {05}},\ \bibinfo {pages} {046} (\bibinfo
  {year} {2004})},\ \Eprint {http://arxiv.org/abs/hep-lat/0401016}
  {arXiv:hep-lat/0401016} \BibitemShut {NoStop}%
\bibitem [{\citenamefont {Fodor}\ and\ \citenamefont
  {Katz}(2004)}]{Fodor:2004nz}%
  \BibitemOpen
  \bibfield  {author} {\bibinfo {author} {\bibfnamefont {Z.}~\bibnamefont
  {Fodor}}\ and\ \bibinfo {author} {\bibfnamefont {S.~D.}\ \bibnamefont
  {Katz}},\ }\href {\doibase 10.1088/1126-6708/2004/04/050} {\bibfield
  {journal} {\bibinfo  {journal} {JHEP}\ }\textbf {\bibinfo {volume} {04}},\
  \bibinfo {pages} {050} (\bibinfo {year} {2004})},\ \Eprint
  {http://arxiv.org/abs/hep-lat/0402006} {arXiv:hep-lat/0402006} \BibitemShut
  {NoStop}%
\bibitem [{\citenamefont {Fodor}\ \emph {et~al.}(2007)\citenamefont {Fodor},
  \citenamefont {Katz},\ and\ \citenamefont {Schmidt}}]{Fodor:2007vv}%
  \BibitemOpen
  \bibfield  {author} {\bibinfo {author} {\bibfnamefont {Z.}~\bibnamefont
  {Fodor}}, \bibinfo {author} {\bibfnamefont {S.~D.}\ \bibnamefont {Katz}}, \
  and\ \bibinfo {author} {\bibfnamefont {C.}~\bibnamefont {Schmidt}},\ }\href
  {\doibase 10.1088/1126-6708/2007/03/121} {\bibfield  {journal} {\bibinfo
  {journal} {JHEP}\ }\textbf {\bibinfo {volume} {03}},\ \bibinfo {pages} {121}
  (\bibinfo {year} {2007})},\ \Eprint {http://arxiv.org/abs/hep-lat/0701022}
  {arXiv:hep-lat/0701022} \BibitemShut {NoStop}%
\bibitem [{\citenamefont {Alexandru}\ \emph {et~al.}(2015)\citenamefont
  {Alexandru}, \citenamefont {Gattringer}, \citenamefont {Schadler},
  \citenamefont {Splittorff},\ and\ \citenamefont
  {Verbaarschot}}]{Alexandru:2014hga}%
  \BibitemOpen
  \bibfield  {author} {\bibinfo {author} {\bibfnamefont {A.}~\bibnamefont
  {Alexandru}}, \bibinfo {author} {\bibfnamefont {C.}~\bibnamefont
  {Gattringer}}, \bibinfo {author} {\bibfnamefont {H.~P.}\ \bibnamefont
  {Schadler}}, \bibinfo {author} {\bibfnamefont {K.}~\bibnamefont
  {Splittorff}}, \ and\ \bibinfo {author} {\bibfnamefont {J.~J.~M.}\
  \bibnamefont {Verbaarschot}},\ }\href {\doibase 10.1103/PhysRevD.91.074501}
  {\bibfield  {journal} {\bibinfo  {journal} {Phys. Rev. D}\ }\textbf {\bibinfo
  {volume} {91}},\ \bibinfo {pages} {074501} (\bibinfo {year} {2015})},\
  \Eprint {http://arxiv.org/abs/1411.4143} {arXiv:1411.4143 [hep-lat]}
  \BibitemShut {NoStop}%
\bibitem [{\citenamefont {Alexandru}\ \emph {et~al.}(2005)\citenamefont
  {Alexandru}, \citenamefont {Faber}, \citenamefont {Horvath},\ and\
  \citenamefont {Liu}}]{Alexandru:2005ix}%
  \BibitemOpen
  \bibfield  {author} {\bibinfo {author} {\bibfnamefont {A.}~\bibnamefont
  {Alexandru}}, \bibinfo {author} {\bibfnamefont {M.}~\bibnamefont {Faber}},
  \bibinfo {author} {\bibfnamefont {I.}~\bibnamefont {Horvath}}, \ and\
  \bibinfo {author} {\bibfnamefont {K.-F.}\ \bibnamefont {Liu}},\ }\href
  {\doibase 10.1103/PhysRevD.72.114513} {\bibfield  {journal} {\bibinfo
  {journal} {Phys. Rev. D}\ }\textbf {\bibinfo {volume} {72}},\ \bibinfo
  {pages} {114513} (\bibinfo {year} {2005})},\ \Eprint
  {http://arxiv.org/abs/hep-lat/0507020} {arXiv:hep-lat/0507020} \BibitemShut
  {NoStop}%
\bibitem [{\citenamefont {Kratochvila}\ and\ \citenamefont
  {de~Forcrand}(2006)}]{Kratochvila:2005mk}%
  \BibitemOpen
  \bibfield  {author} {\bibinfo {author} {\bibfnamefont {S.}~\bibnamefont
  {Kratochvila}}\ and\ \bibinfo {author} {\bibfnamefont {P.}~\bibnamefont
  {de~Forcrand}},\ }\href {\doibase 10.22323/1.020.0167} {\bibfield  {journal}
  {\bibinfo  {journal} {PoS}\ }\textbf {\bibinfo {volume} {LAT2005}},\ \bibinfo
  {pages} {167} (\bibinfo {year} {2006})},\ \Eprint
  {http://arxiv.org/abs/hep-lat/0509143} {arXiv:hep-lat/0509143} \BibitemShut
  {NoStop}%
\bibitem [{\citenamefont {Ejiri}(2008)}]{Ejiri:2008xt}%
  \BibitemOpen
  \bibfield  {author} {\bibinfo {author} {\bibfnamefont {S.}~\bibnamefont
  {Ejiri}},\ }\href {\doibase 10.1103/PhysRevD.78.074507} {\bibfield  {journal}
  {\bibinfo  {journal} {Phys. Rev. D}\ }\textbf {\bibinfo {volume} {78}},\
  \bibinfo {pages} {074507} (\bibinfo {year} {2008})},\ \Eprint
  {http://arxiv.org/abs/0804.3227} {arXiv:0804.3227 [hep-lat]} \BibitemShut
  {NoStop}%
\bibitem [{\citenamefont {Giordano}\ \emph {et~al.}(2020)\citenamefont
  {Giordano}, \citenamefont {Kapas}, \citenamefont {Katz}, \citenamefont
  {Nogradi},\ and\ \citenamefont {Pasztor}}]{Giordano:2020roi}%
  \BibitemOpen
  \bibfield  {author} {\bibinfo {author} {\bibfnamefont {M.}~\bibnamefont
  {Giordano}}, \bibinfo {author} {\bibfnamefont {K.}~\bibnamefont {Kapas}},
  \bibinfo {author} {\bibfnamefont {S.~D.}\ \bibnamefont {Katz}}, \bibinfo
  {author} {\bibfnamefont {D.}~\bibnamefont {Nogradi}}, \ and\ \bibinfo
  {author} {\bibfnamefont {A.}~\bibnamefont {Pasztor}},\ }\href {\doibase
  10.1007/JHEP05(2020)088} {\bibfield  {journal} {\bibinfo  {journal} {JHEP}\
  }\textbf {\bibinfo {volume} {05}},\ \bibinfo {pages} {088} (\bibinfo {year}
  {2020})},\ \Eprint {http://arxiv.org/abs/2004.10800} {arXiv:2004.10800
  [hep-lat]} \BibitemShut {NoStop}%
\bibitem [{\citenamefont {Borsanyi}\ \emph {et~al.}(2022)\citenamefont
  {Borsanyi}, \citenamefont {Fodor}, \citenamefont {Giordano}, \citenamefont
  {Katz}, \citenamefont {Nogradi}, \citenamefont {Pasztor},\ and\ \citenamefont
  {Wong}}]{Borsanyi:2021hbk}%
  \BibitemOpen
  \bibfield  {author} {\bibinfo {author} {\bibfnamefont {S.}~\bibnamefont
  {Borsanyi}}, \bibinfo {author} {\bibfnamefont {Z.}~\bibnamefont {Fodor}},
  \bibinfo {author} {\bibfnamefont {M.}~\bibnamefont {Giordano}}, \bibinfo
  {author} {\bibfnamefont {S.~D.}\ \bibnamefont {Katz}}, \bibinfo {author}
  {\bibfnamefont {D.}~\bibnamefont {Nogradi}}, \bibinfo {author} {\bibfnamefont
  {A.}~\bibnamefont {Pasztor}}, \ and\ \bibinfo {author} {\bibfnamefont
  {C.~H.}\ \bibnamefont {Wong}},\ }\href {\doibase
  10.1103/PhysRevD.105.L051506} {\bibfield  {journal} {\bibinfo  {journal}
  {Phys. Rev. D}\ }\textbf {\bibinfo {volume} {105}},\ \bibinfo {pages}
  {L051506} (\bibinfo {year} {2022})},\ \Eprint
  {http://arxiv.org/abs/2108.09213} {arXiv:2108.09213 [hep-lat]} \BibitemShut
  {NoStop}%
\bibitem [{\citenamefont {Seiler}\ \emph {et~al.}(2013)\citenamefont {Seiler},
  \citenamefont {Sexty},\ and\ \citenamefont {Stamatescu}}]{Seiler:2012wz}%
  \BibitemOpen
  \bibfield  {author} {\bibinfo {author} {\bibfnamefont {E.}~\bibnamefont
  {Seiler}}, \bibinfo {author} {\bibfnamefont {D.}~\bibnamefont {Sexty}}, \
  and\ \bibinfo {author} {\bibfnamefont {I.-O.}\ \bibnamefont {Stamatescu}},\
  }\href {\doibase 10.1016/j.physletb.2013.04.062} {\bibfield  {journal}
  {\bibinfo  {journal} {Phys. Lett. B}\ }\textbf {\bibinfo {volume} {723}},\
  \bibinfo {pages} {213} (\bibinfo {year} {2013})},\ \Eprint
  {http://arxiv.org/abs/1211.3709} {arXiv:1211.3709 [hep-lat]} \BibitemShut
  {NoStop}%
\bibitem [{\citenamefont {Sexty}(2014)}]{Sexty:2013ica}%
  \BibitemOpen
  \bibfield  {author} {\bibinfo {author} {\bibfnamefont {D.}~\bibnamefont
  {Sexty}},\ }\href {\doibase 10.1016/j.physletb.2014.01.019} {\bibfield
  {journal} {\bibinfo  {journal} {Phys. Lett. B}\ }\textbf {\bibinfo {volume}
  {729}},\ \bibinfo {pages} {108} (\bibinfo {year} {2014})},\ \Eprint
  {http://arxiv.org/abs/1307.7748} {arXiv:1307.7748 [hep-lat]} \BibitemShut
  {NoStop}%
\bibitem [{\citenamefont {Sexty}(2019)}]{Sexty:2019vqx}%
  \BibitemOpen
  \bibfield  {author} {\bibinfo {author} {\bibfnamefont {D.}~\bibnamefont
  {Sexty}},\ }\href {\doibase 10.1103/PhysRevD.100.074503} {\bibfield
  {journal} {\bibinfo  {journal} {Phys. Rev. D}\ }\textbf {\bibinfo {volume}
  {100}},\ \bibinfo {pages} {074503} (\bibinfo {year} {2019})},\ \Eprint
  {http://arxiv.org/abs/1907.08712} {arXiv:1907.08712 [hep-lat]} \BibitemShut
  {NoStop}%
\bibitem [{\citenamefont {Attanasio}\ and\ \citenamefont
  {J\"ager}(2019)}]{Attanasio:2018rtq}%
  \BibitemOpen
  \bibfield  {author} {\bibinfo {author} {\bibfnamefont {F.}~\bibnamefont
  {Attanasio}}\ and\ \bibinfo {author} {\bibfnamefont {B.}~\bibnamefont
  {J\"ager}},\ }\href {\doibase 10.1140/epjc/s10052-018-6512-7} {\bibfield
  {journal} {\bibinfo  {journal} {Eur. Phys. J. C}\ }\textbf {\bibinfo {volume}
  {79}},\ \bibinfo {pages} {16} (\bibinfo {year} {2019})},\ \Eprint
  {http://arxiv.org/abs/1808.04400} {arXiv:1808.04400 [hep-lat]} \BibitemShut
  {NoStop}%
\bibitem [{\citenamefont {Attanasio}\ \emph {et~al.}(2022)\citenamefont
  {Attanasio}, \citenamefont {J\"ager},\ and\ \citenamefont
  {Ziegler}}]{Attanasio:2022mjd}%
  \BibitemOpen
  \bibfield  {author} {\bibinfo {author} {\bibfnamefont {F.}~\bibnamefont
  {Attanasio}}, \bibinfo {author} {\bibfnamefont {B.}~\bibnamefont {J\"ager}},
  \ and\ \bibinfo {author} {\bibfnamefont {F.~P.~G.}\ \bibnamefont {Ziegler}},\
  }\href@noop {} {\  (\bibinfo {year} {2022})},\ \Eprint
  {http://arxiv.org/abs/2203.13144} {arXiv:2203.13144 [hep-lat]} \BibitemShut
  {NoStop}%
\bibitem [{\citenamefont {Hebeler}\ \emph {et~al.}(2015)\citenamefont
  {Hebeler}, \citenamefont {Holt}, \citenamefont {Menendez},\ and\
  \citenamefont {Schwenk}}]{Hebeler:2015hla}%
  \BibitemOpen
  \bibfield  {author} {\bibinfo {author} {\bibfnamefont {K.}~\bibnamefont
  {Hebeler}}, \bibinfo {author} {\bibfnamefont {J.~D.}\ \bibnamefont {Holt}},
  \bibinfo {author} {\bibfnamefont {J.}~\bibnamefont {Menendez}}, \ and\
  \bibinfo {author} {\bibfnamefont {A.}~\bibnamefont {Schwenk}},\ }\href
  {\doibase 10.1146/annurev-nucl-102313-025446} {\bibfield  {journal} {\bibinfo
   {journal} {Ann. Rev. Nucl. Part. Sci.}\ }\textbf {\bibinfo {volume} {65}},\
  \bibinfo {pages} {457} (\bibinfo {year} {2015})},\ \Eprint
  {http://arxiv.org/abs/1508.06893} {arXiv:1508.06893 [nucl-th]} \BibitemShut
  {NoStop}%
\bibitem [{\citenamefont {Stroberg}\ \emph {et~al.}(2019)\citenamefont
  {Stroberg}, \citenamefont {Bogner}, \citenamefont {Hergert},\ and\
  \citenamefont {Holt}}]{Stroberg:2019mxo}%
  \BibitemOpen
  \bibfield  {author} {\bibinfo {author} {\bibfnamefont {S.~R.}\ \bibnamefont
  {Stroberg}}, \bibinfo {author} {\bibfnamefont {S.~K.}\ \bibnamefont
  {Bogner}}, \bibinfo {author} {\bibfnamefont {H.}~\bibnamefont {Hergert}}, \
  and\ \bibinfo {author} {\bibfnamefont {J.~D.}\ \bibnamefont {Holt}},\ }\href
  {\doibase 10.1146/annurev-nucl-101917-021120} {\bibfield  {journal} {\bibinfo
   {journal} {Ann. Rev. Nucl. Part. Sci.}\ }\textbf {\bibinfo {volume} {69}},\
  \bibinfo {pages} {307} (\bibinfo {year} {2019})},\ \Eprint
  {http://arxiv.org/abs/1902.06154} {arXiv:1902.06154 [nucl-th]} \BibitemShut
  {NoStop}%
\bibitem [{\citenamefont {Hergert}(2020)}]{Hergert:2020bxy}%
  \BibitemOpen
  \bibfield  {author} {\bibinfo {author} {\bibfnamefont {H.}~\bibnamefont
  {Hergert}},\ }\href {\doibase 10.3389/fphy.2020.00379} {\bibfield  {journal}
  {\bibinfo  {journal} {Front. in Phys.}\ }\textbf {\bibinfo {volume} {8}},\
  \bibinfo {pages} {379} (\bibinfo {year} {2020})},\ \Eprint
  {http://arxiv.org/abs/2008.05061} {arXiv:2008.05061 [nucl-th]} \BibitemShut
  {NoStop}%
\bibitem [{\citenamefont {Carlson}\ \emph {et~al.}(2015)\citenamefont
  {Carlson}, \citenamefont {Gandolfi}, \citenamefont {Pederiva}, \citenamefont
  {Pieper}, \citenamefont {Schiavilla}, \citenamefont {Schmidt},\ and\
  \citenamefont {Wiringa}}]{Carlson_2015}%
  \BibitemOpen
  \bibfield  {author} {\bibinfo {author} {\bibfnamefont {J.}~\bibnamefont
  {Carlson}}, \bibinfo {author} {\bibfnamefont {S.}~\bibnamefont {Gandolfi}},
  \bibinfo {author} {\bibfnamefont {F.}~\bibnamefont {Pederiva}}, \bibinfo
  {author} {\bibfnamefont {S.~C.}\ \bibnamefont {Pieper}}, \bibinfo {author}
  {\bibfnamefont {R.}~\bibnamefont {Schiavilla}}, \bibinfo {author}
  {\bibfnamefont {K.}~\bibnamefont {Schmidt}}, \ and\ \bibinfo {author}
  {\bibfnamefont {R.}~\bibnamefont {Wiringa}},\ }\href {\doibase
  10.1103/revmodphys.87.1067} {\bibfield  {journal} {\bibinfo  {journal}
  {Reviews of Modern Physics}\ }\textbf {\bibinfo {volume} {87}},\ \bibinfo
  {pages} {1067} (\bibinfo {year} {2015})}\BibitemShut {NoStop}%
\bibitem [{\citenamefont {Gandolfi}\ \emph {et~al.}(2020)\citenamefont
  {Gandolfi}, \citenamefont {Lonardoni}, \citenamefont {Lovato},\ and\
  \citenamefont {Piarulli}}]{Gandolfi:2020pbj}%
  \BibitemOpen
  \bibfield  {author} {\bibinfo {author} {\bibfnamefont {S.}~\bibnamefont
  {Gandolfi}}, \bibinfo {author} {\bibfnamefont {D.}~\bibnamefont {Lonardoni}},
  \bibinfo {author} {\bibfnamefont {A.}~\bibnamefont {Lovato}}, \ and\ \bibinfo
  {author} {\bibfnamefont {M.}~\bibnamefont {Piarulli}},\ }\href {\doibase
  10.3389/fphy.2020.00117} {\bibfield  {journal} {\bibinfo  {journal} {Front.
  in Phys.}\ }\textbf {\bibinfo {volume} {8}},\ \bibinfo {pages} {117}
  (\bibinfo {year} {2020})},\ \Eprint {http://arxiv.org/abs/2001.01374}
  {arXiv:2001.01374 [nucl-th]} \BibitemShut {NoStop}%
\bibitem [{\citenamefont {Piarulli}\ and\ \citenamefont
  {Tews}(2020)}]{Piarulli:2019cqu}%
  \BibitemOpen
  \bibfield  {author} {\bibinfo {author} {\bibfnamefont {M.}~\bibnamefont
  {Piarulli}}\ and\ \bibinfo {author} {\bibfnamefont {I.}~\bibnamefont
  {Tews}},\ }\href {\doibase 10.3389/fphy.2019.00245} {\bibfield  {journal}
  {\bibinfo  {journal} {Front. in Phys.}\ }\textbf {\bibinfo {volume} {7}},\
  \bibinfo {pages} {245} (\bibinfo {year} {2020})},\ \Eprint
  {http://arxiv.org/abs/2002.00032} {arXiv:2002.00032 [nucl-th]} \BibitemShut
  {NoStop}%
\bibitem [{\citenamefont {Epelbaum}\ \emph {et~al.}(2015)\citenamefont
  {Epelbaum}, \citenamefont {Krebs},\ and\ \citenamefont
  {Mei\ss{}ner}}]{Epelbaum:2014efa}%
  \BibitemOpen
  \bibfield  {author} {\bibinfo {author} {\bibfnamefont {E.}~\bibnamefont
  {Epelbaum}}, \bibinfo {author} {\bibfnamefont {H.}~\bibnamefont {Krebs}}, \
  and\ \bibinfo {author} {\bibfnamefont {U.~G.}\ \bibnamefont {Mei\ss{}ner}},\
  }\href {\doibase 10.1140/epja/i2015-15053-8} {\bibfield  {journal} {\bibinfo
  {journal} {Eur. Phys. J. A}\ }\textbf {\bibinfo {volume} {51}},\ \bibinfo
  {pages} {53} (\bibinfo {year} {2015})},\ \Eprint
  {http://arxiv.org/abs/1412.0142} {arXiv:1412.0142 [nucl-th]} \BibitemShut
  {NoStop}%
\bibitem [{\citenamefont {Gezerlis}\ \emph {et~al.}(2013)\citenamefont
  {Gezerlis}, \citenamefont {Tews}, \citenamefont {Epelbaum}, \citenamefont
  {Gandolfi}, \citenamefont {Hebeler}, \citenamefont {Nogga},\ and\
  \citenamefont {Schwenk}}]{Gezerlis_2013}%
  \BibitemOpen
  \bibfield  {author} {\bibinfo {author} {\bibfnamefont {A.}~\bibnamefont
  {Gezerlis}}, \bibinfo {author} {\bibfnamefont {I.}~\bibnamefont {Tews}},
  \bibinfo {author} {\bibfnamefont {E.}~\bibnamefont {Epelbaum}}, \bibinfo
  {author} {\bibfnamefont {S.}~\bibnamefont {Gandolfi}}, \bibinfo {author}
  {\bibfnamefont {K.}~\bibnamefont {Hebeler}}, \bibinfo {author} {\bibfnamefont
  {A.}~\bibnamefont {Nogga}}, \ and\ \bibinfo {author} {\bibfnamefont
  {A.}~\bibnamefont {Schwenk}},\ }\href {\doibase
  10.1103/physrevlett.111.032501} {\bibfield  {journal} {\bibinfo  {journal}
  {Physical Review Letters}\ }\textbf {\bibinfo {volume} {111}} (\bibinfo
  {year} {2013}),\ 10.1103/physrevlett.111.032501}\BibitemShut {NoStop}%
\bibitem [{\citenamefont {Gezerlis}\ \emph {et~al.}(2014)\citenamefont
  {Gezerlis}, \citenamefont {Tews}, \citenamefont {Epelbaum}, \citenamefont
  {Freunek}, \citenamefont {Gandolfi}, \citenamefont {Hebeler}, \citenamefont
  {Nogga},\ and\ \citenamefont {Schwenk}}]{Gezerlis_2014}%
  \BibitemOpen
  \bibfield  {author} {\bibinfo {author} {\bibfnamefont {A.}~\bibnamefont
  {Gezerlis}}, \bibinfo {author} {\bibfnamefont {I.}~\bibnamefont {Tews}},
  \bibinfo {author} {\bibfnamefont {E.}~\bibnamefont {Epelbaum}}, \bibinfo
  {author} {\bibfnamefont {M.}~\bibnamefont {Freunek}}, \bibinfo {author}
  {\bibfnamefont {S.}~\bibnamefont {Gandolfi}}, \bibinfo {author}
  {\bibfnamefont {K.}~\bibnamefont {Hebeler}}, \bibinfo {author} {\bibfnamefont
  {A.}~\bibnamefont {Nogga}}, \ and\ \bibinfo {author} {\bibfnamefont
  {A.}~\bibnamefont {Schwenk}},\ }\href {\doibase 10.1103/physrevc.90.054323}
  {\bibfield  {journal} {\bibinfo  {journal} {Physical Review C}\ }\textbf
  {\bibinfo {volume} {90}} (\bibinfo {year} {2014}),\
  10.1103/physrevc.90.054323}\BibitemShut {NoStop}%
\bibitem [{\citenamefont {Lynn}\ \emph {et~al.}(2017)\citenamefont {Lynn},
  \citenamefont {Tews}, \citenamefont {Carlson}, \citenamefont {Gandolfi},
  \citenamefont {Gezerlis}, \citenamefont {Schmidt},\ and\ \citenamefont
  {Schwenk}}]{Lynn_2017}%
  \BibitemOpen
  \bibfield  {author} {\bibinfo {author} {\bibfnamefont {J.~E.}\ \bibnamefont
  {Lynn}}, \bibinfo {author} {\bibfnamefont {I.}~\bibnamefont {Tews}}, \bibinfo
  {author} {\bibfnamefont {J.}~\bibnamefont {Carlson}}, \bibinfo {author}
  {\bibfnamefont {S.}~\bibnamefont {Gandolfi}}, \bibinfo {author}
  {\bibfnamefont {A.}~\bibnamefont {Gezerlis}}, \bibinfo {author}
  {\bibfnamefont {K.~E.}\ \bibnamefont {Schmidt}}, \ and\ \bibinfo {author}
  {\bibfnamefont {A.}~\bibnamefont {Schwenk}},\ }\href {\doibase
  10.1103/physrevc.96.054007} {\bibfield  {journal} {\bibinfo  {journal}
  {Physical Review C}\ }\textbf {\bibinfo {volume} {96}} (\bibinfo {year}
  {2017}),\ 10.1103/physrevc.96.054007}\BibitemShut {NoStop}%
\bibitem [{\citenamefont {Tews}\ \emph
  {et~al.}(2021{\natexlab{a}})\citenamefont {Tews}, \citenamefont {Lonardoni},\
  and\ \citenamefont {Gandolfi}}]{Tews_2021}%
  \BibitemOpen
  \bibfield  {author} {\bibinfo {author} {\bibfnamefont {I.}~\bibnamefont
  {Tews}}, \bibinfo {author} {\bibfnamefont {D.}~\bibnamefont {Lonardoni}}, \
  and\ \bibinfo {author} {\bibfnamefont {S.}~\bibnamefont {Gandolfi}},\ }\href
  {\doibase 10.1007/s00601-021-01687-0} {\bibfield  {journal} {\bibinfo
  {journal} {Few-Body Systems}\ }\textbf {\bibinfo {volume} {62}} (\bibinfo
  {year} {2021}{\natexlab{a}}),\ 10.1007/s00601-021-01687-0}\BibitemShut
  {NoStop}%
\bibitem [{\citenamefont {Piarulli}\ \emph {et~al.}(2015)\citenamefont
  {Piarulli}, \citenamefont {Girlanda}, \citenamefont {Schiavilla},
  \citenamefont {Navarro~P\'erez}, \citenamefont {Amaro},\ and\ \citenamefont
  {Ruiz~Arriola}}]{Piarulli:2014bda}%
  \BibitemOpen
  \bibfield  {author} {\bibinfo {author} {\bibfnamefont {M.}~\bibnamefont
  {Piarulli}}, \bibinfo {author} {\bibfnamefont {L.}~\bibnamefont {Girlanda}},
  \bibinfo {author} {\bibfnamefont {R.}~\bibnamefont {Schiavilla}}, \bibinfo
  {author} {\bibfnamefont {R.}~\bibnamefont {Navarro~P\'erez}}, \bibinfo
  {author} {\bibfnamefont {J.~E.}\ \bibnamefont {Amaro}}, \ and\ \bibinfo
  {author} {\bibfnamefont {E.}~\bibnamefont {Ruiz~Arriola}},\ }\href {\doibase
  10.1103/PhysRevC.91.024003} {\bibfield  {journal} {\bibinfo  {journal} {Phys.
  Rev. C}\ }\textbf {\bibinfo {volume} {91}},\ \bibinfo {pages} {024003}
  (\bibinfo {year} {2015})},\ \Eprint {http://arxiv.org/abs/1412.6446}
  {arXiv:1412.6446 [nucl-th]} \BibitemShut {NoStop}%
\bibitem [{\citenamefont {Piarulli}\ \emph {et~al.}(2016)\citenamefont
  {Piarulli}, \citenamefont {Girlanda}, \citenamefont {Schiavilla},
  \citenamefont {Kievsky}, \citenamefont {Lovato}, \citenamefont {Marcucci},
  \citenamefont {Pieper}, \citenamefont {Viviani},\ and\ \citenamefont
  {Wiringa}}]{Piarulli:2016vel}%
  \BibitemOpen
  \bibfield  {author} {\bibinfo {author} {\bibfnamefont {M.}~\bibnamefont
  {Piarulli}}, \bibinfo {author} {\bibfnamefont {L.}~\bibnamefont {Girlanda}},
  \bibinfo {author} {\bibfnamefont {R.}~\bibnamefont {Schiavilla}}, \bibinfo
  {author} {\bibfnamefont {A.}~\bibnamefont {Kievsky}}, \bibinfo {author}
  {\bibfnamefont {A.}~\bibnamefont {Lovato}}, \bibinfo {author} {\bibfnamefont
  {L.~E.}\ \bibnamefont {Marcucci}}, \bibinfo {author} {\bibfnamefont {S.~C.}\
  \bibnamefont {Pieper}}, \bibinfo {author} {\bibfnamefont {M.}~\bibnamefont
  {Viviani}}, \ and\ \bibinfo {author} {\bibfnamefont {R.~B.}\ \bibnamefont
  {Wiringa}},\ }\href {\doibase 10.1103/PhysRevC.94.054007} {\bibfield
  {journal} {\bibinfo  {journal} {Phys. Rev. C}\ }\textbf {\bibinfo {volume}
  {94}},\ \bibinfo {pages} {054007} (\bibinfo {year} {2016})},\ \Eprint
  {http://arxiv.org/abs/1606.06335} {arXiv:1606.06335 [nucl-th]} \BibitemShut
  {NoStop}%
\bibitem [{\citenamefont {Baroni}\ \emph {et~al.}(2018)\citenamefont {Baroni}
  \emph {et~al.}}]{Baroni:2018fdn}%
  \BibitemOpen
  \bibfield  {author} {\bibinfo {author} {\bibfnamefont {A.}~\bibnamefont
  {Baroni}} \emph {et~al.},\ }\href {\doibase 10.1103/PhysRevC.98.044003}
  {\bibfield  {journal} {\bibinfo  {journal} {Phys. Rev. C}\ }\textbf {\bibinfo
  {volume} {98}},\ \bibinfo {pages} {044003} (\bibinfo {year} {2018})},\
  \Eprint {http://arxiv.org/abs/1806.10245} {arXiv:1806.10245 [nucl-th]}
  \BibitemShut {NoStop}%
\bibitem [{\citenamefont {Lonardoni}\ \emph {et~al.}(2018)\citenamefont
  {Lonardoni}, \citenamefont {Carlson}, \citenamefont {Gandolfi}, \citenamefont
  {Lynn}, \citenamefont {Schmidt}, \citenamefont {Schwenk},\ and\ \citenamefont
  {Wang}}]{Lonardoni_2018}%
  \BibitemOpen
  \bibfield  {author} {\bibinfo {author} {\bibfnamefont {D.}~\bibnamefont
  {Lonardoni}}, \bibinfo {author} {\bibfnamefont {J.}~\bibnamefont {Carlson}},
  \bibinfo {author} {\bibfnamefont {S.}~\bibnamefont {Gandolfi}}, \bibinfo
  {author} {\bibfnamefont {J.}~\bibnamefont {Lynn}}, \bibinfo {author}
  {\bibfnamefont {K.}~\bibnamefont {Schmidt}}, \bibinfo {author} {\bibfnamefont
  {A.}~\bibnamefont {Schwenk}}, \ and\ \bibinfo {author} {\bibfnamefont
  {X.}~\bibnamefont {Wang}},\ }\href {\doibase 10.1103/physrevlett.120.122502}
  {\bibfield  {journal} {\bibinfo  {journal} {Physical Review Letters}\
  }\textbf {\bibinfo {volume} {120}} (\bibinfo {year} {2018}),\
  10.1103/physrevlett.120.122502}\BibitemShut {NoStop}%
\bibitem [{\citenamefont {Lonardoni}\ \emph {et~al.}(2020)\citenamefont
  {Lonardoni}, \citenamefont {Tews}, \citenamefont {Gandolfi},\ and\
  \citenamefont {Carlson}}]{Lonardoni_2020}%
  \BibitemOpen
  \bibfield  {author} {\bibinfo {author} {\bibfnamefont {D.}~\bibnamefont
  {Lonardoni}}, \bibinfo {author} {\bibfnamefont {I.}~\bibnamefont {Tews}},
  \bibinfo {author} {\bibfnamefont {S.}~\bibnamefont {Gandolfi}}, \ and\
  \bibinfo {author} {\bibfnamefont {J.}~\bibnamefont {Carlson}},\ }\href
  {\doibase 10.1103/physrevresearch.2.022033} {\bibfield  {journal} {\bibinfo
  {journal} {Physical Review Research}\ }\textbf {\bibinfo {volume} {2}}
  (\bibinfo {year} {2020}),\ 10.1103/physrevresearch.2.022033}\BibitemShut
  {NoStop}%
\bibitem [{\citenamefont {Piarulli}\ \emph {et~al.}(2018)\citenamefont
  {Piarulli} \emph {et~al.}}]{Piarulli:2017dwd}%
  \BibitemOpen
  \bibfield  {author} {\bibinfo {author} {\bibfnamefont {M.}~\bibnamefont
  {Piarulli}} \emph {et~al.},\ }\href {\doibase 10.1103/PhysRevLett.120.052503}
  {\bibfield  {journal} {\bibinfo  {journal} {Phys. Rev. Lett.}\ }\textbf
  {\bibinfo {volume} {120}},\ \bibinfo {pages} {052503} (\bibinfo {year}
  {2018})},\ \Eprint {http://arxiv.org/abs/1707.02883} {arXiv:1707.02883
  [nucl-th]} \BibitemShut {NoStop}%
\bibitem [{\citenamefont {King}\ \emph {et~al.}(2020)\citenamefont {King},
  \citenamefont {Andreoli}, \citenamefont {Pastore}, \citenamefont {Piarulli},
  \citenamefont {Schiavilla}, \citenamefont {Wiringa}, \citenamefont
  {Carlson},\ and\ \citenamefont {Gandolfi}}]{King:2020wmp}%
  \BibitemOpen
  \bibfield  {author} {\bibinfo {author} {\bibfnamefont {G.~B.}\ \bibnamefont
  {King}}, \bibinfo {author} {\bibfnamefont {L.}~\bibnamefont {Andreoli}},
  \bibinfo {author} {\bibfnamefont {S.}~\bibnamefont {Pastore}}, \bibinfo
  {author} {\bibfnamefont {M.}~\bibnamefont {Piarulli}}, \bibinfo {author}
  {\bibfnamefont {R.}~\bibnamefont {Schiavilla}}, \bibinfo {author}
  {\bibfnamefont {R.~B.}\ \bibnamefont {Wiringa}}, \bibinfo {author}
  {\bibfnamefont {J.}~\bibnamefont {Carlson}}, \ and\ \bibinfo {author}
  {\bibfnamefont {S.}~\bibnamefont {Gandolfi}},\ }\href {\doibase
  10.1103/PhysRevC.102.025501} {\bibfield  {journal} {\bibinfo  {journal}
  {Phys. Rev. C}\ }\textbf {\bibinfo {volume} {102}},\ \bibinfo {pages}
  {025501} (\bibinfo {year} {2020})},\ \Eprint
  {http://arxiv.org/abs/2004.05263} {arXiv:2004.05263 [nucl-th]} \BibitemShut
  {NoStop}%
\bibitem [{\citenamefont {Lynn}\ \emph {et~al.}(2016)\citenamefont {Lynn},
  \citenamefont {Tews}, \citenamefont {Carlson}, \citenamefont {Gandolfi},
  \citenamefont {Gezerlis}, \citenamefont {Schmidt},\ and\ \citenamefont
  {Schwenk}}]{Lynn:2015jua}%
  \BibitemOpen
  \bibfield  {author} {\bibinfo {author} {\bibfnamefont {J.~E.}\ \bibnamefont
  {Lynn}}, \bibinfo {author} {\bibfnamefont {I.}~\bibnamefont {Tews}}, \bibinfo
  {author} {\bibfnamefont {J.}~\bibnamefont {Carlson}}, \bibinfo {author}
  {\bibfnamefont {S.}~\bibnamefont {Gandolfi}}, \bibinfo {author}
  {\bibfnamefont {A.}~\bibnamefont {Gezerlis}}, \bibinfo {author}
  {\bibfnamefont {K.~E.}\ \bibnamefont {Schmidt}}, \ and\ \bibinfo {author}
  {\bibfnamefont {A.}~\bibnamefont {Schwenk}},\ }\href {\doibase
  10.1103/PhysRevLett.116.062501} {\bibfield  {journal} {\bibinfo  {journal}
  {Phys. Rev. Lett.}\ }\textbf {\bibinfo {volume} {116}},\ \bibinfo {pages}
  {062501} (\bibinfo {year} {2016})},\ \Eprint
  {http://arxiv.org/abs/1509.03470} {arXiv:1509.03470 [nucl-th]} \BibitemShut
  {NoStop}%
\bibitem [{\citenamefont {Al-Mamun}\ \emph {et~al.}(2021)\citenamefont
  {Al-Mamun}, \citenamefont {Steiner}, \citenamefont {Nättilä}, \citenamefont
  {Lange}, \citenamefont {O'Shaughnessy}, \citenamefont {Tews}, \citenamefont
  {Gandolfi}, \citenamefont {Heinke},\ and\ \citenamefont
  {Han}}]{Al_Mamun_2021}%
  \BibitemOpen
  \bibfield  {author} {\bibinfo {author} {\bibfnamefont {M.}~\bibnamefont
  {Al-Mamun}}, \bibinfo {author} {\bibfnamefont {A.~W.}\ \bibnamefont
  {Steiner}}, \bibinfo {author} {\bibfnamefont {J.}~\bibnamefont {Nättilä}},
  \bibinfo {author} {\bibfnamefont {J.}~\bibnamefont {Lange}}, \bibinfo
  {author} {\bibfnamefont {R.}~\bibnamefont {O'Shaughnessy}}, \bibinfo {author}
  {\bibfnamefont {I.}~\bibnamefont {Tews}}, \bibinfo {author} {\bibfnamefont
  {S.}~\bibnamefont {Gandolfi}}, \bibinfo {author} {\bibfnamefont
  {C.}~\bibnamefont {Heinke}}, \ and\ \bibinfo {author} {\bibfnamefont
  {S.}~\bibnamefont {Han}},\ }\href {\doibase 10.1103/physrevlett.126.061101}
  {\bibfield  {journal} {\bibinfo  {journal} {Physical Review Letters}\
  }\textbf {\bibinfo {volume} {126}} (\bibinfo {year} {2021}),\
  10.1103/physrevlett.126.061101}\BibitemShut {NoStop}%
\bibitem [{\citenamefont {Dietrich}\ \emph {et~al.}(2020)\citenamefont
  {Dietrich}, \citenamefont {Coughlin}, \citenamefont {Pang}, \citenamefont
  {Bulla}, \citenamefont {Heinzel}, \citenamefont {Issa}, \citenamefont
  {Tews},\ and\ \citenamefont {Antier}}]{Dietrich:2020efo}%
  \BibitemOpen
  \bibfield  {author} {\bibinfo {author} {\bibfnamefont {T.}~\bibnamefont
  {Dietrich}}, \bibinfo {author} {\bibfnamefont {M.~W.}\ \bibnamefont
  {Coughlin}}, \bibinfo {author} {\bibfnamefont {P.~T.~H.}\ \bibnamefont
  {Pang}}, \bibinfo {author} {\bibfnamefont {M.}~\bibnamefont {Bulla}},
  \bibinfo {author} {\bibfnamefont {J.}~\bibnamefont {Heinzel}}, \bibinfo
  {author} {\bibfnamefont {L.}~\bibnamefont {Issa}}, \bibinfo {author}
  {\bibfnamefont {I.}~\bibnamefont {Tews}}, \ and\ \bibinfo {author}
  {\bibfnamefont {S.}~\bibnamefont {Antier}},\ }\href {\doibase
  10.1126/science.abb4317} {\bibfield  {journal} {\bibinfo  {journal}
  {Science}\ }\textbf {\bibinfo {volume} {370}},\ \bibinfo {pages} {1450}
  (\bibinfo {year} {2020})},\ \Eprint {http://arxiv.org/abs/2002.11355}
  {arXiv:2002.11355 [astro-ph.HE]} \BibitemShut {NoStop}%
\bibitem [{\citenamefont {Tichai}\ \emph {et~al.}(2020)\citenamefont {Tichai},
  \citenamefont {Roth},\ and\ \citenamefont {Duguet}}]{Tichai:2020dna}%
  \BibitemOpen
  \bibfield  {author} {\bibinfo {author} {\bibfnamefont {A.}~\bibnamefont
  {Tichai}}, \bibinfo {author} {\bibfnamefont {R.}~\bibnamefont {Roth}}, \ and\
  \bibinfo {author} {\bibfnamefont {T.}~\bibnamefont {Duguet}},\ }\href
  {\doibase 10.3389/fphy.2020.00164} {\bibfield  {journal} {\bibinfo  {journal}
  {Front. in Phys.}\ }\textbf {\bibinfo {volume} {8}},\ \bibinfo {pages} {164}
  (\bibinfo {year} {2020})},\ \Eprint {http://arxiv.org/abs/2001.10433}
  {arXiv:2001.10433 [nucl-th]} \BibitemShut {NoStop}%
\bibitem [{\citenamefont {Bogner}\ \emph {et~al.}(2010)\citenamefont {Bogner},
  \citenamefont {Furnstahl},\ and\ \citenamefont {Schwenk}}]{Bogner:2009bt}%
  \BibitemOpen
  \bibfield  {author} {\bibinfo {author} {\bibfnamefont {S.~K.}\ \bibnamefont
  {Bogner}}, \bibinfo {author} {\bibfnamefont {R.~J.}\ \bibnamefont
  {Furnstahl}}, \ and\ \bibinfo {author} {\bibfnamefont {A.}~\bibnamefont
  {Schwenk}},\ }\href {\doibase 10.1016/j.ppnp.2010.03.001} {\bibfield
  {journal} {\bibinfo  {journal} {Prog. Part. Nucl. Phys.}\ }\textbf {\bibinfo
  {volume} {65}},\ \bibinfo {pages} {94} (\bibinfo {year} {2010})},\ \Eprint
  {http://arxiv.org/abs/0912.3688} {arXiv:0912.3688 [nucl-th]} \BibitemShut
  {NoStop}%
\bibitem [{\citenamefont {Arthuis}\ \emph {et~al.}(2021)\citenamefont
  {Arthuis}, \citenamefont {Tichai}, \citenamefont {Ripoche},\ and\
  \citenamefont {Duguet}}]{Arthuis:2020tjz}%
  \BibitemOpen
  \bibfield  {author} {\bibinfo {author} {\bibfnamefont {P.}~\bibnamefont
  {Arthuis}}, \bibinfo {author} {\bibfnamefont {A.}~\bibnamefont {Tichai}},
  \bibinfo {author} {\bibfnamefont {J.}~\bibnamefont {Ripoche}}, \ and\
  \bibinfo {author} {\bibfnamefont {T.}~\bibnamefont {Duguet}},\ }\href
  {\doibase 10.1016/j.cpc.2020.107677} {\bibfield  {journal} {\bibinfo
  {journal} {Comput. Phys. Commun.}\ }\textbf {\bibinfo {volume} {261}},\
  \bibinfo {pages} {107677} (\bibinfo {year} {2021})},\ \Eprint
  {http://arxiv.org/abs/2007.01661} {arXiv:2007.01661 [nucl-th]} \BibitemShut
  {NoStop}%
\bibitem [{\citenamefont {Drischler}\ \emph {et~al.}(2019)\citenamefont
  {Drischler}, \citenamefont {Hebeler},\ and\ \citenamefont
  {Schwenk}}]{Drischler:2017wtt}%
  \BibitemOpen
  \bibfield  {author} {\bibinfo {author} {\bibfnamefont {C.}~\bibnamefont
  {Drischler}}, \bibinfo {author} {\bibfnamefont {K.}~\bibnamefont {Hebeler}},
  \ and\ \bibinfo {author} {\bibfnamefont {A.}~\bibnamefont {Schwenk}},\ }\href
  {\doibase 10.1103/PhysRevLett.122.042501} {\bibfield  {journal} {\bibinfo
  {journal} {Phys. Rev. Lett.}\ }\textbf {\bibinfo {volume} {122}},\ \bibinfo
  {pages} {042501} (\bibinfo {year} {2019})},\ \Eprint
  {http://arxiv.org/abs/1710.08220} {arXiv:1710.08220 [nucl-th]} \BibitemShut
  {NoStop}%
\bibitem [{\citenamefont {Drischler}\ \emph {et~al.}(2014)\citenamefont
  {Drischler}, \citenamefont {Soma},\ and\ \citenamefont
  {Schwenk}}]{Drischler:2013iza}%
  \BibitemOpen
  \bibfield  {author} {\bibinfo {author} {\bibfnamefont {C.}~\bibnamefont
  {Drischler}}, \bibinfo {author} {\bibfnamefont {V.}~\bibnamefont {Soma}}, \
  and\ \bibinfo {author} {\bibfnamefont {A.}~\bibnamefont {Schwenk}},\ }\href
  {\doibase 10.1103/PhysRevC.89.025806} {\bibfield  {journal} {\bibinfo
  {journal} {Phys. Rev. C}\ }\textbf {\bibinfo {volume} {89}},\ \bibinfo
  {pages} {025806} (\bibinfo {year} {2014})},\ \Eprint
  {http://arxiv.org/abs/1310.5627} {arXiv:1310.5627 [nucl-th]} \BibitemShut
  {NoStop}%
\bibitem [{\citenamefont {Drischler}\ \emph {et~al.}(2016)\citenamefont
  {Drischler}, \citenamefont {Hebeler},\ and\ \citenamefont
  {Schwenk}}]{Drischler:2015eba}%
  \BibitemOpen
  \bibfield  {author} {\bibinfo {author} {\bibfnamefont {C.}~\bibnamefont
  {Drischler}}, \bibinfo {author} {\bibfnamefont {K.}~\bibnamefont {Hebeler}},
  \ and\ \bibinfo {author} {\bibfnamefont {A.}~\bibnamefont {Schwenk}},\ }\href
  {\doibase 10.1103/PhysRevC.93.054314} {\bibfield  {journal} {\bibinfo
  {journal} {Phys. Rev. C}\ }\textbf {\bibinfo {volume} {93}},\ \bibinfo
  {pages} {054314} (\bibinfo {year} {2016})},\ \Eprint
  {http://arxiv.org/abs/1510.06728} {arXiv:1510.06728 [nucl-th]} \BibitemShut
  {NoStop}%
\bibitem [{\citenamefont {Wellenhofer}(2017)}]{Wellenhofer:2017qla}%
  \BibitemOpen
  \bibfield  {author} {\bibinfo {author} {\bibfnamefont {C.}~\bibnamefont
  {Wellenhofer}},\ }\emph {\bibinfo {title} {{Isospin-Asymmetry Dependence of
  the Thermodynamic Nuclear Equation of State in Many-Body Perturbation
  Theory}}},\ \href@noop {} {Ph.D. thesis},\ \bibinfo  {school} {Munich, Tech.
  U.} (\bibinfo {year} {2017}),\ \Eprint {http://arxiv.org/abs/1707.09222}
  {arXiv:1707.09222 [nucl-th]} \BibitemShut {NoStop}%
\bibitem [{\citenamefont {Wen}\ and\ \citenamefont {Holt}(2021)}]{Wen:2020nqs}%
  \BibitemOpen
  \bibfield  {author} {\bibinfo {author} {\bibfnamefont {P.}~\bibnamefont
  {Wen}}\ and\ \bibinfo {author} {\bibfnamefont {J.~W.}\ \bibnamefont {Holt}},\
  }\href {\doibase 10.1103/PhysRevC.103.064002} {\bibfield  {journal} {\bibinfo
   {journal} {Phys. Rev. C}\ }\textbf {\bibinfo {volume} {103}},\ \bibinfo
  {pages} {064002} (\bibinfo {year} {2021})},\ \Eprint
  {http://arxiv.org/abs/2012.02163} {arXiv:2012.02163 [nucl-th]} \BibitemShut
  {NoStop}%
\bibitem [{\citenamefont {Somasundaram}\ \emph {et~al.}(2021)\citenamefont
  {Somasundaram}, \citenamefont {Drischler}, \citenamefont {Tews},\ and\
  \citenamefont {Margueron}}]{Somasundaram:2020chb}%
  \BibitemOpen
  \bibfield  {author} {\bibinfo {author} {\bibfnamefont {R.}~\bibnamefont
  {Somasundaram}}, \bibinfo {author} {\bibfnamefont {C.}~\bibnamefont
  {Drischler}}, \bibinfo {author} {\bibfnamefont {I.}~\bibnamefont {Tews}}, \
  and\ \bibinfo {author} {\bibfnamefont {J.}~\bibnamefont {Margueron}},\ }\href
  {\doibase 10.1103/PhysRevC.103.045803} {\bibfield  {journal} {\bibinfo
  {journal} {Phys. Rev. C}\ }\textbf {\bibinfo {volume} {103}},\ \bibinfo
  {pages} {045803} (\bibinfo {year} {2021})},\ \Eprint
  {http://arxiv.org/abs/2009.04737} {arXiv:2009.04737 [nucl-th]} \BibitemShut
  {NoStop}%
\bibitem [{\citenamefont {Du}\ \emph {et~al.}(2022)\citenamefont {Du},
  \citenamefont {Steiner},\ and\ \citenamefont {Holt}}]{Du:2021rhq}%
  \BibitemOpen
  \bibfield  {author} {\bibinfo {author} {\bibfnamefont {X.}~\bibnamefont
  {Du}}, \bibinfo {author} {\bibfnamefont {A.~W.}\ \bibnamefont {Steiner}}, \
  and\ \bibinfo {author} {\bibfnamefont {J.~W.}\ \bibnamefont {Holt}},\ }\href
  {\doibase 10.1103/PhysRevC.105.035803} {\bibfield  {journal} {\bibinfo
  {journal} {Phys. Rev. C}\ }\textbf {\bibinfo {volume} {105}},\ \bibinfo
  {pages} {035803} (\bibinfo {year} {2022})},\ \Eprint
  {http://arxiv.org/abs/2107.06697} {arXiv:2107.06697 [nucl-th]} \BibitemShut
  {NoStop}%
\bibitem [{\citenamefont {Whitehead}\ \emph {et~al.}(2021)\citenamefont
  {Whitehead}, \citenamefont {Lim},\ and\ \citenamefont
  {Holt}}]{Whitehead:2020wwb}%
  \BibitemOpen
  \bibfield  {author} {\bibinfo {author} {\bibfnamefont {T.~R.}\ \bibnamefont
  {Whitehead}}, \bibinfo {author} {\bibfnamefont {Y.}~\bibnamefont {Lim}}, \
  and\ \bibinfo {author} {\bibfnamefont {J.~W.}\ \bibnamefont {Holt}},\ }\href
  {\doibase 10.1103/PhysRevLett.127.182502} {\bibfield  {journal} {\bibinfo
  {journal} {Phys. Rev. Lett.}\ }\textbf {\bibinfo {volume} {127}},\ \bibinfo
  {pages} {182502} (\bibinfo {year} {2021})},\ \Eprint
  {http://arxiv.org/abs/2009.08436} {arXiv:2009.08436 [nucl-th]} \BibitemShut
  {NoStop}%
\bibitem [{\citenamefont {Holt}\ and\ \citenamefont
  {Whitehead}(2022)}]{Holt:2022piv}%
  \BibitemOpen
  \bibfield  {author} {\bibinfo {author} {\bibfnamefont {J.~W.}\ \bibnamefont
  {Holt}}\ and\ \bibinfo {author} {\bibfnamefont {T.~R.}\ \bibnamefont
  {Whitehead}},\ }\href@noop {} {\  (\bibinfo {year} {2022})},\ \Eprint
  {http://arxiv.org/abs/2201.13404} {arXiv:2201.13404 [nucl-th]} \BibitemShut
  {NoStop}%
\bibitem [{\citenamefont {Hebborn}\ \emph {et~al.}(2022)\citenamefont {Hebborn}
  \emph {et~al.}}]{Hebborn:2022vzm}%
  \BibitemOpen
  \bibfield  {author} {\bibinfo {author} {\bibfnamefont {C.}~\bibnamefont
  {Hebborn}} \emph {et~al.},\ }\href@noop {} {\  (\bibinfo {year} {2022})},\
  \Eprint {http://arxiv.org/abs/2210.07293} {arXiv:2210.07293 [nucl-th]}
  \BibitemShut {NoStop}%
\bibitem [{\citenamefont {Freedman}\ and\ \citenamefont
  {McLerran}(1977)}]{Freedman:1976ub}%
  \BibitemOpen
  \bibfield  {author} {\bibinfo {author} {\bibfnamefont {B.~A.}\ \bibnamefont
  {Freedman}}\ and\ \bibinfo {author} {\bibfnamefont {L.~D.}\ \bibnamefont
  {McLerran}},\ }\href {\doibase 10.1103/PhysRevD.16.1169} {\bibfield
  {journal} {\bibinfo  {journal} {Phys. Rev. D}\ }\textbf {\bibinfo {volume}
  {16}},\ \bibinfo {pages} {1169} (\bibinfo {year} {1977})}\BibitemShut
  {NoStop}%
\bibitem [{\citenamefont {Baluni}(1978)}]{Baluni:1977ms}%
  \BibitemOpen
  \bibfield  {author} {\bibinfo {author} {\bibfnamefont {V.}~\bibnamefont
  {Baluni}},\ }\href {\doibase 10.1103/PhysRevD.17.2092} {\bibfield  {journal}
  {\bibinfo  {journal} {Phys. Rev. D}\ }\textbf {\bibinfo {volume} {17}},\
  \bibinfo {pages} {2092} (\bibinfo {year} {1978})}\BibitemShut {NoStop}%
\bibitem [{\citenamefont {Blaizot}\ \emph {et~al.}(2001)\citenamefont
  {Blaizot}, \citenamefont {Iancu},\ and\ \citenamefont
  {Rebhan}}]{Blaizot:2000fc}%
  \BibitemOpen
  \bibfield  {author} {\bibinfo {author} {\bibfnamefont {J.~P.}\ \bibnamefont
  {Blaizot}}, \bibinfo {author} {\bibfnamefont {E.}~\bibnamefont {Iancu}}, \
  and\ \bibinfo {author} {\bibfnamefont {A.}~\bibnamefont {Rebhan}},\ }\href
  {\doibase 10.1103/PhysRevD.63.065003} {\bibfield  {journal} {\bibinfo
  {journal} {Phys. Rev. D}\ }\textbf {\bibinfo {volume} {63}},\ \bibinfo
  {pages} {065003} (\bibinfo {year} {2001})},\ \Eprint
  {http://arxiv.org/abs/hep-ph/0005003} {arXiv:hep-ph/0005003} \BibitemShut
  {NoStop}%
\bibitem [{\citenamefont {Fraga}\ \emph {et~al.}(2001)\citenamefont {Fraga},
  \citenamefont {Pisarski},\ and\ \citenamefont
  {Schaffner-Bielich}}]{Fraga:2001id}%
  \BibitemOpen
  \bibfield  {author} {\bibinfo {author} {\bibfnamefont {E.~S.}\ \bibnamefont
  {Fraga}}, \bibinfo {author} {\bibfnamefont {R.~D.}\ \bibnamefont {Pisarski}},
  \ and\ \bibinfo {author} {\bibfnamefont {J.}~\bibnamefont
  {Schaffner-Bielich}},\ }\href {\doibase 10.1103/PhysRevD.63.121702}
  {\bibfield  {journal} {\bibinfo  {journal} {Phys. Rev. D}\ }\textbf {\bibinfo
  {volume} {63}},\ \bibinfo {pages} {121702} (\bibinfo {year} {2001})},\
  \Eprint {http://arxiv.org/abs/hep-ph/0101143} {arXiv:hep-ph/0101143}
  \BibitemShut {NoStop}%
\bibitem [{\citenamefont {Vuorinen}(2003)}]{Vuorinen:2003fs}%
  \BibitemOpen
  \bibfield  {author} {\bibinfo {author} {\bibfnamefont {A.}~\bibnamefont
  {Vuorinen}},\ }\href {\doibase 10.1103/PhysRevD.68.054017} {\bibfield
  {journal} {\bibinfo  {journal} {Phys. Rev. D}\ }\textbf {\bibinfo {volume}
  {68}},\ \bibinfo {pages} {054017} (\bibinfo {year} {2003})},\ \Eprint
  {http://arxiv.org/abs/hep-ph/0305183} {arXiv:hep-ph/0305183} \BibitemShut
  {NoStop}%
\bibitem [{\citenamefont {Freedman}\ and\ \citenamefont
  {McLerran}(1978)}]{Freedman:1977gz}%
  \BibitemOpen
  \bibfield  {author} {\bibinfo {author} {\bibfnamefont {B.}~\bibnamefont
  {Freedman}}\ and\ \bibinfo {author} {\bibfnamefont {L.~D.}\ \bibnamefont
  {McLerran}},\ }\href {\doibase 10.1103/PhysRevD.17.1109} {\bibfield
  {journal} {\bibinfo  {journal} {Phys. Rev. D}\ }\textbf {\bibinfo {volume}
  {17}},\ \bibinfo {pages} {1109} (\bibinfo {year} {1978})}\BibitemShut
  {NoStop}%
\bibitem [{\citenamefont {Farhi}\ and\ \citenamefont
  {Jaffe}(1984)}]{Farhi:1984qu}%
  \BibitemOpen
  \bibfield  {author} {\bibinfo {author} {\bibfnamefont {E.}~\bibnamefont
  {Farhi}}\ and\ \bibinfo {author} {\bibfnamefont {R.~L.}\ \bibnamefont
  {Jaffe}},\ }\href {\doibase 10.1103/PhysRevD.30.2379} {\bibfield  {journal}
  {\bibinfo  {journal} {Phys. Rev. D}\ }\textbf {\bibinfo {volume} {30}},\
  \bibinfo {pages} {2379} (\bibinfo {year} {1984})}\BibitemShut {NoStop}%
\bibitem [{\citenamefont {Fraga}\ and\ \citenamefont
  {Romatschke}(2005)}]{Fraga:2004gz}%
  \BibitemOpen
  \bibfield  {author} {\bibinfo {author} {\bibfnamefont {E.~S.}\ \bibnamefont
  {Fraga}}\ and\ \bibinfo {author} {\bibfnamefont {P.}~\bibnamefont
  {Romatschke}},\ }\href {\doibase 10.1103/PhysRevD.71.105014} {\bibfield
  {journal} {\bibinfo  {journal} {Phys. Rev. D}\ }\textbf {\bibinfo {volume}
  {71}},\ \bibinfo {pages} {105014} (\bibinfo {year} {2005})},\ \Eprint
  {http://arxiv.org/abs/hep-ph/0412298} {arXiv:hep-ph/0412298} \BibitemShut
  {NoStop}%
\bibitem [{\citenamefont {Kurkela}\ \emph {et~al.}(2010)\citenamefont
  {Kurkela}, \citenamefont {Romatschke},\ and\ \citenamefont
  {Vuorinen}}]{Kurkela:2009gj}%
  \BibitemOpen
  \bibfield  {author} {\bibinfo {author} {\bibfnamefont {A.}~\bibnamefont
  {Kurkela}}, \bibinfo {author} {\bibfnamefont {P.}~\bibnamefont {Romatschke}},
  \ and\ \bibinfo {author} {\bibfnamefont {A.}~\bibnamefont {Vuorinen}},\
  }\href {\doibase 10.1103/PhysRevD.81.105021} {\bibfield  {journal} {\bibinfo
  {journal} {Phys. Rev. D}\ }\textbf {\bibinfo {volume} {81}},\ \bibinfo
  {pages} {105021} (\bibinfo {year} {2010})},\ \Eprint
  {http://arxiv.org/abs/0912.1856} {arXiv:0912.1856 [hep-ph]} \BibitemShut
  {NoStop}%
\bibitem [{\citenamefont {Kurkela}\ and\ \citenamefont
  {Vuorinen}(2016)}]{Kurkela:2016was}%
  \BibitemOpen
  \bibfield  {author} {\bibinfo {author} {\bibfnamefont {A.}~\bibnamefont
  {Kurkela}}\ and\ \bibinfo {author} {\bibfnamefont {A.}~\bibnamefont
  {Vuorinen}},\ }\href {\doibase 10.1103/PhysRevLett.117.042501} {\bibfield
  {journal} {\bibinfo  {journal} {Phys. Rev. Lett.}\ }\textbf {\bibinfo
  {volume} {117}},\ \bibinfo {pages} {042501} (\bibinfo {year} {2016})},\
  \Eprint {http://arxiv.org/abs/1603.00750} {arXiv:1603.00750 [hep-ph]}
  \BibitemShut {NoStop}%
\bibitem [{\citenamefont {Gorda}\ and\ \citenamefont
  {S\"appi}(2022)}]{Gorda:2021gha}%
  \BibitemOpen
  \bibfield  {author} {\bibinfo {author} {\bibfnamefont {T.}~\bibnamefont
  {Gorda}}\ and\ \bibinfo {author} {\bibfnamefont {S.}~\bibnamefont
  {S\"appi}},\ }\href {\doibase 10.1103/PhysRevD.105.114005} {\bibfield
  {journal} {\bibinfo  {journal} {Phys. Rev. D}\ }\textbf {\bibinfo {volume}
  {105}},\ \bibinfo {pages} {114005} (\bibinfo {year} {2022})},\ \Eprint
  {http://arxiv.org/abs/2112.11472} {arXiv:2112.11472 [hep-ph]} \BibitemShut
  {NoStop}%
\bibitem [{\citenamefont {Gorda}\ \emph {et~al.}(2018)\citenamefont {Gorda},
  \citenamefont {Kurkela}, \citenamefont {Romatschke}, \citenamefont
  {S\"appi},\ and\ \citenamefont {Vuorinen}}]{Gorda:2018gpy}%
  \BibitemOpen
  \bibfield  {author} {\bibinfo {author} {\bibfnamefont {T.}~\bibnamefont
  {Gorda}}, \bibinfo {author} {\bibfnamefont {A.}~\bibnamefont {Kurkela}},
  \bibinfo {author} {\bibfnamefont {P.}~\bibnamefont {Romatschke}}, \bibinfo
  {author} {\bibfnamefont {M.}~\bibnamefont {S\"appi}}, \ and\ \bibinfo
  {author} {\bibfnamefont {A.}~\bibnamefont {Vuorinen}},\ }\href {\doibase
  10.1103/PhysRevLett.121.202701} {\bibfield  {journal} {\bibinfo  {journal}
  {Phys. Rev. Lett.}\ }\textbf {\bibinfo {volume} {121}},\ \bibinfo {pages}
  {202701} (\bibinfo {year} {2018})},\ \Eprint
  {http://arxiv.org/abs/1807.04120} {arXiv:1807.04120 [hep-ph]} \BibitemShut
  {NoStop}%
\bibitem [{\citenamefont {Gorda}\ \emph {et~al.}(2021)\citenamefont {Gorda},
  \citenamefont {Kurkela}, \citenamefont {Paatelainen}, \citenamefont
  {S\"appi},\ and\ \citenamefont {Vuorinen}}]{Gorda:2021znl}%
  \BibitemOpen
  \bibfield  {author} {\bibinfo {author} {\bibfnamefont {T.}~\bibnamefont
  {Gorda}}, \bibinfo {author} {\bibfnamefont {A.}~\bibnamefont {Kurkela}},
  \bibinfo {author} {\bibfnamefont {R.}~\bibnamefont {Paatelainen}}, \bibinfo
  {author} {\bibfnamefont {S.}~\bibnamefont {S\"appi}}, \ and\ \bibinfo
  {author} {\bibfnamefont {A.}~\bibnamefont {Vuorinen}},\ }\href {\doibase
  10.1103/PhysRevLett.127.162003} {\bibfield  {journal} {\bibinfo  {journal}
  {Phys. Rev. Lett.}\ }\textbf {\bibinfo {volume} {127}},\ \bibinfo {pages}
  {162003} (\bibinfo {year} {2021})},\ \Eprint
  {http://arxiv.org/abs/2103.05658} {arXiv:2103.05658 [hep-ph]} \BibitemShut
  {NoStop}%
\bibitem [{\citenamefont {Andersen}\ and\ \citenamefont
  {Strickland}(2002)}]{Andersen:2002jz}%
  \BibitemOpen
  \bibfield  {author} {\bibinfo {author} {\bibfnamefont {J.~O.}\ \bibnamefont
  {Andersen}}\ and\ \bibinfo {author} {\bibfnamefont {M.}~\bibnamefont
  {Strickland}},\ }\href {\doibase 10.1103/PhysRevD.66.105001} {\bibfield
  {journal} {\bibinfo  {journal} {Phys. Rev. D}\ }\textbf {\bibinfo {volume}
  {66}},\ \bibinfo {pages} {105001} (\bibinfo {year} {2002})},\ \Eprint
  {http://arxiv.org/abs/hep-ph/0206196} {arXiv:hep-ph/0206196} \BibitemShut
  {NoStop}%
\bibitem [{\citenamefont {Fujimoto}\ and\ \citenamefont
  {Fukushima}(2022)}]{Fujimoto:2020tjc}%
  \BibitemOpen
  \bibfield  {author} {\bibinfo {author} {\bibfnamefont {Y.}~\bibnamefont
  {Fujimoto}}\ and\ \bibinfo {author} {\bibfnamefont {K.}~\bibnamefont
  {Fukushima}},\ }\href {\doibase 10.1103/PhysRevD.105.014025} {\bibfield
  {journal} {\bibinfo  {journal} {Phys. Rev. D}\ }\textbf {\bibinfo {volume}
  {105}},\ \bibinfo {pages} {014025} (\bibinfo {year} {2022})},\ \Eprint
  {http://arxiv.org/abs/2011.10891} {arXiv:2011.10891 [hep-ph]} \BibitemShut
  {NoStop}%
\bibitem [{\citenamefont {Fernandez}\ and\ \citenamefont
  {Kneur}(2021)}]{Fernandez:2021jfr}%
  \BibitemOpen
  \bibfield  {author} {\bibinfo {author} {\bibfnamefont {L.}~\bibnamefont
  {Fernandez}}\ and\ \bibinfo {author} {\bibfnamefont {J.-L.}\ \bibnamefont
  {Kneur}},\ }\href@noop {} {\  (\bibinfo {year} {2021})},\ \Eprint
  {http://arxiv.org/abs/2109.02410} {arXiv:2109.02410 [hep-ph]} \BibitemShut
  {NoStop}%
\bibitem [{\citenamefont {Kurkela}\ \emph {et~al.}(2014)\citenamefont
  {Kurkela}, \citenamefont {Fraga}, \citenamefont {Schaffner-Bielich},\ and\
  \citenamefont {Vuorinen}}]{Kurkela:2014vha}%
  \BibitemOpen
  \bibfield  {author} {\bibinfo {author} {\bibfnamefont {A.}~\bibnamefont
  {Kurkela}}, \bibinfo {author} {\bibfnamefont {E.~S.}\ \bibnamefont {Fraga}},
  \bibinfo {author} {\bibfnamefont {J.}~\bibnamefont {Schaffner-Bielich}}, \
  and\ \bibinfo {author} {\bibfnamefont {A.}~\bibnamefont {Vuorinen}},\ }\href
  {\doibase 10.1088/0004-637X/789/2/127} {\bibfield  {journal} {\bibinfo
  {journal} {Astrophys. J.}\ }\textbf {\bibinfo {volume} {789}},\ \bibinfo
  {pages} {127} (\bibinfo {year} {2014})},\ \Eprint
  {http://arxiv.org/abs/1402.6618} {arXiv:1402.6618 [astro-ph.HE]} \BibitemShut
  {NoStop}%
\bibitem [{\citenamefont {Gorda}\ \emph {et~al.}(2022)\citenamefont {Gorda},
  \citenamefont {Komoltsev},\ and\ \citenamefont {Kurkela}}]{Gorda:2022jvk}%
  \BibitemOpen
  \bibfield  {author} {\bibinfo {author} {\bibfnamefont {T.}~\bibnamefont
  {Gorda}}, \bibinfo {author} {\bibfnamefont {O.}~\bibnamefont {Komoltsev}}, \
  and\ \bibinfo {author} {\bibfnamefont {A.}~\bibnamefont {Kurkela}},\
  }\href@noop {} {\  (\bibinfo {year} {2022})},\ \Eprint
  {http://arxiv.org/abs/2204.11877} {arXiv:2204.11877 [nucl-th]} \BibitemShut
  {NoStop}%
\bibitem [{\citenamefont {Somasundaram}\ \emph {et~al.}(2022)\citenamefont
  {Somasundaram}, \citenamefont {Tews},\ and\ \citenamefont
  {Margueron}}]{Somasundaram:2022ztm}%
  \BibitemOpen
  \bibfield  {author} {\bibinfo {author} {\bibfnamefont {R.}~\bibnamefont
  {Somasundaram}}, \bibinfo {author} {\bibfnamefont {I.}~\bibnamefont {Tews}},
  \ and\ \bibinfo {author} {\bibfnamefont {J.}~\bibnamefont {Margueron}},\
  }\href@noop {} {\  (\bibinfo {year} {2022})},\ \Eprint
  {http://arxiv.org/abs/2204.14039} {arXiv:2204.14039 [nucl-th]} \BibitemShut
  {NoStop}%
\bibitem [{\citenamefont {Ecker}\ and\ \citenamefont
  {Rezzolla}(2022)}]{Ecker:2022dlg}%
  \BibitemOpen
  \bibfield  {author} {\bibinfo {author} {\bibfnamefont {C.}~\bibnamefont
  {Ecker}}\ and\ \bibinfo {author} {\bibfnamefont {L.}~\bibnamefont
  {Rezzolla}},\ }\href@noop {} {\  (\bibinfo {year} {2022})},\ \Eprint
  {http://arxiv.org/abs/2209.08101} {arXiv:2209.08101 [astro-ph.HE]}
  \BibitemShut {NoStop}%
\bibitem [{\citenamefont {Altiparmak}\ \emph {et~al.}(2022)\citenamefont
  {Altiparmak}, \citenamefont {Ecker},\ and\ \citenamefont
  {Rezzolla}}]{Altiparmak:2022bke}%
  \BibitemOpen
  \bibfield  {author} {\bibinfo {author} {\bibfnamefont {S.}~\bibnamefont
  {Altiparmak}}, \bibinfo {author} {\bibfnamefont {C.}~\bibnamefont {Ecker}}, \
  and\ \bibinfo {author} {\bibfnamefont {L.}~\bibnamefont {Rezzolla}},\
  }\href@noop {} {\  (\bibinfo {year} {2022})},\ \Eprint
  {http://arxiv.org/abs/2203.14974} {arXiv:2203.14974 [astro-ph.HE]}
  \BibitemShut {NoStop}%
\bibitem [{\citenamefont {Parotto}\ \emph {et~al.}(2020)\citenamefont
  {Parotto}, \citenamefont {Bluhm}, \citenamefont {Mroczek}, \citenamefont
  {Nahrgang}, \citenamefont {Noronha-Hostler}, \citenamefont {Rajagopal},
  \citenamefont {Ratti}, \citenamefont {Sch\"afer},\ and\ \citenamefont
  {Stephanov}}]{Parotto:2018pwx}%
  \BibitemOpen
  \bibfield  {author} {\bibinfo {author} {\bibfnamefont {P.}~\bibnamefont
  {Parotto}}, \bibinfo {author} {\bibfnamefont {M.}~\bibnamefont {Bluhm}},
  \bibinfo {author} {\bibfnamefont {D.}~\bibnamefont {Mroczek}}, \bibinfo
  {author} {\bibfnamefont {M.}~\bibnamefont {Nahrgang}}, \bibinfo {author}
  {\bibfnamefont {J.}~\bibnamefont {Noronha-Hostler}}, \bibinfo {author}
  {\bibfnamefont {K.}~\bibnamefont {Rajagopal}}, \bibinfo {author}
  {\bibfnamefont {C.}~\bibnamefont {Ratti}}, \bibinfo {author} {\bibfnamefont
  {T.}~\bibnamefont {Sch\"afer}}, \ and\ \bibinfo {author} {\bibfnamefont
  {M.}~\bibnamefont {Stephanov}},\ }\href {\doibase
  10.1103/PhysRevC.101.034901} {\bibfield  {journal} {\bibinfo  {journal}
  {Phys. Rev. C}\ }\textbf {\bibinfo {volume} {101}},\ \bibinfo {pages}
  {034901} (\bibinfo {year} {2020})},\ \Eprint
  {http://arxiv.org/abs/1805.05249} {arXiv:1805.05249 [hep-ph]} \BibitemShut
  {NoStop}%
\bibitem [{\citenamefont {Karthein}\ \emph {et~al.}(2021)\citenamefont
  {Karthein}, \citenamefont {Mroczek}, \citenamefont {Nava~Acuna},
  \citenamefont {Noronha-Hostler}, \citenamefont {Parotto}, \citenamefont
  {Price},\ and\ \citenamefont {Ratti}}]{Karthein:2021nxe}%
  \BibitemOpen
  \bibfield  {author} {\bibinfo {author} {\bibfnamefont {J.~M.}\ \bibnamefont
  {Karthein}}, \bibinfo {author} {\bibfnamefont {D.}~\bibnamefont {Mroczek}},
  \bibinfo {author} {\bibfnamefont {A.~R.}\ \bibnamefont {Nava~Acuna}},
  \bibinfo {author} {\bibfnamefont {J.}~\bibnamefont {Noronha-Hostler}},
  \bibinfo {author} {\bibfnamefont {P.}~\bibnamefont {Parotto}}, \bibinfo
  {author} {\bibfnamefont {D.~R.~P.}\ \bibnamefont {Price}}, \ and\ \bibinfo
  {author} {\bibfnamefont {C.}~\bibnamefont {Ratti}},\ }\href {\doibase
  10.1140/epjp/s13360-021-01615-5} {\bibfield  {journal} {\bibinfo  {journal}
  {Eur. Phys. J. Plus}\ }\textbf {\bibinfo {volume} {136}},\ \bibinfo {pages}
  {621} (\bibinfo {year} {2021})},\ \Eprint {http://arxiv.org/abs/2103.08146}
  {arXiv:2103.08146 [hep-ph]} \BibitemShut {NoStop}%
\bibitem [{\citenamefont {Shen}(2021)}]{Shen:2020gef}%
  \BibitemOpen
  \bibfield  {author} {\bibinfo {author} {\bibfnamefont {C.}~\bibnamefont
  {Shen}},\ }\href {\doibase 10.1016/j.nuclphysa.2020.121788} {\bibfield
  {journal} {\bibinfo  {journal} {Nucl. Phys. A}\ }\textbf {\bibinfo {volume}
  {1005}},\ \bibinfo {pages} {121788} (\bibinfo {year} {2021})},\ \Eprint
  {http://arxiv.org/abs/2001.11858} {arXiv:2001.11858 [nucl-th]} \BibitemShut
  {NoStop}%
\bibitem [{\citenamefont {Dore}\ \emph {et~al.}(2022)\citenamefont {Dore},
  \citenamefont {Karthein}, \citenamefont {Long}, \citenamefont {Mroczek},
  \citenamefont {Noronha-Hostler}, \citenamefont {Parotto}, \citenamefont
  {Ratti},\ and\ \citenamefont {Yamauchi}}]{Dore:2022qyz}%
  \BibitemOpen
  \bibfield  {author} {\bibinfo {author} {\bibfnamefont {T.}~\bibnamefont
  {Dore}}, \bibinfo {author} {\bibfnamefont {J.~M.}\ \bibnamefont {Karthein}},
  \bibinfo {author} {\bibfnamefont {I.}~\bibnamefont {Long}}, \bibinfo {author}
  {\bibfnamefont {D.}~\bibnamefont {Mroczek}}, \bibinfo {author} {\bibfnamefont
  {J.}~\bibnamefont {Noronha-Hostler}}, \bibinfo {author} {\bibfnamefont
  {P.}~\bibnamefont {Parotto}}, \bibinfo {author} {\bibfnamefont
  {C.}~\bibnamefont {Ratti}}, \ and\ \bibinfo {author} {\bibfnamefont
  {Y.}~\bibnamefont {Yamauchi}},\ }\href@noop {} {\  (\bibinfo {year}
  {2022})},\ \Eprint {http://arxiv.org/abs/2207.04086} {arXiv:2207.04086
  [nucl-th]} \BibitemShut {NoStop}%
\bibitem [{\citenamefont {Rougemont}\ \emph {et~al.}(2017)\citenamefont
  {Rougemont}, \citenamefont {Critelli}, \citenamefont {Noronha-Hostler},
  \citenamefont {Noronha},\ and\ \citenamefont {Ratti}}]{Rougemont:2017tlu}%
  \BibitemOpen
  \bibfield  {author} {\bibinfo {author} {\bibfnamefont {R.}~\bibnamefont
  {Rougemont}}, \bibinfo {author} {\bibfnamefont {R.}~\bibnamefont {Critelli}},
  \bibinfo {author} {\bibfnamefont {J.}~\bibnamefont {Noronha-Hostler}},
  \bibinfo {author} {\bibfnamefont {J.}~\bibnamefont {Noronha}}, \ and\
  \bibinfo {author} {\bibfnamefont {C.}~\bibnamefont {Ratti}},\ }\href
  {\doibase 10.1103/PhysRevD.96.014032} {\bibfield  {journal} {\bibinfo
  {journal} {Phys. Rev. D}\ }\textbf {\bibinfo {volume} {96}},\ \bibinfo
  {pages} {014032} (\bibinfo {year} {2017})},\ \Eprint
  {http://arxiv.org/abs/1704.05558} {arXiv:1704.05558 [hep-ph]} \BibitemShut
  {NoStop}%
\bibitem [{\citenamefont {Critelli}\ \emph {et~al.}(2017)\citenamefont
  {Critelli}, \citenamefont {Noronha}, \citenamefont {Noronha-Hostler},
  \citenamefont {Portillo}, \citenamefont {Ratti},\ and\ \citenamefont
  {Rougemont}}]{Critelli:2017oub}%
  \BibitemOpen
  \bibfield  {author} {\bibinfo {author} {\bibfnamefont {R.}~\bibnamefont
  {Critelli}}, \bibinfo {author} {\bibfnamefont {J.}~\bibnamefont {Noronha}},
  \bibinfo {author} {\bibfnamefont {J.}~\bibnamefont {Noronha-Hostler}},
  \bibinfo {author} {\bibfnamefont {I.}~\bibnamefont {Portillo}}, \bibinfo
  {author} {\bibfnamefont {C.}~\bibnamefont {Ratti}}, \ and\ \bibinfo {author}
  {\bibfnamefont {R.}~\bibnamefont {Rougemont}},\ }\href {\doibase
  10.1103/PhysRevD.96.096026} {\bibfield  {journal} {\bibinfo  {journal} {Phys.
  Rev. D}\ }\textbf {\bibinfo {volume} {96}},\ \bibinfo {pages} {096026}
  (\bibinfo {year} {2017})},\ \Eprint {http://arxiv.org/abs/1706.00455}
  {arXiv:1706.00455 [nucl-th]} \BibitemShut {NoStop}%
\bibitem [{\citenamefont {Grefa}\ \emph {et~al.}(2021)\citenamefont {Grefa},
  \citenamefont {Noronha}, \citenamefont {Noronha-Hostler}, \citenamefont
  {Portillo}, \citenamefont {Ratti},\ and\ \citenamefont
  {Rougemont}}]{Grefa:2021qvt}%
  \BibitemOpen
  \bibfield  {author} {\bibinfo {author} {\bibfnamefont {J.}~\bibnamefont
  {Grefa}}, \bibinfo {author} {\bibfnamefont {J.}~\bibnamefont {Noronha}},
  \bibinfo {author} {\bibfnamefont {J.}~\bibnamefont {Noronha-Hostler}},
  \bibinfo {author} {\bibfnamefont {I.}~\bibnamefont {Portillo}}, \bibinfo
  {author} {\bibfnamefont {C.}~\bibnamefont {Ratti}}, \ and\ \bibinfo {author}
  {\bibfnamefont {R.}~\bibnamefont {Rougemont}},\ }\href {\doibase
  10.1103/PhysRevD.104.034002} {\bibfield  {journal} {\bibinfo  {journal}
  {Phys. Rev. D}\ }\textbf {\bibinfo {volume} {104}},\ \bibinfo {pages}
  {034002} (\bibinfo {year} {2021})},\ \Eprint
  {http://arxiv.org/abs/2102.12042} {arXiv:2102.12042 [nucl-th]} \BibitemShut
  {NoStop}%
\bibitem [{\citenamefont {Grefa}\ \emph {et~al.}(2022)\citenamefont {Grefa},
  \citenamefont {Hippert}, \citenamefont {Noronha}, \citenamefont
  {Noronha-Hostler}, \citenamefont {Portillo}, \citenamefont {Ratti},\ and\
  \citenamefont {Rougemont}}]{Grefa:2022sav}%
  \BibitemOpen
  \bibfield  {author} {\bibinfo {author} {\bibfnamefont {J.}~\bibnamefont
  {Grefa}}, \bibinfo {author} {\bibfnamefont {M.}~\bibnamefont {Hippert}},
  \bibinfo {author} {\bibfnamefont {J.}~\bibnamefont {Noronha}}, \bibinfo
  {author} {\bibfnamefont {J.}~\bibnamefont {Noronha-Hostler}}, \bibinfo
  {author} {\bibfnamefont {I.}~\bibnamefont {Portillo}}, \bibinfo {author}
  {\bibfnamefont {C.}~\bibnamefont {Ratti}}, \ and\ \bibinfo {author}
  {\bibfnamefont {R.}~\bibnamefont {Rougemont}},\ }\href {\doibase
  10.1103/PhysRevD.106.034024} {\bibfield  {journal} {\bibinfo  {journal}
  {Phys. Rev. D}\ }\textbf {\bibinfo {volume} {106}},\ \bibinfo {pages}
  {034024} (\bibinfo {year} {2022})},\ \Eprint
  {http://arxiv.org/abs/2203.00139} {arXiv:2203.00139 [nucl-th]} \BibitemShut
  {NoStop}%
\bibitem [{\citenamefont {Dexheimer}\ and\ \citenamefont
  {Schramm}(2010)}]{Dexheimer:2009hi}%
  \BibitemOpen
  \bibfield  {author} {\bibinfo {author} {\bibfnamefont {V.~A.}\ \bibnamefont
  {Dexheimer}}\ and\ \bibinfo {author} {\bibfnamefont {S.}~\bibnamefont
  {Schramm}},\ }\href {\doibase 10.1103/PhysRevC.81.045201} {\bibfield
  {journal} {\bibinfo  {journal} {Phys. Rev. C}\ }\textbf {\bibinfo {volume}
  {81}},\ \bibinfo {pages} {045201} (\bibinfo {year} {2010})},\ \Eprint
  {http://arxiv.org/abs/0901.1748} {arXiv:0901.1748 [astro-ph.SR]} \BibitemShut
  {NoStop}%
\bibitem [{\citenamefont {Motornenko}\ \emph {et~al.}(2020)\citenamefont
  {Motornenko}, \citenamefont {Steinheimer}, \citenamefont {Vovchenko},
  \citenamefont {Schramm},\ and\ \citenamefont
  {Stoecker}}]{Motornenko:2019arp}%
  \BibitemOpen
  \bibfield  {author} {\bibinfo {author} {\bibfnamefont {A.}~\bibnamefont
  {Motornenko}}, \bibinfo {author} {\bibfnamefont {J.}~\bibnamefont
  {Steinheimer}}, \bibinfo {author} {\bibfnamefont {V.}~\bibnamefont
  {Vovchenko}}, \bibinfo {author} {\bibfnamefont {S.}~\bibnamefont {Schramm}},
  \ and\ \bibinfo {author} {\bibfnamefont {H.}~\bibnamefont {Stoecker}},\
  }\href {\doibase 10.1103/PhysRevC.101.034904} {\bibfield  {journal} {\bibinfo
   {journal} {Phys. Rev. C}\ }\textbf {\bibinfo {volume} {101}},\ \bibinfo
  {pages} {034904} (\bibinfo {year} {2020})},\ \Eprint
  {http://arxiv.org/abs/1905.00866} {arXiv:1905.00866 [hep-ph]} \BibitemShut
  {NoStop}%
\bibitem [{\citenamefont {Chan}\ \emph {et~al.}(2022)\citenamefont {Chan},
  \citenamefont {DeBoer}, \citenamefont {Furnstahl}, \citenamefont {Liyanage},
  \citenamefont {Nunes}, \citenamefont {Odell}, \citenamefont {Phillips},
  \citenamefont {Plumlee}, \citenamefont {Semposki}, \citenamefont {S\"urer},\
  and\ \citenamefont {Wild}}]{bandframework}%
  \BibitemOpen
  \bibfield  {author} {\bibinfo {author} {\bibfnamefont {M.~Y.-H.}\
  \bibnamefont {Chan}}, \bibinfo {author} {\bibfnamefont {R.~J.}\ \bibnamefont
  {DeBoer}}, \bibinfo {author} {\bibfnamefont {R.~J.}\ \bibnamefont
  {Furnstahl}}, \bibinfo {author} {\bibfnamefont {D.}~\bibnamefont {Liyanage}},
  \bibinfo {author} {\bibfnamefont {F.~M.}\ \bibnamefont {Nunes}}, \bibinfo
  {author} {\bibfnamefont {D.}~\bibnamefont {Odell}}, \bibinfo {author}
  {\bibfnamefont {D.~R.}\ \bibnamefont {Phillips}}, \bibinfo {author}
  {\bibfnamefont {M.}~\bibnamefont {Plumlee}}, \bibinfo {author} {\bibfnamefont
  {A.~C.}\ \bibnamefont {Semposki}}, \bibinfo {author} {\bibfnamefont
  {O.}~\bibnamefont {S\"urer}}, \ and\ \bibinfo {author} {\bibfnamefont
  {S.~M.}\ \bibnamefont {Wild}},\ }\href
  {https://github.com/bandframework/bandframework} {\emph {\bibinfo {title}
  {{BANDFramework: An} Open-Source Framework for {B}ayesian Analysis of Nuclear
  Dynamics}}},\ \bibinfo {type} {Tech. Rep.}\ \bibinfo {number} {Version
  0.2.0}\ (\bibinfo {year} {2022})\BibitemShut {NoStop}%
\bibitem [{\citenamefont {Most}\ \emph
  {et~al.}(2022{\natexlab{a}})\citenamefont {Most}, \citenamefont {Motornenko},
  \citenamefont {Steinheimer}, \citenamefont {Dexheimer}, \citenamefont
  {Hanauske}, \citenamefont {Rezzolla},\ and\ \citenamefont
  {Stoecker}}]{Most:2022wgo}%
  \BibitemOpen
  \bibfield  {author} {\bibinfo {author} {\bibfnamefont {E.~R.}\ \bibnamefont
  {Most}}, \bibinfo {author} {\bibfnamefont {A.}~\bibnamefont {Motornenko}},
  \bibinfo {author} {\bibfnamefont {J.}~\bibnamefont {Steinheimer}}, \bibinfo
  {author} {\bibfnamefont {V.}~\bibnamefont {Dexheimer}}, \bibinfo {author}
  {\bibfnamefont {M.}~\bibnamefont {Hanauske}}, \bibinfo {author}
  {\bibfnamefont {L.}~\bibnamefont {Rezzolla}}, \ and\ \bibinfo {author}
  {\bibfnamefont {H.}~\bibnamefont {Stoecker}},\ }\href@noop {} {\  (\bibinfo
  {year} {2022}{\natexlab{a}})},\ \Eprint {http://arxiv.org/abs/2201.13150}
  {arXiv:2201.13150 [nucl-th]} \BibitemShut {NoStop}%
\bibitem [{\citenamefont {Alford}\ \emph {et~al.}(2013)\citenamefont {Alford},
  \citenamefont {Han},\ and\ \citenamefont {Prakash}}]{Alford:2013aca}%
  \BibitemOpen
  \bibfield  {author} {\bibinfo {author} {\bibfnamefont {M.~G.}\ \bibnamefont
  {Alford}}, \bibinfo {author} {\bibfnamefont {S.}~\bibnamefont {Han}}, \ and\
  \bibinfo {author} {\bibfnamefont {M.}~\bibnamefont {Prakash}},\ }\href
  {\doibase 10.1103/PhysRevD.88.083013} {\bibfield  {journal} {\bibinfo
  {journal} {Phys. Rev. D}\ }\textbf {\bibinfo {volume} {88}},\ \bibinfo
  {pages} {083013} (\bibinfo {year} {2013})},\ \Eprint
  {http://arxiv.org/abs/1302.4732} {arXiv:1302.4732 [astro-ph.SR]} \BibitemShut
  {NoStop}%
\bibitem [{\citenamefont {Benic}\ \emph {et~al.}(2015)\citenamefont {Benic},
  \citenamefont {Blaschke}, \citenamefont {Alvarez-Castillo}, \citenamefont
  {Fischer},\ and\ \citenamefont {Typel}}]{Benic:2014jia}%
  \BibitemOpen
  \bibfield  {author} {\bibinfo {author} {\bibfnamefont {S.}~\bibnamefont
  {Benic}}, \bibinfo {author} {\bibfnamefont {D.}~\bibnamefont {Blaschke}},
  \bibinfo {author} {\bibfnamefont {D.~E.}\ \bibnamefont {Alvarez-Castillo}},
  \bibinfo {author} {\bibfnamefont {T.}~\bibnamefont {Fischer}}, \ and\
  \bibinfo {author} {\bibfnamefont {S.}~\bibnamefont {Typel}},\ }\href
  {\doibase 10.1051/0004-6361/201425318} {\bibfield  {journal} {\bibinfo
  {journal} {Astron. Astrophys.}\ }\textbf {\bibinfo {volume} {577}},\ \bibinfo
  {pages} {A40} (\bibinfo {year} {2015})},\ \Eprint
  {http://arxiv.org/abs/1411.2856} {arXiv:1411.2856 [astro-ph.HE]} \BibitemShut
  {NoStop}%
\bibitem [{\citenamefont {Dexheimer}\ \emph {et~al.}(2015)\citenamefont
  {Dexheimer}, \citenamefont {Negreiros},\ and\ \citenamefont
  {Schramm}}]{Dexheimer:2014pea}%
  \BibitemOpen
  \bibfield  {author} {\bibinfo {author} {\bibfnamefont {V.}~\bibnamefont
  {Dexheimer}}, \bibinfo {author} {\bibfnamefont {R.}~\bibnamefont
  {Negreiros}}, \ and\ \bibinfo {author} {\bibfnamefont {S.}~\bibnamefont
  {Schramm}},\ }\href {\doibase 10.1103/PhysRevC.91.055808} {\bibfield
  {journal} {\bibinfo  {journal} {Phys. Rev. C}\ }\textbf {\bibinfo {volume}
  {91}},\ \bibinfo {pages} {055808} (\bibinfo {year} {2015})},\ \Eprint
  {http://arxiv.org/abs/1411.4623} {arXiv:1411.4623 [astro-ph.HE]} \BibitemShut
  {NoStop}%
\bibitem [{\citenamefont {Sagert}\ \emph {et~al.}(2009)\citenamefont {Sagert},
  \citenamefont {Fischer}, \citenamefont {Hempel}, \citenamefont {Pagliara},
  \citenamefont {Schaffner-Bielich}, \citenamefont {Mezzacappa}, \citenamefont
  {Thielemann},\ and\ \citenamefont {Liebendorfer}}]{Sagert:2008ka}%
  \BibitemOpen
  \bibfield  {author} {\bibinfo {author} {\bibfnamefont {I.}~\bibnamefont
  {Sagert}}, \bibinfo {author} {\bibfnamefont {T.}~\bibnamefont {Fischer}},
  \bibinfo {author} {\bibfnamefont {M.}~\bibnamefont {Hempel}}, \bibinfo
  {author} {\bibfnamefont {G.}~\bibnamefont {Pagliara}}, \bibinfo {author}
  {\bibfnamefont {J.}~\bibnamefont {Schaffner-Bielich}}, \bibinfo {author}
  {\bibfnamefont {A.}~\bibnamefont {Mezzacappa}}, \bibinfo {author}
  {\bibfnamefont {F.~K.}\ \bibnamefont {Thielemann}}, \ and\ \bibinfo {author}
  {\bibfnamefont {M.}~\bibnamefont {Liebendorfer}},\ }\href {\doibase
  10.1103/PhysRevLett.102.081101} {\bibfield  {journal} {\bibinfo  {journal}
  {Phys. Rev. Lett.}\ }\textbf {\bibinfo {volume} {102}},\ \bibinfo {pages}
  {081101} (\bibinfo {year} {2009})},\ \Eprint {http://arxiv.org/abs/0809.4225}
  {arXiv:0809.4225 [astro-ph]} \BibitemShut {NoStop}%
\bibitem [{\citenamefont {Fischer}\ \emph {et~al.}(2018)\citenamefont
  {Fischer}, \citenamefont {Bastian}, \citenamefont {Wu}, \citenamefont
  {Baklanov}, \citenamefont {Sorokina}, \citenamefont {Blinnikov},
  \citenamefont {Typel}, \citenamefont {Kl\"ahn},\ and\ \citenamefont
  {Blaschke}}]{Fischer:2017lag}%
  \BibitemOpen
  \bibfield  {author} {\bibinfo {author} {\bibfnamefont {T.}~\bibnamefont
  {Fischer}}, \bibinfo {author} {\bibfnamefont {N.-U.~F.}\ \bibnamefont
  {Bastian}}, \bibinfo {author} {\bibfnamefont {M.-R.}\ \bibnamefont {Wu}},
  \bibinfo {author} {\bibfnamefont {P.}~\bibnamefont {Baklanov}}, \bibinfo
  {author} {\bibfnamefont {E.}~\bibnamefont {Sorokina}}, \bibinfo {author}
  {\bibfnamefont {S.}~\bibnamefont {Blinnikov}}, \bibinfo {author}
  {\bibfnamefont {S.}~\bibnamefont {Typel}}, \bibinfo {author} {\bibfnamefont
  {T.}~\bibnamefont {Kl\"ahn}}, \ and\ \bibinfo {author} {\bibfnamefont
  {D.~B.}\ \bibnamefont {Blaschke}},\ }\href {\doibase
  10.1038/s41550-018-0583-0} {\bibfield  {journal} {\bibinfo  {journal} {Nature
  Astron.}\ }\textbf {\bibinfo {volume} {2}},\ \bibinfo {pages} {980} (\bibinfo
  {year} {2018})},\ \Eprint {http://arxiv.org/abs/1712.08788} {arXiv:1712.08788
  [astro-ph.HE]} \BibitemShut {NoStop}%
\bibitem [{\citenamefont {Bauswein}\ \emph {et~al.}(2019)\citenamefont
  {Bauswein}, \citenamefont {Bastian}, \citenamefont {Blaschke}, \citenamefont
  {Chatziioannou}, \citenamefont {Clark}, \citenamefont {Fischer},\ and\
  \citenamefont {Oertel}}]{Bauswein:2018bma}%
  \BibitemOpen
  \bibfield  {author} {\bibinfo {author} {\bibfnamefont {A.}~\bibnamefont
  {Bauswein}}, \bibinfo {author} {\bibfnamefont {N.-U.~F.}\ \bibnamefont
  {Bastian}}, \bibinfo {author} {\bibfnamefont {D.~B.}\ \bibnamefont
  {Blaschke}}, \bibinfo {author} {\bibfnamefont {K.}~\bibnamefont
  {Chatziioannou}}, \bibinfo {author} {\bibfnamefont {J.~A.}\ \bibnamefont
  {Clark}}, \bibinfo {author} {\bibfnamefont {T.}~\bibnamefont {Fischer}}, \
  and\ \bibinfo {author} {\bibfnamefont {M.}~\bibnamefont {Oertel}},\ }\href
  {\doibase 10.1103/PhysRevLett.122.061102} {\bibfield  {journal} {\bibinfo
  {journal} {Phys. Rev. Lett.}\ }\textbf {\bibinfo {volume} {122}},\ \bibinfo
  {pages} {061102} (\bibinfo {year} {2019})},\ \Eprint
  {http://arxiv.org/abs/1809.01116} {arXiv:1809.01116 [astro-ph.HE]}
  \BibitemShut {NoStop}%
\bibitem [{\citenamefont {Bernhard}\ \emph {et~al.}(2019)\citenamefont
  {Bernhard}, \citenamefont {Moreland},\ and\ \citenamefont
  {Bass}}]{Bernhard:2019bmu}%
  \BibitemOpen
  \bibfield  {author} {\bibinfo {author} {\bibfnamefont {J.~E.}\ \bibnamefont
  {Bernhard}}, \bibinfo {author} {\bibfnamefont {J.~S.}\ \bibnamefont
  {Moreland}}, \ and\ \bibinfo {author} {\bibfnamefont {S.~A.}\ \bibnamefont
  {Bass}},\ }\href {\doibase 10.1038/s41567-019-0611-8} {\bibfield  {journal}
  {\bibinfo  {journal} {Nature Phys.}\ }\textbf {\bibinfo {volume} {15}},\
  \bibinfo {pages} {1113} (\bibinfo {year} {2019})}\BibitemShut {NoStop}%
\bibitem [{\citenamefont {Soloveva}\ \emph {et~al.}(2020)\citenamefont
  {Soloveva}, \citenamefont {Moreau},\ and\ \citenamefont
  {Bratkovskaya}}]{Soloveva:2019xph}%
  \BibitemOpen
  \bibfield  {author} {\bibinfo {author} {\bibfnamefont {O.}~\bibnamefont
  {Soloveva}}, \bibinfo {author} {\bibfnamefont {P.}~\bibnamefont {Moreau}}, \
  and\ \bibinfo {author} {\bibfnamefont {E.}~\bibnamefont {Bratkovskaya}},\
  }\href {\doibase 10.1103/PhysRevC.101.045203} {\bibfield  {journal} {\bibinfo
   {journal} {Phys. Rev. C}\ }\textbf {\bibinfo {volume} {101}},\ \bibinfo
  {pages} {045203} (\bibinfo {year} {2020})},\ \Eprint
  {http://arxiv.org/abs/1911.08547} {arXiv:1911.08547 [nucl-th]} \BibitemShut
  {NoStop}%
\bibitem [{\citenamefont {Rose}\ \emph {et~al.}(2018)\citenamefont {Rose},
  \citenamefont {Torres-Rincon}, \citenamefont {Sch\"afer}, \citenamefont
  {Oliinychenko},\ and\ \citenamefont {Petersen}}]{Rose:2017bjz}%
  \BibitemOpen
  \bibfield  {author} {\bibinfo {author} {\bibfnamefont {J.~B.}\ \bibnamefont
  {Rose}}, \bibinfo {author} {\bibfnamefont {J.~M.}\ \bibnamefont
  {Torres-Rincon}}, \bibinfo {author} {\bibfnamefont {A.}~\bibnamefont
  {Sch\"afer}}, \bibinfo {author} {\bibfnamefont {D.~R.}\ \bibnamefont
  {Oliinychenko}}, \ and\ \bibinfo {author} {\bibfnamefont {H.}~\bibnamefont
  {Petersen}},\ }\href {\doibase 10.1103/PhysRevC.97.055204} {\bibfield
  {journal} {\bibinfo  {journal} {Phys. Rev. C}\ }\textbf {\bibinfo {volume}
  {97}},\ \bibinfo {pages} {055204} (\bibinfo {year} {2018})},\ \Eprint
  {http://arxiv.org/abs/1709.03826} {arXiv:1709.03826 [nucl-th]} \BibitemShut
  {NoStop}%
\bibitem [{\citenamefont {McLaughlin}\ \emph {et~al.}(2022)\citenamefont
  {McLaughlin}, \citenamefont {Rose}, \citenamefont {Dore}, \citenamefont
  {Parotto}, \citenamefont {Ratti},\ and\ \citenamefont
  {Noronha-Hostler}}]{McLaughlin:2021dph}%
  \BibitemOpen
  \bibfield  {author} {\bibinfo {author} {\bibfnamefont {E.}~\bibnamefont
  {McLaughlin}}, \bibinfo {author} {\bibfnamefont {J.}~\bibnamefont {Rose}},
  \bibinfo {author} {\bibfnamefont {T.}~\bibnamefont {Dore}}, \bibinfo {author}
  {\bibfnamefont {P.}~\bibnamefont {Parotto}}, \bibinfo {author} {\bibfnamefont
  {C.}~\bibnamefont {Ratti}}, \ and\ \bibinfo {author} {\bibfnamefont
  {J.}~\bibnamefont {Noronha-Hostler}},\ }\href {\doibase
  10.1103/PhysRevC.105.024903} {\bibfield  {journal} {\bibinfo  {journal}
  {Phys. Rev. C}\ }\textbf {\bibinfo {volume} {105}},\ \bibinfo {pages}
  {024903} (\bibinfo {year} {2022})},\ \Eprint
  {http://arxiv.org/abs/2103.02090} {arXiv:2103.02090 [nucl-th]} \BibitemShut
  {NoStop}%
\bibitem [{\citenamefont {Greif}\ \emph {et~al.}(2018)\citenamefont {Greif},
  \citenamefont {Fotakis}, \citenamefont {Denicol},\ and\ \citenamefont
  {Greiner}}]{Greif:2017byw}%
  \BibitemOpen
  \bibfield  {author} {\bibinfo {author} {\bibfnamefont {M.}~\bibnamefont
  {Greif}}, \bibinfo {author} {\bibfnamefont {J.~A.}\ \bibnamefont {Fotakis}},
  \bibinfo {author} {\bibfnamefont {G.~S.}\ \bibnamefont {Denicol}}, \ and\
  \bibinfo {author} {\bibfnamefont {C.}~\bibnamefont {Greiner}},\ }\href
  {\doibase 10.1103/PhysRevLett.120.242301} {\bibfield  {journal} {\bibinfo
  {journal} {Phys. Rev. Lett.}\ }\textbf {\bibinfo {volume} {120}},\ \bibinfo
  {pages} {242301} (\bibinfo {year} {2018})},\ \Eprint
  {http://arxiv.org/abs/1711.08680} {arXiv:1711.08680 [hep-ph]} \BibitemShut
  {NoStop}%
\bibitem [{\citenamefont {Almaalol}\ \emph
  {et~al.}(2022{\natexlab{b}})\citenamefont {Almaalol}, \citenamefont {Dore},\
  and\ \citenamefont {Noronha-Hostler}}]{Almaalol:2022pjc}%
  \BibitemOpen
  \bibfield  {author} {\bibinfo {author} {\bibfnamefont {D.}~\bibnamefont
  {Almaalol}}, \bibinfo {author} {\bibfnamefont {T.}~\bibnamefont {Dore}}, \
  and\ \bibinfo {author} {\bibfnamefont {J.}~\bibnamefont {Noronha-Hostler}},\
  }\href@noop {} {\  (\bibinfo {year} {2022}{\natexlab{b}})},\ \Eprint
  {http://arxiv.org/abs/2209.11210} {arXiv:2209.11210 [hep-th]} \BibitemShut
  {NoStop}%
\bibitem [{\citenamefont {Alford}\ \emph {et~al.}(2018)\citenamefont {Alford},
  \citenamefont {Bovard}, \citenamefont {Hanauske}, \citenamefont {Rezzolla},\
  and\ \citenamefont {Schwenzer}}]{Alford:2017rxf}%
  \BibitemOpen
  \bibfield  {author} {\bibinfo {author} {\bibfnamefont {M.~G.}\ \bibnamefont
  {Alford}}, \bibinfo {author} {\bibfnamefont {L.}~\bibnamefont {Bovard}},
  \bibinfo {author} {\bibfnamefont {M.}~\bibnamefont {Hanauske}}, \bibinfo
  {author} {\bibfnamefont {L.}~\bibnamefont {Rezzolla}}, \ and\ \bibinfo
  {author} {\bibfnamefont {K.}~\bibnamefont {Schwenzer}},\ }\href {\doibase
  10.1103/PhysRevLett.120.041101} {\bibfield  {journal} {\bibinfo  {journal}
  {Phys. Rev. Lett.}\ }\textbf {\bibinfo {volume} {120}},\ \bibinfo {pages}
  {041101} (\bibinfo {year} {2018})},\ \Eprint
  {http://arxiv.org/abs/1707.09475} {arXiv:1707.09475 [gr-qc]} \BibitemShut
  {NoStop}%
\bibitem [{\citenamefont {Most}\ \emph {et~al.}(2021)\citenamefont {Most},
  \citenamefont {Harris}, \citenamefont {Plumberg}, \citenamefont {Alford},
  \citenamefont {Noronha}, \citenamefont {Noronha-Hostler}, \citenamefont
  {Pretorius}, \citenamefont {Witek},\ and\ \citenamefont
  {Yunes}}]{Most:2021zvc}%
  \BibitemOpen
  \bibfield  {author} {\bibinfo {author} {\bibfnamefont {E.~R.}\ \bibnamefont
  {Most}}, \bibinfo {author} {\bibfnamefont {S.~P.}\ \bibnamefont {Harris}},
  \bibinfo {author} {\bibfnamefont {C.}~\bibnamefont {Plumberg}}, \bibinfo
  {author} {\bibfnamefont {M.~G.}\ \bibnamefont {Alford}}, \bibinfo {author}
  {\bibfnamefont {J.}~\bibnamefont {Noronha}}, \bibinfo {author} {\bibfnamefont
  {J.}~\bibnamefont {Noronha-Hostler}}, \bibinfo {author} {\bibfnamefont
  {F.}~\bibnamefont {Pretorius}}, \bibinfo {author} {\bibfnamefont
  {H.}~\bibnamefont {Witek}}, \ and\ \bibinfo {author} {\bibfnamefont
  {N.}~\bibnamefont {Yunes}},\ }\href {\doibase 10.1093/mnras/stab2793}
  {\bibfield  {journal} {\bibinfo  {journal} {Mon. Not. Roy. Astron. Soc.}\
  }\textbf {\bibinfo {volume} {509}},\ \bibinfo {pages} {1096} (\bibinfo {year}
  {2021})},\ \Eprint {http://arxiv.org/abs/2107.05094} {arXiv:2107.05094
  [astro-ph.HE]} \BibitemShut {NoStop}%
\bibitem [{\citenamefont {Most}\ \emph
  {et~al.}(2022{\natexlab{b}})\citenamefont {Most}, \citenamefont {Haber},
  \citenamefont {Harris}, \citenamefont {Zhang}, \citenamefont {Alford},\ and\
  \citenamefont {Noronha}}]{Most:2022yhe}%
  \BibitemOpen
  \bibfield  {author} {\bibinfo {author} {\bibfnamefont {E.~R.}\ \bibnamefont
  {Most}}, \bibinfo {author} {\bibfnamefont {A.}~\bibnamefont {Haber}},
  \bibinfo {author} {\bibfnamefont {S.~P.}\ \bibnamefont {Harris}}, \bibinfo
  {author} {\bibfnamefont {Z.}~\bibnamefont {Zhang}}, \bibinfo {author}
  {\bibfnamefont {M.~G.}\ \bibnamefont {Alford}}, \ and\ \bibinfo {author}
  {\bibfnamefont {J.}~\bibnamefont {Noronha}},\ }\href@noop {} {\  (\bibinfo
  {year} {2022}{\natexlab{b}})},\ \Eprint {http://arxiv.org/abs/2207.00442}
  {arXiv:2207.00442 [astro-ph.HE]} \BibitemShut {NoStop}%
\bibitem [{\citenamefont {Yakovlev}\ and\ \citenamefont
  {Pethick}(2004)}]{Yakovlev:2004iq}%
  \BibitemOpen
  \bibfield  {author} {\bibinfo {author} {\bibfnamefont {D.~G.}\ \bibnamefont
  {Yakovlev}}\ and\ \bibinfo {author} {\bibfnamefont {C.~J.}\ \bibnamefont
  {Pethick}},\ }\href {\doibase 10.1146/annurev.astro.42.053102.134013}
  {\bibfield  {journal} {\bibinfo  {journal} {Ann. Rev. Astron. Astrophys.}\
  }\textbf {\bibinfo {volume} {42}},\ \bibinfo {pages} {169} (\bibinfo {year}
  {2004})},\ \Eprint {http://arxiv.org/abs/astro-ph/0402143}
  {arXiv:astro-ph/0402143} \BibitemShut {NoStop}%
\bibitem [{\citenamefont {Alford}\ and\ \citenamefont
  {Harris}(2019)}]{Alford:2019qtm}%
  \BibitemOpen
  \bibfield  {author} {\bibinfo {author} {\bibfnamefont {M.~G.}\ \bibnamefont
  {Alford}}\ and\ \bibinfo {author} {\bibfnamefont {S.~P.}\ \bibnamefont
  {Harris}},\ }\href {\doibase 10.1103/PhysRevC.100.035803} {\bibfield
  {journal} {\bibinfo  {journal} {Phys. Rev. C}\ }\textbf {\bibinfo {volume}
  {100}},\ \bibinfo {pages} {035803} (\bibinfo {year} {2019})},\ \Eprint
  {http://arxiv.org/abs/1907.03795} {arXiv:1907.03795 [nucl-th]} \BibitemShut
  {NoStop}%
\bibitem [{\citenamefont {Alford}\ \emph {et~al.}(2019)\citenamefont {Alford},
  \citenamefont {Harutyunyan},\ and\ \citenamefont
  {Sedrakian}}]{Alford:2019kdw}%
  \BibitemOpen
  \bibfield  {author} {\bibinfo {author} {\bibfnamefont {M.}~\bibnamefont
  {Alford}}, \bibinfo {author} {\bibfnamefont {A.}~\bibnamefont {Harutyunyan}},
  \ and\ \bibinfo {author} {\bibfnamefont {A.}~\bibnamefont {Sedrakian}},\
  }\href {\doibase 10.1103/PhysRevD.100.103021} {\bibfield  {journal} {\bibinfo
   {journal} {Phys. Rev. D}\ }\textbf {\bibinfo {volume} {100}},\ \bibinfo
  {pages} {103021} (\bibinfo {year} {2019})},\ \Eprint
  {http://arxiv.org/abs/1907.04192} {arXiv:1907.04192 [astro-ph.HE]}
  \BibitemShut {NoStop}%
\bibitem [{\citenamefont {Alford}\ \emph {et~al.}(2020)\citenamefont {Alford},
  \citenamefont {Harutyunyan},\ and\ \citenamefont
  {Sedrakian}}]{Alford:2020lla}%
  \BibitemOpen
  \bibfield  {author} {\bibinfo {author} {\bibfnamefont {M.}~\bibnamefont
  {Alford}}, \bibinfo {author} {\bibfnamefont {A.}~\bibnamefont {Harutyunyan}},
  \ and\ \bibinfo {author} {\bibfnamefont {A.}~\bibnamefont {Sedrakian}},\
  }\href {\doibase 10.3390/particles3020034} {\bibfield  {journal} {\bibinfo
  {journal} {Particles}\ }\textbf {\bibinfo {volume} {3}},\ \bibinfo {pages}
  {500} (\bibinfo {year} {2020})},\ \Eprint {http://arxiv.org/abs/2006.07975}
  {arXiv:2006.07975 [nucl-th]} \BibitemShut {NoStop}%
\bibitem [{\citenamefont {Alford}\ \emph {et~al.}(2021)\citenamefont {Alford},
  \citenamefont {Harutyunyan},\ and\ \citenamefont
  {Sedrakian}}]{Alford:2021lpp}%
  \BibitemOpen
  \bibfield  {author} {\bibinfo {author} {\bibfnamefont {M.}~\bibnamefont
  {Alford}}, \bibinfo {author} {\bibfnamefont {A.}~\bibnamefont {Harutyunyan}},
  \ and\ \bibinfo {author} {\bibfnamefont {A.}~\bibnamefont {Sedrakian}},\
  }\href {\doibase 10.1103/PhysRevD.104.103027} {\bibfield  {journal} {\bibinfo
   {journal} {Phys. Rev. D}\ }\textbf {\bibinfo {volume} {104}},\ \bibinfo
  {pages} {103027} (\bibinfo {year} {2021})},\ \Eprint
  {http://arxiv.org/abs/2108.07523} {arXiv:2108.07523 [astro-ph.HE]}
  \BibitemShut {NoStop}%
\bibitem [{\citenamefont {Alford}\ \emph {et~al.}(2022)\citenamefont {Alford},
  \citenamefont {Harutyunyan},\ and\ \citenamefont
  {Sedrakian}}]{Alford:2022ufz}%
  \BibitemOpen
  \bibfield  {author} {\bibinfo {author} {\bibfnamefont {M.}~\bibnamefont
  {Alford}}, \bibinfo {author} {\bibfnamefont {A.}~\bibnamefont {Harutyunyan}},
  \ and\ \bibinfo {author} {\bibfnamefont {A.}~\bibnamefont {Sedrakian}},\
  }\href {\doibase 10.3390/particles5030029} {\bibfield  {journal} {\bibinfo
  {journal} {Particles}\ }\textbf {\bibinfo {volume} {5}},\ \bibinfo {pages}
  {361} (\bibinfo {year} {2022})},\ \Eprint {http://arxiv.org/abs/2209.04717}
  {arXiv:2209.04717 [astro-ph.HE]} \BibitemShut {NoStop}%
\bibitem [{\citenamefont {Alford}\ and\ \citenamefont
  {Haber}(2021)}]{Alford:2020pld}%
  \BibitemOpen
  \bibfield  {author} {\bibinfo {author} {\bibfnamefont {M.~G.}\ \bibnamefont
  {Alford}}\ and\ \bibinfo {author} {\bibfnamefont {A.}~\bibnamefont {Haber}},\
  }\href {\doibase 10.1103/PhysRevC.103.045810} {\bibfield  {journal} {\bibinfo
   {journal} {Phys. Rev. C}\ }\textbf {\bibinfo {volume} {103}},\ \bibinfo
  {pages} {045810} (\bibinfo {year} {2021})},\ \Eprint
  {http://arxiv.org/abs/2009.05181} {arXiv:2009.05181 [nucl-th]} \BibitemShut
  {NoStop}%
\bibitem [{\citenamefont {Perego}\ \emph {et~al.}(2019)\citenamefont {Perego},
  \citenamefont {Bernuzzi},\ and\ \citenamefont {Radice}}]{Perego:2019adq}%
  \BibitemOpen
  \bibfield  {author} {\bibinfo {author} {\bibfnamefont {A.}~\bibnamefont
  {Perego}}, \bibinfo {author} {\bibfnamefont {S.}~\bibnamefont {Bernuzzi}}, \
  and\ \bibinfo {author} {\bibfnamefont {D.}~\bibnamefont {Radice}},\ }\href
  {\doibase 10.1140/epja/i2019-12810-7} {\bibfield  {journal} {\bibinfo
  {journal} {Eur. Phys. J. A}\ }\textbf {\bibinfo {volume} {55}},\ \bibinfo
  {pages} {124} (\bibinfo {year} {2019})},\ \Eprint
  {http://arxiv.org/abs/1903.07898} {arXiv:1903.07898 [gr-qc]} \BibitemShut
  {NoStop}%
\bibitem [{\citenamefont {Zappa}\ \emph {et~al.}(2022)\citenamefont {Zappa},
  \citenamefont {Bernuzzi}, \citenamefont {Radice},\ and\ \citenamefont
  {Perego}}]{Zappa:2022rpd}%
  \BibitemOpen
  \bibfield  {author} {\bibinfo {author} {\bibfnamefont {F.}~\bibnamefont
  {Zappa}}, \bibinfo {author} {\bibfnamefont {S.}~\bibnamefont {Bernuzzi}},
  \bibinfo {author} {\bibfnamefont {D.}~\bibnamefont {Radice}}, \ and\ \bibinfo
  {author} {\bibfnamefont {A.}~\bibnamefont {Perego}},\ }\href@noop {} {\
  (\bibinfo {year} {2022})},\ \Eprint {http://arxiv.org/abs/2210.11491}
  {arXiv:2210.11491 [astro-ph.HE]} \BibitemShut {NoStop}%
\bibitem [{\citenamefont {McLerran}\ and\ \citenamefont
  {Pisarski}(2007)}]{McLerran:2007qj}%
  \BibitemOpen
  \bibfield  {author} {\bibinfo {author} {\bibfnamefont {L.}~\bibnamefont
  {McLerran}}\ and\ \bibinfo {author} {\bibfnamefont {R.~D.}\ \bibnamefont
  {Pisarski}},\ }\href {\doibase 10.1016/j.nuclphysa.2007.08.013} {\bibfield
  {journal} {\bibinfo  {journal} {Nucl. Phys. A}\ }\textbf {\bibinfo {volume}
  {796}},\ \bibinfo {pages} {83} (\bibinfo {year} {2007})},\ \Eprint
  {http://arxiv.org/abs/0706.2191} {arXiv:0706.2191 [hep-ph]} \BibitemShut
  {NoStop}%
\bibitem [{\citenamefont {McLerran}\ and\ \citenamefont
  {Reddy}(2019)}]{McLerran:2018hbz}%
  \BibitemOpen
  \bibfield  {author} {\bibinfo {author} {\bibfnamefont {L.}~\bibnamefont
  {McLerran}}\ and\ \bibinfo {author} {\bibfnamefont {S.}~\bibnamefont
  {Reddy}},\ }\href {\doibase 10.1103/PhysRevLett.122.122701} {\bibfield
  {journal} {\bibinfo  {journal} {Phys. Rev. Lett.}\ }\textbf {\bibinfo
  {volume} {122}},\ \bibinfo {pages} {122701} (\bibinfo {year} {2019})},\
  \Eprint {http://arxiv.org/abs/1811.12503} {arXiv:1811.12503 [nucl-th]}
  \BibitemShut {NoStop}%
\bibitem [{\citenamefont {Pisarski}\ \emph {et~al.}(2019)\citenamefont
  {Pisarski}, \citenamefont {Skokov},\ and\ \citenamefont
  {Tsvelik}}]{Pisarski:2018bct}%
  \BibitemOpen
  \bibfield  {author} {\bibinfo {author} {\bibfnamefont {R.~D.}\ \bibnamefont
  {Pisarski}}, \bibinfo {author} {\bibfnamefont {V.~V.}\ \bibnamefont
  {Skokov}}, \ and\ \bibinfo {author} {\bibfnamefont {A.~M.}\ \bibnamefont
  {Tsvelik}},\ }\href {\doibase 10.1103/PhysRevD.99.074025} {\bibfield
  {journal} {\bibinfo  {journal} {Phys. Rev. D}\ }\textbf {\bibinfo {volume}
  {99}},\ \bibinfo {pages} {074025} (\bibinfo {year} {2019})},\ \Eprint
  {http://arxiv.org/abs/1801.08156} {arXiv:1801.08156 [hep-ph]} \BibitemShut
  {NoStop}%
\bibitem [{\citenamefont {Pisarski}\ \emph {et~al.}(2020)\citenamefont
  {Pisarski}, \citenamefont {Tsvelik},\ and\ \citenamefont
  {Valgushev}}]{Pisarski:2020dnx}%
  \BibitemOpen
  \bibfield  {author} {\bibinfo {author} {\bibfnamefont {R.~D.}\ \bibnamefont
  {Pisarski}}, \bibinfo {author} {\bibfnamefont {A.~M.}\ \bibnamefont
  {Tsvelik}}, \ and\ \bibinfo {author} {\bibfnamefont {S.}~\bibnamefont
  {Valgushev}},\ }\href {\doibase 10.1103/PhysRevD.102.016015} {\bibfield
  {journal} {\bibinfo  {journal} {Phys. Rev. D}\ }\textbf {\bibinfo {volume}
  {102}},\ \bibinfo {pages} {016015} (\bibinfo {year} {2020})},\ \Eprint
  {http://arxiv.org/abs/2005.10259} {arXiv:2005.10259 [hep-ph]} \BibitemShut
  {NoStop}%
\bibitem [{\citenamefont {Lajer}\ \emph {et~al.}(2022)\citenamefont {Lajer},
  \citenamefont {Konik}, \citenamefont {Pisarski},\ and\ \citenamefont
  {Tsvelik}}]{Lajer:2021kcz}%
  \BibitemOpen
  \bibfield  {author} {\bibinfo {author} {\bibfnamefont {M.}~\bibnamefont
  {Lajer}}, \bibinfo {author} {\bibfnamefont {R.~M.}\ \bibnamefont {Konik}},
  \bibinfo {author} {\bibfnamefont {R.~D.}\ \bibnamefont {Pisarski}}, \ and\
  \bibinfo {author} {\bibfnamefont {A.~M.}\ \bibnamefont {Tsvelik}},\ }\href
  {\doibase 10.1103/PhysRevD.105.054035} {\bibfield  {journal} {\bibinfo
  {journal} {Phys. Rev. D}\ }\textbf {\bibinfo {volume} {105}},\ \bibinfo
  {pages} {054035} (\bibinfo {year} {2022})},\ \Eprint
  {http://arxiv.org/abs/2112.10238} {arXiv:2112.10238 [hep-th]} \BibitemShut
  {NoStop}%
\bibitem [{\citenamefont {Baym}\ \emph {et~al.}(2018)\citenamefont {Baym},
  \citenamefont {Hatsuda}, \citenamefont {Kojo}, \citenamefont {Powell},
  \citenamefont {Song},\ and\ \citenamefont {Takatsuka}}]{Baym:2017whm}%
  \BibitemOpen
  \bibfield  {author} {\bibinfo {author} {\bibfnamefont {G.}~\bibnamefont
  {Baym}}, \bibinfo {author} {\bibfnamefont {T.}~\bibnamefont {Hatsuda}},
  \bibinfo {author} {\bibfnamefont {T.}~\bibnamefont {Kojo}}, \bibinfo {author}
  {\bibfnamefont {P.~D.}\ \bibnamefont {Powell}}, \bibinfo {author}
  {\bibfnamefont {Y.}~\bibnamefont {Song}}, \ and\ \bibinfo {author}
  {\bibfnamefont {T.}~\bibnamefont {Takatsuka}},\ }\href {\doibase
  10.1088/1361-6633/aaae14} {\bibfield  {journal} {\bibinfo  {journal} {Rept.
  Prog. Phys.}\ }\textbf {\bibinfo {volume} {81}},\ \bibinfo {pages} {056902}
  (\bibinfo {year} {2018})},\ \Eprint {http://arxiv.org/abs/1707.04966}
  {arXiv:1707.04966 [astro-ph.HE]} \BibitemShut {NoStop}%
\bibitem [{\citenamefont {Alford}\ \emph {et~al.}(2008)\citenamefont {Alford},
  \citenamefont {Schmitt}, \citenamefont {Rajagopal},\ and\ \citenamefont
  {Sch\"afer}}]{Alford:2007xm}%
  \BibitemOpen
  \bibfield  {author} {\bibinfo {author} {\bibfnamefont {M.~G.}\ \bibnamefont
  {Alford}}, \bibinfo {author} {\bibfnamefont {A.}~\bibnamefont {Schmitt}},
  \bibinfo {author} {\bibfnamefont {K.}~\bibnamefont {Rajagopal}}, \ and\
  \bibinfo {author} {\bibfnamefont {T.}~\bibnamefont {Sch\"afer}},\ }\href
  {\doibase 10.1103/RevModPhys.80.1455} {\bibfield  {journal} {\bibinfo
  {journal} {Rev. Mod. Phys.}\ }\textbf {\bibinfo {volume} {80}},\ \bibinfo
  {pages} {1455} (\bibinfo {year} {2008})},\ \Eprint
  {http://arxiv.org/abs/0709.4635} {arXiv:0709.4635 [hep-ph]} \BibitemShut
  {NoStop}%
\bibitem [{\citenamefont {Kojo}\ \emph
  {et~al.}(2010{\natexlab{a}})\citenamefont {Kojo}, \citenamefont {Hidaka},
  \citenamefont {McLerran},\ and\ \citenamefont {Pisarski}}]{Kojo:2009ha}%
  \BibitemOpen
  \bibfield  {author} {\bibinfo {author} {\bibfnamefont {T.}~\bibnamefont
  {Kojo}}, \bibinfo {author} {\bibfnamefont {Y.}~\bibnamefont {Hidaka}},
  \bibinfo {author} {\bibfnamefont {L.}~\bibnamefont {McLerran}}, \ and\
  \bibinfo {author} {\bibfnamefont {R.~D.}\ \bibnamefont {Pisarski}},\ }\href
  {\doibase 10.1016/j.nuclphysa.2010.05.053} {\bibfield  {journal} {\bibinfo
  {journal} {Nucl. Phys. A}\ }\textbf {\bibinfo {volume} {843}},\ \bibinfo
  {pages} {37} (\bibinfo {year} {2010}{\natexlab{a}})},\ \Eprint
  {http://arxiv.org/abs/0912.3800} {arXiv:0912.3800 [hep-ph]} \BibitemShut
  {NoStop}%
\bibitem [{\citenamefont {Kojo}\ \emph
  {et~al.}(2010{\natexlab{b}})\citenamefont {Kojo}, \citenamefont {Pisarski},\
  and\ \citenamefont {Tsvelik}}]{Kojo:2010fe}%
  \BibitemOpen
  \bibfield  {author} {\bibinfo {author} {\bibfnamefont {T.}~\bibnamefont
  {Kojo}}, \bibinfo {author} {\bibfnamefont {R.~D.}\ \bibnamefont {Pisarski}},
  \ and\ \bibinfo {author} {\bibfnamefont {A.~M.}\ \bibnamefont {Tsvelik}},\
  }\href {\doibase 10.1103/PhysRevD.82.074015} {\bibfield  {journal} {\bibinfo
  {journal} {Phys. Rev. D}\ }\textbf {\bibinfo {volume} {82}},\ \bibinfo
  {pages} {074015} (\bibinfo {year} {2010}{\natexlab{b}})},\ \Eprint
  {http://arxiv.org/abs/1007.0248} {arXiv:1007.0248 [hep-ph]} \BibitemShut
  {NoStop}%
\bibitem [{\citenamefont {Kojo}\ \emph {et~al.}(2012)\citenamefont {Kojo},
  \citenamefont {Hidaka}, \citenamefont {Fukushima}, \citenamefont {McLerran},\
  and\ \citenamefont {Pisarski}}]{Kojo:2011cn}%
  \BibitemOpen
  \bibfield  {author} {\bibinfo {author} {\bibfnamefont {T.}~\bibnamefont
  {Kojo}}, \bibinfo {author} {\bibfnamefont {Y.}~\bibnamefont {Hidaka}},
  \bibinfo {author} {\bibfnamefont {K.}~\bibnamefont {Fukushima}}, \bibinfo
  {author} {\bibfnamefont {L.~D.}\ \bibnamefont {McLerran}}, \ and\ \bibinfo
  {author} {\bibfnamefont {R.~D.}\ \bibnamefont {Pisarski}},\ }\href {\doibase
  10.1016/j.nuclphysa.2011.11.007} {\bibfield  {journal} {\bibinfo  {journal}
  {Nucl. Phys. A}\ }\textbf {\bibinfo {volume} {875}},\ \bibinfo {pages} {94}
  (\bibinfo {year} {2012})},\ \Eprint {http://arxiv.org/abs/1107.2124}
  {arXiv:1107.2124 [hep-ph]} \BibitemShut {NoStop}%
\bibitem [{\citenamefont {Pisarski}\ and\ \citenamefont
  {Rennecke}(2021)}]{Pisarski:2021qof}%
  \BibitemOpen
  \bibfield  {author} {\bibinfo {author} {\bibfnamefont {R.~D.}\ \bibnamefont
  {Pisarski}}\ and\ \bibinfo {author} {\bibfnamefont {F.}~\bibnamefont
  {Rennecke}},\ }\href {\doibase 10.1103/PhysRevLett.127.152302} {\bibfield
  {journal} {\bibinfo  {journal} {Phys. Rev. Lett.}\ }\textbf {\bibinfo
  {volume} {127}},\ \bibinfo {pages} {152302} (\bibinfo {year} {2021})},\
  \Eprint {http://arxiv.org/abs/2103.06890} {arXiv:2103.06890 [hep-ph]}
  \BibitemShut {NoStop}%
\bibitem [{\citenamefont {Page}\ \emph {et~al.}(2011)\citenamefont {Page},
  \citenamefont {Prakash}, \citenamefont {Lattimer},\ and\ \citenamefont
  {Steiner}}]{Page:2010aw}%
  \BibitemOpen
  \bibfield  {author} {\bibinfo {author} {\bibfnamefont {D.}~\bibnamefont
  {Page}}, \bibinfo {author} {\bibfnamefont {M.}~\bibnamefont {Prakash}},
  \bibinfo {author} {\bibfnamefont {J.~M.}\ \bibnamefont {Lattimer}}, \ and\
  \bibinfo {author} {\bibfnamefont {A.~W.}\ \bibnamefont {Steiner}},\ }\href
  {\doibase 10.1103/PhysRevLett.106.081101} {\bibfield  {journal} {\bibinfo
  {journal} {Phys. Rev. Lett.}\ }\textbf {\bibinfo {volume} {106}},\ \bibinfo
  {pages} {081101} (\bibinfo {year} {2011})},\ \Eprint
  {http://arxiv.org/abs/1011.6142} {arXiv:1011.6142 [astro-ph.HE]} \BibitemShut
  {NoStop}%
\bibitem [{\citenamefont {Hippert}\ \emph {et~al.}(2021)\citenamefont
  {Hippert}, \citenamefont {Fraga},\ and\ \citenamefont
  {Noronha}}]{Hippert:2021gfs}%
  \BibitemOpen
  \bibfield  {author} {\bibinfo {author} {\bibfnamefont {M.}~\bibnamefont
  {Hippert}}, \bibinfo {author} {\bibfnamefont {E.~S.}\ \bibnamefont {Fraga}},
  \ and\ \bibinfo {author} {\bibfnamefont {J.}~\bibnamefont {Noronha}},\ }\href
  {\doibase 10.1103/PhysRevD.104.034011} {\bibfield  {journal} {\bibinfo
  {journal} {Phys. Rev. D}\ }\textbf {\bibinfo {volume} {104}},\ \bibinfo
  {pages} {034011} (\bibinfo {year} {2021})},\ \Eprint
  {http://arxiv.org/abs/2105.04535} {arXiv:2105.04535 [nucl-th]} \BibitemShut
  {NoStop}%
\bibitem [{\citenamefont {Manuel}\ and\ \citenamefont
  {Llanes-Estrada}(2007)}]{Manuel:2007pz}%
  \BibitemOpen
  \bibfield  {author} {\bibinfo {author} {\bibfnamefont {C.}~\bibnamefont
  {Manuel}}\ and\ \bibinfo {author} {\bibfnamefont {F.~J.}\ \bibnamefont
  {Llanes-Estrada}},\ }\href {\doibase 10.1088/1475-7516/2007/08/001}
  {\bibfield  {journal} {\bibinfo  {journal} {JCAP}\ }\textbf {\bibinfo
  {volume} {08}},\ \bibinfo {pages} {001} (\bibinfo {year} {2007})},\ \Eprint
  {http://arxiv.org/abs/0705.3909} {arXiv:0705.3909 [hep-ph]} \BibitemShut
  {NoStop}%
\bibitem [{\citenamefont {Alford}\ \emph {et~al.}(1999)\citenamefont {Alford},
  \citenamefont {Rajagopal},\ and\ \citenamefont {Wilczek}}]{Alford:1998mk}%
  \BibitemOpen
  \bibfield  {author} {\bibinfo {author} {\bibfnamefont {M.~G.}\ \bibnamefont
  {Alford}}, \bibinfo {author} {\bibfnamefont {K.}~\bibnamefont {Rajagopal}}, \
  and\ \bibinfo {author} {\bibfnamefont {F.}~\bibnamefont {Wilczek}},\ }\href
  {\doibase 10.1016/S0550-3213(98)00668-3} {\bibfield  {journal} {\bibinfo
  {journal} {Nucl. Phys. B}\ }\textbf {\bibinfo {volume} {537}},\ \bibinfo
  {pages} {443} (\bibinfo {year} {1999})},\ \Eprint
  {http://arxiv.org/abs/hep-ph/9804403} {arXiv:hep-ph/9804403} \BibitemShut
  {NoStop}%
\bibitem [{\citenamefont {Alford}\ \emph {et~al.}(2005)\citenamefont {Alford},
  \citenamefont {Kouvaris},\ and\ \citenamefont {Rajagopal}}]{Alford:2004hz}%
  \BibitemOpen
  \bibfield  {author} {\bibinfo {author} {\bibfnamefont {M.}~\bibnamefont
  {Alford}}, \bibinfo {author} {\bibfnamefont {C.}~\bibnamefont {Kouvaris}}, \
  and\ \bibinfo {author} {\bibfnamefont {K.}~\bibnamefont {Rajagopal}},\ }\href
  {\doibase 10.1103/PhysRevD.71.054009} {\bibfield  {journal} {\bibinfo
  {journal} {Phys. Rev. D}\ }\textbf {\bibinfo {volume} {71}},\ \bibinfo
  {pages} {054009} (\bibinfo {year} {2005})},\ \Eprint
  {http://arxiv.org/abs/hep-ph/0406137} {arXiv:hep-ph/0406137} \BibitemShut
  {NoStop}%
\bibitem [{\citenamefont {Alford}\ \emph {et~al.}(2001)\citenamefont {Alford},
  \citenamefont {Bowers},\ and\ \citenamefont {Rajagopal}}]{Alford:2000ze}%
  \BibitemOpen
  \bibfield  {author} {\bibinfo {author} {\bibfnamefont {M.~G.}\ \bibnamefont
  {Alford}}, \bibinfo {author} {\bibfnamefont {J.~A.}\ \bibnamefont {Bowers}},
  \ and\ \bibinfo {author} {\bibfnamefont {K.}~\bibnamefont {Rajagopal}},\
  }\href {\doibase 10.1103/PhysRevD.63.074016} {\bibfield  {journal} {\bibinfo
  {journal} {Phys. Rev. D}\ }\textbf {\bibinfo {volume} {63}},\ \bibinfo
  {pages} {074016} (\bibinfo {year} {2001})},\ \Eprint
  {http://arxiv.org/abs/hep-ph/0008208} {arXiv:hep-ph/0008208} \BibitemShut
  {NoStop}%
\bibitem [{\citenamefont {Bowers}\ and\ \citenamefont
  {Rajagopal}(2002)}]{Bowers:2002xr}%
  \BibitemOpen
  \bibfield  {author} {\bibinfo {author} {\bibfnamefont {J.~A.}\ \bibnamefont
  {Bowers}}\ and\ \bibinfo {author} {\bibfnamefont {K.}~\bibnamefont
  {Rajagopal}},\ }\href {\doibase 10.1103/PhysRevD.66.065002} {\bibfield
  {journal} {\bibinfo  {journal} {Phys. Rev. D}\ }\textbf {\bibinfo {volume}
  {66}},\ \bibinfo {pages} {065002} (\bibinfo {year} {2002})},\ \Eprint
  {http://arxiv.org/abs/hep-ph/0204079} {arXiv:hep-ph/0204079} \BibitemShut
  {NoStop}%
\bibitem [{\citenamefont {Mannarelli}\ \emph {et~al.}(2007)\citenamefont
  {Mannarelli}, \citenamefont {Rajagopal},\ and\ \citenamefont
  {Sharma}}]{Mannarelli:2007bs}%
  \BibitemOpen
  \bibfield  {author} {\bibinfo {author} {\bibfnamefont {M.}~\bibnamefont
  {Mannarelli}}, \bibinfo {author} {\bibfnamefont {K.}~\bibnamefont
  {Rajagopal}}, \ and\ \bibinfo {author} {\bibfnamefont {R.}~\bibnamefont
  {Sharma}},\ }\href {\doibase 10.1103/PhysRevD.76.074026} {\bibfield
  {journal} {\bibinfo  {journal} {Phys. Rev. D}\ }\textbf {\bibinfo {volume}
  {76}},\ \bibinfo {pages} {074026} (\bibinfo {year} {2007})},\ \Eprint
  {http://arxiv.org/abs/hep-ph/0702021} {arXiv:hep-ph/0702021} \BibitemShut
  {NoStop}%
\bibitem [{\citenamefont {Bedaque}\ and\ \citenamefont
  {Sch\"afer}(2002)}]{Bedaque:2001je}%
  \BibitemOpen
  \bibfield  {author} {\bibinfo {author} {\bibfnamefont {P.~F.}\ \bibnamefont
  {Bedaque}}\ and\ \bibinfo {author} {\bibfnamefont {T.}~\bibnamefont
  {Sch\"afer}},\ }\href {\doibase 10.1016/S0375-9474(01)01272-6} {\bibfield
  {journal} {\bibinfo  {journal} {Nucl. Phys. A}\ }\textbf {\bibinfo {volume}
  {697}},\ \bibinfo {pages} {802} (\bibinfo {year} {2002})},\ \Eprint
  {http://arxiv.org/abs/hep-ph/0105150} {arXiv:hep-ph/0105150} \BibitemShut
  {NoStop}%
\bibitem [{\citenamefont {Annala}\ \emph {et~al.}(2020)\citenamefont {Annala},
  \citenamefont {Gorda}, \citenamefont {Kurkela}, \citenamefont {N\"attil\"a},\
  and\ \citenamefont {Vuorinen}}]{Annala:2019puf}%
  \BibitemOpen
  \bibfield  {author} {\bibinfo {author} {\bibfnamefont {E.}~\bibnamefont
  {Annala}}, \bibinfo {author} {\bibfnamefont {T.}~\bibnamefont {Gorda}},
  \bibinfo {author} {\bibfnamefont {A.}~\bibnamefont {Kurkela}}, \bibinfo
  {author} {\bibfnamefont {J.}~\bibnamefont {N\"attil\"a}}, \ and\ \bibinfo
  {author} {\bibfnamefont {A.}~\bibnamefont {Vuorinen}},\ }\href {\doibase
  10.1038/s41567-020-0914-9} {\bibfield  {journal} {\bibinfo  {journal} {Nature
  Phys.}\ }\textbf {\bibinfo {volume} {16}},\ \bibinfo {pages} {907} (\bibinfo
  {year} {2020})},\ \Eprint {http://arxiv.org/abs/1903.09121} {arXiv:1903.09121
  [astro-ph.HE]} \BibitemShut {NoStop}%
\bibitem [{\citenamefont {Fujimoto}\ \emph
  {et~al.}(2022{\natexlab{a}})\citenamefont {Fujimoto}, \citenamefont
  {Fukushima}, \citenamefont {McLerran},\ and\ \citenamefont
  {Praszalowicz}}]{Fujimoto:2022ohj}%
  \BibitemOpen
  \bibfield  {author} {\bibinfo {author} {\bibfnamefont {Y.}~\bibnamefont
  {Fujimoto}}, \bibinfo {author} {\bibfnamefont {K.}~\bibnamefont {Fukushima}},
  \bibinfo {author} {\bibfnamefont {L.~D.}\ \bibnamefont {McLerran}}, \ and\
  \bibinfo {author} {\bibfnamefont {M.}~\bibnamefont {Praszalowicz}},\
  }\href@noop {} {\  (\bibinfo {year} {2022}{\natexlab{a}})},\ \Eprint
  {http://arxiv.org/abs/2207.06753} {arXiv:2207.06753 [nucl-th]} \BibitemShut
  {NoStop}%
\bibitem [{\citenamefont {Tan}\ \emph {et~al.}(2022{\natexlab{a}})\citenamefont
  {Tan}, \citenamefont {Dexheimer}, \citenamefont {Noronha-Hostler},\ and\
  \citenamefont {Yunes}}]{Tan:2021nat}%
  \BibitemOpen
  \bibfield  {author} {\bibinfo {author} {\bibfnamefont {H.}~\bibnamefont
  {Tan}}, \bibinfo {author} {\bibfnamefont {V.}~\bibnamefont {Dexheimer}},
  \bibinfo {author} {\bibfnamefont {J.}~\bibnamefont {Noronha-Hostler}}, \ and\
  \bibinfo {author} {\bibfnamefont {N.}~\bibnamefont {Yunes}},\ }\href
  {\doibase 10.1103/PhysRevLett.128.161101} {\bibfield  {journal} {\bibinfo
  {journal} {Phys. Rev. Lett.}\ }\textbf {\bibinfo {volume} {128}},\ \bibinfo
  {pages} {161101} (\bibinfo {year} {2022}{\natexlab{a}})},\ \Eprint
  {http://arxiv.org/abs/2111.10260} {arXiv:2111.10260 [astro-ph.HE]}
  \BibitemShut {NoStop}%
\bibitem [{\citenamefont {Alba}\ \emph {et~al.}(2018)\citenamefont {Alba},
  \citenamefont {Mantovani~Sarti}, \citenamefont {Noronha}, \citenamefont
  {Noronha-Hostler}, \citenamefont {Parotto}, \citenamefont
  {Portillo~Vazquez},\ and\ \citenamefont {Ratti}}]{Alba:2017hhe}%
  \BibitemOpen
  \bibfield  {author} {\bibinfo {author} {\bibfnamefont {P.}~\bibnamefont
  {Alba}}, \bibinfo {author} {\bibfnamefont {V.}~\bibnamefont
  {Mantovani~Sarti}}, \bibinfo {author} {\bibfnamefont {J.}~\bibnamefont
  {Noronha}}, \bibinfo {author} {\bibfnamefont {J.}~\bibnamefont
  {Noronha-Hostler}}, \bibinfo {author} {\bibfnamefont {P.}~\bibnamefont
  {Parotto}}, \bibinfo {author} {\bibfnamefont {I.}~\bibnamefont
  {Portillo~Vazquez}}, \ and\ \bibinfo {author} {\bibfnamefont
  {C.}~\bibnamefont {Ratti}},\ }\href {\doibase 10.1103/PhysRevC.98.034909}
  {\bibfield  {journal} {\bibinfo  {journal} {Phys. Rev. C}\ }\textbf {\bibinfo
  {volume} {98}},\ \bibinfo {pages} {034909} (\bibinfo {year} {2018})},\
  \Eprint {http://arxiv.org/abs/1711.05207} {arXiv:1711.05207 [nucl-th]}
  \BibitemShut {NoStop}%
\bibitem [{\citenamefont {Schenke}\ \emph {et~al.}(2020)\citenamefont
  {Schenke}, \citenamefont {Shen},\ and\ \citenamefont
  {Tribedy}}]{Schenke:2020mbo}%
  \BibitemOpen
  \bibfield  {author} {\bibinfo {author} {\bibfnamefont {B.}~\bibnamefont
  {Schenke}}, \bibinfo {author} {\bibfnamefont {C.}~\bibnamefont {Shen}}, \
  and\ \bibinfo {author} {\bibfnamefont {P.}~\bibnamefont {Tribedy}},\ }\href
  {\doibase 10.1103/PhysRevC.102.044905} {\bibfield  {journal} {\bibinfo
  {journal} {Phys. Rev. C}\ }\textbf {\bibinfo {volume} {102}},\ \bibinfo
  {pages} {044905} (\bibinfo {year} {2020})},\ \Eprint
  {http://arxiv.org/abs/2005.14682} {arXiv:2005.14682 [nucl-th]} \BibitemShut
  {NoStop}%
\bibitem [{\citenamefont {Everett}\ \emph {et~al.}(2021)\citenamefont {Everett}
  \emph {et~al.}}]{JETSCAPE:2020mzn}%
  \BibitemOpen
  \bibfield  {author} {\bibinfo {author} {\bibfnamefont {D.}~\bibnamefont
  {Everett}} \emph {et~al.} (\bibinfo {collaboration} {JETSCAPE}),\ }\href
  {\doibase 10.1103/PhysRevC.103.054904} {\bibfield  {journal} {\bibinfo
  {journal} {Phys. Rev. C}\ }\textbf {\bibinfo {volume} {103}},\ \bibinfo
  {pages} {054904} (\bibinfo {year} {2021})},\ \Eprint
  {http://arxiv.org/abs/2011.01430} {arXiv:2011.01430 [hep-ph]} \BibitemShut
  {NoStop}%
\bibitem [{\citenamefont {Nijs}\ \emph {et~al.}(2021)\citenamefont {Nijs},
  \citenamefont {van~der Schee}, \citenamefont {G\"ursoy},\ and\ \citenamefont
  {Snellings}}]{Nijs:2020ors}%
  \BibitemOpen
  \bibfield  {author} {\bibinfo {author} {\bibfnamefont {G.}~\bibnamefont
  {Nijs}}, \bibinfo {author} {\bibfnamefont {W.}~\bibnamefont {van~der Schee}},
  \bibinfo {author} {\bibfnamefont {U.}~\bibnamefont {G\"ursoy}}, \ and\
  \bibinfo {author} {\bibfnamefont {R.}~\bibnamefont {Snellings}},\ }\href
  {\doibase 10.1103/PhysRevLett.126.202301} {\bibfield  {journal} {\bibinfo
  {journal} {Phys. Rev. Lett.}\ }\textbf {\bibinfo {volume} {126}},\ \bibinfo
  {pages} {202301} (\bibinfo {year} {2021})},\ \Eprint
  {http://arxiv.org/abs/2010.15130} {arXiv:2010.15130 [nucl-th]} \BibitemShut
  {NoStop}%
\bibitem [{\citenamefont {Stephanov}\ \emph {et~al.}(1999)\citenamefont
  {Stephanov}, \citenamefont {Rajagopal},\ and\ \citenamefont
  {Shuryak}}]{Stephanov:1999zu}%
  \BibitemOpen
  \bibfield  {author} {\bibinfo {author} {\bibfnamefont {M.~A.}\ \bibnamefont
  {Stephanov}}, \bibinfo {author} {\bibfnamefont {K.}~\bibnamefont
  {Rajagopal}}, \ and\ \bibinfo {author} {\bibfnamefont {E.~V.}\ \bibnamefont
  {Shuryak}},\ }\href {\doibase 10.1103/PhysRevD.60.114028} {\bibfield
  {journal} {\bibinfo  {journal} {Phys. Rev. D}\ }\textbf {\bibinfo {volume}
  {60}},\ \bibinfo {pages} {114028} (\bibinfo {year} {1999})},\ \Eprint
  {http://arxiv.org/abs/hep-ph/9903292} {arXiv:hep-ph/9903292} \BibitemShut
  {NoStop}%
\bibitem [{\citenamefont {Hatta}\ and\ \citenamefont
  {Stephanov}(2003)}]{Hatta:2003wn}%
  \BibitemOpen
  \bibfield  {author} {\bibinfo {author} {\bibfnamefont {Y.}~\bibnamefont
  {Hatta}}\ and\ \bibinfo {author} {\bibfnamefont {M.~A.}\ \bibnamefont
  {Stephanov}},\ }\href {\doibase 10.1103/PhysRevLett.91.102003} {\bibfield
  {journal} {\bibinfo  {journal} {Phys. Rev. Lett.}\ }\textbf {\bibinfo
  {volume} {91}},\ \bibinfo {pages} {102003} (\bibinfo {year} {2003})},\
  \bibinfo {note} {[Erratum: Phys.Rev.Lett. 91, 129901 (2003)]},\ \Eprint
  {http://arxiv.org/abs/hep-ph/0302002} {arXiv:hep-ph/0302002} \BibitemShut
  {NoStop}%
\bibitem [{\citenamefont {Stephanov}(2009)}]{Stephanov:2008qz}%
  \BibitemOpen
  \bibfield  {author} {\bibinfo {author} {\bibfnamefont {M.~A.}\ \bibnamefont
  {Stephanov}},\ }\href {\doibase 10.1103/PhysRevLett.102.032301} {\bibfield
  {journal} {\bibinfo  {journal} {Phys. Rev. Lett.}\ }\textbf {\bibinfo
  {volume} {102}},\ \bibinfo {pages} {032301} (\bibinfo {year} {2009})},\
  \Eprint {http://arxiv.org/abs/0809.3450} {arXiv:0809.3450 [hep-ph]}
  \BibitemShut {NoStop}%
\bibitem [{\citenamefont {Athanasiou}\ \emph {et~al.}(2010)\citenamefont
  {Athanasiou}, \citenamefont {Rajagopal},\ and\ \citenamefont
  {Stephanov}}]{Athanasiou:2010kw}%
  \BibitemOpen
  \bibfield  {author} {\bibinfo {author} {\bibfnamefont {C.}~\bibnamefont
  {Athanasiou}}, \bibinfo {author} {\bibfnamefont {K.}~\bibnamefont
  {Rajagopal}}, \ and\ \bibinfo {author} {\bibfnamefont {M.}~\bibnamefont
  {Stephanov}},\ }\href {\doibase 10.1103/PhysRevD.82.074008} {\bibfield
  {journal} {\bibinfo  {journal} {Phys. Rev. D}\ }\textbf {\bibinfo {volume}
  {82}},\ \bibinfo {pages} {074008} (\bibinfo {year} {2010})},\ \Eprint
  {http://arxiv.org/abs/1006.4636} {arXiv:1006.4636 [hep-ph]} \BibitemShut
  {NoStop}%
\bibitem [{\citenamefont {Stephanov}(2011)}]{Stephanov:2011pb}%
  \BibitemOpen
  \bibfield  {author} {\bibinfo {author} {\bibfnamefont {M.~A.}\ \bibnamefont
  {Stephanov}},\ }\href {\doibase 10.1103/PhysRevLett.107.052301} {\bibfield
  {journal} {\bibinfo  {journal} {Phys. Rev. Lett.}\ }\textbf {\bibinfo
  {volume} {107}},\ \bibinfo {pages} {052301} (\bibinfo {year} {2011})},\
  \Eprint {http://arxiv.org/abs/1104.1627} {arXiv:1104.1627 [hep-ph]}
  \BibitemShut {NoStop}%
\bibitem [{\citenamefont {Brewer}\ \emph {et~al.}(2018)\citenamefont {Brewer},
  \citenamefont {Mukherjee}, \citenamefont {Rajagopal},\ and\ \citenamefont
  {Yin}}]{Brewer:2018abr}%
  \BibitemOpen
  \bibfield  {author} {\bibinfo {author} {\bibfnamefont {J.}~\bibnamefont
  {Brewer}}, \bibinfo {author} {\bibfnamefont {S.}~\bibnamefont {Mukherjee}},
  \bibinfo {author} {\bibfnamefont {K.}~\bibnamefont {Rajagopal}}, \ and\
  \bibinfo {author} {\bibfnamefont {Y.}~\bibnamefont {Yin}},\ }\href {\doibase
  10.1103/PhysRevC.98.061901} {\bibfield  {journal} {\bibinfo  {journal} {Phys.
  Rev. C}\ }\textbf {\bibinfo {volume} {98}},\ \bibinfo {pages} {061901}
  (\bibinfo {year} {2018})},\ \Eprint {http://arxiv.org/abs/1804.10215}
  {arXiv:1804.10215 [hep-ph]} \BibitemShut {NoStop}%
\bibitem [{\citenamefont {Mroczek}\ \emph {et~al.}(2021)\citenamefont
  {Mroczek}, \citenamefont {Nava~Acuna}, \citenamefont {Noronha-Hostler},
  \citenamefont {Parotto}, \citenamefont {Ratti},\ and\ \citenamefont
  {Stephanov}}]{Mroczek:2020rpm}%
  \BibitemOpen
  \bibfield  {author} {\bibinfo {author} {\bibfnamefont {D.}~\bibnamefont
  {Mroczek}}, \bibinfo {author} {\bibfnamefont {A.~R.}\ \bibnamefont
  {Nava~Acuna}}, \bibinfo {author} {\bibfnamefont {J.}~\bibnamefont
  {Noronha-Hostler}}, \bibinfo {author} {\bibfnamefont {P.}~\bibnamefont
  {Parotto}}, \bibinfo {author} {\bibfnamefont {C.}~\bibnamefont {Ratti}}, \
  and\ \bibinfo {author} {\bibfnamefont {M.~A.}\ \bibnamefont {Stephanov}},\
  }\href {\doibase 10.1103/PhysRevC.103.034901} {\bibfield  {journal} {\bibinfo
   {journal} {Phys. Rev. C}\ }\textbf {\bibinfo {volume} {103}},\ \bibinfo
  {pages} {034901} (\bibinfo {year} {2021})},\ \Eprint
  {http://arxiv.org/abs/2008.04022} {arXiv:2008.04022 [nucl-th]} \BibitemShut
  {NoStop}%
\bibitem [{\citenamefont {Vovchenko}\ \emph {et~al.}(2022)\citenamefont
  {Vovchenko}, \citenamefont {Koch},\ and\ \citenamefont
  {Shen}}]{Vovchenko:2021kxx}%
  \BibitemOpen
  \bibfield  {author} {\bibinfo {author} {\bibfnamefont {V.}~\bibnamefont
  {Vovchenko}}, \bibinfo {author} {\bibfnamefont {V.}~\bibnamefont {Koch}}, \
  and\ \bibinfo {author} {\bibfnamefont {C.}~\bibnamefont {Shen}},\ }\href
  {\doibase 10.1103/PhysRevC.105.014904} {\bibfield  {journal} {\bibinfo
  {journal} {Phys. Rev. C}\ }\textbf {\bibinfo {volume} {105}},\ \bibinfo
  {pages} {014904} (\bibinfo {year} {2022})},\ \Eprint
  {http://arxiv.org/abs/2107.00163} {arXiv:2107.00163 [hep-ph]} \BibitemShut
  {NoStop}%
\bibitem [{\citenamefont {Berdnikov}\ and\ \citenamefont
  {Rajagopal}(2000)}]{Berdnikov:1999ph}%
  \BibitemOpen
  \bibfield  {author} {\bibinfo {author} {\bibfnamefont {B.}~\bibnamefont
  {Berdnikov}}\ and\ \bibinfo {author} {\bibfnamefont {K.}~\bibnamefont
  {Rajagopal}},\ }\href {\doibase 10.1103/PhysRevD.61.105017} {\bibfield
  {journal} {\bibinfo  {journal} {Phys. Rev. D}\ }\textbf {\bibinfo {volume}
  {61}},\ \bibinfo {pages} {105017} (\bibinfo {year} {2000})},\ \Eprint
  {http://arxiv.org/abs/hep-ph/9912274} {arXiv:hep-ph/9912274} \BibitemShut
  {NoStop}%
\bibitem [{\citenamefont {Stephanov}\ and\ \citenamefont
  {Yin}(2018)}]{Stephanov:2017ghc}%
  \BibitemOpen
  \bibfield  {author} {\bibinfo {author} {\bibfnamefont {M.}~\bibnamefont
  {Stephanov}}\ and\ \bibinfo {author} {\bibfnamefont {Y.}~\bibnamefont
  {Yin}},\ }\href {\doibase 10.1103/PhysRevD.98.036006} {\bibfield  {journal}
  {\bibinfo  {journal} {Phys. Rev. D}\ }\textbf {\bibinfo {volume} {98}},\
  \bibinfo {pages} {036006} (\bibinfo {year} {2018})},\ \Eprint
  {http://arxiv.org/abs/1712.10305} {arXiv:1712.10305 [nucl-th]} \BibitemShut
  {NoStop}%
\bibitem [{\citenamefont {Rajagopal}\ \emph {et~al.}(2020)\citenamefont
  {Rajagopal}, \citenamefont {Ridgway}, \citenamefont {Weller},\ and\
  \citenamefont {Yin}}]{Rajagopal:2019xwg}%
  \BibitemOpen
  \bibfield  {author} {\bibinfo {author} {\bibfnamefont {K.}~\bibnamefont
  {Rajagopal}}, \bibinfo {author} {\bibfnamefont {G.}~\bibnamefont {Ridgway}},
  \bibinfo {author} {\bibfnamefont {R.}~\bibnamefont {Weller}}, \ and\ \bibinfo
  {author} {\bibfnamefont {Y.}~\bibnamefont {Yin}},\ }\href {\doibase
  10.1103/PhysRevD.102.094025} {\bibfield  {journal} {\bibinfo  {journal}
  {Phys. Rev. D}\ }\textbf {\bibinfo {volume} {102}},\ \bibinfo {pages}
  {094025} (\bibinfo {year} {2020})},\ \Eprint
  {http://arxiv.org/abs/1908.08539} {arXiv:1908.08539 [hep-ph]} \BibitemShut
  {NoStop}%
\bibitem [{\citenamefont {Du}\ \emph {et~al.}(2020)\citenamefont {Du},
  \citenamefont {Heinz}, \citenamefont {Rajagopal},\ and\ \citenamefont
  {Yin}}]{Du:2020bxp}%
  \BibitemOpen
  \bibfield  {author} {\bibinfo {author} {\bibfnamefont {L.}~\bibnamefont
  {Du}}, \bibinfo {author} {\bibfnamefont {U.}~\bibnamefont {Heinz}}, \bibinfo
  {author} {\bibfnamefont {K.}~\bibnamefont {Rajagopal}}, \ and\ \bibinfo
  {author} {\bibfnamefont {Y.}~\bibnamefont {Yin}},\ }\href {\doibase
  10.1103/PhysRevC.102.054911} {\bibfield  {journal} {\bibinfo  {journal}
  {Phys. Rev. C}\ }\textbf {\bibinfo {volume} {102}},\ \bibinfo {pages}
  {054911} (\bibinfo {year} {2020})},\ \Eprint
  {http://arxiv.org/abs/2004.02719} {arXiv:2004.02719 [nucl-th]} \BibitemShut
  {NoStop}%
\bibitem [{\citenamefont {Pradeep}\ \emph {et~al.}(2022)\citenamefont
  {Pradeep}, \citenamefont {Rajagopal}, \citenamefont {Stephanov},\ and\
  \citenamefont {Yin}}]{Pradeep:2022mkf}%
  \BibitemOpen
  \bibfield  {author} {\bibinfo {author} {\bibfnamefont {M.}~\bibnamefont
  {Pradeep}}, \bibinfo {author} {\bibfnamefont {K.}~\bibnamefont {Rajagopal}},
  \bibinfo {author} {\bibfnamefont {M.}~\bibnamefont {Stephanov}}, \ and\
  \bibinfo {author} {\bibfnamefont {Y.}~\bibnamefont {Yin}},\ }\href {\doibase
  10.1103/PhysRevD.106.036017} {\bibfield  {journal} {\bibinfo  {journal}
  {Phys. Rev. D}\ }\textbf {\bibinfo {volume} {106}},\ \bibinfo {pages}
  {036017} (\bibinfo {year} {2022})},\ \Eprint
  {http://arxiv.org/abs/2204.00639} {arXiv:2204.00639 [hep-ph]} \BibitemShut
  {NoStop}%
\bibitem [{\citenamefont {An}\ \emph {et~al.}(2019)\citenamefont {An},
  \citenamefont {Basar}, \citenamefont {Stephanov},\ and\ \citenamefont
  {Yee}}]{An:2019osr}%
  \BibitemOpen
  \bibfield  {author} {\bibinfo {author} {\bibfnamefont {X.}~\bibnamefont
  {An}}, \bibinfo {author} {\bibfnamefont {G.}~\bibnamefont {Basar}}, \bibinfo
  {author} {\bibfnamefont {M.}~\bibnamefont {Stephanov}}, \ and\ \bibinfo
  {author} {\bibfnamefont {H.-U.}\ \bibnamefont {Yee}},\ }\href {\doibase
  10.1103/PhysRevC.100.024910} {\bibfield  {journal} {\bibinfo  {journal}
  {Phys. Rev. C}\ }\textbf {\bibinfo {volume} {100}},\ \bibinfo {pages}
  {024910} (\bibinfo {year} {2019})},\ \Eprint
  {http://arxiv.org/abs/1902.09517} {arXiv:1902.09517 [hep-th]} \BibitemShut
  {NoStop}%
\bibitem [{\citenamefont {An}\ \emph {et~al.}(2020)\citenamefont {An},
  \citenamefont {Ba\c{s}ar}, \citenamefont {Stephanov},\ and\ \citenamefont
  {Yee}}]{An:2019csj}%
  \BibitemOpen
  \bibfield  {author} {\bibinfo {author} {\bibfnamefont {X.}~\bibnamefont
  {An}}, \bibinfo {author} {\bibfnamefont {G.}~\bibnamefont {Ba\c{s}ar}},
  \bibinfo {author} {\bibfnamefont {M.}~\bibnamefont {Stephanov}}, \ and\
  \bibinfo {author} {\bibfnamefont {H.-U.}\ \bibnamefont {Yee}},\ }\href
  {\doibase 10.1103/PhysRevC.102.034901} {\bibfield  {journal} {\bibinfo
  {journal} {Phys. Rev. C}\ }\textbf {\bibinfo {volume} {102}},\ \bibinfo
  {pages} {034901} (\bibinfo {year} {2020})},\ \Eprint
  {http://arxiv.org/abs/1912.13456} {arXiv:1912.13456 [hep-th]} \BibitemShut
  {NoStop}%
\bibitem [{\citenamefont {An}\ \emph {et~al.}(2021)\citenamefont {An},
  \citenamefont {Ba\c{s}ar}, \citenamefont {Stephanov},\ and\ \citenamefont
  {Yee}}]{An:2020vri}%
  \BibitemOpen
  \bibfield  {author} {\bibinfo {author} {\bibfnamefont {X.}~\bibnamefont
  {An}}, \bibinfo {author} {\bibfnamefont {G.}~\bibnamefont {Ba\c{s}ar}},
  \bibinfo {author} {\bibfnamefont {M.}~\bibnamefont {Stephanov}}, \ and\
  \bibinfo {author} {\bibfnamefont {H.-U.}\ \bibnamefont {Yee}},\ }\href
  {\doibase 10.1103/PhysRevLett.127.072301} {\bibfield  {journal} {\bibinfo
  {journal} {Phys. Rev. Lett.}\ }\textbf {\bibinfo {volume} {127}},\ \bibinfo
  {pages} {072301} (\bibinfo {year} {2021})},\ \Eprint
  {http://arxiv.org/abs/2009.10742} {arXiv:2009.10742 [hep-th]} \BibitemShut
  {NoStop}%
\bibitem [{\citenamefont {Noronha-Hostler}\ \emph {et~al.}(2019)\citenamefont
  {Noronha-Hostler}, \citenamefont {Parotto}, \citenamefont {Ratti},\ and\
  \citenamefont {Stafford}}]{Noronha-Hostler:2019ayj}%
  \BibitemOpen
  \bibfield  {author} {\bibinfo {author} {\bibfnamefont {J.}~\bibnamefont
  {Noronha-Hostler}}, \bibinfo {author} {\bibfnamefont {P.}~\bibnamefont
  {Parotto}}, \bibinfo {author} {\bibfnamefont {C.}~\bibnamefont {Ratti}}, \
  and\ \bibinfo {author} {\bibfnamefont {J.~M.}\ \bibnamefont {Stafford}},\
  }\href {\doibase 10.1103/PhysRevC.100.064910} {\bibfield  {journal} {\bibinfo
   {journal} {Phys. Rev. C}\ }\textbf {\bibinfo {volume} {100}},\ \bibinfo
  {pages} {064910} (\bibinfo {year} {2019})},\ \Eprint
  {http://arxiv.org/abs/1902.06723} {arXiv:1902.06723 [hep-ph]} \BibitemShut
  {NoStop}%
\bibitem [{\citenamefont {Monnai}\ \emph {et~al.}(2019)\citenamefont {Monnai},
  \citenamefont {Schenke},\ and\ \citenamefont {Shen}}]{Monnai:2019hkn}%
  \BibitemOpen
  \bibfield  {author} {\bibinfo {author} {\bibfnamefont {A.}~\bibnamefont
  {Monnai}}, \bibinfo {author} {\bibfnamefont {B.}~\bibnamefont {Schenke}}, \
  and\ \bibinfo {author} {\bibfnamefont {C.}~\bibnamefont {Shen}},\ }\href
  {\doibase 10.1103/PhysRevC.100.024907} {\bibfield  {journal} {\bibinfo
  {journal} {Phys. Rev. C}\ }\textbf {\bibinfo {volume} {100}},\ \bibinfo
  {pages} {024907} (\bibinfo {year} {2019})},\ \Eprint
  {http://arxiv.org/abs/1902.05095} {arXiv:1902.05095 [nucl-th]} \BibitemShut
  {NoStop}%
\bibitem [{\citenamefont {Shen}\ and\ \citenamefont
  {Schenke}(2018)}]{Shen:2017bsr}%
  \BibitemOpen
  \bibfield  {author} {\bibinfo {author} {\bibfnamefont {C.}~\bibnamefont
  {Shen}}\ and\ \bibinfo {author} {\bibfnamefont {B.}~\bibnamefont {Schenke}},\
  }\href {\doibase 10.1103/PhysRevC.97.024907} {\bibfield  {journal} {\bibinfo
  {journal} {Phys. Rev. C}\ }\textbf {\bibinfo {volume} {97}},\ \bibinfo
  {pages} {024907} (\bibinfo {year} {2018})},\ \Eprint
  {http://arxiv.org/abs/1710.00881} {arXiv:1710.00881 [nucl-th]} \BibitemShut
  {NoStop}%
\bibitem [{\citenamefont {Martinez}\ \emph
  {et~al.}(2019{\natexlab{a}})\citenamefont {Martinez}, \citenamefont
  {Sievert}, \citenamefont {Wertepny},\ and\ \citenamefont
  {Noronha-Hostler}}]{Martinez:2019jbu}%
  \BibitemOpen
  \bibfield  {author} {\bibinfo {author} {\bibfnamefont {M.}~\bibnamefont
  {Martinez}}, \bibinfo {author} {\bibfnamefont {M.~D.}\ \bibnamefont
  {Sievert}}, \bibinfo {author} {\bibfnamefont {D.~E.}\ \bibnamefont
  {Wertepny}}, \ and\ \bibinfo {author} {\bibfnamefont {J.}~\bibnamefont
  {Noronha-Hostler}},\ }\href@noop {} {\  (\bibinfo {year}
  {2019}{\natexlab{a}})},\ \Eprint {http://arxiv.org/abs/1911.10272}
  {arXiv:1911.10272 [nucl-th]} \BibitemShut {NoStop}%
\bibitem [{\citenamefont {Carzon}\ \emph {et~al.}(2022)\citenamefont {Carzon},
  \citenamefont {Martinez}, \citenamefont {Sievert}, \citenamefont {Wertepny},\
  and\ \citenamefont {Noronha-Hostler}}]{Carzon:2019qja}%
  \BibitemOpen
  \bibfield  {author} {\bibinfo {author} {\bibfnamefont {P.}~\bibnamefont
  {Carzon}}, \bibinfo {author} {\bibfnamefont {M.}~\bibnamefont {Martinez}},
  \bibinfo {author} {\bibfnamefont {M.~D.}\ \bibnamefont {Sievert}}, \bibinfo
  {author} {\bibfnamefont {D.~E.}\ \bibnamefont {Wertepny}}, \ and\ \bibinfo
  {author} {\bibfnamefont {J.}~\bibnamefont {Noronha-Hostler}},\ }\href
  {\doibase 10.1103/PhysRevC.105.034908} {\bibfield  {journal} {\bibinfo
  {journal} {Phys. Rev. C}\ }\textbf {\bibinfo {volume} {105}},\ \bibinfo
  {pages} {034908} (\bibinfo {year} {2022})},\ \Eprint
  {http://arxiv.org/abs/1911.12454} {arXiv:1911.12454 [nucl-th]} \BibitemShut
  {NoStop}%
\bibitem [{\citenamefont {Inghirami}\ and\ \citenamefont
  {Elfner}(2022)}]{Inghirami:2022afu}%
  \BibitemOpen
  \bibfield  {author} {\bibinfo {author} {\bibfnamefont {G.}~\bibnamefont
  {Inghirami}}\ and\ \bibinfo {author} {\bibfnamefont {H.}~\bibnamefont
  {Elfner}},\ }\href {\doibase 10.1140/epjc/s10052-022-10718-x} {\bibfield
  {journal} {\bibinfo  {journal} {Eur. Phys. J. C}\ }\textbf {\bibinfo {volume}
  {82}},\ \bibinfo {pages} {796} (\bibinfo {year} {2022})},\ \Eprint
  {http://arxiv.org/abs/2201.05934} {arXiv:2201.05934 [hep-ph]} \BibitemShut
  {NoStop}%
\bibitem [{\citenamefont {Pratt}\ and\ \citenamefont
  {Torrieri}(2010)}]{Pratt:2010jt}%
  \BibitemOpen
  \bibfield  {author} {\bibinfo {author} {\bibfnamefont {S.}~\bibnamefont
  {Pratt}}\ and\ \bibinfo {author} {\bibfnamefont {G.}~\bibnamefont
  {Torrieri}},\ }\href {\doibase 10.1103/PhysRevC.82.044901} {\bibfield
  {journal} {\bibinfo  {journal} {Phys. Rev. C}\ }\textbf {\bibinfo {volume}
  {82}},\ \bibinfo {pages} {044901} (\bibinfo {year} {2010})},\ \Eprint
  {http://arxiv.org/abs/1003.0413} {arXiv:1003.0413 [nucl-th]} \BibitemShut
  {NoStop}%
\bibitem [{\citenamefont {Alqahtani}\ \emph
  {et~al.}(2017{\natexlab{a}})\citenamefont {Alqahtani}, \citenamefont
  {Nopoush},\ and\ \citenamefont {Strickland}}]{Alqahtani:2016rth}%
  \BibitemOpen
  \bibfield  {author} {\bibinfo {author} {\bibfnamefont {M.}~\bibnamefont
  {Alqahtani}}, \bibinfo {author} {\bibfnamefont {M.}~\bibnamefont {Nopoush}},
  \ and\ \bibinfo {author} {\bibfnamefont {M.}~\bibnamefont {Strickland}},\
  }\href {\doibase 10.1103/PhysRevC.95.034906} {\bibfield  {journal} {\bibinfo
  {journal} {Phys. Rev. C}\ }\textbf {\bibinfo {volume} {95}},\ \bibinfo
  {pages} {034906} (\bibinfo {year} {2017}{\natexlab{a}})},\ \Eprint
  {http://arxiv.org/abs/1605.02101} {arXiv:1605.02101 [nucl-th]} \BibitemShut
  {NoStop}%
\bibitem [{\citenamefont {Alqahtani}\ \emph
  {et~al.}(2017{\natexlab{b}})\citenamefont {Alqahtani}, \citenamefont
  {Nopoush}, \citenamefont {Ryblewski},\ and\ \citenamefont
  {Strickland}}]{Alqahtani:2017tnq}%
  \BibitemOpen
  \bibfield  {author} {\bibinfo {author} {\bibfnamefont {M.}~\bibnamefont
  {Alqahtani}}, \bibinfo {author} {\bibfnamefont {M.}~\bibnamefont {Nopoush}},
  \bibinfo {author} {\bibfnamefont {R.}~\bibnamefont {Ryblewski}}, \ and\
  \bibinfo {author} {\bibfnamefont {M.}~\bibnamefont {Strickland}},\ }\href
  {\doibase 10.1103/PhysRevC.96.044910} {\bibfield  {journal} {\bibinfo
  {journal} {Phys. Rev. C}\ }\textbf {\bibinfo {volume} {96}},\ \bibinfo
  {pages} {044910} (\bibinfo {year} {2017}{\natexlab{b}})},\ \Eprint
  {http://arxiv.org/abs/1705.10191} {arXiv:1705.10191 [nucl-th]} \BibitemShut
  {NoStop}%
\bibitem [{\citenamefont {Alqahtani}\ \emph {et~al.}(2018)\citenamefont
  {Alqahtani}, \citenamefont {Nopoush},\ and\ \citenamefont
  {Strickland}}]{Alqahtani:2017mhy}%
  \BibitemOpen
  \bibfield  {author} {\bibinfo {author} {\bibfnamefont {M.}~\bibnamefont
  {Alqahtani}}, \bibinfo {author} {\bibfnamefont {M.}~\bibnamefont {Nopoush}},
  \ and\ \bibinfo {author} {\bibfnamefont {M.}~\bibnamefont {Strickland}},\
  }\href {\doibase 10.1016/j.ppnp.2018.05.004} {\bibfield  {journal} {\bibinfo
  {journal} {Prog. Part. Nucl. Phys.}\ }\textbf {\bibinfo {volume} {101}},\
  \bibinfo {pages} {204} (\bibinfo {year} {2018})},\ \Eprint
  {http://arxiv.org/abs/1712.03282} {arXiv:1712.03282 [nucl-th]} \BibitemShut
  {NoStop}%
\bibitem [{\citenamefont {McNelis}\ and\ \citenamefont
  {Heinz}(2021)}]{McNelis:2021acu}%
  \BibitemOpen
  \bibfield  {author} {\bibinfo {author} {\bibfnamefont {M.}~\bibnamefont
  {McNelis}}\ and\ \bibinfo {author} {\bibfnamefont {U.}~\bibnamefont
  {Heinz}},\ }\href {\doibase 10.1103/PhysRevC.103.064903} {\bibfield
  {journal} {\bibinfo  {journal} {Phys. Rev. C}\ }\textbf {\bibinfo {volume}
  {103}},\ \bibinfo {pages} {064903} (\bibinfo {year} {2021})},\ \Eprint
  {http://arxiv.org/abs/2103.03401} {arXiv:2103.03401 [nucl-th]} \BibitemShut
  {NoStop}%
\bibitem [{\citenamefont {Oliinychenko}\ and\ \citenamefont
  {Koch}(2019)}]{Oliinychenko:2019zfk}%
  \BibitemOpen
  \bibfield  {author} {\bibinfo {author} {\bibfnamefont {D.}~\bibnamefont
  {Oliinychenko}}\ and\ \bibinfo {author} {\bibfnamefont {V.}~\bibnamefont
  {Koch}},\ }\href {\doibase 10.1103/PhysRevLett.123.182302} {\bibfield
  {journal} {\bibinfo  {journal} {Phys. Rev. Lett.}\ }\textbf {\bibinfo
  {volume} {123}},\ \bibinfo {pages} {182302} (\bibinfo {year} {2019})},\
  \Eprint {http://arxiv.org/abs/1902.09775} {arXiv:1902.09775 [hep-ph]}
  \BibitemShut {NoStop}%
\bibitem [{\citenamefont {Vovchenko}(2022)}]{Vovchenko:2022syc}%
  \BibitemOpen
  \bibfield  {author} {\bibinfo {author} {\bibfnamefont {V.}~\bibnamefont
  {Vovchenko}},\ }\href@noop {} {\  (\bibinfo {year} {2022})},\ \Eprint
  {http://arxiv.org/abs/2208.13693} {arXiv:2208.13693 [hep-ph]} \BibitemShut
  {NoStop}%
\bibitem [{\citenamefont {Adamczyk}\ \emph {et~al.}(2017)\citenamefont
  {Adamczyk} \emph {et~al.}}]{STAR:2017ckg}%
  \BibitemOpen
  \bibfield  {author} {\bibinfo {author} {\bibfnamefont {L.}~\bibnamefont
  {Adamczyk}} \emph {et~al.} (\bibinfo {collaboration} {STAR}),\ }\href
  {\doibase 10.1038/nature23004} {\bibfield  {journal} {\bibinfo  {journal}
  {Nature}\ }\textbf {\bibinfo {volume} {548}},\ \bibinfo {pages} {62}
  (\bibinfo {year} {2017})},\ \Eprint {http://arxiv.org/abs/1701.06657}
  {arXiv:1701.06657 [nucl-ex]} \BibitemShut {NoStop}%
\bibitem [{\citenamefont {Florkowski}\ \emph
  {et~al.}(2018{\natexlab{a}})\citenamefont {Florkowski}, \citenamefont
  {Friman}, \citenamefont {Jaiswal},\ and\ \citenamefont
  {Speranza}}]{Florkowski:2017ruc}%
  \BibitemOpen
  \bibfield  {author} {\bibinfo {author} {\bibfnamefont {W.}~\bibnamefont
  {Florkowski}}, \bibinfo {author} {\bibfnamefont {B.}~\bibnamefont {Friman}},
  \bibinfo {author} {\bibfnamefont {A.}~\bibnamefont {Jaiswal}}, \ and\
  \bibinfo {author} {\bibfnamefont {E.}~\bibnamefont {Speranza}},\ }\href
  {\doibase 10.1103/PhysRevC.97.041901} {\bibfield  {journal} {\bibinfo
  {journal} {Phys. Rev. C}\ }\textbf {\bibinfo {volume} {97}},\ \bibinfo
  {pages} {041901} (\bibinfo {year} {2018}{\natexlab{a}})},\ \Eprint
  {http://arxiv.org/abs/1705.00587} {arXiv:1705.00587 [nucl-th]} \BibitemShut
  {NoStop}%
\bibitem [{\citenamefont {Hattori}\ \emph {et~al.}(2019)\citenamefont
  {Hattori}, \citenamefont {Hongo}, \citenamefont {Huang}, \citenamefont
  {Matsuo},\ and\ \citenamefont {Taya}}]{Hattori:2019lfp}%
  \BibitemOpen
  \bibfield  {author} {\bibinfo {author} {\bibfnamefont {K.}~\bibnamefont
  {Hattori}}, \bibinfo {author} {\bibfnamefont {M.}~\bibnamefont {Hongo}},
  \bibinfo {author} {\bibfnamefont {X.-G.}\ \bibnamefont {Huang}}, \bibinfo
  {author} {\bibfnamefont {M.}~\bibnamefont {Matsuo}}, \ and\ \bibinfo {author}
  {\bibfnamefont {H.}~\bibnamefont {Taya}},\ }\href {\doibase
  10.1016/j.physletb.2019.05.040} {\bibfield  {journal} {\bibinfo  {journal}
  {Phys. Lett. B}\ }\textbf {\bibinfo {volume} {795}},\ \bibinfo {pages} {100}
  (\bibinfo {year} {2019})},\ \Eprint {http://arxiv.org/abs/1901.06615}
  {arXiv:1901.06615 [hep-th]} \BibitemShut {NoStop}%
\bibitem [{\citenamefont {Bhadury}\ \emph {et~al.}(2021)\citenamefont
  {Bhadury}, \citenamefont {Florkowski}, \citenamefont {Jaiswal}, \citenamefont
  {Kumar},\ and\ \citenamefont {Ryblewski}}]{Bhadury:2020puc}%
  \BibitemOpen
  \bibfield  {author} {\bibinfo {author} {\bibfnamefont {S.}~\bibnamefont
  {Bhadury}}, \bibinfo {author} {\bibfnamefont {W.}~\bibnamefont {Florkowski}},
  \bibinfo {author} {\bibfnamefont {A.}~\bibnamefont {Jaiswal}}, \bibinfo
  {author} {\bibfnamefont {A.}~\bibnamefont {Kumar}}, \ and\ \bibinfo {author}
  {\bibfnamefont {R.}~\bibnamefont {Ryblewski}},\ }\href {\doibase
  10.1016/j.physletb.2021.136096} {\bibfield  {journal} {\bibinfo  {journal}
  {Phys. Lett. B}\ }\textbf {\bibinfo {volume} {814}},\ \bibinfo {pages}
  {136096} (\bibinfo {year} {2021})},\ \Eprint
  {http://arxiv.org/abs/2002.03937} {arXiv:2002.03937 [hep-ph]} \BibitemShut
  {NoStop}%
\bibitem [{\citenamefont {Weickgenannt}\ \emph {et~al.}(2021)\citenamefont
  {Weickgenannt}, \citenamefont {Speranza}, \citenamefont {Sheng},
  \citenamefont {Wang},\ and\ \citenamefont {Rischke}}]{Weickgenannt:2020aaf}%
  \BibitemOpen
  \bibfield  {author} {\bibinfo {author} {\bibfnamefont {N.}~\bibnamefont
  {Weickgenannt}}, \bibinfo {author} {\bibfnamefont {E.}~\bibnamefont
  {Speranza}}, \bibinfo {author} {\bibfnamefont {X.-l.}\ \bibnamefont {Sheng}},
  \bibinfo {author} {\bibfnamefont {Q.}~\bibnamefont {Wang}}, \ and\ \bibinfo
  {author} {\bibfnamefont {D.~H.}\ \bibnamefont {Rischke}},\ }\href {\doibase
  10.1103/PhysRevLett.127.052301} {\bibfield  {journal} {\bibinfo  {journal}
  {Phys. Rev. Lett.}\ }\textbf {\bibinfo {volume} {127}},\ \bibinfo {pages}
  {052301} (\bibinfo {year} {2021})},\ \Eprint
  {http://arxiv.org/abs/2005.01506} {arXiv:2005.01506 [hep-ph]} \BibitemShut
  {NoStop}%
\bibitem [{\citenamefont {Hongo}\ \emph {et~al.}(2021)\citenamefont {Hongo},
  \citenamefont {Huang}, \citenamefont {Kaminski}, \citenamefont {Stephanov},\
  and\ \citenamefont {Yee}}]{Hongo:2021ona}%
  \BibitemOpen
  \bibfield  {author} {\bibinfo {author} {\bibfnamefont {M.}~\bibnamefont
  {Hongo}}, \bibinfo {author} {\bibfnamefont {X.-G.}\ \bibnamefont {Huang}},
  \bibinfo {author} {\bibfnamefont {M.}~\bibnamefont {Kaminski}}, \bibinfo
  {author} {\bibfnamefont {M.}~\bibnamefont {Stephanov}}, \ and\ \bibinfo
  {author} {\bibfnamefont {H.-U.}\ \bibnamefont {Yee}},\ }\href {\doibase
  10.1007/JHEP11(2021)150} {\bibfield  {journal} {\bibinfo  {journal} {JHEP}\
  }\textbf {\bibinfo {volume} {11}},\ \bibinfo {pages} {150} (\bibinfo {year}
  {2021})},\ \Eprint {http://arxiv.org/abs/2107.14231} {arXiv:2107.14231
  [hep-th]} \BibitemShut {NoStop}%
\bibitem [{\citenamefont {Weickgenannt}\ \emph {et~al.}(2022)\citenamefont
  {Weickgenannt}, \citenamefont {Wagner}, \citenamefont {Speranza},\ and\
  \citenamefont {Rischke}}]{Weickgenannt:2022zxs}%
  \BibitemOpen
  \bibfield  {author} {\bibinfo {author} {\bibfnamefont {N.}~\bibnamefont
  {Weickgenannt}}, \bibinfo {author} {\bibfnamefont {D.}~\bibnamefont
  {Wagner}}, \bibinfo {author} {\bibfnamefont {E.}~\bibnamefont {Speranza}}, \
  and\ \bibinfo {author} {\bibfnamefont {D.}~\bibnamefont {Rischke}},\
  }\href@noop {} {\  (\bibinfo {year} {2022})},\ \Eprint
  {http://arxiv.org/abs/2203.04766} {arXiv:2203.04766 [nucl-th]} \BibitemShut
  {NoStop}%
\bibitem [{\citenamefont {Li}(2000)}]{Li:2000bj}%
  \BibitemOpen
  \bibfield  {author} {\bibinfo {author} {\bibfnamefont {B.-A.}\ \bibnamefont
  {Li}},\ }\href {\doibase 10.1103/PhysRevLett.85.4221} {\bibfield  {journal}
  {\bibinfo  {journal} {Phys. Rev. Lett.}\ }\textbf {\bibinfo {volume} {85}},\
  \bibinfo {pages} {4221} (\bibinfo {year} {2000})},\ \Eprint
  {http://arxiv.org/abs/nucl-th/0009069} {arXiv:nucl-th/0009069} \BibitemShut
  {NoStop}%
\bibitem [{\citenamefont {Li}\ \emph {et~al.}(2014)\citenamefont {Li},
  \citenamefont {Ramos}, \citenamefont {Verde},\ and\ \citenamefont
  {Vidana}}]{Li:2014oda}%
  \BibitemOpen
  \bibfield  {author} {\bibinfo {author} {\bibfnamefont {B.-A.}\ \bibnamefont
  {Li}}, \bibinfo {author} {\bibfnamefont {A.}~\bibnamefont {Ramos}}, \bibinfo
  {author} {\bibfnamefont {G.}~\bibnamefont {Verde}}, \ and\ \bibinfo {author}
  {\bibfnamefont {I.}~\bibnamefont {Vidana}},\ }\href {\doibase
  10.1140/epja/i2014-14009-x} {\bibfield  {journal} {\bibinfo  {journal} {Eur.
  Phys. J. A}\ }\textbf {\bibinfo {volume} {50}},\ \bibinfo {pages} {9}
  (\bibinfo {year} {2014})}\BibitemShut {NoStop}%
\bibitem [{\citenamefont {Colonna}(2020)}]{Colonna:2020euy}%
  \BibitemOpen
  \bibfield  {author} {\bibinfo {author} {\bibfnamefont {M.}~\bibnamefont
  {Colonna}},\ }\href {\doibase 10.1016/j.ppnp.2020.103775} {\  (\bibinfo
  {year} {2020}),\ 10.1016/j.ppnp.2020.103775},\ \Eprint
  {http://arxiv.org/abs/2003.02500} {arXiv:2003.02500 [nucl-th]} \BibitemShut
  {NoStop}%
\bibitem [{\citenamefont {Xu}(2019)}]{Xu:2019hqg}%
  \BibitemOpen
  \bibfield  {author} {\bibinfo {author} {\bibfnamefont {J.}~\bibnamefont
  {Xu}},\ }\href {\doibase 10.1016/j.ppnp.2019.02.009} {\bibfield  {journal}
  {\bibinfo  {journal} {Prog. Part. Nucl. Phys.}\ }\textbf {\bibinfo {volume}
  {106}},\ \bibinfo {pages} {312} (\bibinfo {year} {2019})},\ \Eprint
  {http://arxiv.org/abs/1904.00131} {arXiv:1904.00131 [nucl-th]} \BibitemShut
  {NoStop}%
\bibitem [{\citenamefont {Russotto}\ \emph {et~al.}(2011)\citenamefont
  {Russotto} \emph {et~al.}}]{Russotto:2011hq}%
  \BibitemOpen
  \bibfield  {author} {\bibinfo {author} {\bibfnamefont {P.}~\bibnamefont
  {Russotto}} \emph {et~al.},\ }\href {\doibase 10.1016/j.physletb.2011.02.033}
  {\bibfield  {journal} {\bibinfo  {journal} {Phys. Lett. B}\ }\textbf
  {\bibinfo {volume} {697}},\ \bibinfo {pages} {471} (\bibinfo {year}
  {2011})},\ \Eprint {http://arxiv.org/abs/1101.2361} {arXiv:1101.2361
  [nucl-ex]} \BibitemShut {NoStop}%
\bibitem [{\citenamefont {Cozma}(2011)}]{Cozma:2011nr}%
  \BibitemOpen
  \bibfield  {author} {\bibinfo {author} {\bibfnamefont {M.~D.}\ \bibnamefont
  {Cozma}},\ }\href {\doibase 10.1016/j.physletb.2011.05.002} {\bibfield
  {journal} {\bibinfo  {journal} {Phys. Lett. B}\ }\textbf {\bibinfo {volume}
  {700}},\ \bibinfo {pages} {139} (\bibinfo {year} {2011})},\ \Eprint
  {http://arxiv.org/abs/1102.2728} {arXiv:1102.2728 [nucl-th]} \BibitemShut
  {NoStop}%
\bibitem [{\citenamefont {Giordano}\ \emph {et~al.}(2010)\citenamefont
  {Giordano}, \citenamefont {Colonna}, \citenamefont {Di~Toro}, \citenamefont
  {Greco},\ and\ \citenamefont {Rizzo}}]{Giordano:2010pv}%
  \BibitemOpen
  \bibfield  {author} {\bibinfo {author} {\bibfnamefont {V.}~\bibnamefont
  {Giordano}}, \bibinfo {author} {\bibfnamefont {M.}~\bibnamefont {Colonna}},
  \bibinfo {author} {\bibfnamefont {M.}~\bibnamefont {Di~Toro}}, \bibinfo
  {author} {\bibfnamefont {V.}~\bibnamefont {Greco}}, \ and\ \bibinfo {author}
  {\bibfnamefont {J.}~\bibnamefont {Rizzo}},\ }\href {\doibase
  10.1103/PhysRevC.81.044611} {\bibfield  {journal} {\bibinfo  {journal} {Phys.
  Rev. C}\ }\textbf {\bibinfo {volume} {81}},\ \bibinfo {pages} {044611}
  (\bibinfo {year} {2010})},\ \Eprint {http://arxiv.org/abs/1001.4961}
  {arXiv:1001.4961 [nucl-th]} \BibitemShut {NoStop}%
\bibitem [{\citenamefont {Fuchs}\ \emph {et~al.}(2001)\citenamefont {Fuchs},
  \citenamefont {Faessler}, \citenamefont {Zabrodin},\ and\ \citenamefont
  {Zheng}}]{Fuchs:2000kp}%
  \BibitemOpen
  \bibfield  {author} {\bibinfo {author} {\bibfnamefont {C.}~\bibnamefont
  {Fuchs}}, \bibinfo {author} {\bibfnamefont {A.}~\bibnamefont {Faessler}},
  \bibinfo {author} {\bibfnamefont {E.}~\bibnamefont {Zabrodin}}, \ and\
  \bibinfo {author} {\bibfnamefont {Y.-M.}\ \bibnamefont {Zheng}},\ }\href
  {\doibase 10.1103/PhysRevLett.86.1974} {\bibfield  {journal} {\bibinfo
  {journal} {Phys. Rev. Lett.}\ }\textbf {\bibinfo {volume} {86}},\ \bibinfo
  {pages} {1974} (\bibinfo {year} {2001})},\ \Eprint
  {http://arxiv.org/abs/nucl-th/0011102} {arXiv:nucl-th/0011102} \BibitemShut
  {NoStop}%
\bibitem [{\citenamefont {Li}(2002)}]{Li:2002qx}%
  \BibitemOpen
  \bibfield  {author} {\bibinfo {author} {\bibfnamefont {B.-A.}\ \bibnamefont
  {Li}},\ }\href {\doibase 10.1103/PhysRevLett.88.192701} {\bibfield  {journal}
  {\bibinfo  {journal} {Phys. Rev. Lett.}\ }\textbf {\bibinfo {volume} {88}},\
  \bibinfo {pages} {192701} (\bibinfo {year} {2002})},\ \Eprint
  {http://arxiv.org/abs/nucl-th/0205002} {arXiv:nucl-th/0205002} \BibitemShut
  {NoStop}%
\bibitem [{\citenamefont {Xiao}\ \emph {et~al.}(2009)\citenamefont {Xiao},
  \citenamefont {Li}, \citenamefont {Chen}, \citenamefont {Yong},\ and\
  \citenamefont {Zhang}}]{Xiao:2008vm}%
  \BibitemOpen
  \bibfield  {author} {\bibinfo {author} {\bibfnamefont {Z.}~\bibnamefont
  {Xiao}}, \bibinfo {author} {\bibfnamefont {B.-A.}\ \bibnamefont {Li}},
  \bibinfo {author} {\bibfnamefont {L.-W.}\ \bibnamefont {Chen}}, \bibinfo
  {author} {\bibfnamefont {G.-C.}\ \bibnamefont {Yong}}, \ and\ \bibinfo
  {author} {\bibfnamefont {M.}~\bibnamefont {Zhang}},\ }\href {\doibase
  10.1103/PhysRevLett.102.062502} {\bibfield  {journal} {\bibinfo  {journal}
  {Phys. Rev. Lett.}\ }\textbf {\bibinfo {volume} {102}},\ \bibinfo {pages}
  {062502} (\bibinfo {year} {2009})},\ \Eprint {http://arxiv.org/abs/0808.0186}
  {arXiv:0808.0186 [nucl-th]} \BibitemShut {NoStop}%
\bibitem [{\citenamefont {Yong}\ \emph {et~al.}(2022)\citenamefont {Yong},
  \citenamefont {Li}, \citenamefont {Xiao},\ and\ \citenamefont
  {Lin}}]{Yong:2022pyb}%
  \BibitemOpen
  \bibfield  {author} {\bibinfo {author} {\bibfnamefont {G.-C.}\ \bibnamefont
  {Yong}}, \bibinfo {author} {\bibfnamefont {B.-A.}\ \bibnamefont {Li}},
  \bibinfo {author} {\bibfnamefont {Z.-G.}\ \bibnamefont {Xiao}}, \ and\
  \bibinfo {author} {\bibfnamefont {Z.-W.}\ \bibnamefont {Lin}},\ }\href
  {\doibase 10.1103/PhysRevC.106.024902} {\bibfield  {journal} {\bibinfo
  {journal} {Phys. Rev. C}\ }\textbf {\bibinfo {volume} {106}},\ \bibinfo
  {pages} {024902} (\bibinfo {year} {2022})},\ \Eprint
  {http://arxiv.org/abs/2206.10766} {arXiv:2206.10766 [nucl-th]} \BibitemShut
  {NoStop}%
\bibitem [{\citenamefont {Estee}\ \emph {et~al.}(2021)\citenamefont {Estee}
  \emph {et~al.}}]{SRIT:2021gcy}%
  \BibitemOpen
  \bibfield  {author} {\bibinfo {author} {\bibfnamefont {J.}~\bibnamefont
  {Estee}} \emph {et~al.} (\bibinfo {collaboration} {SpiRIT}),\ }\href
  {\doibase 10.1103/PhysRevLett.126.162701} {\bibfield  {journal} {\bibinfo
  {journal} {Phys. Rev. Lett.}\ }\textbf {\bibinfo {volume} {126}},\ \bibinfo
  {pages} {162701} (\bibinfo {year} {2021})},\ \Eprint
  {http://arxiv.org/abs/2103.06861} {arXiv:2103.06861 [nucl-ex]} \BibitemShut
  {NoStop}%
\bibitem [{\citenamefont {Le~F\`evre}\ \emph {et~al.}(2016)\citenamefont
  {Le~F\`evre}, \citenamefont {Leifels}, \citenamefont {Reisdorf},
  \citenamefont {Aichelin},\ and\ \citenamefont {Hartnack}}]{LeFevre:2015paj}%
  \BibitemOpen
  \bibfield  {author} {\bibinfo {author} {\bibfnamefont {A.}~\bibnamefont
  {Le~F\`evre}}, \bibinfo {author} {\bibfnamefont {Y.}~\bibnamefont {Leifels}},
  \bibinfo {author} {\bibfnamefont {W.}~\bibnamefont {Reisdorf}}, \bibinfo
  {author} {\bibfnamefont {J.}~\bibnamefont {Aichelin}}, \ and\ \bibinfo
  {author} {\bibfnamefont {C.}~\bibnamefont {Hartnack}},\ }\href {\doibase
  10.1016/j.nuclphysa.2015.09.015} {\bibfield  {journal} {\bibinfo  {journal}
  {Nucl. Phys. A}\ }\textbf {\bibinfo {volume} {945}},\ \bibinfo {pages} {112}
  (\bibinfo {year} {2016})},\ \Eprint {http://arxiv.org/abs/1501.05246}
  {arXiv:1501.05246 [nucl-ex]} \BibitemShut {NoStop}%
\bibitem [{\citenamefont {Nara}\ \emph {et~al.}(2020)\citenamefont {Nara},
  \citenamefont {Maruyama},\ and\ \citenamefont {Stoecker}}]{Nara:2020ztb}%
  \BibitemOpen
  \bibfield  {author} {\bibinfo {author} {\bibfnamefont {Y.}~\bibnamefont
  {Nara}}, \bibinfo {author} {\bibfnamefont {T.}~\bibnamefont {Maruyama}}, \
  and\ \bibinfo {author} {\bibfnamefont {H.}~\bibnamefont {Stoecker}},\ }\href
  {\doibase 10.1103/PhysRevC.102.024913} {\bibfield  {journal} {\bibinfo
  {journal} {Phys. Rev. C}\ }\textbf {\bibinfo {volume} {102}},\ \bibinfo
  {pages} {024913} (\bibinfo {year} {2020})},\ \Eprint
  {http://arxiv.org/abs/2004.05550} {arXiv:2004.05550 [nucl-th]} \BibitemShut
  {NoStop}%
\bibitem [{\citenamefont {Danielewicz}\ \emph {et~al.}(2002)\citenamefont
  {Danielewicz}, \citenamefont {Lacey},\ and\ \citenamefont
  {Lynch}}]{Danielewicz:2002pu}%
  \BibitemOpen
  \bibfield  {author} {\bibinfo {author} {\bibfnamefont {P.}~\bibnamefont
  {Danielewicz}}, \bibinfo {author} {\bibfnamefont {R.}~\bibnamefont {Lacey}},
  \ and\ \bibinfo {author} {\bibfnamefont {W.~G.}\ \bibnamefont {Lynch}},\
  }\href {\doibase 10.1126/science.1078070} {\bibfield  {journal} {\bibinfo
  {journal} {Science}\ }\textbf {\bibinfo {volume} {298}},\ \bibinfo {pages}
  {1592} (\bibinfo {year} {2002})},\ \Eprint
  {http://arxiv.org/abs/nucl-th/0208016} {arXiv:nucl-th/0208016} \BibitemShut
  {NoStop}%
\bibitem [{\citenamefont {Oliinychenko}\ \emph {et~al.}(2022)\citenamefont
  {Oliinychenko}, \citenamefont {Sorensen}, \citenamefont {Koch},\ and\
  \citenamefont {McLerran}}]{Oliinychenko:2022uvy}%
  \BibitemOpen
  \bibfield  {author} {\bibinfo {author} {\bibfnamefont {D.}~\bibnamefont
  {Oliinychenko}}, \bibinfo {author} {\bibfnamefont {A.}~\bibnamefont
  {Sorensen}}, \bibinfo {author} {\bibfnamefont {V.}~\bibnamefont {Koch}}, \
  and\ \bibinfo {author} {\bibfnamefont {L.}~\bibnamefont {McLerran}},\
  }\href@noop {} {\  (\bibinfo {year} {2022})},\ \Eprint
  {http://arxiv.org/abs/2208.11996} {arXiv:2208.11996 [nucl-th]} \BibitemShut
  {NoStop}%
\bibitem [{\citenamefont {Bedaque}\ and\ \citenamefont
  {Steiner}(2015)}]{Bedaque:2014sqa}%
  \BibitemOpen
  \bibfield  {author} {\bibinfo {author} {\bibfnamefont {P.}~\bibnamefont
  {Bedaque}}\ and\ \bibinfo {author} {\bibfnamefont {A.~W.}\ \bibnamefont
  {Steiner}},\ }\href {\doibase 10.1103/PhysRevLett.114.031103} {\bibfield
  {journal} {\bibinfo  {journal} {Phys. Rev. Lett.}\ }\textbf {\bibinfo
  {volume} {114}},\ \bibinfo {pages} {031103} (\bibinfo {year} {2015})},\
  \Eprint {http://arxiv.org/abs/1408.5116} {arXiv:1408.5116 [nucl-th]}
  \BibitemShut {NoStop}%
\bibitem [{\citenamefont {Fujimoto}\ \emph {et~al.}(2020)\citenamefont
  {Fujimoto}, \citenamefont {Fukushima},\ and\ \citenamefont
  {Murase}}]{Fujimoto:2019hxv}%
  \BibitemOpen
  \bibfield  {author} {\bibinfo {author} {\bibfnamefont {Y.}~\bibnamefont
  {Fujimoto}}, \bibinfo {author} {\bibfnamefont {K.}~\bibnamefont {Fukushima}},
  \ and\ \bibinfo {author} {\bibfnamefont {K.}~\bibnamefont {Murase}},\ }\href
  {\doibase 10.1103/PhysRevD.101.054016} {\bibfield  {journal} {\bibinfo
  {journal} {Phys. Rev. D}\ }\textbf {\bibinfo {volume} {101}},\ \bibinfo
  {pages} {054016} (\bibinfo {year} {2020})},\ \Eprint
  {http://arxiv.org/abs/1903.03400} {arXiv:1903.03400 [nucl-th]} \BibitemShut
  {NoStop}%
\bibitem [{\citenamefont {Tan}\ \emph {et~al.}(2022{\natexlab{b}})\citenamefont
  {Tan}, \citenamefont {Dore}, \citenamefont {Dexheimer}, \citenamefont
  {Noronha-Hostler},\ and\ \citenamefont {Yunes}}]{Tan:2021ahl}%
  \BibitemOpen
  \bibfield  {author} {\bibinfo {author} {\bibfnamefont {H.}~\bibnamefont
  {Tan}}, \bibinfo {author} {\bibfnamefont {T.}~\bibnamefont {Dore}}, \bibinfo
  {author} {\bibfnamefont {V.}~\bibnamefont {Dexheimer}}, \bibinfo {author}
  {\bibfnamefont {J.}~\bibnamefont {Noronha-Hostler}}, \ and\ \bibinfo {author}
  {\bibfnamefont {N.}~\bibnamefont {Yunes}},\ }\href {\doibase
  10.1103/PhysRevD.105.023018} {\bibfield  {journal} {\bibinfo  {journal}
  {Phys. Rev. D}\ }\textbf {\bibinfo {volume} {105}},\ \bibinfo {pages}
  {023018} (\bibinfo {year} {2022}{\natexlab{b}})},\ \Eprint
  {http://arxiv.org/abs/2106.03890} {arXiv:2106.03890 [astro-ph.HE]}
  \BibitemShut {NoStop}%
\bibitem [{\citenamefont {Marczenko}\ \emph {et~al.}(2022)\citenamefont
  {Marczenko}, \citenamefont {McLerran}, \citenamefont {Redlich},\ and\
  \citenamefont {Sasaki}}]{Marczenko:2022jhl}%
  \BibitemOpen
  \bibfield  {author} {\bibinfo {author} {\bibfnamefont {M.}~\bibnamefont
  {Marczenko}}, \bibinfo {author} {\bibfnamefont {L.}~\bibnamefont {McLerran}},
  \bibinfo {author} {\bibfnamefont {K.}~\bibnamefont {Redlich}}, \ and\
  \bibinfo {author} {\bibfnamefont {C.}~\bibnamefont {Sasaki}},\ }\href@noop {}
  {\  (\bibinfo {year} {2022})},\ \Eprint {http://arxiv.org/abs/2207.13059}
  {arXiv:2207.13059 [nucl-th]} \BibitemShut {NoStop}%
\bibitem [{\citenamefont {Sorensen}\ and\ \citenamefont
  {Koch}(2021)}]{Sorensen:2020ygf}%
  \BibitemOpen
  \bibfield  {author} {\bibinfo {author} {\bibfnamefont {A.}~\bibnamefont
  {Sorensen}}\ and\ \bibinfo {author} {\bibfnamefont {V.}~\bibnamefont
  {Koch}},\ }\href {\doibase 10.1103/PhysRevC.104.034904} {\bibfield  {journal}
  {\bibinfo  {journal} {Phys. Rev. C}\ }\textbf {\bibinfo {volume} {104}},\
  \bibinfo {pages} {034904} (\bibinfo {year} {2021})},\ \Eprint
  {http://arxiv.org/abs/2011.06635} {arXiv:2011.06635 [nucl-th]} \BibitemShut
  {NoStop}%
\bibitem [{\citenamefont {Cai}\ and\ \citenamefont {Li}(2022)}]{Cai:2022grw}%
  \BibitemOpen
  \bibfield  {author} {\bibinfo {author} {\bibfnamefont {B.-J.}\ \bibnamefont
  {Cai}}\ and\ \bibinfo {author} {\bibfnamefont {B.-A.}\ \bibnamefont {Li}},\
  }\href@noop {} {\  (\bibinfo {year} {2022})},\ \Eprint
  {http://arxiv.org/abs/2210.10924} {arXiv:2210.10924 [nucl-th]} \BibitemShut
  {NoStop}%
\bibitem [{\citenamefont {Hen}\ \emph {et~al.}(2017)\citenamefont {Hen},
  \citenamefont {Miller}, \citenamefont {Piasetzky},\ and\ \citenamefont
  {Weinstein}}]{Hen:2016kwk}%
  \BibitemOpen
  \bibfield  {author} {\bibinfo {author} {\bibfnamefont {O.}~\bibnamefont
  {Hen}}, \bibinfo {author} {\bibfnamefont {G.~A.}\ \bibnamefont {Miller}},
  \bibinfo {author} {\bibfnamefont {E.}~\bibnamefont {Piasetzky}}, \ and\
  \bibinfo {author} {\bibfnamefont {L.~B.}\ \bibnamefont {Weinstein}},\ }\href
  {\doibase 10.1103/RevModPhys.89.045002} {\bibfield  {journal} {\bibinfo
  {journal} {Rev. Mod. Phys.}\ }\textbf {\bibinfo {volume} {89}},\ \bibinfo
  {pages} {045002} (\bibinfo {year} {2017})},\ \Eprint
  {http://arxiv.org/abs/1611.09748} {arXiv:1611.09748 [nucl-ex]} \BibitemShut
  {NoStop}%
\bibitem [{\citenamefont {Li}\ \emph {et~al.}(2018)\citenamefont {Li},
  \citenamefont {Cai}, \citenamefont {Chen},\ and\ \citenamefont
  {Xu}}]{Li:2018lpy}%
  \BibitemOpen
  \bibfield  {author} {\bibinfo {author} {\bibfnamefont {B.-A.}\ \bibnamefont
  {Li}}, \bibinfo {author} {\bibfnamefont {B.-J.}\ \bibnamefont {Cai}},
  \bibinfo {author} {\bibfnamefont {L.-W.}\ \bibnamefont {Chen}}, \ and\
  \bibinfo {author} {\bibfnamefont {J.}~\bibnamefont {Xu}},\ }\href {\doibase
  10.1016/j.ppnp.2018.01.001} {\bibfield  {journal} {\bibinfo  {journal} {Prog.
  Part. Nucl. Phys.}\ }\textbf {\bibinfo {volume} {99}},\ \bibinfo {pages} {29}
  (\bibinfo {year} {2018})},\ \Eprint {http://arxiv.org/abs/1801.01213}
  {arXiv:1801.01213 [nucl-th]} \BibitemShut {NoStop}%
\bibitem [{\citenamefont {Sorensen}(2021)}]{Sorensen:2021zxd}%
  \BibitemOpen
  \bibfield  {author} {\bibinfo {author} {\bibfnamefont {A.~M.}\ \bibnamefont
  {Sorensen}},\ }\emph {\bibinfo {title} {{Density Functional Equation of State
  and Its Application to the Phenomenology of Heavy-Ion Collisions}}},\
  \href@noop {} {Ph.D. thesis},\ \bibinfo  {school} {UCLA, Los Angeles (main),
  UCLA} (\bibinfo {year} {2021}),\ \Eprint {http://arxiv.org/abs/2109.08105}
  {arXiv:2109.08105 [nucl-th]} \BibitemShut {NoStop}%
\bibitem [{\citenamefont {Wolter}\ \emph {et~al.}(2022)\citenamefont {Wolter}
  \emph {et~al.}}]{TMEP:2022xjg}%
  \BibitemOpen
  \bibfield  {author} {\bibinfo {author} {\bibfnamefont {H.}~\bibnamefont
  {Wolter}} \emph {et~al.} (\bibinfo {collaboration} {TMEP}),\ }\href {\doibase
  10.1016/j.ppnp.2022.103962} {\bibfield  {journal} {\bibinfo  {journal} {Prog.
  Part. Nucl. Phys.}\ }\textbf {\bibinfo {volume} {125}},\ \bibinfo {pages}
  {103962} (\bibinfo {year} {2022})},\ \Eprint
  {http://arxiv.org/abs/2202.06672} {arXiv:2202.06672 [nucl-th]} \BibitemShut
  {NoStop}%
\bibitem [{\citenamefont {Niemi}\ and\ \citenamefont
  {Denicol}(2014)}]{Niemi:2014wta}%
  \BibitemOpen
  \bibfield  {author} {\bibinfo {author} {\bibfnamefont {H.}~\bibnamefont
  {Niemi}}\ and\ \bibinfo {author} {\bibfnamefont {G.~S.}\ \bibnamefont
  {Denicol}},\ }\href@noop {} {\  (\bibinfo {year} {2014})},\ \Eprint
  {http://arxiv.org/abs/1404.7327} {arXiv:1404.7327 [nucl-th]} \BibitemShut
  {NoStop}%
\bibitem [{\citenamefont {Noronha-Hostler}\ \emph
  {et~al.}(2016{\natexlab{a}})\citenamefont {Noronha-Hostler}, \citenamefont
  {Noronha},\ and\ \citenamefont {Gyulassy}}]{Noronha-Hostler:2015coa}%
  \BibitemOpen
  \bibfield  {author} {\bibinfo {author} {\bibfnamefont {J.}~\bibnamefont
  {Noronha-Hostler}}, \bibinfo {author} {\bibfnamefont {J.}~\bibnamefont
  {Noronha}}, \ and\ \bibinfo {author} {\bibfnamefont {M.}~\bibnamefont
  {Gyulassy}},\ }\href {\doibase 10.1103/PhysRevC.93.024909} {\bibfield
  {journal} {\bibinfo  {journal} {Phys. Rev. C}\ }\textbf {\bibinfo {volume}
  {93}},\ \bibinfo {pages} {024909} (\bibinfo {year} {2016}{\natexlab{a}})},\
  \Eprint {http://arxiv.org/abs/1508.02455} {arXiv:1508.02455 [nucl-th]}
  \BibitemShut {NoStop}%
\bibitem [{\citenamefont {Aidala}\ \emph {et~al.}(2019)\citenamefont {Aidala}
  \emph {et~al.}}]{PHENIX:2018lia}%
  \BibitemOpen
  \bibfield  {author} {\bibinfo {author} {\bibfnamefont {C.}~\bibnamefont
  {Aidala}} \emph {et~al.} (\bibinfo {collaboration} {PHENIX}),\ }\href
  {\doibase 10.1038/s41567-018-0360-0} {\bibfield  {journal} {\bibinfo
  {journal} {Nature Phys.}\ }\textbf {\bibinfo {volume} {15}},\ \bibinfo
  {pages} {214} (\bibinfo {year} {2019})},\ \Eprint
  {http://arxiv.org/abs/1805.02973} {arXiv:1805.02973 [nucl-ex]} \BibitemShut
  {NoStop}%
\bibitem [{\citenamefont {Martinez}\ and\ \citenamefont
  {Strickland}(2010)}]{Martinez:2010sc}%
  \BibitemOpen
  \bibfield  {author} {\bibinfo {author} {\bibfnamefont {M.}~\bibnamefont
  {Martinez}}\ and\ \bibinfo {author} {\bibfnamefont {M.}~\bibnamefont
  {Strickland}},\ }\href {\doibase 10.1016/j.nuclphysa.2010.08.011} {\bibfield
  {journal} {\bibinfo  {journal} {Nucl. Phys. A}\ }\textbf {\bibinfo {volume}
  {848}},\ \bibinfo {pages} {183} (\bibinfo {year} {2010})},\ \Eprint
  {http://arxiv.org/abs/1007.0889} {arXiv:1007.0889 [nucl-th]} \BibitemShut
  {NoStop}%
\bibitem [{\citenamefont {Florkowski}\ and\ \citenamefont
  {Ryblewski}(2011)}]{Florkowski:2010cf}%
  \BibitemOpen
  \bibfield  {author} {\bibinfo {author} {\bibfnamefont {W.}~\bibnamefont
  {Florkowski}}\ and\ \bibinfo {author} {\bibfnamefont {R.}~\bibnamefont
  {Ryblewski}},\ }\href {\doibase 10.1103/PhysRevC.83.034907} {\bibfield
  {journal} {\bibinfo  {journal} {Phys. Rev. C}\ }\textbf {\bibinfo {volume}
  {83}},\ \bibinfo {pages} {034907} (\bibinfo {year} {2011})},\ \Eprint
  {http://arxiv.org/abs/1007.0130} {arXiv:1007.0130 [nucl-th]} \BibitemShut
  {NoStop}%
\bibitem [{\citenamefont {Romatschke}\ and\ \citenamefont
  {Romatschke}(2019)}]{Romatschke:2017ejr}%
  \BibitemOpen
  \bibfield  {author} {\bibinfo {author} {\bibfnamefont {P.}~\bibnamefont
  {Romatschke}}\ and\ \bibinfo {author} {\bibfnamefont {U.}~\bibnamefont
  {Romatschke}},\ }\href {\doibase 10.1017/9781108651998} {\emph {\bibinfo
  {title} {{Relativistic Fluid Dynamics In and Out of Equilibrium}}}},\
  Cambridge Monographs on Mathematical Physics\ (\bibinfo  {publisher}
  {Cambridge University Press},\ \bibinfo {year} {2019})\ \Eprint
  {http://arxiv.org/abs/1712.05815} {arXiv:1712.05815 [nucl-th]} \BibitemShut
  {NoStop}%
\bibitem [{\citenamefont {Berges}\ \emph
  {et~al.}(2021{\natexlab{a}})\citenamefont {Berges}, \citenamefont {Heller},
  \citenamefont {Mazeliauskas},\ and\ \citenamefont
  {Venugopalan}}]{Berges:2020fwq}%
  \BibitemOpen
  \bibfield  {author} {\bibinfo {author} {\bibfnamefont {J.}~\bibnamefont
  {Berges}}, \bibinfo {author} {\bibfnamefont {M.~P.}\ \bibnamefont {Heller}},
  \bibinfo {author} {\bibfnamefont {A.}~\bibnamefont {Mazeliauskas}}, \ and\
  \bibinfo {author} {\bibfnamefont {R.}~\bibnamefont {Venugopalan}},\ }\href
  {\doibase 10.1103/RevModPhys.93.035003} {\bibfield  {journal} {\bibinfo
  {journal} {Rev. Mod. Phys.}\ }\textbf {\bibinfo {volume} {93}},\ \bibinfo
  {pages} {035003} (\bibinfo {year} {2021}{\natexlab{a}})},\ \Eprint
  {http://arxiv.org/abs/2005.12299} {arXiv:2005.12299 [hep-th]} \BibitemShut
  {NoStop}%
\bibitem [{\citenamefont {Chesler}\ and\ \citenamefont
  {Yaffe}(2011)}]{Chesler:2010bi}%
  \BibitemOpen
  \bibfield  {author} {\bibinfo {author} {\bibfnamefont {P.~M.}\ \bibnamefont
  {Chesler}}\ and\ \bibinfo {author} {\bibfnamefont {L.~G.}\ \bibnamefont
  {Yaffe}},\ }\href {\doibase 10.1103/PhysRevLett.106.021601} {\bibfield
  {journal} {\bibinfo  {journal} {Phys. Rev. Lett.}\ }\textbf {\bibinfo
  {volume} {106}},\ \bibinfo {pages} {021601} (\bibinfo {year} {2011})},\
  \Eprint {http://arxiv.org/abs/1011.3562} {arXiv:1011.3562 [hep-th]}
  \BibitemShut {NoStop}%
\bibitem [{\citenamefont {Heller}\ \emph
  {et~al.}(2012{\natexlab{a}})\citenamefont {Heller}, \citenamefont {Janik},\
  and\ \citenamefont {Witaszczyk}}]{Heller:2011ju}%
  \BibitemOpen
  \bibfield  {author} {\bibinfo {author} {\bibfnamefont {M.~P.}\ \bibnamefont
  {Heller}}, \bibinfo {author} {\bibfnamefont {R.~A.}\ \bibnamefont {Janik}}, \
  and\ \bibinfo {author} {\bibfnamefont {P.}~\bibnamefont {Witaszczyk}},\
  }\href {\doibase 10.1103/PhysRevLett.108.201602} {\bibfield  {journal}
  {\bibinfo  {journal} {Phys. Rev. Lett.}\ }\textbf {\bibinfo {volume} {108}},\
  \bibinfo {pages} {201602} (\bibinfo {year} {2012}{\natexlab{a}})},\ \Eprint
  {http://arxiv.org/abs/1103.3452} {arXiv:1103.3452 [hep-th]} \BibitemShut
  {NoStop}%
\bibitem [{\citenamefont {Heller}\ \emph
  {et~al.}(2012{\natexlab{b}})\citenamefont {Heller}, \citenamefont {Mateos},
  \citenamefont {van~der Schee},\ and\ \citenamefont
  {Trancanelli}}]{Heller:2012km}%
  \BibitemOpen
  \bibfield  {author} {\bibinfo {author} {\bibfnamefont {M.~P.}\ \bibnamefont
  {Heller}}, \bibinfo {author} {\bibfnamefont {D.}~\bibnamefont {Mateos}},
  \bibinfo {author} {\bibfnamefont {W.}~\bibnamefont {van~der Schee}}, \ and\
  \bibinfo {author} {\bibfnamefont {D.}~\bibnamefont {Trancanelli}},\ }\href
  {\doibase 10.1103/PhysRevLett.108.191601} {\bibfield  {journal} {\bibinfo
  {journal} {Phys. Rev. Lett.}\ }\textbf {\bibinfo {volume} {108}},\ \bibinfo
  {pages} {191601} (\bibinfo {year} {2012}{\natexlab{b}})},\ \Eprint
  {http://arxiv.org/abs/1202.0981} {arXiv:1202.0981 [hep-th]} \BibitemShut
  {NoStop}%
\bibitem [{\citenamefont {van~der Schee}\ \emph {et~al.}(2013)\citenamefont
  {van~der Schee}, \citenamefont {Romatschke},\ and\ \citenamefont
  {Pratt}}]{vanderSchee:2013pia}%
  \BibitemOpen
  \bibfield  {author} {\bibinfo {author} {\bibfnamefont {W.}~\bibnamefont
  {van~der Schee}}, \bibinfo {author} {\bibfnamefont {P.}~\bibnamefont
  {Romatschke}}, \ and\ \bibinfo {author} {\bibfnamefont {S.}~\bibnamefont
  {Pratt}},\ }\href {\doibase 10.1103/PhysRevLett.111.222302} {\bibfield
  {journal} {\bibinfo  {journal} {Phys. Rev. Lett.}\ }\textbf {\bibinfo
  {volume} {111}},\ \bibinfo {pages} {222302} (\bibinfo {year} {2013})},\
  \Eprint {http://arxiv.org/abs/1307.2539} {arXiv:1307.2539 [nucl-th]}
  \BibitemShut {NoStop}%
\bibitem [{\citenamefont {Heller}\ \emph {et~al.}(2013)\citenamefont {Heller},
  \citenamefont {Mateos}, \citenamefont {van~der Schee},\ and\ \citenamefont
  {Triana}}]{Heller:2013oxa}%
  \BibitemOpen
  \bibfield  {author} {\bibinfo {author} {\bibfnamefont {M.~P.}\ \bibnamefont
  {Heller}}, \bibinfo {author} {\bibfnamefont {D.}~\bibnamefont {Mateos}},
  \bibinfo {author} {\bibfnamefont {W.}~\bibnamefont {van~der Schee}}, \ and\
  \bibinfo {author} {\bibfnamefont {M.}~\bibnamefont {Triana}},\ }\href
  {\doibase 10.1007/JHEP09(2013)026} {\bibfield  {journal} {\bibinfo  {journal}
  {JHEP}\ }\textbf {\bibinfo {volume} {09}},\ \bibinfo {pages} {026} (\bibinfo
  {year} {2013})},\ \Eprint {http://arxiv.org/abs/1304.5172} {arXiv:1304.5172
  [hep-th]} \BibitemShut {NoStop}%
\bibitem [{\citenamefont {Casalderrey-Solana}\ \emph
  {et~al.}(2013)\citenamefont {Casalderrey-Solana}, \citenamefont {Heller},
  \citenamefont {Mateos},\ and\ \citenamefont {van~der
  Schee}}]{Casalderrey-Solana:2013aba}%
  \BibitemOpen
  \bibfield  {author} {\bibinfo {author} {\bibfnamefont {J.}~\bibnamefont
  {Casalderrey-Solana}}, \bibinfo {author} {\bibfnamefont {M.~P.}\ \bibnamefont
  {Heller}}, \bibinfo {author} {\bibfnamefont {D.}~\bibnamefont {Mateos}}, \
  and\ \bibinfo {author} {\bibfnamefont {W.}~\bibnamefont {van~der Schee}},\
  }\href {\doibase 10.1103/PhysRevLett.111.181601} {\bibfield  {journal}
  {\bibinfo  {journal} {Phys. Rev. Lett.}\ }\textbf {\bibinfo {volume} {111}},\
  \bibinfo {pages} {181601} (\bibinfo {year} {2013})},\ \Eprint
  {http://arxiv.org/abs/1305.4919} {arXiv:1305.4919 [hep-th]} \BibitemShut
  {NoStop}%
\bibitem [{\citenamefont {Chesler}\ and\ \citenamefont
  {Yaffe}(2014)}]{Chesler:2013lia}%
  \BibitemOpen
  \bibfield  {author} {\bibinfo {author} {\bibfnamefont {P.~M.}\ \bibnamefont
  {Chesler}}\ and\ \bibinfo {author} {\bibfnamefont {L.~G.}\ \bibnamefont
  {Yaffe}},\ }\href {\doibase 10.1007/JHEP07(2014)086} {\bibfield  {journal}
  {\bibinfo  {journal} {JHEP}\ }\textbf {\bibinfo {volume} {07}},\ \bibinfo
  {pages} {086} (\bibinfo {year} {2014})},\ \Eprint
  {http://arxiv.org/abs/1309.1439} {arXiv:1309.1439 [hep-th]} \BibitemShut
  {NoStop}%
\bibitem [{\citenamefont {Chesler}\ and\ \citenamefont
  {Yaffe}(2015)}]{Chesler:2015wra}%
  \BibitemOpen
  \bibfield  {author} {\bibinfo {author} {\bibfnamefont {P.~M.}\ \bibnamefont
  {Chesler}}\ and\ \bibinfo {author} {\bibfnamefont {L.~G.}\ \bibnamefont
  {Yaffe}},\ }\href {\doibase 10.1007/JHEP10(2015)070} {\bibfield  {journal}
  {\bibinfo  {journal} {JHEP}\ }\textbf {\bibinfo {volume} {10}},\ \bibinfo
  {pages} {070} (\bibinfo {year} {2015})},\ \Eprint
  {http://arxiv.org/abs/1501.04644} {arXiv:1501.04644 [hep-th]} \BibitemShut
  {NoStop}%
\bibitem [{\citenamefont {Keegan}\ \emph {et~al.}(2016)\citenamefont {Keegan},
  \citenamefont {Kurkela}, \citenamefont {Romatschke}, \citenamefont {van~der
  Schee},\ and\ \citenamefont {Zhu}}]{Keegan:2015avk}%
  \BibitemOpen
  \bibfield  {author} {\bibinfo {author} {\bibfnamefont {L.}~\bibnamefont
  {Keegan}}, \bibinfo {author} {\bibfnamefont {A.}~\bibnamefont {Kurkela}},
  \bibinfo {author} {\bibfnamefont {P.}~\bibnamefont {Romatschke}}, \bibinfo
  {author} {\bibfnamefont {W.}~\bibnamefont {van~der Schee}}, \ and\ \bibinfo
  {author} {\bibfnamefont {Y.}~\bibnamefont {Zhu}},\ }\href {\doibase
  10.1007/JHEP04(2016)031} {\bibfield  {journal} {\bibinfo  {journal} {JHEP}\
  }\textbf {\bibinfo {volume} {04}},\ \bibinfo {pages} {031} (\bibinfo {year}
  {2016})},\ \Eprint {http://arxiv.org/abs/1512.05347} {arXiv:1512.05347
  [hep-th]} \BibitemShut {NoStop}%
\bibitem [{\citenamefont {Spali\'nski}(2018)}]{Spalinski:2017mel}%
  \BibitemOpen
  \bibfield  {author} {\bibinfo {author} {\bibfnamefont {M.}~\bibnamefont
  {Spali\'nski}},\ }\href {\doibase 10.1016/j.physletb.2017.11.059} {\bibfield
  {journal} {\bibinfo  {journal} {Phys. Lett. B}\ }\textbf {\bibinfo {volume}
  {776}},\ \bibinfo {pages} {468} (\bibinfo {year} {2018})},\ \Eprint
  {http://arxiv.org/abs/1708.01921} {arXiv:1708.01921 [hep-th]} \BibitemShut
  {NoStop}%
\bibitem [{\citenamefont {Denicol}\ \emph
  {et~al.}(2014{\natexlab{a}})\citenamefont {Denicol}, \citenamefont {Heinz},
  \citenamefont {Martinez}, \citenamefont {Noronha},\ and\ \citenamefont
  {Strickland}}]{Denicol:2014xca}%
  \BibitemOpen
  \bibfield  {author} {\bibinfo {author} {\bibfnamefont {G.~S.}\ \bibnamefont
  {Denicol}}, \bibinfo {author} {\bibfnamefont {U.~W.}\ \bibnamefont {Heinz}},
  \bibinfo {author} {\bibfnamefont {M.}~\bibnamefont {Martinez}}, \bibinfo
  {author} {\bibfnamefont {J.}~\bibnamefont {Noronha}}, \ and\ \bibinfo
  {author} {\bibfnamefont {M.}~\bibnamefont {Strickland}},\ }\href {\doibase
  10.1103/PhysRevLett.113.202301} {\bibfield  {journal} {\bibinfo  {journal}
  {Phys. Rev. Lett.}\ }\textbf {\bibinfo {volume} {113}},\ \bibinfo {pages}
  {202301} (\bibinfo {year} {2014}{\natexlab{a}})},\ \Eprint
  {http://arxiv.org/abs/1408.5646} {arXiv:1408.5646 [hep-ph]} \BibitemShut
  {NoStop}%
\bibitem [{\citenamefont {Denicol}\ \emph
  {et~al.}(2014{\natexlab{b}})\citenamefont {Denicol}, \citenamefont {Heinz},
  \citenamefont {Martinez}, \citenamefont {Noronha},\ and\ \citenamefont
  {Strickland}}]{Denicol:2014tha}%
  \BibitemOpen
  \bibfield  {author} {\bibinfo {author} {\bibfnamefont {G.~S.}\ \bibnamefont
  {Denicol}}, \bibinfo {author} {\bibfnamefont {U.~W.}\ \bibnamefont {Heinz}},
  \bibinfo {author} {\bibfnamefont {M.}~\bibnamefont {Martinez}}, \bibinfo
  {author} {\bibfnamefont {J.}~\bibnamefont {Noronha}}, \ and\ \bibinfo
  {author} {\bibfnamefont {M.}~\bibnamefont {Strickland}},\ }\href {\doibase
  10.1103/PhysRevD.90.125026} {\bibfield  {journal} {\bibinfo  {journal} {Phys.
  Rev. D}\ }\textbf {\bibinfo {volume} {90}},\ \bibinfo {pages} {125026}
  (\bibinfo {year} {2014}{\natexlab{b}})},\ \Eprint
  {http://arxiv.org/abs/1408.7048} {arXiv:1408.7048 [hep-ph]} \BibitemShut
  {NoStop}%
\bibitem [{\citenamefont {Kurkela}\ and\ \citenamefont
  {Zhu}(2015)}]{Kurkela:2015qoa}%
  \BibitemOpen
  \bibfield  {author} {\bibinfo {author} {\bibfnamefont {A.}~\bibnamefont
  {Kurkela}}\ and\ \bibinfo {author} {\bibfnamefont {Y.}~\bibnamefont {Zhu}},\
  }\href {\doibase 10.1103/PhysRevLett.115.182301} {\bibfield  {journal}
  {\bibinfo  {journal} {Phys. Rev. Lett.}\ }\textbf {\bibinfo {volume} {115}},\
  \bibinfo {pages} {182301} (\bibinfo {year} {2015})},\ \Eprint
  {http://arxiv.org/abs/1506.06647} {arXiv:1506.06647 [hep-ph]} \BibitemShut
  {NoStop}%
\bibitem [{\citenamefont {Bazow}\ \emph {et~al.}(2016)\citenamefont {Bazow},
  \citenamefont {Denicol}, \citenamefont {Heinz}, \citenamefont {Martinez},\
  and\ \citenamefont {Noronha}}]{Bazow:2015dha}%
  \BibitemOpen
  \bibfield  {author} {\bibinfo {author} {\bibfnamefont {D.}~\bibnamefont
  {Bazow}}, \bibinfo {author} {\bibfnamefont {G.~S.}\ \bibnamefont {Denicol}},
  \bibinfo {author} {\bibfnamefont {U.}~\bibnamefont {Heinz}}, \bibinfo
  {author} {\bibfnamefont {M.}~\bibnamefont {Martinez}}, \ and\ \bibinfo
  {author} {\bibfnamefont {J.}~\bibnamefont {Noronha}},\ }\href {\doibase
  10.1103/PhysRevLett.116.022301} {\bibfield  {journal} {\bibinfo  {journal}
  {Phys. Rev. Lett.}\ }\textbf {\bibinfo {volume} {116}},\ \bibinfo {pages}
  {022301} (\bibinfo {year} {2016})},\ \Eprint
  {http://arxiv.org/abs/1507.07834} {arXiv:1507.07834 [hep-ph]} \BibitemShut
  {NoStop}%
\bibitem [{\citenamefont {Denicol}\ and\ \citenamefont
  {Noronha}(2016)}]{Denicol:2016bjh}%
  \BibitemOpen
  \bibfield  {author} {\bibinfo {author} {\bibfnamefont {G.~S.}\ \bibnamefont
  {Denicol}}\ and\ \bibinfo {author} {\bibfnamefont {J.}~\bibnamefont
  {Noronha}},\ }\href@noop {} {\  (\bibinfo {year} {2016})},\ \Eprint
  {http://arxiv.org/abs/1608.07869} {arXiv:1608.07869 [nucl-th]} \BibitemShut
  {NoStop}%
\bibitem [{\citenamefont {Heller}\ \emph {et~al.}(2018)\citenamefont {Heller},
  \citenamefont {Kurkela}, \citenamefont {Spali\'nski},\ and\ \citenamefont
  {Svensson}}]{Heller:2016rtz}%
  \BibitemOpen
  \bibfield  {author} {\bibinfo {author} {\bibfnamefont {M.~P.}\ \bibnamefont
  {Heller}}, \bibinfo {author} {\bibfnamefont {A.}~\bibnamefont {Kurkela}},
  \bibinfo {author} {\bibfnamefont {M.}~\bibnamefont {Spali\'nski}}, \ and\
  \bibinfo {author} {\bibfnamefont {V.}~\bibnamefont {Svensson}},\ }\href
  {\doibase 10.1103/PhysRevD.97.091503} {\bibfield  {journal} {\bibinfo
  {journal} {Phys. Rev. D}\ }\textbf {\bibinfo {volume} {97}},\ \bibinfo
  {pages} {091503} (\bibinfo {year} {2018})},\ \Eprint
  {http://arxiv.org/abs/1609.04803} {arXiv:1609.04803 [nucl-th]} \BibitemShut
  {NoStop}%
\bibitem [{\citenamefont {Romatschke}(2018)}]{Romatschke:2017vte}%
  \BibitemOpen
  \bibfield  {author} {\bibinfo {author} {\bibfnamefont {P.}~\bibnamefont
  {Romatschke}},\ }\href {\doibase 10.1103/PhysRevLett.120.012301} {\bibfield
  {journal} {\bibinfo  {journal} {Phys. Rev. Lett.}\ }\textbf {\bibinfo
  {volume} {120}},\ \bibinfo {pages} {012301} (\bibinfo {year} {2018})},\
  \Eprint {http://arxiv.org/abs/1704.08699} {arXiv:1704.08699 [hep-th]}
  \BibitemShut {NoStop}%
\bibitem [{\citenamefont {Strickland}(2018)}]{Strickland:2018ayk}%
  \BibitemOpen
  \bibfield  {author} {\bibinfo {author} {\bibfnamefont {M.}~\bibnamefont
  {Strickland}},\ }\href {\doibase 10.1007/JHEP12(2018)128} {\bibfield
  {journal} {\bibinfo  {journal} {JHEP}\ }\textbf {\bibinfo {volume} {12}},\
  \bibinfo {pages} {128} (\bibinfo {year} {2018})},\ \Eprint
  {http://arxiv.org/abs/1809.01200} {arXiv:1809.01200 [nucl-th]} \BibitemShut
  {NoStop}%
\bibitem [{\citenamefont {Kurkela}\ \emph
  {et~al.}(2019{\natexlab{a}})\citenamefont {Kurkela}, \citenamefont
  {Mazeliauskas}, \citenamefont {Paquet}, \citenamefont {Schlichting},\ and\
  \citenamefont {Teaney}}]{Kurkela:2018wud}%
  \BibitemOpen
  \bibfield  {author} {\bibinfo {author} {\bibfnamefont {A.}~\bibnamefont
  {Kurkela}}, \bibinfo {author} {\bibfnamefont {A.}~\bibnamefont
  {Mazeliauskas}}, \bibinfo {author} {\bibfnamefont {J.-F.}\ \bibnamefont
  {Paquet}}, \bibinfo {author} {\bibfnamefont {S.}~\bibnamefont {Schlichting}},
  \ and\ \bibinfo {author} {\bibfnamefont {D.}~\bibnamefont {Teaney}},\ }\href
  {\doibase 10.1103/PhysRevLett.122.122302} {\bibfield  {journal} {\bibinfo
  {journal} {Phys. Rev. Lett.}\ }\textbf {\bibinfo {volume} {122}},\ \bibinfo
  {pages} {122302} (\bibinfo {year} {2019}{\natexlab{a}})},\ \Eprint
  {http://arxiv.org/abs/1805.01604} {arXiv:1805.01604 [hep-ph]} \BibitemShut
  {NoStop}%
\bibitem [{\citenamefont {Kurkela}\ \emph
  {et~al.}(2019{\natexlab{b}})\citenamefont {Kurkela}, \citenamefont
  {Mazeliauskas}, \citenamefont {Paquet}, \citenamefont {Schlichting},\ and\
  \citenamefont {Teaney}}]{Kurkela:2018vqr}%
  \BibitemOpen
  \bibfield  {author} {\bibinfo {author} {\bibfnamefont {A.}~\bibnamefont
  {Kurkela}}, \bibinfo {author} {\bibfnamefont {A.}~\bibnamefont
  {Mazeliauskas}}, \bibinfo {author} {\bibfnamefont {J.-F.}\ \bibnamefont
  {Paquet}}, \bibinfo {author} {\bibfnamefont {S.}~\bibnamefont {Schlichting}},
  \ and\ \bibinfo {author} {\bibfnamefont {D.}~\bibnamefont {Teaney}},\ }\href
  {\doibase 10.1103/PhysRevC.99.034910} {\bibfield  {journal} {\bibinfo
  {journal} {Phys. Rev. C}\ }\textbf {\bibinfo {volume} {99}},\ \bibinfo
  {pages} {034910} (\bibinfo {year} {2019}{\natexlab{b}})},\ \Eprint
  {http://arxiv.org/abs/1805.00961} {arXiv:1805.00961 [hep-ph]} \BibitemShut
  {NoStop}%
\bibitem [{\citenamefont {Kurkela}\ \emph {et~al.}(2020)\citenamefont
  {Kurkela}, \citenamefont {van~der Schee}, \citenamefont {Wiedemann},\ and\
  \citenamefont {Wu}}]{Kurkela:2019set}%
  \BibitemOpen
  \bibfield  {author} {\bibinfo {author} {\bibfnamefont {A.}~\bibnamefont
  {Kurkela}}, \bibinfo {author} {\bibfnamefont {W.}~\bibnamefont {van~der
  Schee}}, \bibinfo {author} {\bibfnamefont {U.~A.}\ \bibnamefont {Wiedemann}},
  \ and\ \bibinfo {author} {\bibfnamefont {B.}~\bibnamefont {Wu}},\ }\href
  {\doibase 10.1103/PhysRevLett.124.102301} {\bibfield  {journal} {\bibinfo
  {journal} {Phys. Rev. Lett.}\ }\textbf {\bibinfo {volume} {124}},\ \bibinfo
  {pages} {102301} (\bibinfo {year} {2020})},\ \Eprint
  {http://arxiv.org/abs/1907.08101} {arXiv:1907.08101 [hep-ph]} \BibitemShut
  {NoStop}%
\bibitem [{\citenamefont {Strickland}\ and\ \citenamefont
  {Tantary}(2019)}]{Strickland:2019hff}%
  \BibitemOpen
  \bibfield  {author} {\bibinfo {author} {\bibfnamefont {M.}~\bibnamefont
  {Strickland}}\ and\ \bibinfo {author} {\bibfnamefont {U.}~\bibnamefont
  {Tantary}},\ }\href {\doibase 10.1007/JHEP10(2019)069} {\bibfield  {journal}
  {\bibinfo  {journal} {JHEP}\ }\textbf {\bibinfo {volume} {10}},\ \bibinfo
  {pages} {069} (\bibinfo {year} {2019})},\ \Eprint
  {http://arxiv.org/abs/1903.03145} {arXiv:1903.03145 [hep-ph]} \BibitemShut
  {NoStop}%
\bibitem [{\citenamefont {Denicol}\ and\ \citenamefont
  {Noronha}(2020)}]{Denicol:2019lio}%
  \BibitemOpen
  \bibfield  {author} {\bibinfo {author} {\bibfnamefont {G.~S.}\ \bibnamefont
  {Denicol}}\ and\ \bibinfo {author} {\bibfnamefont {J.}~\bibnamefont
  {Noronha}},\ }\href {\doibase 10.1103/PhysRevLett.124.152301} {\bibfield
  {journal} {\bibinfo  {journal} {Phys. Rev. Lett.}\ }\textbf {\bibinfo
  {volume} {124}},\ \bibinfo {pages} {152301} (\bibinfo {year} {2020})},\
  \Eprint {http://arxiv.org/abs/1908.09957} {arXiv:1908.09957 [nucl-th]}
  \BibitemShut {NoStop}%
\bibitem [{\citenamefont {Almaalol}\ \emph {et~al.}(2020)\citenamefont
  {Almaalol}, \citenamefont {Kurkela},\ and\ \citenamefont
  {Strickland}}]{Almaalol:2020rnu}%
  \BibitemOpen
  \bibfield  {author} {\bibinfo {author} {\bibfnamefont {D.}~\bibnamefont
  {Almaalol}}, \bibinfo {author} {\bibfnamefont {A.}~\bibnamefont {Kurkela}}, \
  and\ \bibinfo {author} {\bibfnamefont {M.}~\bibnamefont {Strickland}},\
  }\href {\doibase 10.1103/PhysRevLett.125.122302} {\bibfield  {journal}
  {\bibinfo  {journal} {Phys. Rev. Lett.}\ }\textbf {\bibinfo {volume} {125}},\
  \bibinfo {pages} {122302} (\bibinfo {year} {2020})},\ \Eprint
  {http://arxiv.org/abs/2004.05195} {arXiv:2004.05195 [hep-ph]} \BibitemShut
  {NoStop}%
\bibitem [{\citenamefont {Jaiswal}\ \emph {et~al.}(2022)\citenamefont
  {Jaiswal}, \citenamefont {Blaizot}, \citenamefont {Bhalerao}, \citenamefont
  {Chen}, \citenamefont {Jaiswal},\ and\ \citenamefont
  {Yan}}]{Jaiswal:2022udf}%
  \BibitemOpen
  \bibfield  {author} {\bibinfo {author} {\bibfnamefont {S.}~\bibnamefont
  {Jaiswal}}, \bibinfo {author} {\bibfnamefont {J.-P.}\ \bibnamefont
  {Blaizot}}, \bibinfo {author} {\bibfnamefont {R.~S.}\ \bibnamefont
  {Bhalerao}}, \bibinfo {author} {\bibfnamefont {Z.}~\bibnamefont {Chen}},
  \bibinfo {author} {\bibfnamefont {A.}~\bibnamefont {Jaiswal}}, \ and\
  \bibinfo {author} {\bibfnamefont {L.}~\bibnamefont {Yan}},\ }\href {\doibase
  10.1103/PhysRevC.106.044912} {\bibfield  {journal} {\bibinfo  {journal}
  {Phys. Rev. C}\ }\textbf {\bibinfo {volume} {106}},\ \bibinfo {pages}
  {044912} (\bibinfo {year} {2022})},\ \Eprint
  {http://arxiv.org/abs/2208.02750} {arXiv:2208.02750 [nucl-th]} \BibitemShut
  {NoStop}%
\bibitem [{\citenamefont {Alalawi}\ and\ \citenamefont
  {Strickland}(2022)}]{Alalawi:2022pmg}%
  \BibitemOpen
  \bibfield  {author} {\bibinfo {author} {\bibfnamefont {H.}~\bibnamefont
  {Alalawi}}\ and\ \bibinfo {author} {\bibfnamefont {M.}~\bibnamefont
  {Strickland}},\ }\href@noop {} {\  (\bibinfo {year} {2022})},\ \Eprint
  {http://arxiv.org/abs/2210.00658} {arXiv:2210.00658 [hep-ph]} \BibitemShut
  {NoStop}%
\bibitem [{\citenamefont {Mullins}\ \emph {et~al.}(2022)\citenamefont
  {Mullins}, \citenamefont {Denicol},\ and\ \citenamefont
  {Noronha}}]{Mullins:2022fbx}%
  \BibitemOpen
  \bibfield  {author} {\bibinfo {author} {\bibfnamefont {N.}~\bibnamefont
  {Mullins}}, \bibinfo {author} {\bibfnamefont {G.~S.}\ \bibnamefont
  {Denicol}}, \ and\ \bibinfo {author} {\bibfnamefont {J.}~\bibnamefont
  {Noronha}},\ }\href {\doibase 10.1103/PhysRevD.106.056024} {\bibfield
  {journal} {\bibinfo  {journal} {Phys. Rev. D}\ }\textbf {\bibinfo {volume}
  {106}},\ \bibinfo {pages} {056024} (\bibinfo {year} {2022})},\ \Eprint
  {http://arxiv.org/abs/2207.07786} {arXiv:2207.07786 [hep-ph]} \BibitemShut
  {NoStop}%
\bibitem [{\citenamefont {Heller}\ and\ \citenamefont
  {Spalinski}(2015)}]{Heller:2015dha}%
  \BibitemOpen
  \bibfield  {author} {\bibinfo {author} {\bibfnamefont {M.~P.}\ \bibnamefont
  {Heller}}\ and\ \bibinfo {author} {\bibfnamefont {M.}~\bibnamefont
  {Spalinski}},\ }\href {\doibase 10.1103/PhysRevLett.115.072501} {\bibfield
  {journal} {\bibinfo  {journal} {Phys. Rev. Lett.}\ }\textbf {\bibinfo
  {volume} {115}},\ \bibinfo {pages} {072501} (\bibinfo {year} {2015})},\
  \Eprint {http://arxiv.org/abs/1503.07514} {arXiv:1503.07514 [hep-th]}
  \BibitemShut {NoStop}%
\bibitem [{\citenamefont {Florkowski}\ \emph
  {et~al.}(2018{\natexlab{b}})\citenamefont {Florkowski}, \citenamefont
  {Heller},\ and\ \citenamefont {Spalinski}}]{Florkowski:2017olj}%
  \BibitemOpen
  \bibfield  {author} {\bibinfo {author} {\bibfnamefont {W.}~\bibnamefont
  {Florkowski}}, \bibinfo {author} {\bibfnamefont {M.~P.}\ \bibnamefont
  {Heller}}, \ and\ \bibinfo {author} {\bibfnamefont {M.}~\bibnamefont
  {Spalinski}},\ }\href {\doibase 10.1088/1361-6633/aaa091} {\bibfield
  {journal} {\bibinfo  {journal} {Rept. Prog. Phys.}\ }\textbf {\bibinfo
  {volume} {81}},\ \bibinfo {pages} {046001} (\bibinfo {year}
  {2018}{\natexlab{b}})},\ \Eprint {http://arxiv.org/abs/1707.02282}
  {arXiv:1707.02282 [hep-ph]} \BibitemShut {NoStop}%
\bibitem [{\citenamefont {Strickland}\ \emph {et~al.}(2018)\citenamefont
  {Strickland}, \citenamefont {Noronha},\ and\ \citenamefont
  {Denicol}}]{Strickland:2017kux}%
  \BibitemOpen
  \bibfield  {author} {\bibinfo {author} {\bibfnamefont {M.}~\bibnamefont
  {Strickland}}, \bibinfo {author} {\bibfnamefont {J.}~\bibnamefont {Noronha}},
  \ and\ \bibinfo {author} {\bibfnamefont {G.}~\bibnamefont {Denicol}},\ }\href
  {\doibase 10.1103/PhysRevD.97.036020} {\bibfield  {journal} {\bibinfo
  {journal} {Phys. Rev. D}\ }\textbf {\bibinfo {volume} {97}},\ \bibinfo
  {pages} {036020} (\bibinfo {year} {2018})},\ \Eprint
  {http://arxiv.org/abs/1709.06644} {arXiv:1709.06644 [nucl-th]} \BibitemShut
  {NoStop}%
\bibitem [{\citenamefont {Bemfica}\ \emph {et~al.}(2018)\citenamefont
  {Bemfica}, \citenamefont {Disconzi},\ and\ \citenamefont
  {Noronha}}]{Bemfica:2017wps}%
  \BibitemOpen
  \bibfield  {author} {\bibinfo {author} {\bibfnamefont {F.~S.}\ \bibnamefont
  {Bemfica}}, \bibinfo {author} {\bibfnamefont {M.~M.}\ \bibnamefont
  {Disconzi}}, \ and\ \bibinfo {author} {\bibfnamefont {J.}~\bibnamefont
  {Noronha}},\ }\href {\doibase 10.1103/PhysRevD.98.104064} {\bibfield
  {journal} {\bibinfo  {journal} {Phys. Rev. D}\ }\textbf {\bibinfo {volume}
  {98}},\ \bibinfo {pages} {104064} (\bibinfo {year} {2018})},\ \Eprint
  {http://arxiv.org/abs/1708.06255} {arXiv:1708.06255 [gr-qc]} \BibitemShut
  {NoStop}%
\bibitem [{\citenamefont {Kovtun}(2019)}]{Kovtun:2019hdm}%
  \BibitemOpen
  \bibfield  {author} {\bibinfo {author} {\bibfnamefont {P.}~\bibnamefont
  {Kovtun}},\ }\href {\doibase 10.1007/JHEP10(2019)034} {\bibfield  {journal}
  {\bibinfo  {journal} {JHEP}\ }\textbf {\bibinfo {volume} {10}},\ \bibinfo
  {pages} {034} (\bibinfo {year} {2019})},\ \Eprint
  {http://arxiv.org/abs/1907.08191} {arXiv:1907.08191 [hep-th]} \BibitemShut
  {NoStop}%
\bibitem [{\citenamefont {Bemfica}\ \emph
  {et~al.}(2019{\natexlab{a}})\citenamefont {Bemfica}, \citenamefont {Bemfica},
  \citenamefont {Disconzi}, \citenamefont {Disconzi}, \citenamefont {Noronha},\
  and\ \citenamefont {Noronha}}]{Bemfica:2019knx}%
  \BibitemOpen
  \bibfield  {author} {\bibinfo {author} {\bibfnamefont {F.~S.}\ \bibnamefont
  {Bemfica}}, \bibinfo {author} {\bibfnamefont {F.~S.}\ \bibnamefont
  {Bemfica}}, \bibinfo {author} {\bibfnamefont {M.~M.}\ \bibnamefont
  {Disconzi}}, \bibinfo {author} {\bibfnamefont {M.~M.}\ \bibnamefont
  {Disconzi}}, \bibinfo {author} {\bibfnamefont {J.}~\bibnamefont {Noronha}}, \
  and\ \bibinfo {author} {\bibfnamefont {J.}~\bibnamefont {Noronha}},\ }\href
  {\doibase 10.1103/PhysRevD.100.104020} {\bibfield  {journal} {\bibinfo
  {journal} {Phys. Rev. D}\ }\textbf {\bibinfo {volume} {100}},\ \bibinfo
  {pages} {104020} (\bibinfo {year} {2019}{\natexlab{a}})},\ \bibinfo {note}
  {[Erratum: Phys.Rev.D 105, 069902 (2022)]},\ \Eprint
  {http://arxiv.org/abs/1907.12695} {arXiv:1907.12695 [gr-qc]} \BibitemShut
  {NoStop}%
\bibitem [{\citenamefont {Hoult}\ and\ \citenamefont
  {Kovtun}(2020)}]{Hoult:2020eho}%
  \BibitemOpen
  \bibfield  {author} {\bibinfo {author} {\bibfnamefont {R.~E.}\ \bibnamefont
  {Hoult}}\ and\ \bibinfo {author} {\bibfnamefont {P.}~\bibnamefont {Kovtun}},\
  }\href {\doibase 10.1007/JHEP06(2020)067} {\bibfield  {journal} {\bibinfo
  {journal} {JHEP}\ }\textbf {\bibinfo {volume} {06}},\ \bibinfo {pages} {067}
  (\bibinfo {year} {2020})},\ \Eprint {http://arxiv.org/abs/2004.04102}
  {arXiv:2004.04102 [hep-th]} \BibitemShut {NoStop}%
\bibitem [{\citenamefont {Bemfica}\ \emph {et~al.}(2022)\citenamefont
  {Bemfica}, \citenamefont {Disconzi},\ and\ \citenamefont
  {Noronha}}]{Bemfica:2020zjp}%
  \BibitemOpen
  \bibfield  {author} {\bibinfo {author} {\bibfnamefont {F.~S.}\ \bibnamefont
  {Bemfica}}, \bibinfo {author} {\bibfnamefont {M.~M.}\ \bibnamefont
  {Disconzi}}, \ and\ \bibinfo {author} {\bibfnamefont {J.}~\bibnamefont
  {Noronha}},\ }\href {\doibase 10.1103/PhysRevX.12.021044} {\bibfield
  {journal} {\bibinfo  {journal} {Phys. Rev. X}\ }\textbf {\bibinfo {volume}
  {12}},\ \bibinfo {pages} {021044} (\bibinfo {year} {2022})},\ \Eprint
  {http://arxiv.org/abs/2009.11388} {arXiv:2009.11388 [gr-qc]} \BibitemShut
  {NoStop}%
\bibitem [{\citenamefont {Noronha}\ \emph {et~al.}(2022)\citenamefont
  {Noronha}, \citenamefont {Spali\'nski},\ and\ \citenamefont
  {Speranza}}]{Noronha:2021syv}%
  \BibitemOpen
  \bibfield  {author} {\bibinfo {author} {\bibfnamefont {J.}~\bibnamefont
  {Noronha}}, \bibinfo {author} {\bibfnamefont {M.}~\bibnamefont
  {Spali\'nski}}, \ and\ \bibinfo {author} {\bibfnamefont {E.}~\bibnamefont
  {Speranza}},\ }\href {\doibase 10.1103/PhysRevLett.128.252302} {\bibfield
  {journal} {\bibinfo  {journal} {Phys. Rev. Lett.}\ }\textbf {\bibinfo
  {volume} {128}},\ \bibinfo {pages} {252302} (\bibinfo {year} {2022})},\
  \Eprint {http://arxiv.org/abs/2105.01034} {arXiv:2105.01034 [nucl-th]}
  \BibitemShut {NoStop}%
\bibitem [{\citenamefont {Gavassino}(2022)}]{Gavassino:2021owo}%
  \BibitemOpen
  \bibfield  {author} {\bibinfo {author} {\bibfnamefont {L.}~\bibnamefont
  {Gavassino}},\ }\href {\doibase 10.1103/PhysRevX.12.041001} {\bibfield
  {journal} {\bibinfo  {journal} {Phys. Rev. X}\ }\textbf {\bibinfo {volume}
  {12}},\ \bibinfo {pages} {041001} (\bibinfo {year} {2022})},\ \Eprint
  {http://arxiv.org/abs/2111.05254} {arXiv:2111.05254 [gr-qc]} \BibitemShut
  {NoStop}%
\bibitem [{\citenamefont {Brewer}\ \emph {et~al.}(2021)\citenamefont {Brewer},
  \citenamefont {Yan},\ and\ \citenamefont {Yin}}]{Brewer:2019oha}%
  \BibitemOpen
  \bibfield  {author} {\bibinfo {author} {\bibfnamefont {J.}~\bibnamefont
  {Brewer}}, \bibinfo {author} {\bibfnamefont {L.}~\bibnamefont {Yan}}, \ and\
  \bibinfo {author} {\bibfnamefont {Y.}~\bibnamefont {Yin}},\ }\href {\doibase
  10.1016/j.physletb.2021.136189} {\bibfield  {journal} {\bibinfo  {journal}
  {Phys. Lett. B}\ }\textbf {\bibinfo {volume} {816}},\ \bibinfo {pages}
  {136189} (\bibinfo {year} {2021})},\ \Eprint
  {http://arxiv.org/abs/1910.00021} {arXiv:1910.00021 [nucl-th]} \BibitemShut
  {NoStop}%
\bibitem [{\citenamefont {Brewer}\ \emph {et~al.}(2022)\citenamefont {Brewer},
  \citenamefont {Scheihing-Hitschfeld},\ and\ \citenamefont
  {Yin}}]{Brewer:2022vkq}%
  \BibitemOpen
  \bibfield  {author} {\bibinfo {author} {\bibfnamefont {J.}~\bibnamefont
  {Brewer}}, \bibinfo {author} {\bibfnamefont {B.}~\bibnamefont
  {Scheihing-Hitschfeld}}, \ and\ \bibinfo {author} {\bibfnamefont
  {Y.}~\bibnamefont {Yin}},\ }\href {\doibase 10.1007/JHEP05(2022)145}
  {\bibfield  {journal} {\bibinfo  {journal} {JHEP}\ }\textbf {\bibinfo
  {volume} {05}},\ \bibinfo {pages} {145} (\bibinfo {year} {2022})},\ \Eprint
  {http://arxiv.org/abs/2203.02427} {arXiv:2203.02427 [hep-ph]} \BibitemShut
  {NoStop}%
\bibitem [{\citenamefont {Israel}\ and\ \citenamefont
  {Stewart}(1979)}]{Israel:1979wp}%
  \BibitemOpen
  \bibfield  {author} {\bibinfo {author} {\bibfnamefont {W.}~\bibnamefont
  {Israel}}\ and\ \bibinfo {author} {\bibfnamefont {J.~M.}\ \bibnamefont
  {Stewart}},\ }\href {\doibase 10.1016/0003-4916(79)90130-1} {\bibfield
  {journal} {\bibinfo  {journal} {Annals Phys.}\ }\textbf {\bibinfo {volume}
  {118}},\ \bibinfo {pages} {341} (\bibinfo {year} {1979})}\BibitemShut
  {NoStop}%
\bibitem [{\citenamefont {Denicol}\ \emph {et~al.}(2012)\citenamefont
  {Denicol}, \citenamefont {Niemi}, \citenamefont {Molnar},\ and\ \citenamefont
  {Rischke}}]{Denicol:2012cn}%
  \BibitemOpen
  \bibfield  {author} {\bibinfo {author} {\bibfnamefont {G.~S.}\ \bibnamefont
  {Denicol}}, \bibinfo {author} {\bibfnamefont {H.}~\bibnamefont {Niemi}},
  \bibinfo {author} {\bibfnamefont {E.}~\bibnamefont {Molnar}}, \ and\ \bibinfo
  {author} {\bibfnamefont {D.~H.}\ \bibnamefont {Rischke}},\ }\href {\doibase
  10.1103/PhysRevD.85.114047} {\bibfield  {journal} {\bibinfo  {journal} {Phys.
  Rev. D}\ }\textbf {\bibinfo {volume} {85}},\ \bibinfo {pages} {114047}
  (\bibinfo {year} {2012})},\ \bibinfo {note} {[Erratum: Phys.Rev.D 91, 039902
  (2015)]},\ \Eprint {http://arxiv.org/abs/1202.4551} {arXiv:1202.4551
  [nucl-th]} \BibitemShut {NoStop}%
\bibitem [{\citenamefont {Chiu}\ and\ \citenamefont
  {Shen}(2021)}]{Chiu:2021muk}%
  \BibitemOpen
  \bibfield  {author} {\bibinfo {author} {\bibfnamefont {C.}~\bibnamefont
  {Chiu}}\ and\ \bibinfo {author} {\bibfnamefont {C.}~\bibnamefont {Shen}},\
  }\href {\doibase 10.1103/PhysRevC.103.064901} {\bibfield  {journal} {\bibinfo
   {journal} {Phys. Rev. C}\ }\textbf {\bibinfo {volume} {103}},\ \bibinfo
  {pages} {064901} (\bibinfo {year} {2021})},\ \Eprint
  {http://arxiv.org/abs/2103.09848} {arXiv:2103.09848 [nucl-th]} \BibitemShut
  {NoStop}%
\bibitem [{\citenamefont {Plumberg}\ \emph {et~al.}(2022)\citenamefont
  {Plumberg}, \citenamefont {Almaalol}, \citenamefont {Dore}, \citenamefont
  {Noronha},\ and\ \citenamefont {Noronha-Hostler}}]{Plumberg:2021bme}%
  \BibitemOpen
  \bibfield  {author} {\bibinfo {author} {\bibfnamefont {C.}~\bibnamefont
  {Plumberg}}, \bibinfo {author} {\bibfnamefont {D.}~\bibnamefont {Almaalol}},
  \bibinfo {author} {\bibfnamefont {T.}~\bibnamefont {Dore}}, \bibinfo {author}
  {\bibfnamefont {J.}~\bibnamefont {Noronha}}, \ and\ \bibinfo {author}
  {\bibfnamefont {J.}~\bibnamefont {Noronha-Hostler}},\ }\href {\doibase
  10.1103/PhysRevC.105.L061901} {\bibfield  {journal} {\bibinfo  {journal}
  {Phys. Rev. C}\ }\textbf {\bibinfo {volume} {105}},\ \bibinfo {pages}
  {L061901} (\bibinfo {year} {2022})},\ \Eprint
  {http://arxiv.org/abs/2103.15889} {arXiv:2103.15889 [nucl-th]} \BibitemShut
  {NoStop}%
\bibitem [{\citenamefont {Niemi}\ \emph {et~al.}(2016)\citenamefont {Niemi},
  \citenamefont {Eskola}, \citenamefont {Paatelainen},\ and\ \citenamefont
  {Tuominen}}]{Niemi:2015voa}%
  \BibitemOpen
  \bibfield  {author} {\bibinfo {author} {\bibfnamefont {H.}~\bibnamefont
  {Niemi}}, \bibinfo {author} {\bibfnamefont {K.~J.}\ \bibnamefont {Eskola}},
  \bibinfo {author} {\bibfnamefont {R.}~\bibnamefont {Paatelainen}}, \ and\
  \bibinfo {author} {\bibfnamefont {K.}~\bibnamefont {Tuominen}},\ }\href
  {\doibase 10.1103/PhysRevC.93.014912} {\bibfield  {journal} {\bibinfo
  {journal} {Phys. Rev. C}\ }\textbf {\bibinfo {volume} {93}},\ \bibinfo
  {pages} {014912} (\bibinfo {year} {2016})},\ \Eprint
  {http://arxiv.org/abs/1511.04296} {arXiv:1511.04296 [hep-ph]} \BibitemShut
  {NoStop}%
\bibitem [{\citenamefont {Noronha-Hostler}\ \emph
  {et~al.}(2016{\natexlab{b}})\citenamefont {Noronha-Hostler}, \citenamefont
  {Luzum},\ and\ \citenamefont {Ollitrault}}]{Noronha-Hostler:2015uye}%
  \BibitemOpen
  \bibfield  {author} {\bibinfo {author} {\bibfnamefont {J.}~\bibnamefont
  {Noronha-Hostler}}, \bibinfo {author} {\bibfnamefont {M.}~\bibnamefont
  {Luzum}}, \ and\ \bibinfo {author} {\bibfnamefont {J.-Y.}\ \bibnamefont
  {Ollitrault}},\ }\href {\doibase 10.1103/PhysRevC.93.034912} {\bibfield
  {journal} {\bibinfo  {journal} {Phys. Rev. C}\ }\textbf {\bibinfo {volume}
  {93}},\ \bibinfo {pages} {034912} (\bibinfo {year} {2016}{\natexlab{b}})},\
  \Eprint {http://arxiv.org/abs/1511.06289} {arXiv:1511.06289 [nucl-th]}
  \BibitemShut {NoStop}%
\bibitem [{\citenamefont {Hirvonen}\ \emph {et~al.}(2022)\citenamefont
  {Hirvonen}, \citenamefont {Eskola},\ and\ \citenamefont
  {Niemi}}]{Hirvonen:2022xfv}%
  \BibitemOpen
  \bibfield  {author} {\bibinfo {author} {\bibfnamefont {H.}~\bibnamefont
  {Hirvonen}}, \bibinfo {author} {\bibfnamefont {K.~J.}\ \bibnamefont
  {Eskola}}, \ and\ \bibinfo {author} {\bibfnamefont {H.}~\bibnamefont
  {Niemi}},\ }\href {\doibase 10.1103/PhysRevC.106.044913} {\bibfield
  {journal} {\bibinfo  {journal} {Phys. Rev. C}\ }\textbf {\bibinfo {volume}
  {106}},\ \bibinfo {pages} {044913} (\bibinfo {year} {2022})}\BibitemShut
  {NoStop}%
\bibitem [{\citenamefont {Bemfica}\ \emph
  {et~al.}(2019{\natexlab{b}})\citenamefont {Bemfica}, \citenamefont
  {Disconzi},\ and\ \citenamefont {Noronha}}]{Bemfica:2019cop}%
  \BibitemOpen
  \bibfield  {author} {\bibinfo {author} {\bibfnamefont {F.~S.}\ \bibnamefont
  {Bemfica}}, \bibinfo {author} {\bibfnamefont {M.~M.}\ \bibnamefont
  {Disconzi}}, \ and\ \bibinfo {author} {\bibfnamefont {J.}~\bibnamefont
  {Noronha}},\ }\href {\doibase 10.1103/PhysRevLett.122.221602} {\bibfield
  {journal} {\bibinfo  {journal} {Phys. Rev. Lett.}\ }\textbf {\bibinfo
  {volume} {122}},\ \bibinfo {pages} {221602} (\bibinfo {year}
  {2019}{\natexlab{b}})},\ \Eprint {http://arxiv.org/abs/1901.06701}
  {arXiv:1901.06701 [gr-qc]} \BibitemShut {NoStop}%
\bibitem [{\citenamefont {Bemfica}\ \emph {et~al.}(2021)\citenamefont
  {Bemfica}, \citenamefont {Disconzi}, \citenamefont {Hoang}, \citenamefont
  {Noronha},\ and\ \citenamefont {Radosz}}]{Bemfica:2020xym}%
  \BibitemOpen
  \bibfield  {author} {\bibinfo {author} {\bibfnamefont {F.~S.}\ \bibnamefont
  {Bemfica}}, \bibinfo {author} {\bibfnamefont {M.~M.}\ \bibnamefont
  {Disconzi}}, \bibinfo {author} {\bibfnamefont {V.}~\bibnamefont {Hoang}},
  \bibinfo {author} {\bibfnamefont {J.}~\bibnamefont {Noronha}}, \ and\
  \bibinfo {author} {\bibfnamefont {M.}~\bibnamefont {Radosz}},\ }\href
  {\doibase 10.1103/PhysRevLett.126.222301} {\bibfield  {journal} {\bibinfo
  {journal} {Phys. Rev. Lett.}\ }\textbf {\bibinfo {volume} {126}},\ \bibinfo
  {pages} {222301} (\bibinfo {year} {2021})},\ \Eprint
  {http://arxiv.org/abs/2005.11632} {arXiv:2005.11632 [hep-th]} \BibitemShut
  {NoStop}%
\bibitem [{\citenamefont {Martinez}\ \emph
  {et~al.}(2019{\natexlab{b}})\citenamefont {Martinez}, \citenamefont
  {Sch\"afer},\ and\ \citenamefont {Skokov}}]{Martinez:2019bsn}%
  \BibitemOpen
  \bibfield  {author} {\bibinfo {author} {\bibfnamefont {M.}~\bibnamefont
  {Martinez}}, \bibinfo {author} {\bibfnamefont {T.}~\bibnamefont {Sch\"afer}},
  \ and\ \bibinfo {author} {\bibfnamefont {V.}~\bibnamefont {Skokov}},\ }\href
  {\doibase 10.1103/PhysRevD.100.074017} {\bibfield  {journal} {\bibinfo
  {journal} {Phys. Rev. D}\ }\textbf {\bibinfo {volume} {100}},\ \bibinfo
  {pages} {074017} (\bibinfo {year} {2019}{\natexlab{b}})},\ \Eprint
  {http://arxiv.org/abs/1906.11306} {arXiv:1906.11306 [hep-ph]} \BibitemShut
  {NoStop}%
\bibitem [{\citenamefont {Monnai}\ \emph {et~al.}(2017)\citenamefont {Monnai},
  \citenamefont {Mukherjee},\ and\ \citenamefont {Yin}}]{Monnai:2016kud}%
  \BibitemOpen
  \bibfield  {author} {\bibinfo {author} {\bibfnamefont {A.}~\bibnamefont
  {Monnai}}, \bibinfo {author} {\bibfnamefont {S.}~\bibnamefont {Mukherjee}}, \
  and\ \bibinfo {author} {\bibfnamefont {Y.}~\bibnamefont {Yin}},\ }\href
  {\doibase 10.1103/PhysRevC.95.034902} {\bibfield  {journal} {\bibinfo
  {journal} {Phys. Rev. C}\ }\textbf {\bibinfo {volume} {95}},\ \bibinfo
  {pages} {034902} (\bibinfo {year} {2017})},\ \Eprint
  {http://arxiv.org/abs/1606.00771} {arXiv:1606.00771 [nucl-th]} \BibitemShut
  {NoStop}%
\bibitem [{\citenamefont {Dore}\ \emph {et~al.}(2020)\citenamefont {Dore},
  \citenamefont {Noronha-Hostler},\ and\ \citenamefont
  {McLaughlin}}]{Dore:2020jye}%
  \BibitemOpen
  \bibfield  {author} {\bibinfo {author} {\bibfnamefont {T.}~\bibnamefont
  {Dore}}, \bibinfo {author} {\bibfnamefont {J.}~\bibnamefont
  {Noronha-Hostler}}, \ and\ \bibinfo {author} {\bibfnamefont {E.}~\bibnamefont
  {McLaughlin}},\ }\href {\doibase 10.1103/PhysRevD.102.074017} {\bibfield
  {journal} {\bibinfo  {journal} {Phys. Rev. D}\ }\textbf {\bibinfo {volume}
  {102}},\ \bibinfo {pages} {074017} (\bibinfo {year} {2020})},\ \Eprint
  {http://arxiv.org/abs/2007.15083} {arXiv:2007.15083 [nucl-th]} \BibitemShut
  {NoStop}%
\bibitem [{\citenamefont {Chattopadhyay}\ \emph {et~al.}(2022)\citenamefont
  {Chattopadhyay}, \citenamefont {Heinz},\ and\ \citenamefont
  {Schaefer}}]{Chattopadhyay:2022sxk}%
  \BibitemOpen
  \bibfield  {author} {\bibinfo {author} {\bibfnamefont {C.}~\bibnamefont
  {Chattopadhyay}}, \bibinfo {author} {\bibfnamefont {U.}~\bibnamefont
  {Heinz}}, \ and\ \bibinfo {author} {\bibfnamefont {T.}~\bibnamefont
  {Schaefer}},\ }\href@noop {} {\  (\bibinfo {year} {2022})},\ \Eprint
  {http://arxiv.org/abs/2209.10483} {arXiv:2209.10483 [hep-ph]} \BibitemShut
  {NoStop}%
\bibitem [{\citenamefont {Pratt}\ \emph {et~al.}(2015)\citenamefont {Pratt},
  \citenamefont {Sangaline}, \citenamefont {Sorensen},\ and\ \citenamefont
  {Wang}}]{Pratt:2015zsa}%
  \BibitemOpen
  \bibfield  {author} {\bibinfo {author} {\bibfnamefont {S.}~\bibnamefont
  {Pratt}}, \bibinfo {author} {\bibfnamefont {E.}~\bibnamefont {Sangaline}},
  \bibinfo {author} {\bibfnamefont {P.}~\bibnamefont {Sorensen}}, \ and\
  \bibinfo {author} {\bibfnamefont {H.}~\bibnamefont {Wang}},\ }\href {\doibase
  10.1103/PhysRevLett.114.202301} {\bibfield  {journal} {\bibinfo  {journal}
  {Phys. Rev. Lett.}\ }\textbf {\bibinfo {volume} {114}},\ \bibinfo {pages}
  {202301} (\bibinfo {year} {2015})},\ \Eprint
  {http://arxiv.org/abs/1501.04042} {arXiv:1501.04042 [nucl-th]} \BibitemShut
  {NoStop}%
\bibitem [{\citenamefont {Reitze}\ \emph {et~al.}(2019)\citenamefont {Reitze}
  \emph {et~al.}}]{Reitze:2019iox}%
  \BibitemOpen
  \bibfield  {author} {\bibinfo {author} {\bibfnamefont {D.}~\bibnamefont
  {Reitze}} \emph {et~al.},\ }\href@noop {} {\bibfield  {journal} {\bibinfo
  {journal} {Bull. Am. Astron. Soc.}\ }\textbf {\bibinfo {volume} {51}},\
  \bibinfo {pages} {035} (\bibinfo {year} {2019})},\ \Eprint
  {http://arxiv.org/abs/1907.04833} {arXiv:1907.04833 [astro-ph.IM]}
  \BibitemShut {NoStop}%
\bibitem [{\citenamefont {Punturo}\ \emph {et~al.}(2010)\citenamefont {Punturo}
  \emph {et~al.}}]{Punturo:2010zz}%
  \BibitemOpen
  \bibfield  {author} {\bibinfo {author} {\bibfnamefont {M.}~\bibnamefont
  {Punturo}} \emph {et~al.},\ }\href {\doibase 10.1088/0264-9381/27/19/194002}
  {\bibfield  {journal} {\bibinfo  {journal} {Class. Quant. Grav.}\ }\textbf
  {\bibinfo {volume} {27}},\ \bibinfo {pages} {194002} (\bibinfo {year}
  {2010})}\BibitemShut {NoStop}%
\bibitem [{\citenamefont {Ackley}\ \emph {et~al.}(2020)\citenamefont {Ackley}
  \emph {et~al.}}]{Ackley:2020atn}%
  \BibitemOpen
  \bibfield  {author} {\bibinfo {author} {\bibfnamefont {K.}~\bibnamefont
  {Ackley}} \emph {et~al.},\ }\href {\doibase 10.1017/pasa.2020.39} {\bibfield
  {journal} {\bibinfo  {journal} {Publ. Astron. Soc. Austral.}\ }\textbf
  {\bibinfo {volume} {37}},\ \bibinfo {pages} {e047} (\bibinfo {year}
  {2020})},\ \Eprint {http://arxiv.org/abs/2007.03128} {arXiv:2007.03128
  [astro-ph.HE]} \BibitemShut {NoStop}%
\bibitem [{\citenamefont {Abbott}\ \emph {et~al.}(2019)\citenamefont {Abbott}
  \emph {et~al.}}]{LIGOScientific:2018hze}%
  \BibitemOpen
  \bibfield  {author} {\bibinfo {author} {\bibfnamefont {B.~P.}\ \bibnamefont
  {Abbott}} \emph {et~al.} (\bibinfo {collaboration} {LIGO Scientific,
  Virgo}),\ }\href {\doibase 10.1103/PhysRevX.9.011001} {\bibfield  {journal}
  {\bibinfo  {journal} {Phys. Rev. X}\ }\textbf {\bibinfo {volume} {9}},\
  \bibinfo {pages} {011001} (\bibinfo {year} {2019})},\ \Eprint
  {http://arxiv.org/abs/1805.11579} {arXiv:1805.11579 [gr-qc]} \BibitemShut
  {NoStop}%
\bibitem [{\citenamefont {Carson}\ \emph {et~al.}(2019)\citenamefont {Carson},
  \citenamefont {Chatziioannou}, \citenamefont {Haster}, \citenamefont {Yagi},\
  and\ \citenamefont {Yunes}}]{Carson:2019rjx}%
  \BibitemOpen
  \bibfield  {author} {\bibinfo {author} {\bibfnamefont {Z.}~\bibnamefont
  {Carson}}, \bibinfo {author} {\bibfnamefont {K.}~\bibnamefont
  {Chatziioannou}}, \bibinfo {author} {\bibfnamefont {C.-J.}\ \bibnamefont
  {Haster}}, \bibinfo {author} {\bibfnamefont {K.}~\bibnamefont {Yagi}}, \ and\
  \bibinfo {author} {\bibfnamefont {N.}~\bibnamefont {Yunes}},\ }\href
  {\doibase 10.1103/PhysRevD.99.083016} {\bibfield  {journal} {\bibinfo
  {journal} {Phys. Rev. D}\ }\textbf {\bibinfo {volume} {99}},\ \bibinfo
  {pages} {083016} (\bibinfo {year} {2019})},\ \Eprint
  {http://arxiv.org/abs/1903.03909} {arXiv:1903.03909 [gr-qc]} \BibitemShut
  {NoStop}%
\bibitem [{\citenamefont {Hammond}\ \emph {et~al.}(2022)\citenamefont
  {Hammond}, \citenamefont {Hawke},\ and\ \citenamefont
  {Andersson}}]{Hammond:2022uua}%
  \BibitemOpen
  \bibfield  {author} {\bibinfo {author} {\bibfnamefont {P.}~\bibnamefont
  {Hammond}}, \bibinfo {author} {\bibfnamefont {I.}~\bibnamefont {Hawke}}, \
  and\ \bibinfo {author} {\bibfnamefont {N.}~\bibnamefont {Andersson}},\
  }\href@noop {} {\  (\bibinfo {year} {2022})},\ \Eprint
  {http://arxiv.org/abs/2205.11377} {arXiv:2205.11377 [astro-ph.HE]}
  \BibitemShut {NoStop}%
\bibitem [{\citenamefont {Riley}\ \emph {et~al.}(2019)\citenamefont {Riley}
  \emph {et~al.}}]{Riley:2019yda}%
  \BibitemOpen
  \bibfield  {author} {\bibinfo {author} {\bibfnamefont {T.~E.}\ \bibnamefont
  {Riley}} \emph {et~al.},\ }\href {\doibase 10.3847/2041-8213/ab481c}
  {\bibfield  {journal} {\bibinfo  {journal} {Astrophys. J. Lett.}\ }\textbf
  {\bibinfo {volume} {887}},\ \bibinfo {pages} {L21} (\bibinfo {year}
  {2019})},\ \Eprint {http://arxiv.org/abs/1912.05702} {arXiv:1912.05702
  [astro-ph.HE]} \BibitemShut {NoStop}%
\bibitem [{\citenamefont {Miller}\ \emph {et~al.}(2019)\citenamefont {Miller}
  \emph {et~al.}}]{Miller:2019cac}%
  \BibitemOpen
  \bibfield  {author} {\bibinfo {author} {\bibfnamefont {M.~C.}\ \bibnamefont
  {Miller}} \emph {et~al.},\ }\href {\doibase 10.3847/2041-8213/ab50c5}
  {\bibfield  {journal} {\bibinfo  {journal} {Astrophys. J. Lett.}\ }\textbf
  {\bibinfo {volume} {887}},\ \bibinfo {pages} {L24} (\bibinfo {year}
  {2019})},\ \Eprint {http://arxiv.org/abs/1912.05705} {arXiv:1912.05705
  [astro-ph.HE]} \BibitemShut {NoStop}%
\bibitem [{\citenamefont {Catuneanu}\ \emph {et~al.}(2013)\citenamefont
  {Catuneanu}, \citenamefont {Heinke}, \citenamefont {Sivakoff}, \citenamefont
  {Ho},\ and\ \citenamefont {Servillat}}]{2013ApJ...764..145C}%
  \BibitemOpen
  \bibfield  {author} {\bibinfo {author} {\bibfnamefont {A.}~\bibnamefont
  {Catuneanu}}, \bibinfo {author} {\bibfnamefont {C.~O.}\ \bibnamefont
  {Heinke}}, \bibinfo {author} {\bibfnamefont {G.~R.}\ \bibnamefont
  {Sivakoff}}, \bibinfo {author} {\bibfnamefont {W.~C.~G.}\ \bibnamefont {Ho}},
  \ and\ \bibinfo {author} {\bibfnamefont {M.}~\bibnamefont {Servillat}},\
  }\href {\doibase 10.1088/0004-637X/764/2/145} {\bibfield  {journal} {\bibinfo
   {journal} {Astrophys. J.}\ }\textbf {\bibinfo {volume} {764}},\ \bibinfo
  {pages} {145} (\bibinfo {year} {2013})},\ \Eprint
  {http://arxiv.org/abs/1301.3768} {arXiv:1301.3768 [astro-ph.HE]} \BibitemShut
  {NoStop}%
\bibitem [{\citenamefont {Servillat}\ \emph {et~al.}(2012)\citenamefont
  {Servillat}, \citenamefont {Heinke}, \citenamefont {Ho}, \citenamefont
  {Grindlay}, \citenamefont {Hong}, \citenamefont {Berg},\ and\ \citenamefont
  {Bogdanov}}]{2012MNRAS.423.1556S}%
  \BibitemOpen
  \bibfield  {author} {\bibinfo {author} {\bibfnamefont {M.}~\bibnamefont
  {Servillat}}, \bibinfo {author} {\bibfnamefont {C.~O.}\ \bibnamefont
  {Heinke}}, \bibinfo {author} {\bibfnamefont {W.~C.~G.}\ \bibnamefont {Ho}},
  \bibinfo {author} {\bibfnamefont {J.~E.}\ \bibnamefont {Grindlay}}, \bibinfo
  {author} {\bibfnamefont {J.}~\bibnamefont {Hong}}, \bibinfo {author}
  {\bibfnamefont {M.~v.~d.}\ \bibnamefont {Berg}}, \ and\ \bibinfo {author}
  {\bibfnamefont {S.}~\bibnamefont {Bogdanov}},\ }\href {\doibase
  10.1111/j.1365-2966.2012.20976.x} {\bibfield  {journal} {\bibinfo  {journal}
  {Mon. Not. Roy. Astron. Soc.}\ }\textbf {\bibinfo {volume} {423}},\ \bibinfo
  {pages} {1556} (\bibinfo {year} {2012})},\ \Eprint
  {http://arxiv.org/abs/1203.5807} {arXiv:1203.5807 [astro-ph.HE]} \BibitemShut
  {NoStop}%
\bibitem [{\citenamefont {Torres-Rivas}\ \emph {et~al.}(2019)\citenamefont
  {Torres-Rivas}, \citenamefont {Chatziioannou}, \citenamefont {Bauswein},\
  and\ \citenamefont {Clark}}]{Torres-Rivas:2018svp}%
  \BibitemOpen
  \bibfield  {author} {\bibinfo {author} {\bibfnamefont {A.}~\bibnamefont
  {Torres-Rivas}}, \bibinfo {author} {\bibfnamefont {K.}~\bibnamefont
  {Chatziioannou}}, \bibinfo {author} {\bibfnamefont {A.}~\bibnamefont
  {Bauswein}}, \ and\ \bibinfo {author} {\bibfnamefont {J.~A.}\ \bibnamefont
  {Clark}},\ }\href {\doibase 10.1103/PhysRevD.99.044014} {\bibfield  {journal}
  {\bibinfo  {journal} {Phys. Rev. D}\ }\textbf {\bibinfo {volume} {99}},\
  \bibinfo {pages} {044014} (\bibinfo {year} {2019})},\ \Eprint
  {http://arxiv.org/abs/1811.08931} {arXiv:1811.08931 [gr-qc]} \BibitemShut
  {NoStop}%
\bibitem [{\citenamefont {Dietrich}\ \emph {et~al.}(2021)\citenamefont
  {Dietrich}, \citenamefont {Hinderer},\ and\ \citenamefont
  {Samajdar}}]{Dietrich:2020eud}%
  \BibitemOpen
  \bibfield  {author} {\bibinfo {author} {\bibfnamefont {T.}~\bibnamefont
  {Dietrich}}, \bibinfo {author} {\bibfnamefont {T.}~\bibnamefont {Hinderer}},
  \ and\ \bibinfo {author} {\bibfnamefont {A.}~\bibnamefont {Samajdar}},\
  }\href {\doibase 10.1007/s10714-020-02751-6} {\bibfield  {journal} {\bibinfo
  {journal} {Gen. Rel. Grav.}\ }\textbf {\bibinfo {volume} {53}},\ \bibinfo
  {pages} {27} (\bibinfo {year} {2021})},\ \Eprint
  {http://arxiv.org/abs/2004.02527} {arXiv:2004.02527 [gr-qc]} \BibitemShut
  {NoStop}%
\bibitem [{\citenamefont {Farr}\ \emph {et~al.}(2016)\citenamefont {Farr} \emph
  {et~al.}}]{Farr:2015lna}%
  \BibitemOpen
  \bibfield  {author} {\bibinfo {author} {\bibfnamefont {B.}~\bibnamefont
  {Farr}} \emph {et~al.},\ }\href {\doibase 10.3847/0004-637X/825/2/116}
  {\bibfield  {journal} {\bibinfo  {journal} {Astrophys. J.}\ }\textbf
  {\bibinfo {volume} {825}},\ \bibinfo {pages} {116} (\bibinfo {year}
  {2016})},\ \Eprint {http://arxiv.org/abs/1508.05336} {arXiv:1508.05336
  [astro-ph.HE]} \BibitemShut {NoStop}%
\bibitem [{\citenamefont {Ng}\ \emph {et~al.}(2018)\citenamefont {Ng},
  \citenamefont {Vitale}, \citenamefont {Zimmerman}, \citenamefont
  {Chatziioannou}, \citenamefont {Gerosa},\ and\ \citenamefont
  {Haster}}]{Ng:2018neg}%
  \BibitemOpen
  \bibfield  {author} {\bibinfo {author} {\bibfnamefont {K.~K.~Y.}\
  \bibnamefont {Ng}}, \bibinfo {author} {\bibfnamefont {S.}~\bibnamefont
  {Vitale}}, \bibinfo {author} {\bibfnamefont {A.}~\bibnamefont {Zimmerman}},
  \bibinfo {author} {\bibfnamefont {K.}~\bibnamefont {Chatziioannou}}, \bibinfo
  {author} {\bibfnamefont {D.}~\bibnamefont {Gerosa}}, \ and\ \bibinfo {author}
  {\bibfnamefont {C.-J.}\ \bibnamefont {Haster}},\ }\href {\doibase
  10.1103/PhysRevD.98.083007} {\bibfield  {journal} {\bibinfo  {journal} {Phys.
  Rev. D}\ }\textbf {\bibinfo {volume} {98}},\ \bibinfo {pages} {083007}
  (\bibinfo {year} {2018})},\ \Eprint {http://arxiv.org/abs/1805.03046}
  {arXiv:1805.03046 [gr-qc]} \BibitemShut {NoStop}%
\bibitem [{\citenamefont {Abbott}\ \emph
  {et~al.}(2020{\natexlab{a}})\citenamefont {Abbott} \emph
  {et~al.}}]{LIGOScientific:2020aai}%
  \BibitemOpen
  \bibfield  {author} {\bibinfo {author} {\bibfnamefont {B.~P.}\ \bibnamefont
  {Abbott}} \emph {et~al.} (\bibinfo {collaboration} {LIGO Scientific,
  Virgo}),\ }\href {\doibase 10.3847/2041-8213/ab75f5} {\bibfield  {journal}
  {\bibinfo  {journal} {Astrophys. J. Lett.}\ }\textbf {\bibinfo {volume}
  {892}},\ \bibinfo {pages} {L3} (\bibinfo {year} {2020}{\natexlab{a}})},\
  \Eprint {http://arxiv.org/abs/2001.01761} {arXiv:2001.01761 [astro-ph.HE]}
  \BibitemShut {NoStop}%
\bibitem [{\citenamefont {Lattimer}(2012)}]{Lattimer:2012nd}%
  \BibitemOpen
  \bibfield  {author} {\bibinfo {author} {\bibfnamefont {J.~M.}\ \bibnamefont
  {Lattimer}},\ }\href {\doibase 10.1146/annurev-nucl-102711-095018} {\bibfield
   {journal} {\bibinfo  {journal} {Ann. Rev. Nucl. Part. Sci.}\ }\textbf
  {\bibinfo {volume} {62}},\ \bibinfo {pages} {485} (\bibinfo {year} {2012})},\
  \Eprint {http://arxiv.org/abs/1305.3510} {arXiv:1305.3510 [nucl-th]}
  \BibitemShut {NoStop}%
\bibitem [{\citenamefont {Abbott}\ \emph
  {et~al.}(2020{\natexlab{b}})\citenamefont {Abbott} \emph
  {et~al.}}]{LIGOScientific:2020zkf}%
  \BibitemOpen
  \bibfield  {author} {\bibinfo {author} {\bibfnamefont {R.}~\bibnamefont
  {Abbott}} \emph {et~al.} (\bibinfo {collaboration} {LIGO Scientific,
  Virgo}),\ }\href {\doibase 10.3847/2041-8213/ab960f} {\bibfield  {journal}
  {\bibinfo  {journal} {Astrophys. J. Lett.}\ }\textbf {\bibinfo {volume}
  {896}},\ \bibinfo {pages} {L44} (\bibinfo {year} {2020}{\natexlab{b}})},\
  \Eprint {http://arxiv.org/abs/2006.12611} {arXiv:2006.12611 [astro-ph.HE]}
  \BibitemShut {NoStop}%
\bibitem [{\citenamefont {Chatziioannou}(2022)}]{Chatziioannou:2021tdi}%
  \BibitemOpen
  \bibfield  {author} {\bibinfo {author} {\bibfnamefont {K.}~\bibnamefont
  {Chatziioannou}},\ }\href {\doibase 10.1103/PhysRevD.105.084021} {\bibfield
  {journal} {\bibinfo  {journal} {Phys. Rev. D}\ }\textbf {\bibinfo {volume}
  {105}},\ \bibinfo {pages} {084021} (\bibinfo {year} {2022})},\ \Eprint
  {http://arxiv.org/abs/2108.12368} {arXiv:2108.12368 [gr-qc]} \BibitemShut
  {NoStop}%
\bibitem [{\citenamefont {Annala}\ \emph {et~al.}(2018)\citenamefont {Annala},
  \citenamefont {Gorda}, \citenamefont {Kurkela},\ and\ \citenamefont
  {Vuorinen}}]{Annala:2017llu}%
  \BibitemOpen
  \bibfield  {author} {\bibinfo {author} {\bibfnamefont {E.}~\bibnamefont
  {Annala}}, \bibinfo {author} {\bibfnamefont {T.}~\bibnamefont {Gorda}},
  \bibinfo {author} {\bibfnamefont {A.}~\bibnamefont {Kurkela}}, \ and\
  \bibinfo {author} {\bibfnamefont {A.}~\bibnamefont {Vuorinen}},\ }\href
  {\doibase 10.1103/PhysRevLett.120.172703} {\bibfield  {journal} {\bibinfo
  {journal} {Phys. Rev. Lett.}\ }\textbf {\bibinfo {volume} {120}},\ \bibinfo
  {pages} {172703} (\bibinfo {year} {2018})},\ \Eprint
  {http://arxiv.org/abs/1711.02644} {arXiv:1711.02644 [astro-ph.HE]}
  \BibitemShut {NoStop}%
\bibitem [{\citenamefont {Bauswein}\ \emph {et~al.}(2017)\citenamefont
  {Bauswein}, \citenamefont {Just}, \citenamefont {Janka},\ and\ \citenamefont
  {Stergioulas}}]{Bauswein:2017vtn}%
  \BibitemOpen
  \bibfield  {author} {\bibinfo {author} {\bibfnamefont {A.}~\bibnamefont
  {Bauswein}}, \bibinfo {author} {\bibfnamefont {O.}~\bibnamefont {Just}},
  \bibinfo {author} {\bibfnamefont {H.-T.}\ \bibnamefont {Janka}}, \ and\
  \bibinfo {author} {\bibfnamefont {N.}~\bibnamefont {Stergioulas}},\ }\href
  {\doibase 10.3847/2041-8213/aa9994} {\bibfield  {journal} {\bibinfo
  {journal} {Astrophys. J. Lett.}\ }\textbf {\bibinfo {volume} {850}},\
  \bibinfo {pages} {L34} (\bibinfo {year} {2017})},\ \Eprint
  {http://arxiv.org/abs/1710.06843} {arXiv:1710.06843 [astro-ph.HE]}
  \BibitemShut {NoStop}%
\bibitem [{\citenamefont {De}\ \emph {et~al.}(2018)\citenamefont {De},
  \citenamefont {Finstad}, \citenamefont {Lattimer}, \citenamefont {Brown},
  \citenamefont {Berger},\ and\ \citenamefont {Biwer}}]{De:2018uhw}%
  \BibitemOpen
  \bibfield  {author} {\bibinfo {author} {\bibfnamefont {S.}~\bibnamefont
  {De}}, \bibinfo {author} {\bibfnamefont {D.}~\bibnamefont {Finstad}},
  \bibinfo {author} {\bibfnamefont {J.~M.}\ \bibnamefont {Lattimer}}, \bibinfo
  {author} {\bibfnamefont {D.~A.}\ \bibnamefont {Brown}}, \bibinfo {author}
  {\bibfnamefont {E.}~\bibnamefont {Berger}}, \ and\ \bibinfo {author}
  {\bibfnamefont {C.~M.}\ \bibnamefont {Biwer}},\ }\href {\doibase
  10.1103/PhysRevLett.121.091102} {\bibfield  {journal} {\bibinfo  {journal}
  {Phys. Rev. Lett.}\ }\textbf {\bibinfo {volume} {121}},\ \bibinfo {pages}
  {091102} (\bibinfo {year} {2018})},\ \bibinfo {note} {[Erratum:
  Phys.Rev.Lett. 121, 259902 (2018)]},\ \Eprint
  {http://arxiv.org/abs/1804.08583} {arXiv:1804.08583 [astro-ph.HE]}
  \BibitemShut {NoStop}%
\bibitem [{\citenamefont {Most}\ \emph {et~al.}(2018)\citenamefont {Most},
  \citenamefont {Weih}, \citenamefont {Rezzolla},\ and\ \citenamefont
  {Schaffner-Bielich}}]{Most:2018hfd}%
  \BibitemOpen
  \bibfield  {author} {\bibinfo {author} {\bibfnamefont {E.~R.}\ \bibnamefont
  {Most}}, \bibinfo {author} {\bibfnamefont {L.~R.}\ \bibnamefont {Weih}},
  \bibinfo {author} {\bibfnamefont {L.}~\bibnamefont {Rezzolla}}, \ and\
  \bibinfo {author} {\bibfnamefont {J.}~\bibnamefont {Schaffner-Bielich}},\
  }\href {\doibase 10.1103/PhysRevLett.120.261103} {\bibfield  {journal}
  {\bibinfo  {journal} {Phys. Rev. Lett.}\ }\textbf {\bibinfo {volume} {120}},\
  \bibinfo {pages} {261103} (\bibinfo {year} {2018})},\ \Eprint
  {http://arxiv.org/abs/1803.00549} {arXiv:1803.00549 [gr-qc]} \BibitemShut
  {NoStop}%
\bibitem [{\citenamefont {Raithel}\ \emph {et~al.}(2018)\citenamefont
  {Raithel}, \citenamefont {\"Ozel},\ and\ \citenamefont
  {Psaltis}}]{Raithel:2018ncd}%
  \BibitemOpen
  \bibfield  {author} {\bibinfo {author} {\bibfnamefont {C.}~\bibnamefont
  {Raithel}}, \bibinfo {author} {\bibfnamefont {F.}~\bibnamefont {\"Ozel}}, \
  and\ \bibinfo {author} {\bibfnamefont {D.}~\bibnamefont {Psaltis}},\ }\href
  {\doibase 10.3847/2041-8213/aabcbf} {\bibfield  {journal} {\bibinfo
  {journal} {Astrophys. J. Lett.}\ }\textbf {\bibinfo {volume} {857}},\
  \bibinfo {pages} {L23} (\bibinfo {year} {2018})},\ \Eprint
  {http://arxiv.org/abs/1803.07687} {arXiv:1803.07687 [astro-ph.HE]}
  \BibitemShut {NoStop}%
\bibitem [{\citenamefont {Abbott}\ \emph {et~al.}(2018)\citenamefont {Abbott}
  \emph {et~al.}}]{LIGOScientific:2018cki}%
  \BibitemOpen
  \bibfield  {author} {\bibinfo {author} {\bibfnamefont {B.~P.}\ \bibnamefont
  {Abbott}} \emph {et~al.} (\bibinfo {collaboration} {LIGO Scientific,
  Virgo}),\ }\href {\doibase 10.1103/PhysRevLett.121.161101} {\bibfield
  {journal} {\bibinfo  {journal} {Phys. Rev. Lett.}\ }\textbf {\bibinfo
  {volume} {121}},\ \bibinfo {pages} {161101} (\bibinfo {year} {2018})},\
  \Eprint {http://arxiv.org/abs/1805.11581} {arXiv:1805.11581 [gr-qc]}
  \BibitemShut {NoStop}%
\bibitem [{\citenamefont {Raithel}\ \emph
  {et~al.}(2021{\natexlab{a}})\citenamefont {Raithel}, \citenamefont {Ozel},\
  and\ \citenamefont {Psaltis}}]{Raithel:2020vvg}%
  \BibitemOpen
  \bibfield  {author} {\bibinfo {author} {\bibfnamefont {C.}~\bibnamefont
  {Raithel}}, \bibinfo {author} {\bibfnamefont {F.}~\bibnamefont {Ozel}}, \
  and\ \bibinfo {author} {\bibfnamefont {D.}~\bibnamefont {Psaltis}},\ }\href
  {\doibase 10.3847/1538-4357/abd3a4} {\bibfield  {journal} {\bibinfo
  {journal} {Astrophys. J.}\ }\textbf {\bibinfo {volume} {908}},\ \bibinfo
  {pages} {103} (\bibinfo {year} {2021}{\natexlab{a}})},\ \Eprint
  {http://arxiv.org/abs/2004.00656} {arXiv:2004.00656 [astro-ph.HE]}
  \BibitemShut {NoStop}%
\bibitem [{\citenamefont {Chatziioannou}\ and\ \citenamefont
  {Han}(2020)}]{Chatziioannou:2019yko}%
  \BibitemOpen
  \bibfield  {author} {\bibinfo {author} {\bibfnamefont {K.}~\bibnamefont
  {Chatziioannou}}\ and\ \bibinfo {author} {\bibfnamefont {S.}~\bibnamefont
  {Han}},\ }\href {\doibase 10.1103/PhysRevD.101.044019} {\bibfield  {journal}
  {\bibinfo  {journal} {Phys. Rev. D}\ }\textbf {\bibinfo {volume} {101}},\
  \bibinfo {pages} {044019} (\bibinfo {year} {2020})},\ \Eprint
  {http://arxiv.org/abs/1911.07091} {arXiv:1911.07091 [gr-qc]} \BibitemShut
  {NoStop}%
\bibitem [{\citenamefont {Pang}\ \emph {et~al.}(2020)\citenamefont {Pang},
  \citenamefont {Dietrich}, \citenamefont {Tews},\ and\ \citenamefont {Van
  Den~Broeck}}]{Pang:2020ilf}%
  \BibitemOpen
  \bibfield  {author} {\bibinfo {author} {\bibfnamefont {P.~T.~H.}\
  \bibnamefont {Pang}}, \bibinfo {author} {\bibfnamefont {T.}~\bibnamefont
  {Dietrich}}, \bibinfo {author} {\bibfnamefont {I.}~\bibnamefont {Tews}}, \
  and\ \bibinfo {author} {\bibfnamefont {C.}~\bibnamefont {Van Den~Broeck}},\
  }\href {\doibase 10.1103/PhysRevResearch.2.033514} {\bibfield  {journal}
  {\bibinfo  {journal} {Phys. Rev. Res.}\ }\textbf {\bibinfo {volume} {2}},\
  \bibinfo {pages} {033514} (\bibinfo {year} {2020})},\ \Eprint
  {http://arxiv.org/abs/2006.14936} {arXiv:2006.14936 [astro-ph.HE]}
  \BibitemShut {NoStop}%
\bibitem [{\citenamefont {Tan}\ \emph {et~al.}(2020)\citenamefont {Tan},
  \citenamefont {Noronha-Hostler},\ and\ \citenamefont {Yunes}}]{Tan:2020ics}%
  \BibitemOpen
  \bibfield  {author} {\bibinfo {author} {\bibfnamefont {H.}~\bibnamefont
  {Tan}}, \bibinfo {author} {\bibfnamefont {J.}~\bibnamefont
  {Noronha-Hostler}}, \ and\ \bibinfo {author} {\bibfnamefont {N.}~\bibnamefont
  {Yunes}},\ }\href {\doibase 10.1103/PhysRevLett.125.261104} {\bibfield
  {journal} {\bibinfo  {journal} {Phys. Rev. Lett.}\ }\textbf {\bibinfo
  {volume} {125}},\ \bibinfo {pages} {261104} (\bibinfo {year} {2020})},\
  \Eprint {http://arxiv.org/abs/2006.16296} {arXiv:2006.16296 [astro-ph.HE]}
  \BibitemShut {NoStop}%
\bibitem [{\citenamefont {Jayasinghe}\ \emph {et~al.}(2021)\citenamefont
  {Jayasinghe} \emph {et~al.}}]{Jayasinghe:2021uqb}%
  \BibitemOpen
  \bibfield  {author} {\bibinfo {author} {\bibfnamefont {T.}~\bibnamefont
  {Jayasinghe}} \emph {et~al.},\ }\href {\doibase 10.1093/mnras/stab907}
  {\bibfield  {journal} {\bibinfo  {journal} {Mon. Not. Roy. Astron. Soc.}\
  }\textbf {\bibinfo {volume} {504}},\ \bibinfo {pages} {2577} (\bibinfo {year}
  {2021})},\ \Eprint {http://arxiv.org/abs/2101.02212} {arXiv:2101.02212
  [astro-ph.SR]} \BibitemShut {NoStop}%
\bibitem [{\citenamefont {van Kerkwijk}\ \emph {et~al.}(2011)\citenamefont {van
  Kerkwijk}, \citenamefont {Breton},\ and\ \citenamefont
  {Kulkarni}}]{vanKerkwijk:2010mt}%
  \BibitemOpen
  \bibfield  {author} {\bibinfo {author} {\bibfnamefont {M.~H.}\ \bibnamefont
  {van Kerkwijk}}, \bibinfo {author} {\bibfnamefont {R.}~\bibnamefont
  {Breton}}, \ and\ \bibinfo {author} {\bibfnamefont {S.~R.}\ \bibnamefont
  {Kulkarni}},\ }\href {\doibase 10.1088/0004-637X/728/2/95} {\bibfield
  {journal} {\bibinfo  {journal} {Astrophys. J.}\ }\textbf {\bibinfo {volume}
  {728}},\ \bibinfo {pages} {95} (\bibinfo {year} {2011})},\ \Eprint
  {http://arxiv.org/abs/1009.5427} {arXiv:1009.5427 [astro-ph.HE]} \BibitemShut
  {NoStop}%
\bibitem [{\citenamefont {Romani}\ \emph {et~al.}(2022)\citenamefont {Romani},
  \citenamefont {Kandel}, \citenamefont {Filippenko}, \citenamefont {Brink},\
  and\ \citenamefont {Zheng}}]{Romani:2022jhd}%
  \BibitemOpen
  \bibfield  {author} {\bibinfo {author} {\bibfnamefont {R.~W.}\ \bibnamefont
  {Romani}}, \bibinfo {author} {\bibfnamefont {D.}~\bibnamefont {Kandel}},
  \bibinfo {author} {\bibfnamefont {A.~V.}\ \bibnamefont {Filippenko}},
  \bibinfo {author} {\bibfnamefont {T.~G.}\ \bibnamefont {Brink}}, \ and\
  \bibinfo {author} {\bibfnamefont {W.}~\bibnamefont {Zheng}},\ }\href
  {\doibase 10.3847/2041-8213/ac8007} {\bibfield  {journal} {\bibinfo
  {journal} {Astrophys. J. Lett.}\ }\textbf {\bibinfo {volume} {934}},\
  \bibinfo {pages} {L18} (\bibinfo {year} {2022})},\ \Eprint
  {http://arxiv.org/abs/2207.05124} {arXiv:2207.05124 [astro-ph.HE]}
  \BibitemShut {NoStop}%
\bibitem [{\citenamefont {Tews}\ \emph
  {et~al.}(2021{\natexlab{b}})\citenamefont {Tews}, \citenamefont {Pang},
  \citenamefont {Dietrich}, \citenamefont {Coughlin}, \citenamefont {Antier},
  \citenamefont {Bulla}, \citenamefont {Heinzel},\ and\ \citenamefont
  {Issa}}]{Tews:2020ylw}%
  \BibitemOpen
  \bibfield  {author} {\bibinfo {author} {\bibfnamefont {I.}~\bibnamefont
  {Tews}}, \bibinfo {author} {\bibfnamefont {P.~T.~H.}\ \bibnamefont {Pang}},
  \bibinfo {author} {\bibfnamefont {T.}~\bibnamefont {Dietrich}}, \bibinfo
  {author} {\bibfnamefont {M.~W.}\ \bibnamefont {Coughlin}}, \bibinfo {author}
  {\bibfnamefont {S.}~\bibnamefont {Antier}}, \bibinfo {author} {\bibfnamefont
  {M.}~\bibnamefont {Bulla}}, \bibinfo {author} {\bibfnamefont
  {J.}~\bibnamefont {Heinzel}}, \ and\ \bibinfo {author} {\bibfnamefont
  {L.}~\bibnamefont {Issa}},\ }\href {\doibase 10.3847/2041-8213/abdaae}
  {\bibfield  {journal} {\bibinfo  {journal} {Astrophys. J. Lett.}\ }\textbf
  {\bibinfo {volume} {908}},\ \bibinfo {pages} {L1} (\bibinfo {year}
  {2021}{\natexlab{b}})},\ \Eprint {http://arxiv.org/abs/2007.06057}
  {arXiv:2007.06057 [astro-ph.HE]} \BibitemShut {NoStop}%
\bibitem [{\citenamefont {Margalit}\ and\ \citenamefont
  {Metzger}(2017)}]{Margalit:2017dij}%
  \BibitemOpen
  \bibfield  {author} {\bibinfo {author} {\bibfnamefont {B.}~\bibnamefont
  {Margalit}}\ and\ \bibinfo {author} {\bibfnamefont {B.~D.}\ \bibnamefont
  {Metzger}},\ }\href {\doibase 10.3847/2041-8213/aa991c} {\bibfield  {journal}
  {\bibinfo  {journal} {Astrophys. J. Lett.}\ }\textbf {\bibinfo {volume}
  {850}},\ \bibinfo {pages} {L19} (\bibinfo {year} {2017})},\ \Eprint
  {http://arxiv.org/abs/1710.05938} {arXiv:1710.05938 [astro-ph.HE]}
  \BibitemShut {NoStop}%
\bibitem [{\citenamefont {Rezzolla}\ \emph {et~al.}(2018)\citenamefont
  {Rezzolla}, \citenamefont {Most},\ and\ \citenamefont
  {Weih}}]{Rezzolla:2017aly}%
  \BibitemOpen
  \bibfield  {author} {\bibinfo {author} {\bibfnamefont {L.}~\bibnamefont
  {Rezzolla}}, \bibinfo {author} {\bibfnamefont {E.~R.}\ \bibnamefont {Most}},
  \ and\ \bibinfo {author} {\bibfnamefont {L.~R.}\ \bibnamefont {Weih}},\
  }\href {\doibase 10.3847/2041-8213/aaa401} {\bibfield  {journal} {\bibinfo
  {journal} {Astrophys. J. Lett.}\ }\textbf {\bibinfo {volume} {852}},\
  \bibinfo {pages} {L25} (\bibinfo {year} {2018})},\ \Eprint
  {http://arxiv.org/abs/1711.00314} {arXiv:1711.00314 [astro-ph.HE]}
  \BibitemShut {NoStop}%
\bibitem [{\citenamefont {Shibata}\ \emph {et~al.}(2019)\citenamefont
  {Shibata}, \citenamefont {Zhou}, \citenamefont {Kiuchi},\ and\ \citenamefont
  {Fujibayashi}}]{Shibata:2019ctb}%
  \BibitemOpen
  \bibfield  {author} {\bibinfo {author} {\bibfnamefont {M.}~\bibnamefont
  {Shibata}}, \bibinfo {author} {\bibfnamefont {E.}~\bibnamefont {Zhou}},
  \bibinfo {author} {\bibfnamefont {K.}~\bibnamefont {Kiuchi}}, \ and\ \bibinfo
  {author} {\bibfnamefont {S.}~\bibnamefont {Fujibayashi}},\ }\href {\doibase
  10.1103/PhysRevD.100.023015} {\bibfield  {journal} {\bibinfo  {journal}
  {Phys. Rev. D}\ }\textbf {\bibinfo {volume} {100}},\ \bibinfo {pages}
  {023015} (\bibinfo {year} {2019})},\ \Eprint
  {http://arxiv.org/abs/1905.03656} {arXiv:1905.03656 [astro-ph.HE]}
  \BibitemShut {NoStop}%
\bibitem [{\citenamefont {Bauswein}\ \emph {et~al.}(2010)\citenamefont
  {Bauswein}, \citenamefont {Janka},\ and\ \citenamefont
  {Oechslin}}]{Bauswein:2010dn}%
  \BibitemOpen
  \bibfield  {author} {\bibinfo {author} {\bibfnamefont {A.}~\bibnamefont
  {Bauswein}}, \bibinfo {author} {\bibfnamefont {H.~T.}\ \bibnamefont {Janka}},
  \ and\ \bibinfo {author} {\bibfnamefont {R.}~\bibnamefont {Oechslin}},\
  }\href {\doibase 10.1103/PhysRevD.82.084043} {\bibfield  {journal} {\bibinfo
  {journal} {Phys. Rev. D}\ }\textbf {\bibinfo {volume} {82}},\ \bibinfo
  {pages} {084043} (\bibinfo {year} {2010})},\ \Eprint
  {http://arxiv.org/abs/1006.3315} {arXiv:1006.3315 [astro-ph.SR]} \BibitemShut
  {NoStop}%
\bibitem [{\citenamefont {Kastaun}\ \emph {et~al.}(2016)\citenamefont
  {Kastaun}, \citenamefont {Ciolfi},\ and\ \citenamefont
  {Giacomazzo}}]{Kastaun:2016yaf}%
  \BibitemOpen
  \bibfield  {author} {\bibinfo {author} {\bibfnamefont {W.}~\bibnamefont
  {Kastaun}}, \bibinfo {author} {\bibfnamefont {R.}~\bibnamefont {Ciolfi}}, \
  and\ \bibinfo {author} {\bibfnamefont {B.}~\bibnamefont {Giacomazzo}},\
  }\href {\doibase 10.1103/PhysRevD.94.044060} {\bibfield  {journal} {\bibinfo
  {journal} {Phys. Rev. D}\ }\textbf {\bibinfo {volume} {94}},\ \bibinfo
  {pages} {044060} (\bibinfo {year} {2016})},\ \Eprint
  {http://arxiv.org/abs/1607.02186} {arXiv:1607.02186 [astro-ph.HE]}
  \BibitemShut {NoStop}%
\bibitem [{\citenamefont {Hanauske}\ \emph {et~al.}(2017)\citenamefont
  {Hanauske}, \citenamefont {Takami}, \citenamefont {Bovard}, \citenamefont
  {Rezzolla}, \citenamefont {Font}, \citenamefont {Galeazzi},\ and\
  \citenamefont {St\"ocker}}]{Hanauske:2016gia}%
  \BibitemOpen
  \bibfield  {author} {\bibinfo {author} {\bibfnamefont {M.}~\bibnamefont
  {Hanauske}}, \bibinfo {author} {\bibfnamefont {K.}~\bibnamefont {Takami}},
  \bibinfo {author} {\bibfnamefont {L.}~\bibnamefont {Bovard}}, \bibinfo
  {author} {\bibfnamefont {L.}~\bibnamefont {Rezzolla}}, \bibinfo {author}
  {\bibfnamefont {J.~A.}\ \bibnamefont {Font}}, \bibinfo {author}
  {\bibfnamefont {F.}~\bibnamefont {Galeazzi}}, \ and\ \bibinfo {author}
  {\bibfnamefont {H.}~\bibnamefont {St\"ocker}},\ }\href {\doibase
  10.1103/PhysRevD.96.043004} {\bibfield  {journal} {\bibinfo  {journal} {Phys.
  Rev. D}\ }\textbf {\bibinfo {volume} {96}},\ \bibinfo {pages} {043004}
  (\bibinfo {year} {2017})},\ \Eprint {http://arxiv.org/abs/1611.07152}
  {arXiv:1611.07152 [gr-qc]} \BibitemShut {NoStop}%
\bibitem [{\citenamefont {Endrizzi}\ \emph {et~al.}(2020)\citenamefont
  {Endrizzi}, \citenamefont {Perego}, \citenamefont {Fabbri}, \citenamefont
  {Branca}, \citenamefont {Radice}, \citenamefont {Bernuzzi}, \citenamefont
  {Giacomazzo}, \citenamefont {Pederiva},\ and\ \citenamefont
  {Lovato}}]{Endrizzi:2019trv}%
  \BibitemOpen
  \bibfield  {author} {\bibinfo {author} {\bibfnamefont {A.}~\bibnamefont
  {Endrizzi}}, \bibinfo {author} {\bibfnamefont {A.}~\bibnamefont {Perego}},
  \bibinfo {author} {\bibfnamefont {F.~M.}\ \bibnamefont {Fabbri}}, \bibinfo
  {author} {\bibfnamefont {L.}~\bibnamefont {Branca}}, \bibinfo {author}
  {\bibfnamefont {D.}~\bibnamefont {Radice}}, \bibinfo {author} {\bibfnamefont
  {S.}~\bibnamefont {Bernuzzi}}, \bibinfo {author} {\bibfnamefont
  {B.}~\bibnamefont {Giacomazzo}}, \bibinfo {author} {\bibfnamefont
  {F.}~\bibnamefont {Pederiva}}, \ and\ \bibinfo {author} {\bibfnamefont
  {A.}~\bibnamefont {Lovato}},\ }\href {\doibase
  10.1140/epja/s10050-019-00018-6} {\bibfield  {journal} {\bibinfo  {journal}
  {Eur. Phys. J. A}\ }\textbf {\bibinfo {volume} {56}},\ \bibinfo {pages} {15}
  (\bibinfo {year} {2020})},\ \Eprint {http://arxiv.org/abs/1908.04952}
  {arXiv:1908.04952 [astro-ph.HE]} \BibitemShut {NoStop}%
\bibitem [{\citenamefont {Raithel}\ \emph
  {et~al.}(2021{\natexlab{b}})\citenamefont {Raithel}, \citenamefont
  {Paschalidis},\ and\ \citenamefont {\"Ozel}}]{Raithel:2021hye}%
  \BibitemOpen
  \bibfield  {author} {\bibinfo {author} {\bibfnamefont {C.}~\bibnamefont
  {Raithel}}, \bibinfo {author} {\bibfnamefont {V.}~\bibnamefont
  {Paschalidis}}, \ and\ \bibinfo {author} {\bibfnamefont {F.}~\bibnamefont
  {\"Ozel}},\ }\href {\doibase 10.1103/PhysRevD.104.063016} {\bibfield
  {journal} {\bibinfo  {journal} {Phys. Rev. D}\ }\textbf {\bibinfo {volume}
  {104}},\ \bibinfo {pages} {063016} (\bibinfo {year} {2021}{\natexlab{b}})},\
  \Eprint {http://arxiv.org/abs/2104.07226} {arXiv:2104.07226 [astro-ph.HE]}
  \BibitemShut {NoStop}%
\bibitem [{\citenamefont {Radice}\ \emph {et~al.}(2020)\citenamefont {Radice},
  \citenamefont {Bernuzzi},\ and\ \citenamefont {Perego}}]{Radice:2020ddv}%
  \BibitemOpen
  \bibfield  {author} {\bibinfo {author} {\bibfnamefont {D.}~\bibnamefont
  {Radice}}, \bibinfo {author} {\bibfnamefont {S.}~\bibnamefont {Bernuzzi}}, \
  and\ \bibinfo {author} {\bibfnamefont {A.}~\bibnamefont {Perego}},\ }\href
  {\doibase 10.1146/annurev-nucl-013120-114541} {\bibfield  {journal} {\bibinfo
   {journal} {Ann. Rev. Nucl. Part. Sci.}\ }\textbf {\bibinfo {volume} {70}},\
  \bibinfo {pages} {95} (\bibinfo {year} {2020})},\ \Eprint
  {http://arxiv.org/abs/2002.03863} {arXiv:2002.03863 [astro-ph.HE]}
  \BibitemShut {NoStop}%
\bibitem [{\citenamefont {Most}\ \emph {et~al.}(2020)\citenamefont {Most},
  \citenamefont {Jens~Papenfort}, \citenamefont {Dexheimer}, \citenamefont
  {Hanauske}, \citenamefont {Stoecker},\ and\ \citenamefont
  {Rezzolla}}]{Most:2019onn}%
  \BibitemOpen
  \bibfield  {author} {\bibinfo {author} {\bibfnamefont {E.~R.}\ \bibnamefont
  {Most}}, \bibinfo {author} {\bibfnamefont {L.}~\bibnamefont
  {Jens~Papenfort}}, \bibinfo {author} {\bibfnamefont {V.}~\bibnamefont
  {Dexheimer}}, \bibinfo {author} {\bibfnamefont {M.}~\bibnamefont {Hanauske}},
  \bibinfo {author} {\bibfnamefont {H.}~\bibnamefont {Stoecker}}, \ and\
  \bibinfo {author} {\bibfnamefont {L.}~\bibnamefont {Rezzolla}},\ }\href
  {\doibase 10.1140/epja/s10050-020-00073-4} {\bibfield  {journal} {\bibinfo
  {journal} {Eur. Phys. J. A}\ }\textbf {\bibinfo {volume} {56}},\ \bibinfo
  {pages} {59} (\bibinfo {year} {2020})},\ \Eprint
  {http://arxiv.org/abs/1910.13893} {arXiv:1910.13893 [astro-ph.HE]}
  \BibitemShut {NoStop}%
\bibitem [{\citenamefont {Sekiguchi}\ \emph {et~al.}(2011)\citenamefont
  {Sekiguchi}, \citenamefont {Kiuchi}, \citenamefont {Kyutoku},\ and\
  \citenamefont {Shibata}}]{Sekiguchi:2011mc}%
  \BibitemOpen
  \bibfield  {author} {\bibinfo {author} {\bibfnamefont {Y.}~\bibnamefont
  {Sekiguchi}}, \bibinfo {author} {\bibfnamefont {K.}~\bibnamefont {Kiuchi}},
  \bibinfo {author} {\bibfnamefont {K.}~\bibnamefont {Kyutoku}}, \ and\
  \bibinfo {author} {\bibfnamefont {M.}~\bibnamefont {Shibata}},\ }\href
  {\doibase 10.1103/PhysRevLett.107.211101} {\bibfield  {journal} {\bibinfo
  {journal} {Phys. Rev. Lett.}\ }\textbf {\bibinfo {volume} {107}},\ \bibinfo
  {pages} {211101} (\bibinfo {year} {2011})},\ \Eprint
  {http://arxiv.org/abs/1110.4442} {arXiv:1110.4442 [astro-ph.HE]} \BibitemShut
  {NoStop}%
\bibitem [{\citenamefont {Radice}\ \emph {et~al.}(2017)\citenamefont {Radice},
  \citenamefont {Bernuzzi}, \citenamefont {Del~Pozzo}, \citenamefont
  {Roberts},\ and\ \citenamefont {Ott}}]{Radice:2016rys}%
  \BibitemOpen
  \bibfield  {author} {\bibinfo {author} {\bibfnamefont {D.}~\bibnamefont
  {Radice}}, \bibinfo {author} {\bibfnamefont {S.}~\bibnamefont {Bernuzzi}},
  \bibinfo {author} {\bibfnamefont {W.}~\bibnamefont {Del~Pozzo}}, \bibinfo
  {author} {\bibfnamefont {L.~F.}\ \bibnamefont {Roberts}}, \ and\ \bibinfo
  {author} {\bibfnamefont {C.~D.}\ \bibnamefont {Ott}},\ }\href {\doibase
  10.3847/2041-8213/aa775f} {\bibfield  {journal} {\bibinfo  {journal}
  {Astrophys. J. Lett.}\ }\textbf {\bibinfo {volume} {842}},\ \bibinfo {pages}
  {L10} (\bibinfo {year} {2017})},\ \Eprint {http://arxiv.org/abs/1612.06429}
  {arXiv:1612.06429 [astro-ph.HE]} \BibitemShut {NoStop}%
\bibitem [{\citenamefont {Weih}\ \emph {et~al.}(2020)\citenamefont {Weih},
  \citenamefont {Hanauske},\ and\ \citenamefont {Rezzolla}}]{Weih:2019xvw}%
  \BibitemOpen
  \bibfield  {author} {\bibinfo {author} {\bibfnamefont {L.~R.}\ \bibnamefont
  {Weih}}, \bibinfo {author} {\bibfnamefont {M.}~\bibnamefont {Hanauske}}, \
  and\ \bibinfo {author} {\bibfnamefont {L.}~\bibnamefont {Rezzolla}},\ }\href
  {\doibase 10.1103/PhysRevLett.124.171103} {\bibfield  {journal} {\bibinfo
  {journal} {Phys. Rev. Lett.}\ }\textbf {\bibinfo {volume} {124}},\ \bibinfo
  {pages} {171103} (\bibinfo {year} {2020})},\ \Eprint
  {http://arxiv.org/abs/1912.09340} {arXiv:1912.09340 [gr-qc]} \BibitemShut
  {NoStop}%
\bibitem [{\citenamefont {Blacker}\ \emph {et~al.}(2020)\citenamefont
  {Blacker}, \citenamefont {Bastian}, \citenamefont {Bauswein}, \citenamefont
  {Blaschke}, \citenamefont {Fischer}, \citenamefont {Oertel}, \citenamefont
  {Soultanis},\ and\ \citenamefont {Typel}}]{Blacker:2020nlq}%
  \BibitemOpen
  \bibfield  {author} {\bibinfo {author} {\bibfnamefont {S.}~\bibnamefont
  {Blacker}}, \bibinfo {author} {\bibfnamefont {N.-U.~F.}\ \bibnamefont
  {Bastian}}, \bibinfo {author} {\bibfnamefont {A.}~\bibnamefont {Bauswein}},
  \bibinfo {author} {\bibfnamefont {D.~B.}\ \bibnamefont {Blaschke}}, \bibinfo
  {author} {\bibfnamefont {T.}~\bibnamefont {Fischer}}, \bibinfo {author}
  {\bibfnamefont {M.}~\bibnamefont {Oertel}}, \bibinfo {author} {\bibfnamefont
  {T.}~\bibnamefont {Soultanis}}, \ and\ \bibinfo {author} {\bibfnamefont
  {S.}~\bibnamefont {Typel}},\ }\href {\doibase 10.1103/PhysRevD.102.123023}
  {\bibfield  {journal} {\bibinfo  {journal} {Phys. Rev. D}\ }\textbf {\bibinfo
  {volume} {102}},\ \bibinfo {pages} {123023} (\bibinfo {year} {2020})},\
  \Eprint {http://arxiv.org/abs/2006.03789} {arXiv:2006.03789 [astro-ph.HE]}
  \BibitemShut {NoStop}%
\bibitem [{\citenamefont {Liebling}\ \emph {et~al.}(2021)\citenamefont
  {Liebling}, \citenamefont {Palenzuela},\ and\ \citenamefont
  {Lehner}}]{Liebling:2020dhf}%
  \BibitemOpen
  \bibfield  {author} {\bibinfo {author} {\bibfnamefont {S.~L.}\ \bibnamefont
  {Liebling}}, \bibinfo {author} {\bibfnamefont {C.}~\bibnamefont
  {Palenzuela}}, \ and\ \bibinfo {author} {\bibfnamefont {L.}~\bibnamefont
  {Lehner}},\ }\href {\doibase 10.1088/1361-6382/abf898} {\bibfield  {journal}
  {\bibinfo  {journal} {Class. Quant. Grav.}\ }\textbf {\bibinfo {volume}
  {38}},\ \bibinfo {pages} {115007} (\bibinfo {year} {2021})},\ \Eprint
  {http://arxiv.org/abs/2010.12567} {arXiv:2010.12567 [gr-qc]} \BibitemShut
  {NoStop}%
\bibitem [{\citenamefont {Prakash}\ \emph {et~al.}(2021)\citenamefont
  {Prakash}, \citenamefont {Radice}, \citenamefont {Logoteta}, \citenamefont
  {Perego}, \citenamefont {Nedora}, \citenamefont {Bombaci}, \citenamefont
  {Kashyap}, \citenamefont {Bernuzzi},\ and\ \citenamefont
  {Endrizzi}}]{Prakash:2021wpz}%
  \BibitemOpen
  \bibfield  {author} {\bibinfo {author} {\bibfnamefont {A.}~\bibnamefont
  {Prakash}}, \bibinfo {author} {\bibfnamefont {D.}~\bibnamefont {Radice}},
  \bibinfo {author} {\bibfnamefont {D.}~\bibnamefont {Logoteta}}, \bibinfo
  {author} {\bibfnamefont {A.}~\bibnamefont {Perego}}, \bibinfo {author}
  {\bibfnamefont {V.}~\bibnamefont {Nedora}}, \bibinfo {author} {\bibfnamefont
  {I.}~\bibnamefont {Bombaci}}, \bibinfo {author} {\bibfnamefont
  {R.}~\bibnamefont {Kashyap}}, \bibinfo {author} {\bibfnamefont
  {S.}~\bibnamefont {Bernuzzi}}, \ and\ \bibinfo {author} {\bibfnamefont
  {A.}~\bibnamefont {Endrizzi}},\ }\href {\doibase 10.1103/PhysRevD.104.083029}
  {\bibfield  {journal} {\bibinfo  {journal} {Phys. Rev. D}\ }\textbf {\bibinfo
  {volume} {104}},\ \bibinfo {pages} {083029} (\bibinfo {year} {2021})},\
  \Eprint {http://arxiv.org/abs/2106.07885} {arXiv:2106.07885 [astro-ph.HE]}
  \BibitemShut {NoStop}%
\bibitem [{\citenamefont {Abbott}\ \emph
  {et~al.}(2017{\natexlab{b}})\citenamefont {Abbott} \emph
  {et~al.}}]{LIGOScientific:2017ync}%
  \BibitemOpen
  \bibfield  {author} {\bibinfo {author} {\bibfnamefont {B.~P.}\ \bibnamefont
  {Abbott}} \emph {et~al.} (\bibinfo {collaboration} {LIGO Scientific, Virgo,
  Fermi GBM, INTEGRAL, IceCube, AstroSat Cadmium Zinc Telluride Imager Team,
  IPN, Insight-Hxmt, ANTARES, Swift, AGILE Team, 1M2H Team, Dark Energy Camera
  GW-EM, DES, DLT40, GRAWITA, Fermi-LAT, ATCA, ASKAP, Las Cumbres Observatory
  Group, OzGrav, DWF (Deeper Wider Faster Program), AST3, CAASTRO, VINROUGE,
  MASTER, J-GEM, GROWTH, JAGWAR, CaltechNRAO, TTU-NRAO, NuSTAR, Pan-STARRS,
  MAXI Team, TZAC Consortium, KU, Nordic Optical Telescope, ePESSTO, GROND,
  Texas Tech University, SALT Group, TOROS, BOOTES, MWA, CALET, IKI-GW
  Follow-up, H.E.S.S., LOFAR, LWA, HAWC, Pierre Auger, ALMA, Euro VLBI Team, Pi
  of Sky, Chandra Team at McGill University, DFN, ATLAS Telescopes, High Time
  Resolution Universe Survey, RIMAS, RATIR, SKA South Africa/MeerKAT}),\ }\href
  {\doibase 10.3847/2041-8213/aa91c9} {\bibfield  {journal} {\bibinfo
  {journal} {Astrophys. J. Lett.}\ }\textbf {\bibinfo {volume} {848}},\
  \bibinfo {pages} {L12} (\bibinfo {year} {2017}{\natexlab{b}})},\ \Eprint
  {http://arxiv.org/abs/1710.05833} {arXiv:1710.05833 [astro-ph.HE]}
  \BibitemShut {NoStop}%
\bibitem [{\citenamefont {Kasliwal}\ \emph {et~al.}(2017)\citenamefont
  {Kasliwal} \emph {et~al.}}]{Kasliwal:2017ngb}%
  \BibitemOpen
  \bibfield  {author} {\bibinfo {author} {\bibfnamefont {M.~M.}\ \bibnamefont
  {Kasliwal}} \emph {et~al.},\ }\href {\doibase 10.1126/science.aap9455}
  {\bibfield  {journal} {\bibinfo  {journal} {Science}\ }\textbf {\bibinfo
  {volume} {358}},\ \bibinfo {pages} {1559} (\bibinfo {year} {2017})},\ \Eprint
  {http://arxiv.org/abs/1710.05436} {arXiv:1710.05436 [astro-ph.HE]}
  \BibitemShut {NoStop}%
\bibitem [{\citenamefont {Fujimoto}\ \emph
  {et~al.}(2022{\natexlab{b}})\citenamefont {Fujimoto}, \citenamefont
  {Fukushima}, \citenamefont {Hotokezaka},\ and\ \citenamefont
  {Kyutoku}}]{Fujimoto:2022xhv}%
  \BibitemOpen
  \bibfield  {author} {\bibinfo {author} {\bibfnamefont {Y.}~\bibnamefont
  {Fujimoto}}, \bibinfo {author} {\bibfnamefont {K.}~\bibnamefont {Fukushima}},
  \bibinfo {author} {\bibfnamefont {K.}~\bibnamefont {Hotokezaka}}, \ and\
  \bibinfo {author} {\bibfnamefont {K.}~\bibnamefont {Kyutoku}},\ }\href@noop
  {} {\  (\bibinfo {year} {2022}{\natexlab{b}})},\ \Eprint
  {http://arxiv.org/abs/2205.03882} {arXiv:2205.03882 [astro-ph.HE]}
  \BibitemShut {NoStop}%
\bibitem [{\citenamefont {Bose}\ \emph {et~al.}(2018)\citenamefont {Bose},
  \citenamefont {Chakravarti}, \citenamefont {Rezzolla}, \citenamefont
  {Sathyaprakash},\ and\ \citenamefont {Takami}}]{Bose:2017jvk}%
  \BibitemOpen
  \bibfield  {author} {\bibinfo {author} {\bibfnamefont {S.}~\bibnamefont
  {Bose}}, \bibinfo {author} {\bibfnamefont {K.}~\bibnamefont {Chakravarti}},
  \bibinfo {author} {\bibfnamefont {L.}~\bibnamefont {Rezzolla}}, \bibinfo
  {author} {\bibfnamefont {B.~S.}\ \bibnamefont {Sathyaprakash}}, \ and\
  \bibinfo {author} {\bibfnamefont {K.}~\bibnamefont {Takami}},\ }\href
  {\doibase 10.1103/PhysRevLett.120.031102} {\bibfield  {journal} {\bibinfo
  {journal} {Phys. Rev. Lett.}\ }\textbf {\bibinfo {volume} {120}},\ \bibinfo
  {pages} {031102} (\bibinfo {year} {2018})},\ \Eprint
  {http://arxiv.org/abs/1705.10850} {arXiv:1705.10850 [gr-qc]} \BibitemShut
  {NoStop}%
\bibitem [{\citenamefont {Breschi}\ \emph
  {et~al.}(2022{\natexlab{a}})\citenamefont {Breschi}, \citenamefont {Gamba},
  \citenamefont {Borhanian}, \citenamefont {Carullo},\ and\ \citenamefont
  {Bernuzzi}}]{Breschi:2022ens}%
  \BibitemOpen
  \bibfield  {author} {\bibinfo {author} {\bibfnamefont {M.}~\bibnamefont
  {Breschi}}, \bibinfo {author} {\bibfnamefont {R.}~\bibnamefont {Gamba}},
  \bibinfo {author} {\bibfnamefont {S.}~\bibnamefont {Borhanian}}, \bibinfo
  {author} {\bibfnamefont {G.}~\bibnamefont {Carullo}}, \ and\ \bibinfo
  {author} {\bibfnamefont {S.}~\bibnamefont {Bernuzzi}},\ }\href@noop {} {\
  (\bibinfo {year} {2022}{\natexlab{a}})},\ \Eprint
  {http://arxiv.org/abs/2205.09979} {arXiv:2205.09979 [gr-qc]} \BibitemShut
  {NoStop}%
\bibitem [{\citenamefont {Breschi}\ \emph
  {et~al.}(2022{\natexlab{b}})\citenamefont {Breschi}, \citenamefont
  {Bernuzzi}, \citenamefont {Chakravarti}, \citenamefont {Camilletti},
  \citenamefont {Prakash},\ and\ \citenamefont {Perego}}]{Breschi:2022xnc}%
  \BibitemOpen
  \bibfield  {author} {\bibinfo {author} {\bibfnamefont {M.}~\bibnamefont
  {Breschi}}, \bibinfo {author} {\bibfnamefont {S.}~\bibnamefont {Bernuzzi}},
  \bibinfo {author} {\bibfnamefont {K.}~\bibnamefont {Chakravarti}}, \bibinfo
  {author} {\bibfnamefont {A.}~\bibnamefont {Camilletti}}, \bibinfo {author}
  {\bibfnamefont {A.}~\bibnamefont {Prakash}}, \ and\ \bibinfo {author}
  {\bibfnamefont {A.}~\bibnamefont {Perego}},\ }\href@noop {} {\  (\bibinfo
  {year} {2022}{\natexlab{b}})},\ \Eprint {http://arxiv.org/abs/2205.09112}
  {arXiv:2205.09112 [gr-qc]} \BibitemShut {NoStop}%
\bibitem [{\citenamefont {Wijngaarden}\ \emph {et~al.}(2022)\citenamefont
  {Wijngaarden}, \citenamefont {Chatziioannou}, \citenamefont {Bauswein},
  \citenamefont {Clark},\ and\ \citenamefont {Cornish}}]{Wijngaarden:2022sah}%
  \BibitemOpen
  \bibfield  {author} {\bibinfo {author} {\bibfnamefont {M.}~\bibnamefont
  {Wijngaarden}}, \bibinfo {author} {\bibfnamefont {K.}~\bibnamefont
  {Chatziioannou}}, \bibinfo {author} {\bibfnamefont {A.}~\bibnamefont
  {Bauswein}}, \bibinfo {author} {\bibfnamefont {J.~A.}\ \bibnamefont {Clark}},
  \ and\ \bibinfo {author} {\bibfnamefont {N.~J.}\ \bibnamefont {Cornish}},\
  }\href {\doibase 10.1103/PhysRevD.105.104019} {\bibfield  {journal} {\bibinfo
   {journal} {Phys. Rev. D}\ }\textbf {\bibinfo {volume} {105}},\ \bibinfo
  {pages} {104019} (\bibinfo {year} {2022})},\ \Eprint
  {http://arxiv.org/abs/2202.09382} {arXiv:2202.09382 [gr-qc]} \BibitemShut
  {NoStop}%
\bibitem [{\citenamefont {Puecher}\ \emph {et~al.}(2022)\citenamefont
  {Puecher}, \citenamefont {Dietrich}, \citenamefont {Tsang}, \citenamefont
  {Kalaghatgi}, \citenamefont {Roy}, \citenamefont {Setyawati},\ and\
  \citenamefont {Van Den~Broeck}}]{Puecher:2022oiz}%
  \BibitemOpen
  \bibfield  {author} {\bibinfo {author} {\bibfnamefont {A.}~\bibnamefont
  {Puecher}}, \bibinfo {author} {\bibfnamefont {T.}~\bibnamefont {Dietrich}},
  \bibinfo {author} {\bibfnamefont {K.~W.}\ \bibnamefont {Tsang}}, \bibinfo
  {author} {\bibfnamefont {C.}~\bibnamefont {Kalaghatgi}}, \bibinfo {author}
  {\bibfnamefont {S.}~\bibnamefont {Roy}}, \bibinfo {author} {\bibfnamefont
  {Y.}~\bibnamefont {Setyawati}}, \ and\ \bibinfo {author} {\bibfnamefont
  {C.}~\bibnamefont {Van Den~Broeck}},\ }\href@noop {} {\  (\bibinfo {year}
  {2022})},\ \Eprint {http://arxiv.org/abs/2210.09259} {arXiv:2210.09259
  [gr-qc]} \BibitemShut {NoStop}%
\bibitem [{\citenamefont {Bauswein}\ and\ \citenamefont
  {Janka}(2012)}]{Bauswein:2011tp}%
  \BibitemOpen
  \bibfield  {author} {\bibinfo {author} {\bibfnamefont {A.}~\bibnamefont
  {Bauswein}}\ and\ \bibinfo {author} {\bibfnamefont {H.~T.}\ \bibnamefont
  {Janka}},\ }\href {\doibase 10.1103/PhysRevLett.108.011101} {\bibfield
  {journal} {\bibinfo  {journal} {Phys. Rev. Lett.}\ }\textbf {\bibinfo
  {volume} {108}},\ \bibinfo {pages} {011101} (\bibinfo {year} {2012})},\
  \Eprint {http://arxiv.org/abs/1106.1616} {arXiv:1106.1616 [astro-ph.SR]}
  \BibitemShut {NoStop}%
\bibitem [{\citenamefont {Bernuzzi}\ \emph {et~al.}(2012)\citenamefont
  {Bernuzzi}, \citenamefont {Nagar}, \citenamefont {Thierfelder},\ and\
  \citenamefont {Brugmann}}]{Bernuzzi:2012ci}%
  \BibitemOpen
  \bibfield  {author} {\bibinfo {author} {\bibfnamefont {S.}~\bibnamefont
  {Bernuzzi}}, \bibinfo {author} {\bibfnamefont {A.}~\bibnamefont {Nagar}},
  \bibinfo {author} {\bibfnamefont {M.}~\bibnamefont {Thierfelder}}, \ and\
  \bibinfo {author} {\bibfnamefont {B.}~\bibnamefont {Brugmann}},\ }\href
  {\doibase 10.1103/PhysRevD.86.044030} {\bibfield  {journal} {\bibinfo
  {journal} {Phys. Rev. D}\ }\textbf {\bibinfo {volume} {86}},\ \bibinfo
  {pages} {044030} (\bibinfo {year} {2012})},\ \Eprint
  {http://arxiv.org/abs/1205.3403} {arXiv:1205.3403 [gr-qc]} \BibitemShut
  {NoStop}%
\bibitem [{\citenamefont {Takami}\ \emph {et~al.}(2014)\citenamefont {Takami},
  \citenamefont {Rezzolla},\ and\ \citenamefont {Baiotti}}]{Takami:2014zpa}%
  \BibitemOpen
  \bibfield  {author} {\bibinfo {author} {\bibfnamefont {K.}~\bibnamefont
  {Takami}}, \bibinfo {author} {\bibfnamefont {L.}~\bibnamefont {Rezzolla}}, \
  and\ \bibinfo {author} {\bibfnamefont {L.}~\bibnamefont {Baiotti}},\ }\href
  {\doibase 10.1103/PhysRevLett.113.091104} {\bibfield  {journal} {\bibinfo
  {journal} {Phys. Rev. Lett.}\ }\textbf {\bibinfo {volume} {113}},\ \bibinfo
  {pages} {091104} (\bibinfo {year} {2014})},\ \Eprint
  {http://arxiv.org/abs/1403.5672} {arXiv:1403.5672 [gr-qc]} \BibitemShut
  {NoStop}%
\bibitem [{\citenamefont {Raithel}\ and\ \citenamefont
  {Most}(2022)}]{Raithel:2022orm}%
  \BibitemOpen
  \bibfield  {author} {\bibinfo {author} {\bibfnamefont {C.~A.}\ \bibnamefont
  {Raithel}}\ and\ \bibinfo {author} {\bibfnamefont {E.~R.}\ \bibnamefont
  {Most}},\ }\href {\doibase 10.3847/2041-8213/ac7c75} {\bibfield  {journal}
  {\bibinfo  {journal} {Astrophys. J. Lett.}\ }\textbf {\bibinfo {volume}
  {933}},\ \bibinfo {pages} {L39} (\bibinfo {year} {2022})},\ \Eprint
  {http://arxiv.org/abs/2201.03594} {arXiv:2201.03594 [astro-ph.HE]}
  \BibitemShut {NoStop}%
\bibitem [{\citenamefont {Lattimer}\ \emph {et~al.}(1977)\citenamefont
  {Lattimer}, \citenamefont {Mackie}, \citenamefont {Ravenhall},\ and\
  \citenamefont {Schramm}}]{Lattimer:1977igd}%
  \BibitemOpen
  \bibfield  {author} {\bibinfo {author} {\bibfnamefont {J.~M.}\ \bibnamefont
  {Lattimer}}, \bibinfo {author} {\bibfnamefont {F.}~\bibnamefont {Mackie}},
  \bibinfo {author} {\bibfnamefont {D.~G.}\ \bibnamefont {Ravenhall}}, \ and\
  \bibinfo {author} {\bibfnamefont {D.~N.}\ \bibnamefont {Schramm}},\ }\href
  {\doibase 10.1086/155148} {\bibfield  {journal} {\bibinfo  {journal}
  {Astrophys. J.}\ }\textbf {\bibinfo {volume} {213}},\ \bibinfo {pages} {225}
  (\bibinfo {year} {1977})}\BibitemShut {NoStop}%
\bibitem [{\citenamefont {Metzger}(2020)}]{Metzger:2019zeh}%
  \BibitemOpen
  \bibfield  {author} {\bibinfo {author} {\bibfnamefont {B.~D.}\ \bibnamefont
  {Metzger}},\ }\href {\doibase 10.1007/s41114-019-0024-0} {\bibfield
  {journal} {\bibinfo  {journal} {Living Rev. Rel.}\ }\textbf {\bibinfo
  {volume} {23}},\ \bibinfo {pages} {1} (\bibinfo {year} {2020})},\ \Eprint
  {http://arxiv.org/abs/1910.01617} {arXiv:1910.01617 [astro-ph.HE]}
  \BibitemShut {NoStop}%
\bibitem [{\citenamefont {Radice}\ \emph {et~al.}(2018)\citenamefont {Radice},
  \citenamefont {Perego}, \citenamefont {Zappa},\ and\ \citenamefont
  {Bernuzzi}}]{Radice:2017lry}%
  \BibitemOpen
  \bibfield  {author} {\bibinfo {author} {\bibfnamefont {D.}~\bibnamefont
  {Radice}}, \bibinfo {author} {\bibfnamefont {A.}~\bibnamefont {Perego}},
  \bibinfo {author} {\bibfnamefont {F.}~\bibnamefont {Zappa}}, \ and\ \bibinfo
  {author} {\bibfnamefont {S.}~\bibnamefont {Bernuzzi}},\ }\href {\doibase
  10.3847/2041-8213/aaa402} {\bibfield  {journal} {\bibinfo  {journal}
  {Astrophys. J. Lett.}\ }\textbf {\bibinfo {volume} {852}},\ \bibinfo {pages}
  {L29} (\bibinfo {year} {2018})},\ \Eprint {http://arxiv.org/abs/1711.03647}
  {arXiv:1711.03647 [astro-ph.HE]} \BibitemShut {NoStop}%
\bibitem [{\citenamefont {Coughlin}\ \emph {et~al.}(2018)\citenamefont
  {Coughlin} \emph {et~al.}}]{Coughlin:2018miv}%
  \BibitemOpen
  \bibfield  {author} {\bibinfo {author} {\bibfnamefont {M.~W.}\ \bibnamefont
  {Coughlin}} \emph {et~al.},\ }\href {\doibase 10.1093/mnras/sty2174}
  {\bibfield  {journal} {\bibinfo  {journal} {Mon. Not. Roy. Astron. Soc.}\
  }\textbf {\bibinfo {volume} {480}},\ \bibinfo {pages} {3871} (\bibinfo {year}
  {2018})},\ \Eprint {http://arxiv.org/abs/1805.09371} {arXiv:1805.09371
  [astro-ph.HE]} \BibitemShut {NoStop}%
\bibitem [{\citenamefont {Kiuchi}\ \emph {et~al.}(2019)\citenamefont {Kiuchi},
  \citenamefont {Kyutoku}, \citenamefont {Shibata},\ and\ \citenamefont
  {Taniguchi}}]{Kiuchi:2019lls}%
  \BibitemOpen
  \bibfield  {author} {\bibinfo {author} {\bibfnamefont {K.}~\bibnamefont
  {Kiuchi}}, \bibinfo {author} {\bibfnamefont {K.}~\bibnamefont {Kyutoku}},
  \bibinfo {author} {\bibfnamefont {M.}~\bibnamefont {Shibata}}, \ and\
  \bibinfo {author} {\bibfnamefont {K.}~\bibnamefont {Taniguchi}},\ }\href
  {\doibase 10.3847/2041-8213/ab1e45} {\bibfield  {journal} {\bibinfo
  {journal} {Astrophys. J. Lett.}\ }\textbf {\bibinfo {volume} {876}},\
  \bibinfo {pages} {L31} (\bibinfo {year} {2019})},\ \Eprint
  {http://arxiv.org/abs/1903.01466} {arXiv:1903.01466 [astro-ph.HE]}
  \BibitemShut {NoStop}%
\bibitem [{\citenamefont {Breu}\ and\ \citenamefont
  {Rezzolla}(2016)}]{Breu:2016ufb}%
  \BibitemOpen
  \bibfield  {author} {\bibinfo {author} {\bibfnamefont {C.}~\bibnamefont
  {Breu}}\ and\ \bibinfo {author} {\bibfnamefont {L.}~\bibnamefont
  {Rezzolla}},\ }\href {\doibase 10.1093/mnras/stw575} {\bibfield  {journal}
  {\bibinfo  {journal} {Mon. Not. Roy. Astron. Soc.}\ }\textbf {\bibinfo
  {volume} {459}},\ \bibinfo {pages} {646} (\bibinfo {year} {2016})},\ \Eprint
  {http://arxiv.org/abs/1601.06083} {arXiv:1601.06083 [gr-qc]} \BibitemShut
  {NoStop}%
\bibitem [{\citenamefont {Bozzola}\ \emph {et~al.}(2019)\citenamefont
  {Bozzola}, \citenamefont {Espino}, \citenamefont {Lewin},\ and\ \citenamefont
  {Paschalidis}}]{Bozzola:2019tit}%
  \BibitemOpen
  \bibfield  {author} {\bibinfo {author} {\bibfnamefont {G.}~\bibnamefont
  {Bozzola}}, \bibinfo {author} {\bibfnamefont {P.~L.}\ \bibnamefont {Espino}},
  \bibinfo {author} {\bibfnamefont {C.~D.}\ \bibnamefont {Lewin}}, \ and\
  \bibinfo {author} {\bibfnamefont {V.}~\bibnamefont {Paschalidis}},\ }\href
  {\doibase 10.1140/epja/i2019-12831-2} {\bibfield  {journal} {\bibinfo
  {journal} {Eur. Phys. J. A}\ }\textbf {\bibinfo {volume} {55}},\ \bibinfo
  {pages} {149} (\bibinfo {year} {2019})},\ \Eprint
  {http://arxiv.org/abs/1905.00028} {arXiv:1905.00028 [astro-ph.HE]}
  \BibitemShut {NoStop}%
\bibitem [{\citenamefont {Koliogiannis}\ and\ \citenamefont
  {Moustakidis}(2020)}]{Koliogiannis:2019rvh}%
  \BibitemOpen
  \bibfield  {author} {\bibinfo {author} {\bibfnamefont {P.~S.}\ \bibnamefont
  {Koliogiannis}}\ and\ \bibinfo {author} {\bibfnamefont {C.~C.}\ \bibnamefont
  {Moustakidis}},\ }\href {\doibase 10.1103/PhysRevC.101.015805} {\bibfield
  {journal} {\bibinfo  {journal} {Phys. Rev. C}\ }\textbf {\bibinfo {volume}
  {101}},\ \bibinfo {pages} {015805} (\bibinfo {year} {2020})},\ \Eprint
  {http://arxiv.org/abs/1907.13375} {arXiv:1907.13375 [nucl-th]} \BibitemShut
  {NoStop}%
\bibitem [{\citenamefont {Annala}\ \emph {et~al.}(2022)\citenamefont {Annala},
  \citenamefont {Gorda}, \citenamefont {Katerini}, \citenamefont {Kurkela},
  \citenamefont {N\"attil\"a}, \citenamefont {Paschalidis},\ and\ \citenamefont
  {Vuorinen}}]{Annala:2021gom}%
  \BibitemOpen
  \bibfield  {author} {\bibinfo {author} {\bibfnamefont {E.}~\bibnamefont
  {Annala}}, \bibinfo {author} {\bibfnamefont {T.}~\bibnamefont {Gorda}},
  \bibinfo {author} {\bibfnamefont {E.}~\bibnamefont {Katerini}}, \bibinfo
  {author} {\bibfnamefont {A.}~\bibnamefont {Kurkela}}, \bibinfo {author}
  {\bibfnamefont {J.}~\bibnamefont {N\"attil\"a}}, \bibinfo {author}
  {\bibfnamefont {V.}~\bibnamefont {Paschalidis}}, \ and\ \bibinfo {author}
  {\bibfnamefont {A.}~\bibnamefont {Vuorinen}},\ }\href {\doibase
  10.1103/PhysRevX.12.011058} {\bibfield  {journal} {\bibinfo  {journal} {Phys.
  Rev. X}\ }\textbf {\bibinfo {volume} {12}},\ \bibinfo {pages} {011058}
  (\bibinfo {year} {2022})},\ \Eprint {http://arxiv.org/abs/2105.05132}
  {arXiv:2105.05132 [astro-ph.HE]} \BibitemShut {NoStop}%
\bibitem [{\citenamefont {Baiotti}(2019)}]{Baiotti:2019sew}%
  \BibitemOpen
  \bibfield  {author} {\bibinfo {author} {\bibfnamefont {L.}~\bibnamefont
  {Baiotti}},\ }\href {\doibase 10.1016/j.ppnp.2019.103714} {\bibfield
  {journal} {\bibinfo  {journal} {Prog. Part. Nucl. Phys.}\ }\textbf {\bibinfo
  {volume} {109}},\ \bibinfo {pages} {103714} (\bibinfo {year} {2019})},\
  \Eprint {http://arxiv.org/abs/1907.08534} {arXiv:1907.08534 [astro-ph.HE]}
  \BibitemShut {NoStop}%
\bibitem [{\citenamefont {Ruiz}\ and\ \citenamefont
  {Shapiro}(2017)}]{Ruiz:2017inq}%
  \BibitemOpen
  \bibfield  {author} {\bibinfo {author} {\bibfnamefont {M.}~\bibnamefont
  {Ruiz}}\ and\ \bibinfo {author} {\bibfnamefont {S.~L.}\ \bibnamefont
  {Shapiro}},\ }\href {\doibase 10.1103/PhysRevD.96.084063} {\bibfield
  {journal} {\bibinfo  {journal} {Phys. Rev. D}\ }\textbf {\bibinfo {volume}
  {96}},\ \bibinfo {pages} {084063} (\bibinfo {year} {2017})},\ \Eprint
  {http://arxiv.org/abs/1709.00414} {arXiv:1709.00414 [astro-ph.HE]}
  \BibitemShut {NoStop}%
\bibitem [{\citenamefont {Bernuzzi}\ \emph {et~al.}(2020)\citenamefont
  {Bernuzzi} \emph {et~al.}}]{Bernuzzi:2020txg}%
  \BibitemOpen
  \bibfield  {author} {\bibinfo {author} {\bibfnamefont {S.}~\bibnamefont
  {Bernuzzi}} \emph {et~al.},\ }\href {\doibase 10.1093/mnras/staa1860}
  {\bibfield  {journal} {\bibinfo  {journal} {Mon. Not. Roy. Astron. Soc.}\
  }\textbf {\bibinfo {volume} {497}},\ \bibinfo {pages} {1488} (\bibinfo {year}
  {2020})},\ \Eprint {http://arxiv.org/abs/2003.06015} {arXiv:2003.06015
  [astro-ph.HE]} \BibitemShut {NoStop}%
\bibitem [{\citenamefont {Bauswein}\ \emph {et~al.}(2013)\citenamefont
  {Bauswein}, \citenamefont {Baumgarte},\ and\ \citenamefont
  {Janka}}]{Bauswein:2013jpa}%
  \BibitemOpen
  \bibfield  {author} {\bibinfo {author} {\bibfnamefont {A.}~\bibnamefont
  {Bauswein}}, \bibinfo {author} {\bibfnamefont {T.~W.}\ \bibnamefont
  {Baumgarte}}, \ and\ \bibinfo {author} {\bibfnamefont {H.~T.}\ \bibnamefont
  {Janka}},\ }\href {\doibase 10.1103/PhysRevLett.111.131101} {\bibfield
  {journal} {\bibinfo  {journal} {Phys. Rev. Lett.}\ }\textbf {\bibinfo
  {volume} {111}},\ \bibinfo {pages} {131101} (\bibinfo {year} {2013})},\
  \Eprint {http://arxiv.org/abs/1307.5191} {arXiv:1307.5191 [astro-ph.SR]}
  \BibitemShut {NoStop}%
\bibitem [{\citenamefont {K\"oppel}\ \emph {et~al.}(2019)\citenamefont
  {K\"oppel}, \citenamefont {Bovard},\ and\ \citenamefont
  {Rezzolla}}]{Koppel:2019pys}%
  \BibitemOpen
  \bibfield  {author} {\bibinfo {author} {\bibfnamefont {S.}~\bibnamefont
  {K\"oppel}}, \bibinfo {author} {\bibfnamefont {L.}~\bibnamefont {Bovard}}, \
  and\ \bibinfo {author} {\bibfnamefont {L.}~\bibnamefont {Rezzolla}},\ }\href
  {\doibase 10.3847/2041-8213/ab0210} {\bibfield  {journal} {\bibinfo
  {journal} {Astrophys. J. Lett.}\ }\textbf {\bibinfo {volume} {872}},\
  \bibinfo {pages} {L16} (\bibinfo {year} {2019})},\ \Eprint
  {http://arxiv.org/abs/1901.09977} {arXiv:1901.09977 [gr-qc]} \BibitemShut
  {NoStop}%
\bibitem [{\citenamefont {Bauswein}\ \emph {et~al.}(2020)\citenamefont
  {Bauswein}, \citenamefont {Blacker}, \citenamefont {Vijayan}, \citenamefont
  {Stergioulas}, \citenamefont {Chatziioannou}, \citenamefont {Clark},
  \citenamefont {Bastian}, \citenamefont {Blaschke}, \citenamefont {Cierniak},\
  and\ \citenamefont {Fischer}}]{Bauswein:2020aag}%
  \BibitemOpen
  \bibfield  {author} {\bibinfo {author} {\bibfnamefont {A.}~\bibnamefont
  {Bauswein}}, \bibinfo {author} {\bibfnamefont {S.}~\bibnamefont {Blacker}},
  \bibinfo {author} {\bibfnamefont {V.}~\bibnamefont {Vijayan}}, \bibinfo
  {author} {\bibfnamefont {N.}~\bibnamefont {Stergioulas}}, \bibinfo {author}
  {\bibfnamefont {K.}~\bibnamefont {Chatziioannou}}, \bibinfo {author}
  {\bibfnamefont {J.~A.}\ \bibnamefont {Clark}}, \bibinfo {author}
  {\bibfnamefont {N.-U.~F.}\ \bibnamefont {Bastian}}, \bibinfo {author}
  {\bibfnamefont {D.~B.}\ \bibnamefont {Blaschke}}, \bibinfo {author}
  {\bibfnamefont {M.}~\bibnamefont {Cierniak}}, \ and\ \bibinfo {author}
  {\bibfnamefont {T.}~\bibnamefont {Fischer}},\ }\href {\doibase
  10.1103/PhysRevLett.125.141103} {\bibfield  {journal} {\bibinfo  {journal}
  {Phys. Rev. Lett.}\ }\textbf {\bibinfo {volume} {125}},\ \bibinfo {pages}
  {141103} (\bibinfo {year} {2020})},\ \Eprint
  {http://arxiv.org/abs/2004.00846} {arXiv:2004.00846 [astro-ph.HE]}
  \BibitemShut {NoStop}%
\bibitem [{\citenamefont {Tootle}\ \emph {et~al.}(2021)\citenamefont {Tootle},
  \citenamefont {Papenfort}, \citenamefont {Most},\ and\ \citenamefont
  {Rezzolla}}]{Tootle:2021umi}%
  \BibitemOpen
  \bibfield  {author} {\bibinfo {author} {\bibfnamefont {S.~D.}\ \bibnamefont
  {Tootle}}, \bibinfo {author} {\bibfnamefont {L.~J.}\ \bibnamefont
  {Papenfort}}, \bibinfo {author} {\bibfnamefont {E.~R.}\ \bibnamefont {Most}},
  \ and\ \bibinfo {author} {\bibfnamefont {L.}~\bibnamefont {Rezzolla}},\
  }\href {\doibase 10.3847/2041-8213/ac350d} {\bibfield  {journal} {\bibinfo
  {journal} {Astrophys. J. Lett.}\ }\textbf {\bibinfo {volume} {922}},\
  \bibinfo {pages} {L19} (\bibinfo {year} {2021})},\ \Eprint
  {http://arxiv.org/abs/2109.00940} {arXiv:2109.00940 [gr-qc]} \BibitemShut
  {NoStop}%
\bibitem [{\citenamefont {Kashyap}\ \emph {et~al.}(2022)\citenamefont {Kashyap}
  \emph {et~al.}}]{Kashyap:2021wzs}%
  \BibitemOpen
  \bibfield  {author} {\bibinfo {author} {\bibfnamefont {R.}~\bibnamefont
  {Kashyap}} \emph {et~al.},\ }\href {\doibase 10.1103/PhysRevD.105.103022}
  {\bibfield  {journal} {\bibinfo  {journal} {Phys. Rev. D}\ }\textbf {\bibinfo
  {volume} {105}},\ \bibinfo {pages} {103022} (\bibinfo {year} {2022})},\
  \Eprint {http://arxiv.org/abs/2111.05183} {arXiv:2111.05183 [astro-ph.HE]}
  \BibitemShut {NoStop}%
\bibitem [{\citenamefont {K\"olsch}\ \emph {et~al.}(2022)\citenamefont
  {K\"olsch}, \citenamefont {Dietrich}, \citenamefont {Ujevic},\ and\
  \citenamefont {Bruegmann}}]{Kolsch:2021lub}%
  \BibitemOpen
  \bibfield  {author} {\bibinfo {author} {\bibfnamefont {M.}~\bibnamefont
  {K\"olsch}}, \bibinfo {author} {\bibfnamefont {T.}~\bibnamefont {Dietrich}},
  \bibinfo {author} {\bibfnamefont {M.}~\bibnamefont {Ujevic}}, \ and\ \bibinfo
  {author} {\bibfnamefont {B.}~\bibnamefont {Bruegmann}},\ }\href {\doibase
  10.1103/PhysRevD.106.044026} {\bibfield  {journal} {\bibinfo  {journal}
  {Phys. Rev. D}\ }\textbf {\bibinfo {volume} {106}},\ \bibinfo {pages}
  {044026} (\bibinfo {year} {2022})},\ \Eprint
  {http://arxiv.org/abs/2112.11851} {arXiv:2112.11851 [gr-qc]} \BibitemShut
  {NoStop}%
\bibitem [{\citenamefont {Perego}\ \emph {et~al.}(2022)\citenamefont {Perego},
  \citenamefont {Logoteta}, \citenamefont {Radice}, \citenamefont {Bernuzzi},
  \citenamefont {Kashyap}, \citenamefont {Das}, \citenamefont {Padamata},\ and\
  \citenamefont {Prakash}}]{Perego:2021mkd}%
  \BibitemOpen
  \bibfield  {author} {\bibinfo {author} {\bibfnamefont {A.}~\bibnamefont
  {Perego}}, \bibinfo {author} {\bibfnamefont {D.}~\bibnamefont {Logoteta}},
  \bibinfo {author} {\bibfnamefont {D.}~\bibnamefont {Radice}}, \bibinfo
  {author} {\bibfnamefont {S.}~\bibnamefont {Bernuzzi}}, \bibinfo {author}
  {\bibfnamefont {R.}~\bibnamefont {Kashyap}}, \bibinfo {author} {\bibfnamefont
  {A.}~\bibnamefont {Das}}, \bibinfo {author} {\bibfnamefont {S.}~\bibnamefont
  {Padamata}}, \ and\ \bibinfo {author} {\bibfnamefont {A.}~\bibnamefont
  {Prakash}},\ }\href {\doibase 10.1103/PhysRevLett.129.032701} {\bibfield
  {journal} {\bibinfo  {journal} {Phys. Rev. Lett.}\ }\textbf {\bibinfo
  {volume} {129}},\ \bibinfo {pages} {032701} (\bibinfo {year} {2022})},\
  \Eprint {http://arxiv.org/abs/2112.05864} {arXiv:2112.05864 [astro-ph.HE]}
  \BibitemShut {NoStop}%
\bibitem [{\citenamefont {Ruiz}\ \emph {et~al.}(2018)\citenamefont {Ruiz},
  \citenamefont {Shapiro},\ and\ \citenamefont {Tsokaros}}]{Ruiz:2017due}%
  \BibitemOpen
  \bibfield  {author} {\bibinfo {author} {\bibfnamefont {M.}~\bibnamefont
  {Ruiz}}, \bibinfo {author} {\bibfnamefont {S.~L.}\ \bibnamefont {Shapiro}}, \
  and\ \bibinfo {author} {\bibfnamefont {A.}~\bibnamefont {Tsokaros}},\ }\href
  {\doibase 10.1103/PhysRevD.97.021501} {\bibfield  {journal} {\bibinfo
  {journal} {Phys. Rev. D}\ }\textbf {\bibinfo {volume} {97}},\ \bibinfo
  {pages} {021501} (\bibinfo {year} {2018})},\ \Eprint
  {http://arxiv.org/abs/1711.00473} {arXiv:1711.00473 [astro-ph.HE]}
  \BibitemShut {NoStop}%
\bibitem [{\citenamefont {Ellis}\ \emph
  {et~al.}(2018{\natexlab{a}})\citenamefont {Ellis}, \citenamefont {H\"utsi},
  \citenamefont {Kannike}, \citenamefont {Marzola}, \citenamefont {Raidal},\
  and\ \citenamefont {Vaskonen}}]{Ellis:2018bkr}%
  \BibitemOpen
  \bibfield  {author} {\bibinfo {author} {\bibfnamefont {J.}~\bibnamefont
  {Ellis}}, \bibinfo {author} {\bibfnamefont {G.}~\bibnamefont {H\"utsi}},
  \bibinfo {author} {\bibfnamefont {K.}~\bibnamefont {Kannike}}, \bibinfo
  {author} {\bibfnamefont {L.}~\bibnamefont {Marzola}}, \bibinfo {author}
  {\bibfnamefont {M.}~\bibnamefont {Raidal}}, \ and\ \bibinfo {author}
  {\bibfnamefont {V.}~\bibnamefont {Vaskonen}},\ }\href {\doibase
  10.1103/PhysRevD.97.123007} {\bibfield  {journal} {\bibinfo  {journal} {Phys.
  Rev. D}\ }\textbf {\bibinfo {volume} {97}},\ \bibinfo {pages} {123007}
  (\bibinfo {year} {2018}{\natexlab{a}})},\ \Eprint
  {http://arxiv.org/abs/1804.01418} {arXiv:1804.01418 [astro-ph.CO]}
  \BibitemShut {NoStop}%
\bibitem [{\citenamefont {Deliyergiyev}\ \emph {et~al.}(2019)\citenamefont
  {Deliyergiyev}, \citenamefont {Del~Popolo}, \citenamefont {Tolos},
  \citenamefont {Le~Delliou}, \citenamefont {Lee},\ and\ \citenamefont
  {Burgio}}]{Deliyergiyev:2019vti}%
  \BibitemOpen
  \bibfield  {author} {\bibinfo {author} {\bibfnamefont {M.}~\bibnamefont
  {Deliyergiyev}}, \bibinfo {author} {\bibfnamefont {A.}~\bibnamefont
  {Del~Popolo}}, \bibinfo {author} {\bibfnamefont {L.}~\bibnamefont {Tolos}},
  \bibinfo {author} {\bibfnamefont {M.}~\bibnamefont {Le~Delliou}}, \bibinfo
  {author} {\bibfnamefont {X.}~\bibnamefont {Lee}}, \ and\ \bibinfo {author}
  {\bibfnamefont {F.}~\bibnamefont {Burgio}},\ }\href {\doibase
  10.1103/PhysRevD.99.063015} {\bibfield  {journal} {\bibinfo  {journal} {Phys.
  Rev. D}\ }\textbf {\bibinfo {volume} {99}},\ \bibinfo {pages} {063015}
  (\bibinfo {year} {2019})},\ \Eprint {http://arxiv.org/abs/1903.01183}
  {arXiv:1903.01183 [gr-qc]} \BibitemShut {NoStop}%
\bibitem [{\citenamefont {Kain}(2021)}]{Kain:2021hpk}%
  \BibitemOpen
  \bibfield  {author} {\bibinfo {author} {\bibfnamefont {B.}~\bibnamefont
  {Kain}},\ }\href {\doibase 10.1103/PhysRevD.103.043009} {\bibfield  {journal}
  {\bibinfo  {journal} {Phys. Rev. D}\ }\textbf {\bibinfo {volume} {103}},\
  \bibinfo {pages} {043009} (\bibinfo {year} {2021})},\ \Eprint
  {http://arxiv.org/abs/2102.08257} {arXiv:2102.08257 [gr-qc]} \BibitemShut
  {NoStop}%
\bibitem [{\citenamefont {Rutherford}\ \emph {et~al.}(2022)\citenamefont
  {Rutherford}, \citenamefont {Raaijmakers}, \citenamefont
  {Prescod-Weinstein},\ and\ \citenamefont {Watts}}]{Rutherford:2022xeb}%
  \BibitemOpen
  \bibfield  {author} {\bibinfo {author} {\bibfnamefont {N.}~\bibnamefont
  {Rutherford}}, \bibinfo {author} {\bibfnamefont {G.}~\bibnamefont
  {Raaijmakers}}, \bibinfo {author} {\bibfnamefont {C.}~\bibnamefont
  {Prescod-Weinstein}}, \ and\ \bibinfo {author} {\bibfnamefont
  {A.}~\bibnamefont {Watts}},\ }\href@noop {} {\  (\bibinfo {year} {2022})},\
  \Eprint {http://arxiv.org/abs/2208.03282} {arXiv:2208.03282 [astro-ph.HE]}
  \BibitemShut {NoStop}%
\bibitem [{\citenamefont {Horowitz}\ and\ \citenamefont
  {Reddy}(2019)}]{Horowitz:2019aim}%
  \BibitemOpen
  \bibfield  {author} {\bibinfo {author} {\bibfnamefont {C.~J.}\ \bibnamefont
  {Horowitz}}\ and\ \bibinfo {author} {\bibfnamefont {S.}~\bibnamefont
  {Reddy}},\ }\href {\doibase 10.1103/PhysRevLett.122.071102} {\bibfield
  {journal} {\bibinfo  {journal} {Phys. Rev. Lett.}\ }\textbf {\bibinfo
  {volume} {122}},\ \bibinfo {pages} {071102} (\bibinfo {year} {2019})},\
  \Eprint {http://arxiv.org/abs/1902.04597} {arXiv:1902.04597 [astro-ph.HE]}
  \BibitemShut {NoStop}%
\bibitem [{\citenamefont {Hippert}\ \emph {et~al.}(2022)\citenamefont
  {Hippert}, \citenamefont {Setford}, \citenamefont {Tan}, \citenamefont
  {Curtin}, \citenamefont {Noronha-Hostler},\ and\ \citenamefont
  {Yunes}}]{Hippert:2021fch}%
  \BibitemOpen
  \bibfield  {author} {\bibinfo {author} {\bibfnamefont {M.}~\bibnamefont
  {Hippert}}, \bibinfo {author} {\bibfnamefont {J.}~\bibnamefont {Setford}},
  \bibinfo {author} {\bibfnamefont {H.}~\bibnamefont {Tan}}, \bibinfo {author}
  {\bibfnamefont {D.}~\bibnamefont {Curtin}}, \bibinfo {author} {\bibfnamefont
  {J.}~\bibnamefont {Noronha-Hostler}}, \ and\ \bibinfo {author} {\bibfnamefont
  {N.}~\bibnamefont {Yunes}},\ }\href {\doibase 10.1103/PhysRevD.106.035025}
  {\bibfield  {journal} {\bibinfo  {journal} {Phys. Rev. D}\ }\textbf {\bibinfo
  {volume} {106}},\ \bibinfo {pages} {035025} (\bibinfo {year} {2022})},\
  \Eprint {http://arxiv.org/abs/2103.01965} {arXiv:2103.01965 [astro-ph.HE]}
  \BibitemShut {NoStop}%
\bibitem [{\citenamefont {Yagi}\ and\ \citenamefont
  {Yunes}(2013)}]{Yagi:2013awa}%
  \BibitemOpen
  \bibfield  {author} {\bibinfo {author} {\bibfnamefont {K.}~\bibnamefont
  {Yagi}}\ and\ \bibinfo {author} {\bibfnamefont {N.}~\bibnamefont {Yunes}},\
  }\href {\doibase 10.1103/PhysRevD.88.023009} {\bibfield  {journal} {\bibinfo
  {journal} {Phys. Rev. D}\ }\textbf {\bibinfo {volume} {88}},\ \bibinfo
  {pages} {023009} (\bibinfo {year} {2013})},\ \Eprint
  {http://arxiv.org/abs/1303.1528} {arXiv:1303.1528 [gr-qc]} \BibitemShut
  {NoStop}%
\bibitem [{\citenamefont {Lopes}\ \emph {et~al.}(2022)\citenamefont {Lopes},
  \citenamefont {Farias}, \citenamefont {Dexheimer}, \citenamefont
  {Bandyopadhyay},\ and\ \citenamefont {O.~Ramos}}]{Lopes:2022efy}%
  \BibitemOpen
  \bibfield  {author} {\bibinfo {author} {\bibfnamefont {B.~S.}\ \bibnamefont
  {Lopes}}, \bibinfo {author} {\bibfnamefont {R.~L.~S.}\ \bibnamefont
  {Farias}}, \bibinfo {author} {\bibfnamefont {V.}~\bibnamefont {Dexheimer}},
  \bibinfo {author} {\bibfnamefont {A.}~\bibnamefont {Bandyopadhyay}}, \ and\
  \bibinfo {author} {\bibfnamefont {R.}~\bibnamefont {O.~Ramos}},\ }\href@noop
  {} {\  (\bibinfo {year} {2022})},\ \Eprint {http://arxiv.org/abs/2206.01631}
  {arXiv:2206.01631 [hep-ph]} \BibitemShut {NoStop}%
\bibitem [{\citenamefont {Raffelt}(1996)}]{Raffelt:1996wa}%
  \BibitemOpen
  \bibfield  {author} {\bibinfo {author} {\bibfnamefont {G.~G.}\ \bibnamefont
  {Raffelt}},\ }\href@noop {} {\emph {\bibinfo {title} {{Stars as laboratories
  for fundamental physics}: {The astrophysics of neutrinos, axions, and other
  weakly interacting particles}}}}\ (\bibinfo {year} {1996})\BibitemShut
  {NoStop}%
\bibitem [{\citenamefont {Fortin}\ \emph {et~al.}(2021)\citenamefont {Fortin},
  \citenamefont {Guo}, \citenamefont {Harris}, \citenamefont {Kim},
  \citenamefont {Sinha},\ and\ \citenamefont {Sun}}]{Fortin:2021cog}%
  \BibitemOpen
  \bibfield  {author} {\bibinfo {author} {\bibfnamefont {J.-F.}\ \bibnamefont
  {Fortin}}, \bibinfo {author} {\bibfnamefont {H.-K.}\ \bibnamefont {Guo}},
  \bibinfo {author} {\bibfnamefont {S.~P.}\ \bibnamefont {Harris}}, \bibinfo
  {author} {\bibfnamefont {D.}~\bibnamefont {Kim}}, \bibinfo {author}
  {\bibfnamefont {K.}~\bibnamefont {Sinha}}, \ and\ \bibinfo {author}
  {\bibfnamefont {C.}~\bibnamefont {Sun}},\ }\href {\doibase
  10.1142/S0218271821300020} {\bibfield  {journal} {\bibinfo  {journal} {Int.
  J. Mod. Phys. D}\ }\textbf {\bibinfo {volume} {30}},\ \bibinfo {pages}
  {2130002} (\bibinfo {year} {2021})},\ \Eprint
  {http://arxiv.org/abs/2102.12503} {arXiv:2102.12503 [hep-ph]} \BibitemShut
  {NoStop}%
\bibitem [{\citenamefont {Sedrakian}(2016)}]{Sedrakian:2015krq}%
  \BibitemOpen
  \bibfield  {author} {\bibinfo {author} {\bibfnamefont {A.}~\bibnamefont
  {Sedrakian}},\ }\href {\doibase 10.1103/PhysRevD.93.065044} {\bibfield
  {journal} {\bibinfo  {journal} {Phys. Rev. D}\ }\textbf {\bibinfo {volume}
  {93}},\ \bibinfo {pages} {065044} (\bibinfo {year} {2016})},\ \Eprint
  {http://arxiv.org/abs/1512.07828} {arXiv:1512.07828 [astro-ph.HE]}
  \BibitemShut {NoStop}%
\bibitem [{\citenamefont {Hamaguchi}\ \emph {et~al.}(2018)\citenamefont
  {Hamaguchi}, \citenamefont {Nagata}, \citenamefont {Yanagi},\ and\
  \citenamefont {Zheng}}]{Hamaguchi:2018oqw}%
  \BibitemOpen
  \bibfield  {author} {\bibinfo {author} {\bibfnamefont {K.}~\bibnamefont
  {Hamaguchi}}, \bibinfo {author} {\bibfnamefont {N.}~\bibnamefont {Nagata}},
  \bibinfo {author} {\bibfnamefont {K.}~\bibnamefont {Yanagi}}, \ and\ \bibinfo
  {author} {\bibfnamefont {J.}~\bibnamefont {Zheng}},\ }\href {\doibase
  10.1103/PhysRevD.98.103015} {\bibfield  {journal} {\bibinfo  {journal} {Phys.
  Rev. D}\ }\textbf {\bibinfo {volume} {98}},\ \bibinfo {pages} {103015}
  (\bibinfo {year} {2018})},\ \Eprint {http://arxiv.org/abs/1806.07151}
  {arXiv:1806.07151 [hep-ph]} \BibitemShut {NoStop}%
\bibitem [{\citenamefont {Buschmann}\ \emph {et~al.}(2021)\citenamefont
  {Buschmann}, \citenamefont {Co}, \citenamefont {Dessert},\ and\ \citenamefont
  {Safdi}}]{Buschmann:2019pfp}%
  \BibitemOpen
  \bibfield  {author} {\bibinfo {author} {\bibfnamefont {M.}~\bibnamefont
  {Buschmann}}, \bibinfo {author} {\bibfnamefont {R.~T.}\ \bibnamefont {Co}},
  \bibinfo {author} {\bibfnamefont {C.}~\bibnamefont {Dessert}}, \ and\
  \bibinfo {author} {\bibfnamefont {B.~R.}\ \bibnamefont {Safdi}},\ }\href
  {\doibase 10.1103/PhysRevLett.126.021102} {\bibfield  {journal} {\bibinfo
  {journal} {Phys. Rev. Lett.}\ }\textbf {\bibinfo {volume} {126}},\ \bibinfo
  {pages} {021102} (\bibinfo {year} {2021})},\ \Eprint
  {http://arxiv.org/abs/1910.04164} {arXiv:1910.04164 [hep-ph]} \BibitemShut
  {NoStop}%
\bibitem [{\citenamefont {Giangrandi}\ \emph {et~al.}(2022)\citenamefont
  {Giangrandi}, \citenamefont {Sagun}, \citenamefont {Ivanytskyi},
  \citenamefont {Provid\^encia},\ and\ \citenamefont
  {Dietrich}}]{Giangrandi:2022wht}%
  \BibitemOpen
  \bibfield  {author} {\bibinfo {author} {\bibfnamefont {E.}~\bibnamefont
  {Giangrandi}}, \bibinfo {author} {\bibfnamefont {V.}~\bibnamefont {Sagun}},
  \bibinfo {author} {\bibfnamefont {O.}~\bibnamefont {Ivanytskyi}}, \bibinfo
  {author} {\bibfnamefont {C.}~\bibnamefont {Provid\^encia}}, \ and\ \bibinfo
  {author} {\bibfnamefont {T.}~\bibnamefont {Dietrich}},\ }\href@noop {} {\
  (\bibinfo {year} {2022})},\ \Eprint {http://arxiv.org/abs/2209.10905}
  {arXiv:2209.10905 [astro-ph.HE]} \BibitemShut {NoStop}%
\bibitem [{\citenamefont {Baryakhtar}\ \emph {et~al.}(2022)\citenamefont
  {Baryakhtar} \emph {et~al.}}]{Baryakhtar:2022hbu}%
  \BibitemOpen
  \bibfield  {author} {\bibinfo {author} {\bibfnamefont {M.}~\bibnamefont
  {Baryakhtar}} \emph {et~al.},\ }in\ \href@noop {} {\emph {\bibinfo
  {booktitle} {{2022 Snowmass Summer Study}}}}\ (\bibinfo {year} {2022})\
  \Eprint {http://arxiv.org/abs/2203.07984} {arXiv:2203.07984 [hep-ph]}
  \BibitemShut {NoStop}%
\bibitem [{\citenamefont {Ellis}\ \emph
  {et~al.}(2018{\natexlab{b}})\citenamefont {Ellis}, \citenamefont {Hektor},
  \citenamefont {H\"utsi}, \citenamefont {Kannike}, \citenamefont {Marzola},
  \citenamefont {Raidal},\ and\ \citenamefont {Vaskonen}}]{Ellis:2017jgp}%
  \BibitemOpen
  \bibfield  {author} {\bibinfo {author} {\bibfnamefont {J.}~\bibnamefont
  {Ellis}}, \bibinfo {author} {\bibfnamefont {A.}~\bibnamefont {Hektor}},
  \bibinfo {author} {\bibfnamefont {G.}~\bibnamefont {H\"utsi}}, \bibinfo
  {author} {\bibfnamefont {K.}~\bibnamefont {Kannike}}, \bibinfo {author}
  {\bibfnamefont {L.}~\bibnamefont {Marzola}}, \bibinfo {author} {\bibfnamefont
  {M.}~\bibnamefont {Raidal}}, \ and\ \bibinfo {author} {\bibfnamefont
  {V.}~\bibnamefont {Vaskonen}},\ }\href {\doibase
  10.1016/j.physletb.2018.04.048} {\bibfield  {journal} {\bibinfo  {journal}
  {Phys. Lett. B}\ }\textbf {\bibinfo {volume} {781}},\ \bibinfo {pages} {607}
  (\bibinfo {year} {2018}{\natexlab{b}})},\ \Eprint
  {http://arxiv.org/abs/1710.05540} {arXiv:1710.05540 [astro-ph.CO]}
  \BibitemShut {NoStop}%
\bibitem [{\citenamefont {Bezares}\ \emph {et~al.}(2019)\citenamefont
  {Bezares}, \citenamefont {Vigan\`o},\ and\ \citenamefont
  {Palenzuela}}]{Bezares:2019jcb}%
  \BibitemOpen
  \bibfield  {author} {\bibinfo {author} {\bibfnamefont {M.}~\bibnamefont
  {Bezares}}, \bibinfo {author} {\bibfnamefont {D.}~\bibnamefont {Vigan\`o}}, \
  and\ \bibinfo {author} {\bibfnamefont {C.}~\bibnamefont {Palenzuela}},\
  }\href {\doibase 10.1103/PhysRevD.100.044049} {\bibfield  {journal} {\bibinfo
   {journal} {Phys. Rev. D}\ }\textbf {\bibinfo {volume} {100}},\ \bibinfo
  {pages} {044049} (\bibinfo {year} {2019})},\ \Eprint
  {http://arxiv.org/abs/1905.08551} {arXiv:1905.08551 [gr-qc]} \BibitemShut
  {NoStop}%
\bibitem [{\citenamefont {Zhang}\ \emph {et~al.}(2021)\citenamefont {Zhang},
  \citenamefont {Lyu}, \citenamefont {Huang}, \citenamefont {Johnson},
  \citenamefont {Sagunski}, \citenamefont {Sakellariadou},\ and\ \citenamefont
  {Yang}}]{Zhang:2021mks}%
  \BibitemOpen
  \bibfield  {author} {\bibinfo {author} {\bibfnamefont {J.}~\bibnamefont
  {Zhang}}, \bibinfo {author} {\bibfnamefont {Z.}~\bibnamefont {Lyu}}, \bibinfo
  {author} {\bibfnamefont {J.}~\bibnamefont {Huang}}, \bibinfo {author}
  {\bibfnamefont {M.~C.}\ \bibnamefont {Johnson}}, \bibinfo {author}
  {\bibfnamefont {L.}~\bibnamefont {Sagunski}}, \bibinfo {author}
  {\bibfnamefont {M.}~\bibnamefont {Sakellariadou}}, \ and\ \bibinfo {author}
  {\bibfnamefont {H.}~\bibnamefont {Yang}},\ }\href {\doibase
  10.1103/PhysRevLett.127.161101} {\bibfield  {journal} {\bibinfo  {journal}
  {Phys. Rev. Lett.}\ }\textbf {\bibinfo {volume} {127}},\ \bibinfo {pages}
  {161101} (\bibinfo {year} {2021})},\ \Eprint
  {http://arxiv.org/abs/2105.13963} {arXiv:2105.13963 [hep-ph]} \BibitemShut
  {NoStop}%
\bibitem [{\citenamefont {Dev}\ \emph {et~al.}(2022)\citenamefont {Dev},
  \citenamefont {Fortin}, \citenamefont {Harris}, \citenamefont {Sinha},\ and\
  \citenamefont {Zhang}}]{Dev:2021kje}%
  \BibitemOpen
  \bibfield  {author} {\bibinfo {author} {\bibfnamefont {P.~S.~B.}\
  \bibnamefont {Dev}}, \bibinfo {author} {\bibfnamefont {J.-F.}\ \bibnamefont
  {Fortin}}, \bibinfo {author} {\bibfnamefont {S.~P.}\ \bibnamefont {Harris}},
  \bibinfo {author} {\bibfnamefont {K.}~\bibnamefont {Sinha}}, \ and\ \bibinfo
  {author} {\bibfnamefont {Y.}~\bibnamefont {Zhang}},\ }\href {\doibase
  10.1088/1475-7516/2022/01/006} {\bibfield  {journal} {\bibinfo  {journal}
  {JCAP}\ }\textbf {\bibinfo {volume} {01}},\ \bibinfo {pages} {006} (\bibinfo
  {year} {2022})},\ \Eprint {http://arxiv.org/abs/2111.05852} {arXiv:2111.05852
  [hep-ph]} \BibitemShut {NoStop}%
\bibitem [{\citenamefont {Harris}\ \emph {et~al.}(2020)\citenamefont {Harris},
  \citenamefont {Fortin}, \citenamefont {Sinha},\ and\ \citenamefont
  {Alford}}]{Harris:2020qim}%
  \BibitemOpen
  \bibfield  {author} {\bibinfo {author} {\bibfnamefont {S.~P.}\ \bibnamefont
  {Harris}}, \bibinfo {author} {\bibfnamefont {J.-F.}\ \bibnamefont {Fortin}},
  \bibinfo {author} {\bibfnamefont {K.}~\bibnamefont {Sinha}}, \ and\ \bibinfo
  {author} {\bibfnamefont {M.~G.}\ \bibnamefont {Alford}},\ }\href {\doibase
  10.1088/1475-7516/2020/07/023} {\bibfield  {journal} {\bibinfo  {journal}
  {JCAP}\ }\textbf {\bibinfo {volume} {07}},\ \bibinfo {pages} {023} (\bibinfo
  {year} {2020})},\ \Eprint {http://arxiv.org/abs/2003.09768} {arXiv:2003.09768
  [hep-ph]} \BibitemShut {NoStop}%
\bibitem [{\citenamefont {Diamond}\ and\ \citenamefont
  {Marques-Tavares}(2022)}]{Diamond:2021ekg}%
  \BibitemOpen
  \bibfield  {author} {\bibinfo {author} {\bibfnamefont {M.~D.}\ \bibnamefont
  {Diamond}}\ and\ \bibinfo {author} {\bibfnamefont {G.}~\bibnamefont
  {Marques-Tavares}},\ }\href {\doibase 10.1103/PhysRevLett.128.211101}
  {\bibfield  {journal} {\bibinfo  {journal} {Phys. Rev. Lett.}\ }\textbf
  {\bibinfo {volume} {128}},\ \bibinfo {pages} {211101} (\bibinfo {year}
  {2022})},\ \Eprint {http://arxiv.org/abs/2106.03879} {arXiv:2106.03879
  [hep-ph]} \BibitemShut {NoStop}%
\bibitem [{\citenamefont {Li}\ \emph {et~al.}(2019)\citenamefont {Li},
  \citenamefont {Krastev}, \citenamefont {Wen},\ and\ \citenamefont
  {Zhang}}]{Li:2019xxz}%
  \BibitemOpen
  \bibfield  {author} {\bibinfo {author} {\bibfnamefont {B.-A.}\ \bibnamefont
  {Li}}, \bibinfo {author} {\bibfnamefont {P.~G.}\ \bibnamefont {Krastev}},
  \bibinfo {author} {\bibfnamefont {D.-H.}\ \bibnamefont {Wen}}, \ and\
  \bibinfo {author} {\bibfnamefont {N.-B.}\ \bibnamefont {Zhang}},\ }\href
  {\doibase 10.1140/epja/i2019-12780-8} {\bibfield  {journal} {\bibinfo
  {journal} {Eur. Phys. J. A}\ }\textbf {\bibinfo {volume} {55}},\ \bibinfo
  {pages} {117} (\bibinfo {year} {2019})},\ \Eprint
  {http://arxiv.org/abs/1905.13175} {arXiv:1905.13175 [nucl-th]} \BibitemShut
  {NoStop}%
\bibitem [{\citenamefont {Xu}(2022)}]{Xu:2022ikx}%
  \BibitemOpen
  \bibfield  {author} {\bibinfo {author} {\bibfnamefont {H.}~\bibnamefont {Xu}}
  (\bibinfo {collaboration} {STAR}),\ }in\ \href@noop {} {\emph {\bibinfo
  {booktitle} {{29th International Conference on Ultra-relativistic
  Nucleus-Nucleus Collisions}}}}\ (\bibinfo {year} {2022})\ \Eprint
  {http://arxiv.org/abs/2208.06149} {arXiv:2208.06149 [nucl-ex]} \BibitemShut
  {NoStop}%
\bibitem [{\citenamefont {Raithel}\ and\ \citenamefont
  {Ozel}(2019)}]{Raithel:2019ejc}%
  \BibitemOpen
  \bibfield  {author} {\bibinfo {author} {\bibfnamefont {C.~A.}\ \bibnamefont
  {Raithel}}\ and\ \bibinfo {author} {\bibfnamefont {F.}~\bibnamefont {Ozel}},\
  }\href {\doibase 10.3847/1538-4357/ab48e6} {\  (\bibinfo {year} {2019}),\
  10.3847/1538-4357/ab48e6},\ \Eprint {http://arxiv.org/abs/1908.00018}
  {arXiv:1908.00018 [astro-ph.HE]} \BibitemShut {NoStop}%
\bibitem [{\citenamefont {Li}\ \emph {et~al.}(2021{\natexlab{a}})\citenamefont
  {Li}, \citenamefont {Cai}, \citenamefont {Xie},\ and\ \citenamefont
  {Zhang}}]{Li:2021thg}%
  \BibitemOpen
  \bibfield  {author} {\bibinfo {author} {\bibfnamefont {B.-A.}\ \bibnamefont
  {Li}}, \bibinfo {author} {\bibfnamefont {B.-J.}\ \bibnamefont {Cai}},
  \bibinfo {author} {\bibfnamefont {W.-J.}\ \bibnamefont {Xie}}, \ and\
  \bibinfo {author} {\bibfnamefont {N.-B.}\ \bibnamefont {Zhang}},\ }\href
  {\doibase 10.3390/universe7060182} {\bibfield  {journal} {\bibinfo  {journal}
  {Universe}\ }\textbf {\bibinfo {volume} {7}},\ \bibinfo {pages} {182}
  (\bibinfo {year} {2021}{\natexlab{a}})},\ \Eprint
  {http://arxiv.org/abs/2105.04629} {arXiv:2105.04629 [nucl-th]} \BibitemShut
  {NoStop}%
\bibitem [{\citenamefont {Essick}\ \emph
  {et~al.}(2021{\natexlab{b}})\citenamefont {Essick}, \citenamefont {Tews},
  \citenamefont {Landry},\ and\ \citenamefont {Schwenk}}]{Essick:2021kjb}%
  \BibitemOpen
  \bibfield  {author} {\bibinfo {author} {\bibfnamefont {R.}~\bibnamefont
  {Essick}}, \bibinfo {author} {\bibfnamefont {I.}~\bibnamefont {Tews}},
  \bibinfo {author} {\bibfnamefont {P.}~\bibnamefont {Landry}}, \ and\ \bibinfo
  {author} {\bibfnamefont {A.}~\bibnamefont {Schwenk}},\ }\href {\doibase
  10.1103/PhysRevLett.127.192701} {\bibfield  {journal} {\bibinfo  {journal}
  {Phys. Rev. Lett.}\ }\textbf {\bibinfo {volume} {127}},\ \bibinfo {pages}
  {192701} (\bibinfo {year} {2021}{\natexlab{b}})},\ \Eprint
  {http://arxiv.org/abs/2102.10074} {arXiv:2102.10074 [nucl-th]} \BibitemShut
  {NoStop}%
\bibitem [{\citenamefont {Most}\ and\ \citenamefont
  {Raithel}(2021)}]{Most:2021ktk}%
  \BibitemOpen
  \bibfield  {author} {\bibinfo {author} {\bibfnamefont {E.~R.}\ \bibnamefont
  {Most}}\ and\ \bibinfo {author} {\bibfnamefont {C.~A.}\ \bibnamefont
  {Raithel}},\ }\href {\doibase 10.1103/PhysRevD.104.124012} {\bibfield
  {journal} {\bibinfo  {journal} {Phys. Rev. D}\ }\textbf {\bibinfo {volume}
  {104}},\ \bibinfo {pages} {124012} (\bibinfo {year} {2021})},\ \Eprint
  {http://arxiv.org/abs/2107.06804} {arXiv:2107.06804 [astro-ph.HE]}
  \BibitemShut {NoStop}%
\bibitem [{\citenamefont {Li}\ \emph {et~al.}(2008)\citenamefont {Li},
  \citenamefont {Chen},\ and\ \citenamefont {Ko}}]{LCK08}%
  \BibitemOpen
  \bibfield  {author} {\bibinfo {author} {\bibfnamefont {B.-A.}\ \bibnamefont
  {Li}}, \bibinfo {author} {\bibfnamefont {L.-W.}\ \bibnamefont {Chen}}, \ and\
  \bibinfo {author} {\bibfnamefont {C.~M.}\ \bibnamefont {Ko}},\ }\href
  {\doibase 10.1016/j.physrep.2008.04.005} {\bibfield  {journal} {\bibinfo
  {journal} {Phys. Rept.}\ }\textbf {\bibinfo {volume} {464}},\ \bibinfo
  {pages} {113} (\bibinfo {year} {2008})},\ \Eprint
  {http://arxiv.org/abs/0804.3580} {arXiv:0804.3580 [nucl-th]} \BibitemShut
  {NoStop}%
\bibitem [{FRI(2019)}]{FRIB400}%
  \BibitemOpen
  \href {https://fribusers.org/documents/2019/FRIB400-Upgrade.pdf} {\bibfield
  {journal} {\bibinfo  {journal} {FRIB400 White Paper: The Scientific Case for
  the 400 MeV/u Energy Upgrade of FRIB}\ } (\bibinfo {year}
  {2019})}\BibitemShut {NoStop}%
\bibitem [{\citenamefont {Huth}\ \emph {et~al.}(2022)\citenamefont {Huth} \emph
  {et~al.}}]{Huth:2021bsp}%
  \BibitemOpen
  \bibfield  {author} {\bibinfo {author} {\bibfnamefont {S.}~\bibnamefont
  {Huth}} \emph {et~al.},\ }\href {\doibase 10.1038/s41586-022-04750-w}
  {\bibfield  {journal} {\bibinfo  {journal} {Nature}\ }\textbf {\bibinfo
  {volume} {606}},\ \bibinfo {pages} {276} (\bibinfo {year} {2022})},\ \Eprint
  {http://arxiv.org/abs/2107.06229} {arXiv:2107.06229 [nucl-th]} \BibitemShut
  {NoStop}%
\bibitem [{\citenamefont {Aryal}\ \emph {et~al.}(2020)\citenamefont {Aryal},
  \citenamefont {Constantinou}, \citenamefont {Farias},\ and\ \citenamefont
  {Dexheimer}}]{Aryal:2020ocm}%
  \BibitemOpen
  \bibfield  {author} {\bibinfo {author} {\bibfnamefont {K.}~\bibnamefont
  {Aryal}}, \bibinfo {author} {\bibfnamefont {C.}~\bibnamefont {Constantinou}},
  \bibinfo {author} {\bibfnamefont {R.~L.~S.}\ \bibnamefont {Farias}}, \ and\
  \bibinfo {author} {\bibfnamefont {V.}~\bibnamefont {Dexheimer}},\ }\href
  {\doibase 10.1103/PhysRevD.102.076016} {\bibfield  {journal} {\bibinfo
  {journal} {Phys. Rev. D}\ }\textbf {\bibinfo {volume} {102}},\ \bibinfo
  {pages} {076016} (\bibinfo {year} {2020})},\ \Eprint
  {http://arxiv.org/abs/2004.03039} {arXiv:2004.03039 [nucl-th]} \BibitemShut
  {NoStop}%
\bibitem [{\citenamefont {Costa}\ \emph {et~al.}(2020)\citenamefont {Costa},
  \citenamefont {C\^amara~Pereira},\ and\ \citenamefont
  {Provid\^encia}}]{Costa:2020dgc}%
  \BibitemOpen
  \bibfield  {author} {\bibinfo {author} {\bibfnamefont {P.}~\bibnamefont
  {Costa}}, \bibinfo {author} {\bibfnamefont {R.}~\bibnamefont
  {C\^amara~Pereira}}, \ and\ \bibinfo {author} {\bibfnamefont
  {C.}~\bibnamefont {Provid\^encia}},\ }\href {\doibase
  10.1103/PhysRevD.102.054010} {\bibfield  {journal} {\bibinfo  {journal}
  {Phys. Rev. D}\ }\textbf {\bibinfo {volume} {102}},\ \bibinfo {pages}
  {054010} (\bibinfo {year} {2020})},\ \Eprint
  {http://arxiv.org/abs/2009.01781} {arXiv:2009.01781 [hep-ph]} \BibitemShut
  {NoStop}%
\bibitem [{\citenamefont {Qian}\ \emph {et~al.}(2000)\citenamefont {Qian},
  \citenamefont {Su},\ and\ \citenamefont {Wang}}]{Qian:2000fq}%
  \BibitemOpen
  \bibfield  {author} {\bibinfo {author} {\bibfnamefont {W.~L.}\ \bibnamefont
  {Qian}}, \bibinfo {author} {\bibfnamefont {R.-K.}\ \bibnamefont {Su}}, \ and\
  \bibinfo {author} {\bibfnamefont {P.}~\bibnamefont {Wang}},\ }\href {\doibase
  10.1016/S0370-2693(00)00981-3} {\bibfield  {journal} {\bibinfo  {journal}
  {Phys. Lett. B}\ }\textbf {\bibinfo {volume} {491}},\ \bibinfo {pages} {90}
  (\bibinfo {year} {2000})},\ \Eprint {http://arxiv.org/abs/nucl-th/0008057}
  {arXiv:nucl-th/0008057} \BibitemShut {NoStop}%
\bibitem [{\citenamefont {Hatsuda}\ \emph {et~al.}(2006)\citenamefont
  {Hatsuda}, \citenamefont {Tachibana}, \citenamefont {Yamamoto},\ and\
  \citenamefont {Baym}}]{Hatsuda:2006ps}%
  \BibitemOpen
  \bibfield  {author} {\bibinfo {author} {\bibfnamefont {T.}~\bibnamefont
  {Hatsuda}}, \bibinfo {author} {\bibfnamefont {M.}~\bibnamefont {Tachibana}},
  \bibinfo {author} {\bibfnamefont {N.}~\bibnamefont {Yamamoto}}, \ and\
  \bibinfo {author} {\bibfnamefont {G.}~\bibnamefont {Baym}},\ }\href {\doibase
  10.1103/PhysRevLett.97.122001} {\bibfield  {journal} {\bibinfo  {journal}
  {Phys. Rev. Lett.}\ }\textbf {\bibinfo {volume} {97}},\ \bibinfo {pages}
  {122001} (\bibinfo {year} {2006})},\ \Eprint
  {http://arxiv.org/abs/hep-ph/0605018} {arXiv:hep-ph/0605018} \BibitemShut
  {NoStop}%
\bibitem [{\citenamefont {Danielewicz}\ and\ \citenamefont
  {Kurata-Nishimura}(2022)}]{Danielewicz:2021vqq}%
  \BibitemOpen
  \bibfield  {author} {\bibinfo {author} {\bibfnamefont {P.}~\bibnamefont
  {Danielewicz}}\ and\ \bibinfo {author} {\bibfnamefont {M.}~\bibnamefont
  {Kurata-Nishimura}},\ }\href {\doibase 10.1103/PhysRevC.105.034608}
  {\bibfield  {journal} {\bibinfo  {journal} {Phys. Rev. C}\ }\textbf {\bibinfo
  {volume} {105}},\ \bibinfo {pages} {034608} (\bibinfo {year} {2022})},\
  \Eprint {http://arxiv.org/abs/2109.02626} {arXiv:2109.02626 [nucl-th]}
  \BibitemShut {NoStop}%
\bibitem [{\citenamefont {D'Agostini}(1995)}]{dagostini_multidimensional_1995}%
  \BibitemOpen
  \bibfield  {author} {\bibinfo {author} {\bibfnamefont {G.}~\bibnamefont
  {D'Agostini}},\ }\href {\doibase 10.1016/0168-9002(95)00274-X} {\bibfield
  {journal} {\bibinfo  {journal} {Nucl. Instrum. Methods. Phys. Res. A}\
  }\textbf {\bibinfo {volume} {362}},\ \bibinfo {pages} {487} (\bibinfo {year}
  {1995})}\BibitemShut {NoStop}%
\bibitem [{\citenamefont {Landry}\ \emph {et~al.}(2020)\citenamefont {Landry},
  \citenamefont {Essick},\ and\ \citenamefont
  {Chatziioannou}}]{Landry:2020vaw}%
  \BibitemOpen
  \bibfield  {author} {\bibinfo {author} {\bibfnamefont {P.}~\bibnamefont
  {Landry}}, \bibinfo {author} {\bibfnamefont {R.}~\bibnamefont {Essick}}, \
  and\ \bibinfo {author} {\bibfnamefont {K.}~\bibnamefont {Chatziioannou}},\
  }\href {\doibase 10.1103/PhysRevD.101.123007} {\bibfield  {journal} {\bibinfo
   {journal} {Phys. Rev. D}\ }\textbf {\bibinfo {volume} {101}},\ \bibinfo
  {pages} {123007} (\bibinfo {year} {2020})},\ \Eprint
  {http://arxiv.org/abs/2003.04880} {arXiv:2003.04880 [astro-ph.HE]}
  \BibitemShut {NoStop}%
\bibitem [{\citenamefont {Han}\ \emph {et~al.}(2022)\citenamefont {Han},
  \citenamefont {Tang},\ and\ \citenamefont {Fan}}]{Han:2022sxt}%
  \BibitemOpen
  \bibfield  {author} {\bibinfo {author} {\bibfnamefont {M.-Z.}\ \bibnamefont
  {Han}}, \bibinfo {author} {\bibfnamefont {S.-P.}\ \bibnamefont {Tang}}, \
  and\ \bibinfo {author} {\bibfnamefont {Y.-Z.}\ \bibnamefont {Fan}},\
  }\href@noop {} {\  (\bibinfo {year} {2022})},\ \Eprint
  {http://arxiv.org/abs/2205.03855} {arXiv:2205.03855 [astro-ph.HE]}
  \BibitemShut {NoStop}%
\bibitem [{\citenamefont {Haensel}\ \emph {et~al.}(2006)\citenamefont
  {Haensel}, \citenamefont {Potekhin},\ and\ \citenamefont
  {Yakovlev}}]{haensel2006neutron}%
  \BibitemOpen
  \bibfield  {author} {\bibinfo {author} {\bibfnamefont {P.}~\bibnamefont
  {Haensel}}, \bibinfo {author} {\bibfnamefont {A.}~\bibnamefont {Potekhin}}, \
  and\ \bibinfo {author} {\bibfnamefont {D.}~\bibnamefont {Yakovlev}},\ }\href
  {https://books.google.fr/books?id=iIrj9nfHnesC} {\emph {\bibinfo {title}
  {Neutron Stars 1: Equation of State and Structure}}},\ Astrophysics and Space
  Science Library\ (\bibinfo  {publisher} {Springer New York},\ \bibinfo {year}
  {2006})\BibitemShut {NoStop}%
\bibitem [{\citenamefont {Makishima}\ \emph {et~al.}(2014)\citenamefont
  {Makishima}, \citenamefont {Enoto}, \citenamefont {Hiraga}, \citenamefont
  {Nakano}, \citenamefont {Nakazawa}, \citenamefont {Sakurai}, \citenamefont
  {Sasano},\ and\ \citenamefont {Murakami}}]{Makishima:2014dua}%
  \BibitemOpen
  \bibfield  {author} {\bibinfo {author} {\bibfnamefont {K.}~\bibnamefont
  {Makishima}}, \bibinfo {author} {\bibfnamefont {T.}~\bibnamefont {Enoto}},
  \bibinfo {author} {\bibfnamefont {J.~S.}\ \bibnamefont {Hiraga}}, \bibinfo
  {author} {\bibfnamefont {T.}~\bibnamefont {Nakano}}, \bibinfo {author}
  {\bibfnamefont {K.}~\bibnamefont {Nakazawa}}, \bibinfo {author}
  {\bibfnamefont {S.}~\bibnamefont {Sakurai}}, \bibinfo {author} {\bibfnamefont
  {M.}~\bibnamefont {Sasano}}, \ and\ \bibinfo {author} {\bibfnamefont
  {H.}~\bibnamefont {Murakami}},\ }\href {\doibase
  10.1103/PhysRevLett.112.171102} {\bibfield  {journal} {\bibinfo  {journal}
  {Phys. Rev. Lett.}\ }\textbf {\bibinfo {volume} {112}},\ \bibinfo {pages}
  {171102} (\bibinfo {year} {2014})},\ \Eprint {http://arxiv.org/abs/1404.3705}
  {arXiv:1404.3705 [astro-ph.HE]} \BibitemShut {NoStop}%
\bibitem [{\citenamefont {Dall'Osso}\ \emph {et~al.}(2018)\citenamefont
  {Dall'Osso}, \citenamefont {Stella},\ and\ \citenamefont
  {Palomba}}]{DallOsso:2018dos}%
  \BibitemOpen
  \bibfield  {author} {\bibinfo {author} {\bibfnamefont {S.}~\bibnamefont
  {Dall'Osso}}, \bibinfo {author} {\bibfnamefont {L.}~\bibnamefont {Stella}}, \
  and\ \bibinfo {author} {\bibfnamefont {C.}~\bibnamefont {Palomba}},\ }\href
  {\doibase 10.1093/mnras/sty1706} {\bibfield  {journal} {\bibinfo  {journal}
  {Mon. Not. Roy. Astron. Soc.}\ }\textbf {\bibinfo {volume} {480}},\ \bibinfo
  {pages} {1353} (\bibinfo {year} {2018})},\ \Eprint
  {http://arxiv.org/abs/1806.11164} {arXiv:1806.11164 [astro-ph.HE]}
  \BibitemShut {NoStop}%
\bibitem [{\citenamefont {Dexheimer}\ \emph {et~al.}(2017)\citenamefont
  {Dexheimer}, \citenamefont {Franzon}, \citenamefont {Gomes}, \citenamefont
  {Farias}, \citenamefont {Avancini},\ and\ \citenamefont
  {Schramm}}]{Dexheimer:2016yqu}%
  \BibitemOpen
  \bibfield  {author} {\bibinfo {author} {\bibfnamefont {V.}~\bibnamefont
  {Dexheimer}}, \bibinfo {author} {\bibfnamefont {B.}~\bibnamefont {Franzon}},
  \bibinfo {author} {\bibfnamefont {R.~O.}\ \bibnamefont {Gomes}}, \bibinfo
  {author} {\bibfnamefont {R.~L.~S.}\ \bibnamefont {Farias}}, \bibinfo {author}
  {\bibfnamefont {S.~S.}\ \bibnamefont {Avancini}}, \ and\ \bibinfo {author}
  {\bibfnamefont {S.}~\bibnamefont {Schramm}},\ }\href {\doibase
  10.1016/j.physletb.2017.09.008} {\bibfield  {journal} {\bibinfo  {journal}
  {Phys. Lett. B}\ }\textbf {\bibinfo {volume} {773}},\ \bibinfo {pages} {487}
  (\bibinfo {year} {2017})},\ \Eprint {http://arxiv.org/abs/1612.05795}
  {arXiv:1612.05795 [astro-ph.HE]} \BibitemShut {NoStop}%
\bibitem [{\citenamefont {{Lai}}\ and\ \citenamefont
  {{Shapiro}}(1991)}]{1991ApJ...383..745L}%
  \BibitemOpen
  \bibfield  {author} {\bibinfo {author} {\bibfnamefont {D.}~\bibnamefont
  {{Lai}}}\ and\ \bibinfo {author} {\bibfnamefont {S.~L.}\ \bibnamefont
  {{Shapiro}}},\ }\href {\doibase 10.1086/170831} {\bibfield  {journal}
  {\bibinfo  {journal} {\apj}\ }\textbf {\bibinfo {volume} {383}},\ \bibinfo
  {pages} {745} (\bibinfo {year} {1991})}\BibitemShut {NoStop}%
\bibitem [{\citenamefont {Bonazzola}\ \emph {et~al.}(1993)\citenamefont
  {Bonazzola}, \citenamefont {Gourgoulhon}, \citenamefont {Salgado},\ and\
  \citenamefont {Marck}}]{Bonazzola:1993zz}%
  \BibitemOpen
  \bibfield  {author} {\bibinfo {author} {\bibfnamefont {S.}~\bibnamefont
  {Bonazzola}}, \bibinfo {author} {\bibfnamefont {E.}~\bibnamefont
  {Gourgoulhon}}, \bibinfo {author} {\bibfnamefont {M.}~\bibnamefont
  {Salgado}}, \ and\ \bibinfo {author} {\bibfnamefont {J.~A.}\ \bibnamefont
  {Marck}},\ }\href@noop {} {\bibfield  {journal} {\bibinfo  {journal} {Astron.
  Astrophys.}\ }\textbf {\bibinfo {volume} {278}},\ \bibinfo {pages} {421}
  (\bibinfo {year} {1993})}\BibitemShut {NoStop}%
\bibitem [{\citenamefont {Price}\ and\ \citenamefont
  {Rosswog}(2006)}]{Price:2006fi}%
  \BibitemOpen
  \bibfield  {author} {\bibinfo {author} {\bibfnamefont {D.}~\bibnamefont
  {Price}}\ and\ \bibinfo {author} {\bibfnamefont {S.}~\bibnamefont
  {Rosswog}},\ }\href {\doibase 10.1126/science.1125201} {\bibfield  {journal}
  {\bibinfo  {journal} {Science}\ }\textbf {\bibinfo {volume} {312}},\ \bibinfo
  {pages} {719} (\bibinfo {year} {2006})},\ \Eprint
  {http://arxiv.org/abs/astro-ph/0603845} {arXiv:astro-ph/0603845} \BibitemShut
  {NoStop}%
\bibitem [{\citenamefont {Giacomazzo}\ \emph {et~al.}(2015)\citenamefont
  {Giacomazzo}, \citenamefont {Zrake}, \citenamefont {Duffell}, \citenamefont
  {MacFadyen},\ and\ \citenamefont {Perna}}]{Giacomazzo:2014qba}%
  \BibitemOpen
  \bibfield  {author} {\bibinfo {author} {\bibfnamefont {B.}~\bibnamefont
  {Giacomazzo}}, \bibinfo {author} {\bibfnamefont {J.}~\bibnamefont {Zrake}},
  \bibinfo {author} {\bibfnamefont {P.}~\bibnamefont {Duffell}}, \bibinfo
  {author} {\bibfnamefont {A.~I.}\ \bibnamefont {MacFadyen}}, \ and\ \bibinfo
  {author} {\bibfnamefont {R.}~\bibnamefont {Perna}},\ }\href {\doibase
  10.1088/0004-637X/809/1/39} {\bibfield  {journal} {\bibinfo  {journal}
  {Astrophys. J.}\ }\textbf {\bibinfo {volume} {809}},\ \bibinfo {pages} {39}
  (\bibinfo {year} {2015})},\ \Eprint {http://arxiv.org/abs/1410.0013}
  {arXiv:1410.0013 [astro-ph.HE]} \BibitemShut {NoStop}%
\bibitem [{\citenamefont {Kiuchi}\ \emph {et~al.}(2015)\citenamefont {Kiuchi},
  \citenamefont {Cerd\'a-Dur\'an}, \citenamefont {Kyutoku}, \citenamefont
  {Sekiguchi},\ and\ \citenamefont {Shibata}}]{Kiuchi:2015sga}%
  \BibitemOpen
  \bibfield  {author} {\bibinfo {author} {\bibfnamefont {K.}~\bibnamefont
  {Kiuchi}}, \bibinfo {author} {\bibfnamefont {P.}~\bibnamefont
  {Cerd\'a-Dur\'an}}, \bibinfo {author} {\bibfnamefont {K.}~\bibnamefont
  {Kyutoku}}, \bibinfo {author} {\bibfnamefont {Y.}~\bibnamefont {Sekiguchi}},
  \ and\ \bibinfo {author} {\bibfnamefont {M.}~\bibnamefont {Shibata}},\ }\href
  {\doibase 10.1103/PhysRevD.92.124034} {\bibfield  {journal} {\bibinfo
  {journal} {Phys. Rev. D}\ }\textbf {\bibinfo {volume} {92}},\ \bibinfo
  {pages} {124034} (\bibinfo {year} {2015})},\ \Eprint
  {http://arxiv.org/abs/1509.09205} {arXiv:1509.09205 [astro-ph.HE]}
  \BibitemShut {NoStop}%
\bibitem [{\citenamefont {Kiuchi}\ \emph {et~al.}(2018)\citenamefont {Kiuchi},
  \citenamefont {Kyutoku}, \citenamefont {Sekiguchi},\ and\ \citenamefont
  {Shibata}}]{Kiuchi:2017zzg}%
  \BibitemOpen
  \bibfield  {author} {\bibinfo {author} {\bibfnamefont {K.}~\bibnamefont
  {Kiuchi}}, \bibinfo {author} {\bibfnamefont {K.}~\bibnamefont {Kyutoku}},
  \bibinfo {author} {\bibfnamefont {Y.}~\bibnamefont {Sekiguchi}}, \ and\
  \bibinfo {author} {\bibfnamefont {M.}~\bibnamefont {Shibata}},\ }\href
  {\doibase 10.1103/PhysRevD.97.124039} {\bibfield  {journal} {\bibinfo
  {journal} {Phys. Rev. D}\ }\textbf {\bibinfo {volume} {97}},\ \bibinfo
  {pages} {124039} (\bibinfo {year} {2018})},\ \Eprint
  {http://arxiv.org/abs/1710.01311} {arXiv:1710.01311 [astro-ph.HE]}
  \BibitemShut {NoStop}%
\bibitem [{\citenamefont {Palenzuela}\ \emph {et~al.}(2022)\citenamefont
  {Palenzuela}, \citenamefont {Aguilera-Miret}, \citenamefont {Carrasco},
  \citenamefont {Ciolfi}, \citenamefont {Kalinani}, \citenamefont {Kastaun},
  \citenamefont {Mi\~nano},\ and\ \citenamefont
  {Vigan\`o}}]{Palenzuela:2021gdo}%
  \BibitemOpen
  \bibfield  {author} {\bibinfo {author} {\bibfnamefont {C.}~\bibnamefont
  {Palenzuela}}, \bibinfo {author} {\bibfnamefont {R.}~\bibnamefont
  {Aguilera-Miret}}, \bibinfo {author} {\bibfnamefont {F.}~\bibnamefont
  {Carrasco}}, \bibinfo {author} {\bibfnamefont {R.}~\bibnamefont {Ciolfi}},
  \bibinfo {author} {\bibfnamefont {J.~V.}\ \bibnamefont {Kalinani}}, \bibinfo
  {author} {\bibfnamefont {W.}~\bibnamefont {Kastaun}}, \bibinfo {author}
  {\bibfnamefont {B.}~\bibnamefont {Mi\~nano}}, \ and\ \bibinfo {author}
  {\bibfnamefont {D.}~\bibnamefont {Vigan\`o}},\ }\href {\doibase
  10.1103/PhysRevD.106.023013} {\bibfield  {journal} {\bibinfo  {journal}
  {Phys. Rev. D}\ }\textbf {\bibinfo {volume} {106}},\ \bibinfo {pages}
  {023013} (\bibinfo {year} {2022})},\ \Eprint
  {http://arxiv.org/abs/2112.08413} {arXiv:2112.08413 [gr-qc]} \BibitemShut
  {NoStop}%
\bibitem [{\citenamefont {Aguilera-Miret}\ \emph {et~al.}(2022)\citenamefont
  {Aguilera-Miret}, \citenamefont {Vigan\`o},\ and\ \citenamefont
  {Palenzuela}}]{Aguilera-Miret:2021fre}%
  \BibitemOpen
  \bibfield  {author} {\bibinfo {author} {\bibfnamefont {R.}~\bibnamefont
  {Aguilera-Miret}}, \bibinfo {author} {\bibfnamefont {D.}~\bibnamefont
  {Vigan\`o}}, \ and\ \bibinfo {author} {\bibfnamefont {C.}~\bibnamefont
  {Palenzuela}},\ }\href {\doibase 10.3847/2041-8213/ac50a7} {\bibfield
  {journal} {\bibinfo  {journal} {Astrophys. J. Lett.}\ }\textbf {\bibinfo
  {volume} {926}},\ \bibinfo {pages} {L31} (\bibinfo {year} {2022})},\ \Eprint
  {http://arxiv.org/abs/2112.08406} {arXiv:2112.08406 [gr-qc]} \BibitemShut
  {NoStop}%
\bibitem [{\citenamefont {Fore}\ and\ \citenamefont
  {Reddy}(2020)}]{Fore:2019wib}%
  \BibitemOpen
  \bibfield  {author} {\bibinfo {author} {\bibfnamefont {B.}~\bibnamefont
  {Fore}}\ and\ \bibinfo {author} {\bibfnamefont {S.}~\bibnamefont {Reddy}},\
  }\href {\doibase 10.1103/PhysRevC.101.035809} {\bibfield  {journal} {\bibinfo
   {journal} {Phys. Rev. C}\ }\textbf {\bibinfo {volume} {101}},\ \bibinfo
  {pages} {035809} (\bibinfo {year} {2020})},\ \Eprint
  {http://arxiv.org/abs/1911.02632} {arXiv:1911.02632 [astro-ph.HE]}
  \BibitemShut {NoStop}%
\bibitem [{\citenamefont {Yakovlev}\ \emph {et~al.}(2001)\citenamefont
  {Yakovlev}, \citenamefont {Kaminker}, \citenamefont {Gnedin},\ and\
  \citenamefont {Haensel}}]{Yakovlev:2000jp}%
  \BibitemOpen
  \bibfield  {author} {\bibinfo {author} {\bibfnamefont {D.~G.}\ \bibnamefont
  {Yakovlev}}, \bibinfo {author} {\bibfnamefont {A.~D.}\ \bibnamefont
  {Kaminker}}, \bibinfo {author} {\bibfnamefont {O.~Y.}\ \bibnamefont
  {Gnedin}}, \ and\ \bibinfo {author} {\bibfnamefont {P.}~\bibnamefont
  {Haensel}},\ }\href {\doibase 10.1016/S0370-1573(00)00131-9} {\bibfield
  {journal} {\bibinfo  {journal} {Phys. Rept.}\ }\textbf {\bibinfo {volume}
  {354}},\ \bibinfo {pages} {1} (\bibinfo {year} {2001})},\ \Eprint
  {http://arxiv.org/abs/astro-ph/0012122} {arXiv:astro-ph/0012122} \BibitemShut
  {NoStop}%
\bibitem [{\citenamefont {Arras}\ and\ \citenamefont
  {Weinberg}(2019)}]{Arras:2018fxj}%
  \BibitemOpen
  \bibfield  {author} {\bibinfo {author} {\bibfnamefont {P.}~\bibnamefont
  {Arras}}\ and\ \bibinfo {author} {\bibfnamefont {N.~N.}\ \bibnamefont
  {Weinberg}},\ }\href {\doibase 10.1093/mnras/stz880} {\bibfield  {journal}
  {\bibinfo  {journal} {Mon. Not. Roy. Astron. Soc.}\ }\textbf {\bibinfo
  {volume} {486}},\ \bibinfo {pages} {1424} (\bibinfo {year} {2019})},\ \Eprint
  {http://arxiv.org/abs/1806.04163} {arXiv:1806.04163 [astro-ph.HE]}
  \BibitemShut {NoStop}%
\bibitem [{\citenamefont {Gavassino}\ \emph {et~al.}(2021)\citenamefont
  {Gavassino}, \citenamefont {Antonelli},\ and\ \citenamefont
  {Haskell}}]{Gavassino:2020kwo}%
  \BibitemOpen
  \bibfield  {author} {\bibinfo {author} {\bibfnamefont {L.}~\bibnamefont
  {Gavassino}}, \bibinfo {author} {\bibfnamefont {M.}~\bibnamefont
  {Antonelli}}, \ and\ \bibinfo {author} {\bibfnamefont {B.}~\bibnamefont
  {Haskell}},\ }\href {\doibase 10.1088/1361-6382/abe588} {\bibfield  {journal}
  {\bibinfo  {journal} {Class. Quant. Grav.}\ }\textbf {\bibinfo {volume}
  {38}},\ \bibinfo {pages} {075001} (\bibinfo {year} {2021})},\ \Eprint
  {http://arxiv.org/abs/2003.04609} {arXiv:2003.04609 [gr-qc]} \BibitemShut
  {NoStop}%
\bibitem [{\citenamefont {Pandya}\ \emph
  {et~al.}(2022{\natexlab{a}})\citenamefont {Pandya}, \citenamefont {Most},\
  and\ \citenamefont {Pretorius}}]{Pandya:2022pif}%
  \BibitemOpen
  \bibfield  {author} {\bibinfo {author} {\bibfnamefont {A.}~\bibnamefont
  {Pandya}}, \bibinfo {author} {\bibfnamefont {E.~R.}\ \bibnamefont {Most}}, \
  and\ \bibinfo {author} {\bibfnamefont {F.}~\bibnamefont {Pretorius}},\ }\href
  {\doibase 10.1103/PhysRevD.105.123001} {\bibfield  {journal} {\bibinfo
  {journal} {Phys. Rev. D}\ }\textbf {\bibinfo {volume} {105}},\ \bibinfo
  {pages} {123001} (\bibinfo {year} {2022}{\natexlab{a}})},\ \Eprint
  {http://arxiv.org/abs/2201.12317} {arXiv:2201.12317 [gr-qc]} \BibitemShut
  {NoStop}%
\bibitem [{\citenamefont {Pandya}\ \emph
  {et~al.}(2022{\natexlab{b}})\citenamefont {Pandya}, \citenamefont {Most},\
  and\ \citenamefont {Pretorius}}]{Pandya:2022sff}%
  \BibitemOpen
  \bibfield  {author} {\bibinfo {author} {\bibfnamefont {A.}~\bibnamefont
  {Pandya}}, \bibinfo {author} {\bibfnamefont {E.~R.}\ \bibnamefont {Most}}, \
  and\ \bibinfo {author} {\bibfnamefont {F.}~\bibnamefont {Pretorius}},\
  }\href@noop {} {\  (\bibinfo {year} {2022}{\natexlab{b}})},\ \Eprint
  {http://arxiv.org/abs/2209.09265} {arXiv:2209.09265 [gr-qc]} \BibitemShut
  {NoStop}%
\bibitem [{\citenamefont {Camelio}\ \emph
  {et~al.}(2022{\natexlab{a}})\citenamefont {Camelio}, \citenamefont
  {Gavassino}, \citenamefont {Antonelli}, \citenamefont {Bernuzzi},\ and\
  \citenamefont {Haskell}}]{Camelio:2022ljs}%
  \BibitemOpen
  \bibfield  {author} {\bibinfo {author} {\bibfnamefont {G.}~\bibnamefont
  {Camelio}}, \bibinfo {author} {\bibfnamefont {L.}~\bibnamefont {Gavassino}},
  \bibinfo {author} {\bibfnamefont {M.}~\bibnamefont {Antonelli}}, \bibinfo
  {author} {\bibfnamefont {S.}~\bibnamefont {Bernuzzi}}, \ and\ \bibinfo
  {author} {\bibfnamefont {B.}~\bibnamefont {Haskell}},\ }\href@noop {} {\
  (\bibinfo {year} {2022}{\natexlab{a}})},\ \Eprint
  {http://arxiv.org/abs/2204.11809} {arXiv:2204.11809 [gr-qc]} \BibitemShut
  {NoStop}%
\bibitem [{\citenamefont {Camelio}\ \emph
  {et~al.}(2022{\natexlab{b}})\citenamefont {Camelio}, \citenamefont
  {Gavassino}, \citenamefont {Antonelli}, \citenamefont {Bernuzzi},\ and\
  \citenamefont {Haskell}}]{Camelio:2022fds}%
  \BibitemOpen
  \bibfield  {author} {\bibinfo {author} {\bibfnamefont {G.}~\bibnamefont
  {Camelio}}, \bibinfo {author} {\bibfnamefont {L.}~\bibnamefont {Gavassino}},
  \bibinfo {author} {\bibfnamefont {M.}~\bibnamefont {Antonelli}}, \bibinfo
  {author} {\bibfnamefont {S.}~\bibnamefont {Bernuzzi}}, \ and\ \bibinfo
  {author} {\bibfnamefont {B.}~\bibnamefont {Haskell}},\ }\href@noop {} {\
  (\bibinfo {year} {2022}{\natexlab{b}})},\ \Eprint
  {http://arxiv.org/abs/2204.11810} {arXiv:2204.11810 [gr-qc]} \BibitemShut
  {NoStop}%
\bibitem [{\citenamefont {Foucart}\ \emph {et~al.}(2016)\citenamefont
  {Foucart}, \citenamefont {O'Connor}, \citenamefont {Roberts}, \citenamefont
  {Kidder}, \citenamefont {Pfeiffer},\ and\ \citenamefont
  {Scheel}}]{Foucart:2016rxm}%
  \BibitemOpen
  \bibfield  {author} {\bibinfo {author} {\bibfnamefont {F.}~\bibnamefont
  {Foucart}}, \bibinfo {author} {\bibfnamefont {E.}~\bibnamefont {O'Connor}},
  \bibinfo {author} {\bibfnamefont {L.}~\bibnamefont {Roberts}}, \bibinfo
  {author} {\bibfnamefont {L.~E.}\ \bibnamefont {Kidder}}, \bibinfo {author}
  {\bibfnamefont {H.~P.}\ \bibnamefont {Pfeiffer}}, \ and\ \bibinfo {author}
  {\bibfnamefont {M.~A.}\ \bibnamefont {Scheel}},\ }\href {\doibase
  10.1103/PhysRevD.94.123016} {\bibfield  {journal} {\bibinfo  {journal} {Phys.
  Rev. D}\ }\textbf {\bibinfo {volume} {94}},\ \bibinfo {pages} {123016}
  (\bibinfo {year} {2016})},\ \Eprint {http://arxiv.org/abs/1607.07450}
  {arXiv:1607.07450 [astro-ph.HE]} \BibitemShut {NoStop}%
\bibitem [{\citenamefont {Radice}\ \emph {et~al.}(2022)\citenamefont {Radice},
  \citenamefont {Bernuzzi}, \citenamefont {Perego},\ and\ \citenamefont
  {Haas}}]{Radice:2021jtw}%
  \BibitemOpen
  \bibfield  {author} {\bibinfo {author} {\bibfnamefont {D.}~\bibnamefont
  {Radice}}, \bibinfo {author} {\bibfnamefont {S.}~\bibnamefont {Bernuzzi}},
  \bibinfo {author} {\bibfnamefont {A.}~\bibnamefont {Perego}}, \ and\ \bibinfo
  {author} {\bibfnamefont {R.}~\bibnamefont {Haas}},\ }\href {\doibase
  10.1093/mnras/stac589} {\bibfield  {journal} {\bibinfo  {journal} {Mon. Not.
  Roy. Astron. Soc.}\ }\textbf {\bibinfo {volume} {512}},\ \bibinfo {pages}
  {1499} (\bibinfo {year} {2022})},\ \Eprint {http://arxiv.org/abs/2111.14858}
  {arXiv:2111.14858 [astro-ph.HE]} \BibitemShut {NoStop}%
\bibitem [{\citenamefont {Foucart}\ \emph {et~al.}(2020)\citenamefont
  {Foucart}, \citenamefont {Duez}, \citenamefont {Hebert}, \citenamefont
  {Kidder}, \citenamefont {Pfeiffer},\ and\ \citenamefont
  {Scheel}}]{Foucart:2020qjb}%
  \BibitemOpen
  \bibfield  {author} {\bibinfo {author} {\bibfnamefont {F.}~\bibnamefont
  {Foucart}}, \bibinfo {author} {\bibfnamefont {M.~D.}\ \bibnamefont {Duez}},
  \bibinfo {author} {\bibfnamefont {F.}~\bibnamefont {Hebert}}, \bibinfo
  {author} {\bibfnamefont {L.~E.}\ \bibnamefont {Kidder}}, \bibinfo {author}
  {\bibfnamefont {H.~P.}\ \bibnamefont {Pfeiffer}}, \ and\ \bibinfo {author}
  {\bibfnamefont {M.~A.}\ \bibnamefont {Scheel}},\ }\href {\doibase
  10.3847/2041-8213/abbb87} {\bibfield  {journal} {\bibinfo  {journal}
  {Astrophys. J. Lett.}\ }\textbf {\bibinfo {volume} {902}},\ \bibinfo {pages}
  {L27} (\bibinfo {year} {2020})},\ \Eprint {http://arxiv.org/abs/2008.08089}
  {arXiv:2008.08089 [astro-ph.HE]} \BibitemShut {NoStop}%
\bibitem [{\citenamefont {Troyer}\ and\ \citenamefont
  {Wiese}(2005)}]{troyer2005computational}%
  \BibitemOpen
  \bibfield  {author} {\bibinfo {author} {\bibfnamefont {M.}~\bibnamefont
  {Troyer}}\ and\ \bibinfo {author} {\bibfnamefont {U.-J.}\ \bibnamefont
  {Wiese}},\ }\href@noop {} {\bibfield  {journal} {\bibinfo  {journal}
  {Physical Review Letters}\ }\textbf {\bibinfo {volume} {94}},\ \bibinfo
  {pages} {170201} (\bibinfo {year} {2005})}\BibitemShut {NoStop}%
\bibitem [{\citenamefont {Banuls}\ and\ \citenamefont
  {Cichy}(2020)}]{banuls2020review}%
  \BibitemOpen
  \bibfield  {author} {\bibinfo {author} {\bibfnamefont {M.~C.}\ \bibnamefont
  {Banuls}}\ and\ \bibinfo {author} {\bibfnamefont {K.}~\bibnamefont {Cichy}},\
  }\href@noop {} {\bibfield  {journal} {\bibinfo  {journal} {Reports on
  Progress in Physics}\ }\textbf {\bibinfo {volume} {83}},\ \bibinfo {pages}
  {024401} (\bibinfo {year} {2020})}\BibitemShut {NoStop}%
\bibitem [{\citenamefont {Meurice}\ \emph {et~al.}(2020)\citenamefont
  {Meurice}, \citenamefont {Sakai},\ and\ \citenamefont
  {Unmuth-Yockey}}]{meurice2020tensor}%
  \BibitemOpen
  \bibfield  {author} {\bibinfo {author} {\bibfnamefont {Y.}~\bibnamefont
  {Meurice}}, \bibinfo {author} {\bibfnamefont {R.}~\bibnamefont {Sakai}}, \
  and\ \bibinfo {author} {\bibfnamefont {J.}~\bibnamefont {Unmuth-Yockey}},\
  }\href@noop {} {\bibfield  {journal} {\bibinfo  {journal} {arXiv preprint
  arXiv:2010.06539}\ } (\bibinfo {year} {2020})}\BibitemShut {NoStop}%
\bibitem [{\citenamefont {Meurice}\ \emph {et~al.}(2022)\citenamefont
  {Meurice}, \citenamefont {Osborn}, \citenamefont {Sakai}, \citenamefont
  {Unmuth-Yockey}, \citenamefont {Catterall},\ and\ \citenamefont
  {Somma}}]{meurice2022tensor}%
  \BibitemOpen
  \bibfield  {author} {\bibinfo {author} {\bibfnamefont {Y.}~\bibnamefont
  {Meurice}}, \bibinfo {author} {\bibfnamefont {J.~C.}\ \bibnamefont {Osborn}},
  \bibinfo {author} {\bibfnamefont {R.}~\bibnamefont {Sakai}}, \bibinfo
  {author} {\bibfnamefont {J.}~\bibnamefont {Unmuth-Yockey}}, \bibinfo {author}
  {\bibfnamefont {S.}~\bibnamefont {Catterall}}, \ and\ \bibinfo {author}
  {\bibfnamefont {R.~D.}\ \bibnamefont {Somma}},\ }\href@noop {} {\bibfield
  {journal} {\bibinfo  {journal} {arXiv preprint arXiv:2203.04902}\ } (\bibinfo
  {year} {2022})}\BibitemShut {NoStop}%
\bibitem [{\citenamefont {Lloyd}(1996)}]{lloyd1996universal}%
  \BibitemOpen
  \bibfield  {author} {\bibinfo {author} {\bibfnamefont {S.}~\bibnamefont
  {Lloyd}},\ }\href {\doibase 10.1126/science.273.5278.1073} {\bibfield
  {journal} {\bibinfo  {journal} {Science}\ }\textbf {\bibinfo {volume}
  {273}},\ \bibinfo {pages} {1073} (\bibinfo {year} {1996})}\BibitemShut
  {NoStop}%
\bibitem [{\citenamefont {Jordan}\ \emph {et~al.}(2011)\citenamefont {Jordan},
  \citenamefont {Lee},\ and\ \citenamefont {Preskill}}]{jordan2011quantum}%
  \BibitemOpen
  \bibfield  {author} {\bibinfo {author} {\bibfnamefont {S.~P.}\ \bibnamefont
  {Jordan}}, \bibinfo {author} {\bibfnamefont {K.~S.}\ \bibnamefont {Lee}}, \
  and\ \bibinfo {author} {\bibfnamefont {J.}~\bibnamefont {Preskill}},\
  }\href@noop {} {\bibfield  {journal} {\bibinfo  {journal} {arXiv preprint
  arXiv:1112.4833}\ } (\bibinfo {year} {2011})}\BibitemShut {NoStop}%
\bibitem [{\citenamefont {Shaw}\ \emph {et~al.}(2020)\citenamefont {Shaw},
  \citenamefont {Lougovski}, \citenamefont {Stryker},\ and\ \citenamefont
  {Wiebe}}]{shaw2020quantum}%
  \BibitemOpen
  \bibfield  {author} {\bibinfo {author} {\bibfnamefont {A.~F.}\ \bibnamefont
  {Shaw}}, \bibinfo {author} {\bibfnamefont {P.}~\bibnamefont {Lougovski}},
  \bibinfo {author} {\bibfnamefont {J.~R.}\ \bibnamefont {Stryker}}, \ and\
  \bibinfo {author} {\bibfnamefont {N.}~\bibnamefont {Wiebe}},\ }\href@noop {}
  {\bibfield  {journal} {\bibinfo  {journal} {Quantum}\ }\textbf {\bibinfo
  {volume} {4}},\ \bibinfo {pages} {306} (\bibinfo {year} {2020})}\BibitemShut
  {NoStop}%
\bibitem [{\citenamefont {Byrnes}\ and\ \citenamefont
  {Yamamoto}(2006)}]{Byrnes:2005qx}%
  \BibitemOpen
  \bibfield  {author} {\bibinfo {author} {\bibfnamefont {T.}~\bibnamefont
  {Byrnes}}\ and\ \bibinfo {author} {\bibfnamefont {Y.}~\bibnamefont
  {Yamamoto}},\ }\href {\doibase 10.1103/PhysRevA.73.022328} {\bibfield
  {journal} {\bibinfo  {journal} {Phys. Rev. A}\ }\textbf {\bibinfo {volume}
  {73}},\ \bibinfo {pages} {022328} (\bibinfo {year} {2006})},\ \Eprint
  {http://arxiv.org/abs/quant-ph/0510027} {arXiv:quant-ph/0510027} \BibitemShut
  {NoStop}%
\bibitem [{\citenamefont {Ciavarella}\ \emph {et~al.}(2021)\citenamefont
  {Ciavarella}, \citenamefont {Klco},\ and\ \citenamefont
  {Savage}}]{ciavarella2021trailhead}%
  \BibitemOpen
  \bibfield  {author} {\bibinfo {author} {\bibfnamefont {A.}~\bibnamefont
  {Ciavarella}}, \bibinfo {author} {\bibfnamefont {N.}~\bibnamefont {Klco}}, \
  and\ \bibinfo {author} {\bibfnamefont {M.~J.}\ \bibnamefont {Savage}},\
  }\href@noop {} {\bibfield  {journal} {\bibinfo  {journal} {Physical Review
  D}\ }\textbf {\bibinfo {volume} {103}},\ \bibinfo {pages} {094501} (\bibinfo
  {year} {2021})}\BibitemShut {NoStop}%
\bibitem [{\citenamefont {Lamm}\ \emph {et~al.}(2019)\citenamefont {Lamm},
  \citenamefont {Lawrence},\ and\ \citenamefont {Yamauchi}}]{Lamm:2019bik}%
  \BibitemOpen
  \bibfield  {author} {\bibinfo {author} {\bibfnamefont {H.}~\bibnamefont
  {Lamm}}, \bibinfo {author} {\bibfnamefont {S.}~\bibnamefont {Lawrence}}, \
  and\ \bibinfo {author} {\bibfnamefont {Y.}~\bibnamefont {Yamauchi}} (\bibinfo
  {collaboration} {NuQS}),\ }\href {\doibase 10.1103/PhysRevD.100.034518}
  {\bibfield  {journal} {\bibinfo  {journal} {Phys. Rev. D}\ }\textbf {\bibinfo
  {volume} {100}},\ \bibinfo {pages} {034518} (\bibinfo {year} {2019})},\
  \Eprint {http://arxiv.org/abs/1903.08807} {arXiv:1903.08807 [hep-lat]}
  \BibitemShut {NoStop}%
\bibitem [{\citenamefont {Kan}\ and\ \citenamefont
  {Nam}(2021)}]{kan2021lattice}%
  \BibitemOpen
  \bibfield  {author} {\bibinfo {author} {\bibfnamefont {A.}~\bibnamefont
  {Kan}}\ and\ \bibinfo {author} {\bibfnamefont {Y.}~\bibnamefont {Nam}},\
  }\href {https://arxiv.org/abs/2107.12769} {\bibfield  {journal} {\bibinfo
  {journal} {arXiv:2107.12769}\ } (\bibinfo {year} {2021})}\BibitemShut
  {NoStop}%
\bibitem [{\citenamefont {Banuls}\ \emph {et~al.}(2020)\citenamefont {Banuls},
  \citenamefont {Blatt}, \citenamefont {Catani}, \citenamefont {Celi},
  \citenamefont {Cirac}, \citenamefont {Dalmonte}, \citenamefont {Fallani},
  \citenamefont {Jansen}, \citenamefont {Lewenstein}, \citenamefont
  {Montangero} \emph {et~al.}}]{banuls2020simulating}%
  \BibitemOpen
  \bibfield  {author} {\bibinfo {author} {\bibfnamefont {M.~C.}\ \bibnamefont
  {Banuls}}, \bibinfo {author} {\bibfnamefont {R.}~\bibnamefont {Blatt}},
  \bibinfo {author} {\bibfnamefont {J.}~\bibnamefont {Catani}}, \bibinfo
  {author} {\bibfnamefont {A.}~\bibnamefont {Celi}}, \bibinfo {author}
  {\bibfnamefont {J.~I.}\ \bibnamefont {Cirac}}, \bibinfo {author}
  {\bibfnamefont {M.}~\bibnamefont {Dalmonte}}, \bibinfo {author}
  {\bibfnamefont {L.}~\bibnamefont {Fallani}}, \bibinfo {author} {\bibfnamefont
  {K.}~\bibnamefont {Jansen}}, \bibinfo {author} {\bibfnamefont
  {M.}~\bibnamefont {Lewenstein}}, \bibinfo {author} {\bibfnamefont
  {S.}~\bibnamefont {Montangero}},  \emph {et~al.},\ }\href@noop {} {\bibfield
  {journal} {\bibinfo  {journal} {The European physical journal D}\ }\textbf
  {\bibinfo {volume} {74}},\ \bibinfo {pages} {1} (\bibinfo {year}
  {2020})}\BibitemShut {NoStop}%
\bibitem [{\citenamefont {Klco}\ \emph {et~al.}(2022)\citenamefont {Klco},
  \citenamefont {Roggero},\ and\ \citenamefont {Savage}}]{klco2022standard}%
  \BibitemOpen
  \bibfield  {author} {\bibinfo {author} {\bibfnamefont {N.}~\bibnamefont
  {Klco}}, \bibinfo {author} {\bibfnamefont {A.}~\bibnamefont {Roggero}}, \
  and\ \bibinfo {author} {\bibfnamefont {M.~J.}\ \bibnamefont {Savage}},\
  }\href@noop {} {\bibfield  {journal} {\bibinfo  {journal} {Reports on
  Progress in Physics}\ } (\bibinfo {year} {2022})}\BibitemShut {NoStop}%
\bibitem [{\citenamefont {Lamm}\ \emph {et~al.}(2020)\citenamefont {Lamm},
  \citenamefont {Lawrence}, \citenamefont {Yamauchi}, \citenamefont
  {Collaboration} \emph {et~al.}}]{lamm2020parton}%
  \BibitemOpen
  \bibfield  {author} {\bibinfo {author} {\bibfnamefont {H.}~\bibnamefont
  {Lamm}}, \bibinfo {author} {\bibfnamefont {S.}~\bibnamefont {Lawrence}},
  \bibinfo {author} {\bibfnamefont {Y.}~\bibnamefont {Yamauchi}}, \bibinfo
  {author} {\bibfnamefont {N.}~\bibnamefont {Collaboration}},  \emph {et~al.},\
  }\href@noop {} {\bibfield  {journal} {\bibinfo  {journal} {Physical Review
  Research}\ }\textbf {\bibinfo {volume} {2}},\ \bibinfo {pages} {013272}
  (\bibinfo {year} {2020})}\BibitemShut {NoStop}%
\bibitem [{\citenamefont {Echevarria}\ \emph {et~al.}(2021)\citenamefont
  {Echevarria}, \citenamefont {Egusquiza}, \citenamefont {Rico},\ and\
  \citenamefont {Schnell}}]{echevarria2021quantum}%
  \BibitemOpen
  \bibfield  {author} {\bibinfo {author} {\bibfnamefont {M.}~\bibnamefont
  {Echevarria}}, \bibinfo {author} {\bibfnamefont {I.}~\bibnamefont
  {Egusquiza}}, \bibinfo {author} {\bibfnamefont {E.}~\bibnamefont {Rico}}, \
  and\ \bibinfo {author} {\bibfnamefont {G.}~\bibnamefont {Schnell}},\
  }\href@noop {} {\bibfield  {journal} {\bibinfo  {journal} {Physical Review
  D}\ }\textbf {\bibinfo {volume} {104}},\ \bibinfo {pages} {014512} (\bibinfo
  {year} {2021})}\BibitemShut {NoStop}%
\bibitem [{\citenamefont {Kreshchuk}\ \emph {et~al.}(2021)\citenamefont
  {Kreshchuk}, \citenamefont {Jia}, \citenamefont {Kirby}, \citenamefont
  {Goldstein}, \citenamefont {Vary},\ and\ \citenamefont
  {Love}}]{kreshchuk2021simulating}%
  \BibitemOpen
  \bibfield  {author} {\bibinfo {author} {\bibfnamefont {M.}~\bibnamefont
  {Kreshchuk}}, \bibinfo {author} {\bibfnamefont {S.}~\bibnamefont {Jia}},
  \bibinfo {author} {\bibfnamefont {W.~M.}\ \bibnamefont {Kirby}}, \bibinfo
  {author} {\bibfnamefont {G.}~\bibnamefont {Goldstein}}, \bibinfo {author}
  {\bibfnamefont {J.~P.}\ \bibnamefont {Vary}}, \ and\ \bibinfo {author}
  {\bibfnamefont {P.~J.}\ \bibnamefont {Love}},\ }\href@noop {} {\bibfield
  {journal} {\bibinfo  {journal} {Physical Review A}\ }\textbf {\bibinfo
  {volume} {103}},\ \bibinfo {pages} {062601} (\bibinfo {year}
  {2021})}\BibitemShut {NoStop}%
\bibitem [{\citenamefont {Li}\ \emph {et~al.}(2021{\natexlab{b}})\citenamefont
  {Li}, \citenamefont {Guo}, \citenamefont {Lai}, \citenamefont {Liu},
  \citenamefont {Wang}, \citenamefont {Xing}, \citenamefont {Zhang},\ and\
  \citenamefont {Zhu}}]{li2021partonic}%
  \BibitemOpen
  \bibfield  {author} {\bibinfo {author} {\bibfnamefont {T.}~\bibnamefont
  {Li}}, \bibinfo {author} {\bibfnamefont {X.}~\bibnamefont {Guo}}, \bibinfo
  {author} {\bibfnamefont {W.~K.}\ \bibnamefont {Lai}}, \bibinfo {author}
  {\bibfnamefont {X.}~\bibnamefont {Liu}}, \bibinfo {author} {\bibfnamefont
  {E.}~\bibnamefont {Wang}}, \bibinfo {author} {\bibfnamefont {H.}~\bibnamefont
  {Xing}}, \bibinfo {author} {\bibfnamefont {D.-B.}\ \bibnamefont {Zhang}}, \
  and\ \bibinfo {author} {\bibfnamefont {S.-L.}\ \bibnamefont {Zhu}},\
  }\href@noop {} {\bibfield  {journal} {\bibinfo  {journal} {arXiv preprint
  arXiv:2106.03865}\ } (\bibinfo {year} {2021}{\natexlab{b}})}\BibitemShut
  {NoStop}%
\bibitem [{\citenamefont {P{\'e}rez-Salinas}\ \emph {et~al.}(2021)\citenamefont
  {P{\'e}rez-Salinas}, \citenamefont {Cruz-Martinez}, \citenamefont {Alhajri},\
  and\ \citenamefont {Carrazza}}]{perez2021determining}%
  \BibitemOpen
  \bibfield  {author} {\bibinfo {author} {\bibfnamefont {A.}~\bibnamefont
  {P{\'e}rez-Salinas}}, \bibinfo {author} {\bibfnamefont {J.}~\bibnamefont
  {Cruz-Martinez}}, \bibinfo {author} {\bibfnamefont {A.~A.}\ \bibnamefont
  {Alhajri}}, \ and\ \bibinfo {author} {\bibfnamefont {S.}~\bibnamefont
  {Carrazza}},\ }\href@noop {} {\bibfield  {journal} {\bibinfo  {journal}
  {Physical Review D}\ }\textbf {\bibinfo {volume} {103}},\ \bibinfo {pages}
  {034027} (\bibinfo {year} {2021})}\BibitemShut {NoStop}%
\bibitem [{\citenamefont {Cohen}\ \emph {et~al.}(2021)\citenamefont {Cohen},
  \citenamefont {Lamm}, \citenamefont {Lawrence}, \citenamefont {Yamauchi},
  \citenamefont {Collaboration} \emph {et~al.}}]{cohen2021quantum}%
  \BibitemOpen
  \bibfield  {author} {\bibinfo {author} {\bibfnamefont {T.~D.}\ \bibnamefont
  {Cohen}}, \bibinfo {author} {\bibfnamefont {H.}~\bibnamefont {Lamm}},
  \bibinfo {author} {\bibfnamefont {S.}~\bibnamefont {Lawrence}}, \bibinfo
  {author} {\bibfnamefont {Y.}~\bibnamefont {Yamauchi}}, \bibinfo {author}
  {\bibfnamefont {N.}~\bibnamefont {Collaboration}},  \emph {et~al.},\
  }\href@noop {} {\bibfield  {journal} {\bibinfo  {journal} {Physical Review
  D}\ }\textbf {\bibinfo {volume} {104}},\ \bibinfo {pages} {094514} (\bibinfo
  {year} {2021})}\BibitemShut {NoStop}%
\bibitem [{\citenamefont {De~Jong}\ \emph {et~al.}(2021)\citenamefont
  {De~Jong}, \citenamefont {Metcalf}, \citenamefont {Mulligan}, \citenamefont
  {P\l{}osko\'n}, \citenamefont {Ringer},\ and\ \citenamefont
  {Yao}}]{deJong:2020riy}%
  \BibitemOpen
  \bibfield  {author} {\bibinfo {author} {\bibfnamefont {W.~A.}\ \bibnamefont
  {De~Jong}}, \bibinfo {author} {\bibfnamefont {M.}~\bibnamefont {Metcalf}},
  \bibinfo {author} {\bibfnamefont {J.}~\bibnamefont {Mulligan}}, \bibinfo
  {author} {\bibfnamefont {M.}~\bibnamefont {P\l{}osko\'n}}, \bibinfo {author}
  {\bibfnamefont {F.}~\bibnamefont {Ringer}}, \ and\ \bibinfo {author}
  {\bibfnamefont {X.}~\bibnamefont {Yao}},\ }\href {\doibase
  10.1103/PhysRevD.104.L051501} {\bibfield  {journal} {\bibinfo  {journal}
  {Phys. Rev. D}\ }\textbf {\bibinfo {volume} {104}},\ \bibinfo {pages}
  {051501} (\bibinfo {year} {2021})},\ \Eprint
  {http://arxiv.org/abs/2010.03571} {arXiv:2010.03571 [hep-ph]} \BibitemShut
  {NoStop}%
\bibitem [{\citenamefont {Nachman}\ \emph {et~al.}(2021)\citenamefont
  {Nachman}, \citenamefont {Provasoli}, \citenamefont {de~Jong},\ and\
  \citenamefont {Bauer}}]{nachman2021quantum}%
  \BibitemOpen
  \bibfield  {author} {\bibinfo {author} {\bibfnamefont {B.}~\bibnamefont
  {Nachman}}, \bibinfo {author} {\bibfnamefont {D.}~\bibnamefont {Provasoli}},
  \bibinfo {author} {\bibfnamefont {W.~A.}\ \bibnamefont {de~Jong}}, \ and\
  \bibinfo {author} {\bibfnamefont {C.~W.}\ \bibnamefont {Bauer}},\ }\href@noop
  {} {\bibfield  {journal} {\bibinfo  {journal} {Physical Review Letters}\
  }\textbf {\bibinfo {volume} {126}},\ \bibinfo {pages} {062001} (\bibinfo
  {year} {2021})}\BibitemShut {NoStop}%
\bibitem [{\citenamefont {Yao}(2022)}]{Yao:2022eqm}%
  \BibitemOpen
  \bibfield  {author} {\bibinfo {author} {\bibfnamefont {X.}~\bibnamefont
  {Yao}},\ }\href@noop {} {\  (\bibinfo {year} {2022})},\ \Eprint
  {http://arxiv.org/abs/2205.07902} {arXiv:2205.07902 [hep-ph]} \BibitemShut
  {NoStop}%
\bibitem [{\citenamefont {Czajka}\ \emph {et~al.}(2022)\citenamefont {Czajka},
  \citenamefont {Kang}, \citenamefont {Ma},\ and\ \citenamefont
  {Zhao}}]{czajka2022quantum}%
  \BibitemOpen
  \bibfield  {author} {\bibinfo {author} {\bibfnamefont {A.~M.}\ \bibnamefont
  {Czajka}}, \bibinfo {author} {\bibfnamefont {Z.-B.}\ \bibnamefont {Kang}},
  \bibinfo {author} {\bibfnamefont {H.}~\bibnamefont {Ma}}, \ and\ \bibinfo
  {author} {\bibfnamefont {F.}~\bibnamefont {Zhao}},\ }\href@noop {} {\bibfield
   {journal} {\bibinfo  {journal} {Journal of High Energy Physics}\ }\textbf
  {\bibinfo {volume} {2022}},\ \bibinfo {pages} {1} (\bibinfo {year}
  {2022})}\BibitemShut {NoStop}%
\bibitem [{\citenamefont {Davoudi}\ \emph
  {et~al.}(2022{\natexlab{b}})\citenamefont {Davoudi}, \citenamefont
  {Mueller},\ and\ \citenamefont {Powers}}]{davoudi2022toward}%
  \BibitemOpen
  \bibfield  {author} {\bibinfo {author} {\bibfnamefont {Z.}~\bibnamefont
  {Davoudi}}, \bibinfo {author} {\bibfnamefont {N.}~\bibnamefont {Mueller}}, \
  and\ \bibinfo {author} {\bibfnamefont {C.}~\bibnamefont {Powers}},\
  }\href@noop {} {\bibfield  {journal} {\bibinfo  {journal} {arXiv preprint
  arXiv:2208.13112}\ } (\bibinfo {year} {2022}{\natexlab{b}})}\BibitemShut
  {NoStop}%
\bibitem [{\citenamefont {Riera}\ \emph {et~al.}(2012)\citenamefont {Riera},
  \citenamefont {Gogolin},\ and\ \citenamefont
  {Eisert}}]{riera2012thermalization}%
  \BibitemOpen
  \bibfield  {author} {\bibinfo {author} {\bibfnamefont {A.}~\bibnamefont
  {Riera}}, \bibinfo {author} {\bibfnamefont {C.}~\bibnamefont {Gogolin}}, \
  and\ \bibinfo {author} {\bibfnamefont {J.}~\bibnamefont {Eisert}},\
  }\href@noop {} {\bibfield  {journal} {\bibinfo  {journal} {Physical review
  letters}\ }\textbf {\bibinfo {volume} {108}},\ \bibinfo {pages} {080402}
  (\bibinfo {year} {2012})}\BibitemShut {NoStop}%
\bibitem [{\citenamefont {de~Jong}\ \emph {et~al.}(2022)\citenamefont
  {de~Jong}, \citenamefont {Lee}, \citenamefont {Mulligan}, \citenamefont
  {P\l{}osko\'n}, \citenamefont {Ringer},\ and\ \citenamefont
  {Yao}}]{deJong:2021wsd}%
  \BibitemOpen
  \bibfield  {author} {\bibinfo {author} {\bibfnamefont {W.~A.}\ \bibnamefont
  {de~Jong}}, \bibinfo {author} {\bibfnamefont {K.}~\bibnamefont {Lee}},
  \bibinfo {author} {\bibfnamefont {J.}~\bibnamefont {Mulligan}}, \bibinfo
  {author} {\bibfnamefont {M.}~\bibnamefont {P\l{}osko\'n}}, \bibinfo {author}
  {\bibfnamefont {F.}~\bibnamefont {Ringer}}, \ and\ \bibinfo {author}
  {\bibfnamefont {X.}~\bibnamefont {Yao}},\ }\href {\doibase
  10.1103/PhysRevD.106.054508} {\bibfield  {journal} {\bibinfo  {journal}
  {Phys. Rev. D}\ }\textbf {\bibinfo {volume} {106}},\ \bibinfo {pages}
  {054508} (\bibinfo {year} {2022})},\ \Eprint
  {http://arxiv.org/abs/2106.08394} {arXiv:2106.08394 [quant-ph]} \BibitemShut
  {NoStop}%
\bibitem [{\citenamefont {Kharzeev}\ and\ \citenamefont
  {Tuchin}(2005)}]{kharzeev2005color}%
  \BibitemOpen
  \bibfield  {author} {\bibinfo {author} {\bibfnamefont {D.}~\bibnamefont
  {Kharzeev}}\ and\ \bibinfo {author} {\bibfnamefont {K.}~\bibnamefont
  {Tuchin}},\ }\href@noop {} {\bibfield  {journal} {\bibinfo  {journal}
  {Nuclear Physics A}\ }\textbf {\bibinfo {volume} {753}},\ \bibinfo {pages}
  {316} (\bibinfo {year} {2005})}\BibitemShut {NoStop}%
\bibitem [{\citenamefont {Kharzeev}\ and\ \citenamefont
  {Levin}(2017)}]{kharzeev2017deep}%
  \BibitemOpen
  \bibfield  {author} {\bibinfo {author} {\bibfnamefont {D.~E.}\ \bibnamefont
  {Kharzeev}}\ and\ \bibinfo {author} {\bibfnamefont {E.~M.}\ \bibnamefont
  {Levin}},\ }\href@noop {} {\bibfield  {journal} {\bibinfo  {journal}
  {Physical Review D}\ }\textbf {\bibinfo {volume} {95}},\ \bibinfo {pages}
  {114008} (\bibinfo {year} {2017})}\BibitemShut {NoStop}%
\bibitem [{\citenamefont {Baker}\ and\ \citenamefont
  {Kharzeev}(2018)}]{baker2018thermal}%
  \BibitemOpen
  \bibfield  {author} {\bibinfo {author} {\bibfnamefont {O.}~\bibnamefont
  {Baker}}\ and\ \bibinfo {author} {\bibfnamefont {D.}~\bibnamefont
  {Kharzeev}},\ }\href@noop {} {\bibfield  {journal} {\bibinfo  {journal}
  {Physical Review D}\ }\textbf {\bibinfo {volume} {98}},\ \bibinfo {pages}
  {054007} (\bibinfo {year} {2018})}\BibitemShut {NoStop}%
\bibitem [{\citenamefont {Berges}\ \emph
  {et~al.}(2021{\natexlab{b}})\citenamefont {Berges}, \citenamefont {Heller},
  \citenamefont {Mazeliauskas},\ and\ \citenamefont
  {Venugopalan}}]{berges2021qcd}%
  \BibitemOpen
  \bibfield  {author} {\bibinfo {author} {\bibfnamefont {J.}~\bibnamefont
  {Berges}}, \bibinfo {author} {\bibfnamefont {M.~P.}\ \bibnamefont {Heller}},
  \bibinfo {author} {\bibfnamefont {A.}~\bibnamefont {Mazeliauskas}}, \ and\
  \bibinfo {author} {\bibfnamefont {R.}~\bibnamefont {Venugopalan}},\
  }\href@noop {} {\bibfield  {journal} {\bibinfo  {journal} {Reviews of Modern
  Physics}\ }\textbf {\bibinfo {volume} {93}},\ \bibinfo {pages} {035003}
  (\bibinfo {year} {2021}{\natexlab{b}})}\BibitemShut {NoStop}%
\bibitem [{\citenamefont {Deutsch}(1991)}]{deutsch1991quantum}%
  \BibitemOpen
  \bibfield  {author} {\bibinfo {author} {\bibfnamefont {J.~M.}\ \bibnamefont
  {Deutsch}},\ }\href@noop {} {\bibfield  {journal} {\bibinfo  {journal}
  {Physical review a}\ }\textbf {\bibinfo {volume} {43}},\ \bibinfo {pages}
  {2046} (\bibinfo {year} {1991})}\BibitemShut {NoStop}%
\bibitem [{\citenamefont {Srednicki}(1994)}]{srednicki1994chaos}%
  \BibitemOpen
  \bibfield  {author} {\bibinfo {author} {\bibfnamefont {M.}~\bibnamefont
  {Srednicki}},\ }\href@noop {} {\bibfield  {journal} {\bibinfo  {journal}
  {Physical review e}\ }\textbf {\bibinfo {volume} {50}},\ \bibinfo {pages}
  {888} (\bibinfo {year} {1994})}\BibitemShut {NoStop}%
\bibitem [{\citenamefont {Li}\ and\ \citenamefont
  {Haldane}(2008)}]{li2008entanglement}%
  \BibitemOpen
  \bibfield  {author} {\bibinfo {author} {\bibfnamefont {H.}~\bibnamefont
  {Li}}\ and\ \bibinfo {author} {\bibfnamefont {F.~D.~M.}\ \bibnamefont
  {Haldane}},\ }\href@noop {} {\bibfield  {journal} {\bibinfo  {journal}
  {Physical review letters}\ }\textbf {\bibinfo {volume} {101}},\ \bibinfo
  {pages} {010504} (\bibinfo {year} {2008})}\BibitemShut {NoStop}%
\bibitem [{\citenamefont {Amico}\ \emph {et~al.}(2008)\citenamefont {Amico},
  \citenamefont {Fazio}, \citenamefont {Osterloh},\ and\ \citenamefont
  {Vedral}}]{amico2008entanglement}%
  \BibitemOpen
  \bibfield  {author} {\bibinfo {author} {\bibfnamefont {L.}~\bibnamefont
  {Amico}}, \bibinfo {author} {\bibfnamefont {R.}~\bibnamefont {Fazio}},
  \bibinfo {author} {\bibfnamefont {A.}~\bibnamefont {Osterloh}}, \ and\
  \bibinfo {author} {\bibfnamefont {V.}~\bibnamefont {Vedral}},\ }\href@noop {}
  {\bibfield  {journal} {\bibinfo  {journal} {Reviews of modern physics}\
  }\textbf {\bibinfo {volume} {80}},\ \bibinfo {pages} {517} (\bibinfo {year}
  {2008})}\BibitemShut {NoStop}%
\bibitem [{\citenamefont {Eisert}\ \emph {et~al.}(2010)\citenamefont {Eisert},
  \citenamefont {Cramer},\ and\ \citenamefont {Plenio}}]{eisert2010colloquium}%
  \BibitemOpen
  \bibfield  {author} {\bibinfo {author} {\bibfnamefont {J.}~\bibnamefont
  {Eisert}}, \bibinfo {author} {\bibfnamefont {M.}~\bibnamefont {Cramer}}, \
  and\ \bibinfo {author} {\bibfnamefont {M.~B.}\ \bibnamefont {Plenio}},\
  }\href@noop {} {\bibfield  {journal} {\bibinfo  {journal} {Reviews of modern
  physics}\ }\textbf {\bibinfo {volume} {82}},\ \bibinfo {pages} {277}
  (\bibinfo {year} {2010})}\BibitemShut {NoStop}%
\bibitem [{\citenamefont {Geraedts}\ \emph {et~al.}(2016)\citenamefont
  {Geraedts}, \citenamefont {Nandkishore},\ and\ \citenamefont
  {Regnault}}]{geraedts2016many}%
  \BibitemOpen
  \bibfield  {author} {\bibinfo {author} {\bibfnamefont {S.~D.}\ \bibnamefont
  {Geraedts}}, \bibinfo {author} {\bibfnamefont {R.}~\bibnamefont
  {Nandkishore}}, \ and\ \bibinfo {author} {\bibfnamefont {N.}~\bibnamefont
  {Regnault}},\ }\href@noop {} {\bibfield  {journal} {\bibinfo  {journal}
  {Physical Review B}\ }\textbf {\bibinfo {volume} {93}},\ \bibinfo {pages}
  {174202} (\bibinfo {year} {2016})}\BibitemShut {NoStop}%
\bibitem [{\citenamefont {Kaufman}\ \emph {et~al.}(2016)\citenamefont
  {Kaufman}, \citenamefont {Tai}, \citenamefont {Lukin}, \citenamefont
  {Rispoli}, \citenamefont {Schittko}, \citenamefont {Preiss},\ and\
  \citenamefont {Greiner}}]{kaufman2016quantum}%
  \BibitemOpen
  \bibfield  {author} {\bibinfo {author} {\bibfnamefont {A.~M.}\ \bibnamefont
  {Kaufman}}, \bibinfo {author} {\bibfnamefont {M.~E.}\ \bibnamefont {Tai}},
  \bibinfo {author} {\bibfnamefont {A.}~\bibnamefont {Lukin}}, \bibinfo
  {author} {\bibfnamefont {M.}~\bibnamefont {Rispoli}}, \bibinfo {author}
  {\bibfnamefont {R.}~\bibnamefont {Schittko}}, \bibinfo {author}
  {\bibfnamefont {P.~M.}\ \bibnamefont {Preiss}}, \ and\ \bibinfo {author}
  {\bibfnamefont {M.}~\bibnamefont {Greiner}},\ }\href@noop {} {\bibfield
  {journal} {\bibinfo  {journal} {Science}\ }\textbf {\bibinfo {volume}
  {353}},\ \bibinfo {pages} {794} (\bibinfo {year} {2016})}\BibitemShut
  {NoStop}%
\bibitem [{\citenamefont {Yang}\ \emph {et~al.}(2015)\citenamefont {Yang},
  \citenamefont {Chamon}, \citenamefont {Hamma},\ and\ \citenamefont
  {Mucciolo}}]{yang2015two}%
  \BibitemOpen
  \bibfield  {author} {\bibinfo {author} {\bibfnamefont {Z.-C.}\ \bibnamefont
  {Yang}}, \bibinfo {author} {\bibfnamefont {C.}~\bibnamefont {Chamon}},
  \bibinfo {author} {\bibfnamefont {A.}~\bibnamefont {Hamma}}, \ and\ \bibinfo
  {author} {\bibfnamefont {E.~R.}\ \bibnamefont {Mucciolo}},\ }\href@noop {}
  {\bibfield  {journal} {\bibinfo  {journal} {Physical review letters}\
  }\textbf {\bibinfo {volume} {115}},\ \bibinfo {pages} {267206} (\bibinfo
  {year} {2015})}\BibitemShut {NoStop}%
\bibitem [{\citenamefont {Mueller}\ \emph
  {et~al.}(2022{\natexlab{a}})\citenamefont {Mueller}, \citenamefont {Zache},\
  and\ \citenamefont {Ott}}]{Mueller:2021gxd}%
  \BibitemOpen
  \bibfield  {author} {\bibinfo {author} {\bibfnamefont {N.}~\bibnamefont
  {Mueller}}, \bibinfo {author} {\bibfnamefont {T.~V.}\ \bibnamefont {Zache}},
  \ and\ \bibinfo {author} {\bibfnamefont {R.}~\bibnamefont {Ott}},\ }\href
  {\doibase 10.1103/PhysRevLett.129.011601} {\bibfield  {journal} {\bibinfo
  {journal} {Phys. Rev. Lett.}\ }\textbf {\bibinfo {volume} {129}},\ \bibinfo
  {pages} {011601} (\bibinfo {year} {2022}{\natexlab{a}})},\ \Eprint
  {http://arxiv.org/abs/2107.11416} {arXiv:2107.11416 [quant-ph]} \BibitemShut
  {NoStop}%
\bibitem [{\citenamefont {Halimeh}\ \emph {et~al.}(2022)\citenamefont
  {Halimeh}, \citenamefont {Barbiero}, \citenamefont {Hauke}, \citenamefont
  {Grusdt},\ and\ \citenamefont {Bohrdt}}]{halimeh2022robust}%
  \BibitemOpen
  \bibfield  {author} {\bibinfo {author} {\bibfnamefont {J.~C.}\ \bibnamefont
  {Halimeh}}, \bibinfo {author} {\bibfnamefont {L.}~\bibnamefont {Barbiero}},
  \bibinfo {author} {\bibfnamefont {P.}~\bibnamefont {Hauke}}, \bibinfo
  {author} {\bibfnamefont {F.}~\bibnamefont {Grusdt}}, \ and\ \bibinfo {author}
  {\bibfnamefont {A.}~\bibnamefont {Bohrdt}},\ }\href@noop {} {\bibfield
  {journal} {\bibinfo  {journal} {arXiv preprint arXiv:2203.08828}\ } (\bibinfo
  {year} {2022})}\BibitemShut {NoStop}%
\bibitem [{\citenamefont {Banerjee}\ and\ \citenamefont
  {Sen}(2021)}]{banerjee2021quantum}%
  \BibitemOpen
  \bibfield  {author} {\bibinfo {author} {\bibfnamefont {D.}~\bibnamefont
  {Banerjee}}\ and\ \bibinfo {author} {\bibfnamefont {A.}~\bibnamefont {Sen}},\
  }\href@noop {} {\bibfield  {journal} {\bibinfo  {journal} {Physical Review
  Letters}\ }\textbf {\bibinfo {volume} {126}},\ \bibinfo {pages} {220601}
  (\bibinfo {year} {2021})}\BibitemShut {NoStop}%
\bibitem [{\citenamefont {Brenes}\ \emph {et~al.}(2018)\citenamefont {Brenes},
  \citenamefont {Dalmonte}, \citenamefont {Heyl},\ and\ \citenamefont
  {Scardicchio}}]{brenes2018many}%
  \BibitemOpen
  \bibfield  {author} {\bibinfo {author} {\bibfnamefont {M.}~\bibnamefont
  {Brenes}}, \bibinfo {author} {\bibfnamefont {M.}~\bibnamefont {Dalmonte}},
  \bibinfo {author} {\bibfnamefont {M.}~\bibnamefont {Heyl}}, \ and\ \bibinfo
  {author} {\bibfnamefont {A.}~\bibnamefont {Scardicchio}},\ }\href@noop {}
  {\bibfield  {journal} {\bibinfo  {journal} {Physical review letters}\
  }\textbf {\bibinfo {volume} {120}},\ \bibinfo {pages} {030601} (\bibinfo
  {year} {2018})}\BibitemShut {NoStop}%
\bibitem [{\citenamefont {Dalmonte}\ \emph {et~al.}(2022)\citenamefont
  {Dalmonte}, \citenamefont {Eisler}, \citenamefont {Falconi},\ and\
  \citenamefont {Vermersch}}]{dalmonte2022entanglement}%
  \BibitemOpen
  \bibfield  {author} {\bibinfo {author} {\bibfnamefont {M.}~\bibnamefont
  {Dalmonte}}, \bibinfo {author} {\bibfnamefont {V.}~\bibnamefont {Eisler}},
  \bibinfo {author} {\bibfnamefont {M.}~\bibnamefont {Falconi}}, \ and\
  \bibinfo {author} {\bibfnamefont {B.}~\bibnamefont {Vermersch}},\ }\href@noop
  {} {\bibfield  {journal} {\bibinfo  {journal} {arXiv preprint
  arXiv:2202.05045}\ } (\bibinfo {year} {2022})}\BibitemShut {NoStop}%
\bibitem [{\citenamefont {Kokail}\ \emph
  {et~al.}(2021{\natexlab{a}})\citenamefont {Kokail}, \citenamefont {van
  Bijnen}, \citenamefont {Elben}, \citenamefont {Vermersch},\ and\
  \citenamefont {Zoller}}]{kokail2021entanglement}%
  \BibitemOpen
  \bibfield  {author} {\bibinfo {author} {\bibfnamefont {C.}~\bibnamefont
  {Kokail}}, \bibinfo {author} {\bibfnamefont {R.}~\bibnamefont {van Bijnen}},
  \bibinfo {author} {\bibfnamefont {A.}~\bibnamefont {Elben}}, \bibinfo
  {author} {\bibfnamefont {B.}~\bibnamefont {Vermersch}}, \ and\ \bibinfo
  {author} {\bibfnamefont {P.}~\bibnamefont {Zoller}},\ }\href@noop {}
  {\bibfield  {journal} {\bibinfo  {journal} {Nature Physics}\ }\textbf
  {\bibinfo {volume} {17}},\ \bibinfo {pages} {936} (\bibinfo {year}
  {2021}{\natexlab{a}})}\BibitemShut {NoStop}%
\bibitem [{\citenamefont {Pichler}\ \emph {et~al.}(2016)\citenamefont
  {Pichler}, \citenamefont {Zhu}, \citenamefont {Seif}, \citenamefont
  {Zoller},\ and\ \citenamefont {Hafezi}}]{pichler2016measurement}%
  \BibitemOpen
  \bibfield  {author} {\bibinfo {author} {\bibfnamefont {H.}~\bibnamefont
  {Pichler}}, \bibinfo {author} {\bibfnamefont {G.}~\bibnamefont {Zhu}},
  \bibinfo {author} {\bibfnamefont {A.}~\bibnamefont {Seif}}, \bibinfo {author}
  {\bibfnamefont {P.}~\bibnamefont {Zoller}}, \ and\ \bibinfo {author}
  {\bibfnamefont {M.}~\bibnamefont {Hafezi}},\ }\href@noop {} {\bibfield
  {journal} {\bibinfo  {journal} {Physical Review X}\ }\textbf {\bibinfo
  {volume} {6}},\ \bibinfo {pages} {041033} (\bibinfo {year}
  {2016})}\BibitemShut {NoStop}%
\bibitem [{\citenamefont {Yirka}\ and\ \citenamefont
  {Suba{\c{s}}{\i}}(2021)}]{yirka2021qubit}%
  \BibitemOpen
  \bibfield  {author} {\bibinfo {author} {\bibfnamefont {J.}~\bibnamefont
  {Yirka}}\ and\ \bibinfo {author} {\bibfnamefont {Y.}~\bibnamefont
  {Suba{\c{s}}{\i}}},\ }\href@noop {} {\bibfield  {journal} {\bibinfo
  {journal} {Quantum}\ }\textbf {\bibinfo {volume} {5}},\ \bibinfo {pages}
  {535} (\bibinfo {year} {2021})}\BibitemShut {NoStop}%
\bibitem [{\citenamefont {Kokail}\ \emph
  {et~al.}(2021{\natexlab{b}})\citenamefont {Kokail}, \citenamefont {Sundar},
  \citenamefont {Zache}, \citenamefont {Elben}, \citenamefont {Vermersch},
  \citenamefont {Dalmonte}, \citenamefont {van Bijnen},\ and\ \citenamefont
  {Zoller}}]{kokail2021quantum}%
  \BibitemOpen
  \bibfield  {author} {\bibinfo {author} {\bibfnamefont {C.}~\bibnamefont
  {Kokail}}, \bibinfo {author} {\bibfnamefont {B.}~\bibnamefont {Sundar}},
  \bibinfo {author} {\bibfnamefont {T.~V.}\ \bibnamefont {Zache}}, \bibinfo
  {author} {\bibfnamefont {A.}~\bibnamefont {Elben}}, \bibinfo {author}
  {\bibfnamefont {B.}~\bibnamefont {Vermersch}}, \bibinfo {author}
  {\bibfnamefont {M.}~\bibnamefont {Dalmonte}}, \bibinfo {author}
  {\bibfnamefont {R.}~\bibnamefont {van Bijnen}}, \ and\ \bibinfo {author}
  {\bibfnamefont {P.}~\bibnamefont {Zoller}},\ }\href@noop {} {\bibfield
  {journal} {\bibinfo  {journal} {Physical review letters}\ }\textbf {\bibinfo
  {volume} {127}},\ \bibinfo {pages} {170501} (\bibinfo {year}
  {2021}{\natexlab{b}})}\BibitemShut {NoStop}%
\bibitem [{\citenamefont {Mitra}(2018)}]{mitra2018quantum}%
  \BibitemOpen
  \bibfield  {author} {\bibinfo {author} {\bibfnamefont {A.}~\bibnamefont
  {Mitra}},\ }\href@noop {} {\bibfield  {journal} {\bibinfo  {journal} {Annual
  Review of Condensed Matter Physics}\ }\textbf {\bibinfo {volume} {9}},\
  \bibinfo {pages} {245} (\bibinfo {year} {2018})}\BibitemShut {NoStop}%
\bibitem [{\citenamefont {Canovi}\ \emph {et~al.}(2011)\citenamefont {Canovi},
  \citenamefont {Rossini}, \citenamefont {Fazio}, \citenamefont {Santoro},\
  and\ \citenamefont {Silva}}]{canovi2011quantum}%
  \BibitemOpen
  \bibfield  {author} {\bibinfo {author} {\bibfnamefont {E.}~\bibnamefont
  {Canovi}}, \bibinfo {author} {\bibfnamefont {D.}~\bibnamefont {Rossini}},
  \bibinfo {author} {\bibfnamefont {R.}~\bibnamefont {Fazio}}, \bibinfo
  {author} {\bibfnamefont {G.~E.}\ \bibnamefont {Santoro}}, \ and\ \bibinfo
  {author} {\bibfnamefont {A.}~\bibnamefont {Silva}},\ }\href@noop {}
  {\bibfield  {journal} {\bibinfo  {journal} {Physical Review B}\ }\textbf
  {\bibinfo {volume} {83}},\ \bibinfo {pages} {094431} (\bibinfo {year}
  {2011})}\BibitemShut {NoStop}%
\bibitem [{\citenamefont {Heyl}(2019)}]{heyl2019dynamical}%
  \BibitemOpen
  \bibfield  {author} {\bibinfo {author} {\bibfnamefont {M.}~\bibnamefont
  {Heyl}},\ }\href@noop {} {\bibfield  {journal} {\bibinfo  {journal} {EPL
  (Europhysics Letters)}\ }\textbf {\bibinfo {volume} {125}},\ \bibinfo {pages}
  {26001} (\bibinfo {year} {2019})}\BibitemShut {NoStop}%
\bibitem [{\citenamefont {Zhang}\ \emph {et~al.}(2017)\citenamefont {Zhang},
  \citenamefont {Pagano}, \citenamefont {Hess}, \citenamefont {Kyprianidis},
  \citenamefont {Becker}, \citenamefont {Kaplan}, \citenamefont {Gorshkov},
  \citenamefont {Gong},\ and\ \citenamefont {Monroe}}]{zhang2017observation}%
  \BibitemOpen
  \bibfield  {author} {\bibinfo {author} {\bibfnamefont {J.}~\bibnamefont
  {Zhang}}, \bibinfo {author} {\bibfnamefont {G.}~\bibnamefont {Pagano}},
  \bibinfo {author} {\bibfnamefont {P.~W.}\ \bibnamefont {Hess}}, \bibinfo
  {author} {\bibfnamefont {A.}~\bibnamefont {Kyprianidis}}, \bibinfo {author}
  {\bibfnamefont {P.}~\bibnamefont {Becker}}, \bibinfo {author} {\bibfnamefont
  {H.}~\bibnamefont {Kaplan}}, \bibinfo {author} {\bibfnamefont {A.~V.}\
  \bibnamefont {Gorshkov}}, \bibinfo {author} {\bibfnamefont {Z.-X.}\
  \bibnamefont {Gong}}, \ and\ \bibinfo {author} {\bibfnamefont
  {C.}~\bibnamefont {Monroe}},\ }\href@noop {} {\bibfield  {journal} {\bibinfo
  {journal} {Nature}\ }\textbf {\bibinfo {volume} {551}},\ \bibinfo {pages}
  {601} (\bibinfo {year} {2017})}\BibitemShut {NoStop}%
\bibitem [{\citenamefont {Jurcevic}\ \emph {et~al.}(2017)\citenamefont
  {Jurcevic}, \citenamefont {Shen}, \citenamefont {Hauke}, \citenamefont
  {Maier}, \citenamefont {Brydges}, \citenamefont {Hempel}, \citenamefont
  {Lanyon}, \citenamefont {Heyl}, \citenamefont {Blatt},\ and\ \citenamefont
  {Roos}}]{jurcevic2017direct}%
  \BibitemOpen
  \bibfield  {author} {\bibinfo {author} {\bibfnamefont {P.}~\bibnamefont
  {Jurcevic}}, \bibinfo {author} {\bibfnamefont {H.}~\bibnamefont {Shen}},
  \bibinfo {author} {\bibfnamefont {P.}~\bibnamefont {Hauke}}, \bibinfo
  {author} {\bibfnamefont {C.}~\bibnamefont {Maier}}, \bibinfo {author}
  {\bibfnamefont {T.}~\bibnamefont {Brydges}}, \bibinfo {author} {\bibfnamefont
  {C.}~\bibnamefont {Hempel}}, \bibinfo {author} {\bibfnamefont
  {B.}~\bibnamefont {Lanyon}}, \bibinfo {author} {\bibfnamefont
  {M.}~\bibnamefont {Heyl}}, \bibinfo {author} {\bibfnamefont {R.}~\bibnamefont
  {Blatt}}, \ and\ \bibinfo {author} {\bibfnamefont {C.}~\bibnamefont {Roos}},\
  }\href@noop {} {\bibfield  {journal} {\bibinfo  {journal} {Physical review
  letters}\ }\textbf {\bibinfo {volume} {119}},\ \bibinfo {pages} {080501}
  (\bibinfo {year} {2017})}\BibitemShut {NoStop}%
\bibitem [{\citenamefont {Nie}\ \emph {et~al.}(2019)\citenamefont {Nie},
  \citenamefont {Wei}, \citenamefont {Chen}, \citenamefont {Zhang},
  \citenamefont {Zhao}, \citenamefont {Qiu}, \citenamefont {Tian},
  \citenamefont {Ji}, \citenamefont {Xin}, \citenamefont {Lu},\ and\
  \citenamefont {Li}}]{nie2019experimental}%
  \BibitemOpen
  \bibfield  {author} {\bibinfo {author} {\bibfnamefont {X.}~\bibnamefont
  {Nie}}, \bibinfo {author} {\bibfnamefont {B.-B.}\ \bibnamefont {Wei}},
  \bibinfo {author} {\bibfnamefont {X.}~\bibnamefont {Chen}}, \bibinfo {author}
  {\bibfnamefont {Z.}~\bibnamefont {Zhang}}, \bibinfo {author} {\bibfnamefont
  {X.}~\bibnamefont {Zhao}}, \bibinfo {author} {\bibfnamefont {C.}~\bibnamefont
  {Qiu}}, \bibinfo {author} {\bibfnamefont {Y.}~\bibnamefont {Tian}}, \bibinfo
  {author} {\bibfnamefont {Y.}~\bibnamefont {Ji}}, \bibinfo {author}
  {\bibfnamefont {T.}~\bibnamefont {Xin}}, \bibinfo {author} {\bibfnamefont
  {D.}~\bibnamefont {Lu}}, \ and\ \bibinfo {author} {\bibfnamefont
  {J.}~\bibnamefont {Li}},\ }\href@noop {} {\enquote {\bibinfo {title}
  {Experimental observation of equilibrium and dynamical quantum phase
  transitions via out-of-time-ordered correlators},}\ } (\bibinfo {year}
  {2019}),\ \Eprint {http://arxiv.org/abs/1912.12038} {arXiv:1912.12038
  [quant-ph]} \BibitemShut {NoStop}%
\bibitem [{\citenamefont {Zache}\ \emph {et~al.}(2019)\citenamefont {Zache},
  \citenamefont {Mueller}, \citenamefont {Schneider}, \citenamefont
  {Jendrzejewski}, \citenamefont {Berges},\ and\ \citenamefont
  {Hauke}}]{zache2019dynamical}%
  \BibitemOpen
  \bibfield  {author} {\bibinfo {author} {\bibfnamefont {T.}~\bibnamefont
  {Zache}}, \bibinfo {author} {\bibfnamefont {N.}~\bibnamefont {Mueller}},
  \bibinfo {author} {\bibfnamefont {J.}~\bibnamefont {Schneider}}, \bibinfo
  {author} {\bibfnamefont {F.}~\bibnamefont {Jendrzejewski}}, \bibinfo {author}
  {\bibfnamefont {J.}~\bibnamefont {Berges}}, \ and\ \bibinfo {author}
  {\bibfnamefont {P.}~\bibnamefont {Hauke}},\ }\href@noop {} {\bibfield
  {journal} {\bibinfo  {journal} {Physical review letters}\ }\textbf {\bibinfo
  {volume} {122}},\ \bibinfo {pages} {050403} (\bibinfo {year}
  {2019})}\BibitemShut {NoStop}%
\bibitem [{\citenamefont {Mueller}\ \emph
  {et~al.}(2022{\natexlab{b}})\citenamefont {Mueller}, \citenamefont {Carolan},
  \citenamefont {Connelly}, \citenamefont {Davoudi}, \citenamefont
  {Dumitrescu},\ and\ \citenamefont {Yeter-Aydeniz}}]{mueller2022quantum}%
  \BibitemOpen
  \bibfield  {author} {\bibinfo {author} {\bibfnamefont {N.}~\bibnamefont
  {Mueller}}, \bibinfo {author} {\bibfnamefont {J.~A.}\ \bibnamefont
  {Carolan}}, \bibinfo {author} {\bibfnamefont {A.}~\bibnamefont {Connelly}},
  \bibinfo {author} {\bibfnamefont {Z.}~\bibnamefont {Davoudi}}, \bibinfo
  {author} {\bibfnamefont {E.~F.}\ \bibnamefont {Dumitrescu}}, \ and\ \bibinfo
  {author} {\bibfnamefont {K.}~\bibnamefont {Yeter-Aydeniz}},\ }\href@noop {}
  {\bibfield  {journal} {\bibinfo  {journal} {arXiv preprint arXiv:2210.03089}\
  } (\bibinfo {year} {2022}{\natexlab{b}})}\BibitemShut {NoStop}%
\bibitem [{\citenamefont {Van~Damme}\ \emph {et~al.}(2022)\citenamefont
  {Van~Damme}, \citenamefont {Desaules}, \citenamefont {Papi{\'c}},\ and\
  \citenamefont {Halimeh}}]{van2022anatomy}%
  \BibitemOpen
  \bibfield  {author} {\bibinfo {author} {\bibfnamefont {M.}~\bibnamefont
  {Van~Damme}}, \bibinfo {author} {\bibfnamefont {J.-Y.}\ \bibnamefont
  {Desaules}}, \bibinfo {author} {\bibfnamefont {Z.}~\bibnamefont {Papi{\'c}}},
  \ and\ \bibinfo {author} {\bibfnamefont {J.~C.}\ \bibnamefont {Halimeh}},\
  }\href@noop {} {\bibfield  {journal} {\bibinfo  {journal} {arXiv preprint
  arXiv:2210.02453}\ } (\bibinfo {year} {2022})}\BibitemShut {NoStop}%
\bibitem [{\citenamefont {Bernien}\ \emph {et~al.}(2017)\citenamefont
  {Bernien}, \citenamefont {Schwartz}, \citenamefont {Keesling}, \citenamefont
  {Levine}, \citenamefont {Omran}, \citenamefont {Pichler}, \citenamefont
  {Choi}, \citenamefont {Zibrov}, \citenamefont {Endres}, \citenamefont
  {Greiner} \emph {et~al.}}]{bernien2017probing}%
  \BibitemOpen
  \bibfield  {author} {\bibinfo {author} {\bibfnamefont {H.}~\bibnamefont
  {Bernien}}, \bibinfo {author} {\bibfnamefont {S.}~\bibnamefont {Schwartz}},
  \bibinfo {author} {\bibfnamefont {A.}~\bibnamefont {Keesling}}, \bibinfo
  {author} {\bibfnamefont {H.}~\bibnamefont {Levine}}, \bibinfo {author}
  {\bibfnamefont {A.}~\bibnamefont {Omran}}, \bibinfo {author} {\bibfnamefont
  {H.}~\bibnamefont {Pichler}}, \bibinfo {author} {\bibfnamefont
  {S.}~\bibnamefont {Choi}}, \bibinfo {author} {\bibfnamefont {A.~S.}\
  \bibnamefont {Zibrov}}, \bibinfo {author} {\bibfnamefont {M.}~\bibnamefont
  {Endres}}, \bibinfo {author} {\bibfnamefont {M.}~\bibnamefont {Greiner}},
  \emph {et~al.},\ }\href@noop {} {\bibfield  {journal} {\bibinfo  {journal}
  {Nature}\ }\textbf {\bibinfo {volume} {551}},\ \bibinfo {pages} {579}
  (\bibinfo {year} {2017})}\BibitemShut {NoStop}%
\end{thebibliography}%
\end{document}